\documentclass[a4paper,11pt]{article}
\usepackage[skins]{tcolorbox}
\usepackage{mathtools,slashed}
\usepackage{subcaption}
\usepackage[T1]{fontenc}
\usepackage{slashed,verbatim}
\usepackage[numbers,sort&compress]{natbib}
\usepackage{amsmath}
\usepackage{mathrsfs}
\usepackage{amsbsy}
\usepackage{amssymb}
\usepackage{appendix}
\usepackage{graphicx}
\usepackage[final]{pdfpages}
\usepackage{dcolumn}
\usepackage{bm}
\usepackage{cancel}
\usepackage[colorlinks=true,linktocpage=true,
linkcolor=blue,citecolor=blue]{hyperref}
\usepackage{url}
\definecolor{db5}{cmyk}{0.5,0.5,0,0.5}
\definecolor{mauve}{cmyk}{0.3,0.7,0.1,0.3}
\definecolor{palemauve}{cmyk}{0.3,0.7,0.1,0.0}
\definecolor{pb}{cmyk}{0.4,0.1,0,0.1}
\definecolor{pgreen}{cmyk}{0.4,0.0,0.3,0.0}
\definecolor{pink}{cmyk}{0.0,0.5,0.3,0.0}
\newcommand{\bs}{\begin{slide}}
\newcommand{\es}{\end{slide}}
\newcommand{\tcb}{\textcolor {blue}}
\usepackage[a4paper]{geometry}
\catcode`@11
\def\seceqaa{\@addtoreset{equation}{section}
	\def\theequation{A\arabic{equation}}}
\def\seceqbb{\@addtoreset{equation}{section}
	\def\theequation{B\arabic{equation}}}
\def\seceqcc{\@addtoreset{equation}{section}
	\def\theequation{C\arabic{equation}}}
\def\seceqdd{\@addtoreset{equation}{section}
	\def\theequation{D\arabic{equation}}}
\def\seceqee{\@addtoreset{equation}{section}
	\def\theequation{E\arabic{equation}}}
\def\seceqff{\@addtoreset{equation}{section}
	\def\theequation{F\arabic{equation}}}	
\def\seceqgg{\@addtoreset{equation}{section}
	\def\theequation{G\arabic{equation}}}
\def\seceqhh{\@addtoreset{equation}{section}
	\def\theequation{H\arabic{equation}}}
\def\seceqii{\@addtoreset{equation}{section}
	\def\theequation{H\arabic{equation}}}	
\def\seceqjj{\@addtoreset{equation}{section}
	\def\theequation{H\arabic{equation}}}
\catcode`@11
\usepackage{authblk}
\newcommand{\be}{\begin{eqnarray}}
\newcommand{\ee}{\end{eqnarray}}
\begin{document}
\large
\title{$T_c$, Photoproduction,  Paramagnetic Anisotropic Plasma,  IR Log-Gravitational-DBI Renormalization and $G_2$-Structure Induced (Almost) Contact 3-Structures in Hot Strongly Magnetic MQCD at Intermediate Coupling}
\author{Shivam Singh Kushwah\footnote{email- shivams\_kushwah@ph.iitr.ac.in}~ and ~Aalok Misra\footnote{email- aalok.misra@ph.iitr.ac.in}\vspace{0.1in}\\
Department of Physics,\\
Indian Institute of Technology Roorkee, Roorkee 247667, Uttarakhand, India}
\date{}
\maketitle
\begin{abstract}
{\scriptsize
After obtaining the gauge fields that can be supported on the world-volume of flavor $D6$-branes in the type IIA dual of thermal QCD-like theories at high temperatures and intermediate coupling (the latter incorporated via the inclusion of ${\cal O}(R^4)$ corrections in its ${\cal M}$-theory uplift), combining with the results of \cite{Gopal+Vikas+Aalok}, we show that the deconfinement temperature $T_c$ decreases in the presence of a strong magnetic field as in lattice QCD \cite{decrease-Tc-B}. By working out gauge-invariant fluctuations about the aforementioned world-volume gauge fields, in the (absence and) presence of a strong magnetic field ($B>\left(T_c(B=0)\right)^2$ in $e=1$-units), we obtain the $\frac{\chi}{N^2 T^2\omega}-\omega$ variation ($\chi$ being the spectral function for the in(reaction)-plane photon polarization and $N$ being the number of color $D3$-branes in the parent type IIB dual \cite{metrics} of thermal QCD-like theories). We further obtain a nice agreement with, e.g., bottom-up holographic anisotropic backgrounds in gauged supergravity \cite{photoprod-vs-w_Bneq0}. Implementing Dirichlet boundary condition at the horizon for the world-volume gauge fields, we also demonstrate at the level of EoS that the holographic dual, in principle, could correspond to several scenarios above $T_c$. These include (i)the anisotropic plasma transitioning via a smooth crossover to exotic matter as the universe cools (the converse being prohibited in our setup), (ii) stable wormholes (where we also remark that a resolved conifold near $r=0$ is somewhat like a half Ellis wormhole), and (iii) a paramagnetic pressure/energy-anisotropic plasma. Given that above $T_c$ QGP is expected to be paramagnetic \cite{Bali et al-magnetic-chi}, the third possibility appears to be the preferred one. Generalizing the TOV equations to include angular mass/pressure/energy profiles, we show up to first order in $G$, that it is not possible that the anisotropic plasma leads to the formation of a compact star. En route, we show that the IR renormalization of the DBI action requires a boundary Log-determinant-of-Ricci-tensor counter term. {\it We further conjecture that (i) quantities like photoproduction spectral function, speed of sound (and hence bulk viscosity \cite{Shivam+Aalok_Bulk}), etc. in the absence of a magnetic field that are determined from world-volume gauge field fluctuations that receive ${\cal O}(R^4)$-corrections, if complexified, include a non-analytic-complexified gauge-coupling dependence, and correspond to Contact 3-Structures; (ii) quantities, e.g., pressure/free energy, energy density, etc. in the presence of a strong magnetic field that are determined from world-volume gauge fields that are not ${\cal O}(R^4)$-corrected, if complexified, are analytic in the complexified gauge coupling, and correspond to Almost Contact 3-Structures (AC3S), both induced from the $G_2$ structure of a closed seven-fold - a warped product of the ${\cal M}$-theory circle and a non-Kähler six-fold with the six-fold being a warped product of the thermal circle with a non-Einsteinian deformation of $T^{1,1}$, and (iii) the lack of $N$-path connectedness in the parameter space associated with AC3S and C3S \cite{ACMS}, corresponds therefore to that gauge field fluctuations can not be finite, and in the zero-instanton sector, (type-IIB modular-completion-inspired) ${\cal O}(R^4)$ non-renormalized gauge fields produce ${\cal O}(R^4)$-corrected gauge fluctuations.}
}
\end{abstract}

\tableofcontents

\section{Introduction}
\label{Introduction}

The QCD has an interesting phase diagram, below the deconfinement temperature it contains hadrons and mesons as degrees of freedom, which are the bound states of quarks formed due to their strong interactions mediated by gluons. Above the deconfinement temperature, a new state of matter is predicted called a Quark-Gluon Plasma (QGP) which behaves very close to an ideal fluid \cite{STAR:2005gfr}. In recent years, Relativistic Heavy Ion Collisions (RHIC) experiments have started the quest to detect the QGP, and revealed that QGP is strongly coupled, leading to non-applicability of perturbative methods. AdS/CFT is a proposal to deal with the strongly coupled quantum field theories, where a weakly coupled gravity dual can be constructed for them. The more extended version of the AdS/CFT duality is the gauge/gravity duality, where one can consider non-conformal theories like QCD, resulted in a fruitful way to deal with such strongly coupled systems. QGP is believed to be found in the core of stellar objects called Neutron stars, which can be probed experimentally via gravitational wave data, for a detailed discussion on Neutron stars see \cite{Kovensky:2021kzl}, and for their core supporting quark matter see \cite{Hoyos:2016zke}. Recent heavy ion collision experiments reveal the presence of a strong magnetic field during the early times of production of QGP, and in stellar objects there exists a certain class of Neutron stars, called Magnetors \cite{Duncan:1992hi,Thompson:1995gw,Lopes:2014vva} characterized by the strong magnetic field and low frequency of rotation compared to neutron stars. These findings make the inclusion of magnetic field an interesting probe to study QGP stars. The photon or dilepton production is another aspect which is interesting to explore because of the thermal nature of plasma. Due to small electromagnetic coupling ($\alpha_{em}$) the photon interacts very weakly with plasma, hence is considered to be optically thin. The presence of a strong magnetic field in the early stages of QGP production verified by RHIC experiments \cite{Zhong:2014sua} enhances the rate of photoproduction \cite{Tuchin:2012mf}, and produces anisotropy results in their elliptic flow, see \cite{Bzdak:2012fr}. Hence, the strong magnetic field provides an interesting probe to study photoproduction in QGP within close analogy with RHIC experiments.

The study conducted in this paper can be divided into two parts:
\begin{enumerate}
\item Photo-production in top-down holographic QGP.
\item Study of generalized Equation of State(EoS).
\end{enumerate}

\subsection{Photo-production in top-down holographic QGP}
\label{photo-prod}

Since QGP is a charged medium it will eventually emit photons or dileptons. These radiated thermal photons(say) encode the characteristic features of the medium. Due to the weak nature of electromagnetic interaction, they do not interact with the strong coupling medium, and they significantly provide information about the current-current correlators in the hot QGP produced during heavy ion collision experiments. The differential photon emission rate per unit volume in thermal equilibrium can be written as \cite{Holographic-Photoprod-B}:
\begin{equation}
d\Gamma=\frac{d^{3} k}{2(2\pi)^3}\frac{\chi(k)}{\omega (e^{\beta\omega}-1)},\quad\quad\quad\quad \chi(k)=-2Im[\Sigma_{i=1,2}\varepsilon^{\mu}_{i}\varepsilon^{\nu}_{i} C_{\mu\nu}(k)]
\end{equation}
where, $\chi(k)$ is the trace of the spectral density, $C_{\mu\nu}$ is the retarded current-current correlators, $i=1,2$ denotes the polarization states of the photon.

Depending on the photon momentum, three dynamical classifications of photons as hard, soft, and ultra-soft photons are discussed in \cite{MartinContreras:2016ada}. Depending on the type of photon, the characteristic properties of the plasma, such as electrical conductivity, susceptibility, bulk viscosity, etc. are affected. Due to such features, it becomes interesting to explore the photoproduction from QGP.

\subsection{Study of generalized EoS}
\label{EoS}

In recent years a lot of work has been devoted to exploring the features of the EoS of QCD below de-confinement and above de-confinement where physicists consider the form $p=\omega\epsilon$, where $p$ is pressure density, $E$ is energy density, and $\omega$ is EoS parameter, and $\omega$ is constant. But it seems to be an interesting problem where one considers the EoS parameter is not to be a constant. This generalization of EoS leads to interesting phenomena such as dark matter(DM), dark energy(DE), phantom dark energy(PDE), etc., which makes it an interesting problem to explore.

According to cosmological studies, the Friedman equation can be written as,
\begin{equation}
\label{Friedmaneq}
\frac{\ddot{a}}{a}=-4 \pi (E+3P) 
\end{equation}
where $P$ is the total pressure, $E$ is the energy density, and $a$ is the scale factor that appears in the FRW metric to describe the size of the universe. Consider the generalized EoS $P=P(E)$, where the $\omega(r)$ is the EoS parameter.  Now, from the eq(\ref{Friedmaneq}), one can see that for $\omega(r)<-\frac{1}{3}$, the Universe is expanding i.e. $\ddot{a}>0$. The EOS with $ P =\omega \epsilon $, with $ \omega < -\frac{1}{3}$ corresponds to the Dark Energy. Hence one can say that DE is sourcing the accelerated expansion of the current universe. According to the $\Lambda$CDM model, the universe consists of nearly $70\%$ dark energy. EoS $w=-1$ corresponds to the positive cosmological constant term, $-1<\omega<-\frac{1}{3}$ corresponds to quintessence, and $w<-1$ corresponds to the special kind of dark energy called Phantom Dark Energy(PDE). Phantom energy is sourced by the negative-kinetic-energy-term of phantom(or ghost) scalar field which is introduced by hand in the bottom-up models of gravity, say for $\phi$ as the phantom (or ghost) field \cite{Dzhunushaliev:2010kow},
\begin{equation}
L=-\frac{R}{16\pi G}-\frac{1}{2}\partial_{\mu}\phi\partial^{\mu}\phi-V(\phi)
\end{equation}

But in string theory, a phantom field naturally emerges \cite{Sen:2002nu}. Phantom energy boosts the accelerated expansion of the universe leading to the Big Rip scenario. It also violates the null energy condition and hence becomes a candidate to support the wormholes. 

Wormholes are considered to be the objects that connect the two regions of the same spacetime or two different spacetimes. Wormholes were first introduced by Flamm in 1916 and then by Einstein and Nathan Rosen in 1935\cite{Einstein:1935tc}, then also called the Einstein-Rosen bridge. These are the geometries supported by the general theory of relativity that appear as solutions to Einstein's field equations. Certain energy conditions need to be followed by the ordinary matter (written in terms of stress-energy tensor) and the perfect fluid (written in terms of energy density($E$) and pressure density($P$)) which are based on the restrictions on energy-momentum tensor which are as follows\cite{Kontou:2020bta}:

\begin{itemize}
\item Weak Energy Condition (WEC): $T_{\mu\nu}u^{\mu}u^{\nu}\geq0$, or $\epsilon\geq 0$, $\epsilon+P\geq 0$, which ensure that for a time like observer moving with $4-$ velocity, $u^{\mu}$, the observed energy is always positive.
\item The Null energy condition(NEC):$T_{\mu\nu}k^{\mu}k^{\nu}\geq0$, or $\epsilon+P\geq 0$, which prevents the negative energy-like situation along the null trajectories moving with $4$-velocity vector $k^{\mu}$.
\item The Strong Energy Condition (SEC):$(T_{\mu\nu}-\frac{1}{2}T g_{\mu\nu})u^{\mu}u^{\nu}\geq0$, or $\epsilon+P\geq 0$, $\epsilon+3P\geq 0$, which prevents the matter to have gravitational repulsion, hence avoids the exoticness-like scenario.
\item The Dominant Energy Conditions(DEC): $T_{\mu\nu}u^{\mu}v^{\nu}\geq0$, where $u^{\mu}$, and $v^{\nu}$ are the co-oriented time-like vectors, or $\epsilon\geq |P|$, which prevents the superluminal transport and maintain causality of energy-momentum flow.
\end{itemize}

Wormholes violate the null energy condition (NEC). Traversable wormholes are a special class of wormholes that allow travel from one point of the universe to another or from one universe to another and do not contain an event horizon as it prohibits two-way travel. Exotic matter is considered to be required to stabilize the wormhole as it violates the average null energy condition and weak energy condition. Due to negative energy density, the wormholes get the required gravitational repulsion to stabilize \cite{Morris:1988cz}. It has been argued that the exotic matter is not necessarily required to stabilize them, but ordinary fermionic matter is also sufficient to do so, for example, for traversable wormholes in Einstein-Maxwell-Dirac theory see \cite{Blazquez-Salcedo:2020czn,Maldacena:2018gjk}. In certain conditions of the failure of the negative energy source, wormhole converts into a black hole, and inversely a black hole can also convert into a wormhole if it irradiated with negative energy, resulting in stationary wormholes that could be the final state of radiating black holes \cite{Hayward:1998pp}, and would possibly resolve the information loss paradox as there is no singularity present in a traversable wormhole where information loss could happen. There is another well-known effect in quantum field theory, argued to stabilize the traversable wormholes, is the Casimir effect. A Casimir energy is also the candidate that can support the traversable wormholes \cite{Maldacena:2018gjk} known as Casimir Wormholes, extensively studied in the context of modified theories of gravity, see \cite{Hassan:2022hcb,Sokoliuk:2022jcq}. The creation/annihilation of wormholes is controlled by PDE (or DE). Varying EoS parameter crossing the phantom divide $\omega=-1$ generates excessive radial pressure in the dark stars resulting in opening the tunnel leads to the creation of a wormhole \cite{DeBenedictis:2008qm}. At the spacetime foam level in Euclidean quantum gravity, Wormholes also have some topological implications that arise in higher derivative theories as shown \cite{Tsilioukas:2023tdw} states that topology changes occur due to the formation of the wormhole, later this induced the effective cosmological term purely of topological origin depending on HD-correction-coupling-parameter and density of wormholes, resulting in the DE sector in general time-dependent background. The only example compatible with quantum and classical description is based on traversable wormholes known as $ER=EPR$ conjecture \cite{Maldacena:2013xja} to resolve the EPR paradox. These studies make it very interesting to study the interconnection between ordinary matter (here quark matter), dark energy, phantom energy, and wormholes as they coexist, and play a crucial role in the stellar structure of the universe.

%ma
The rest of the paper is organized as follows. Section \ref{MQGP limit} consists of an introduction to the ${\cal M}$-theoretic uplift of the type-IIA  SYZ (Strominger-Yau-Zaslow) mirror  constructed via  triple T-duality, of the parent type-IIB dual of holographic QCD like theories as developed in \cite{metrics,MQGP,NPB}. Section \ref{O(R4)} discusses the inclusion of ${\cal O}(R^4)$-corrections in the  ${\cal M}$-theoretic uplift and when to go beyond. In section \ref{ACM3S-basics}, we summarize the basics of (Almost) Contact 3-Structures. In section \ref{gauge-fields-B0}, we obtain the flavor D$6$-brane world-volume gauge fields in the absence of magnetic field in the UV and the IR in a self-consistent truncation resulting in non-renormalization of gauge fields at ${\cal O}(R^4)$ (in the zero-instanton sector). The analysis of section \ref{gauge-fields-B0} is repeated in section \ref{gauge-fields-B} but with the inclusion of a strong magnetic field. Section \ref{Tc-large-B} demonstrates that the deconfinement temperature decreases by turning on a (strong) magnetic field. Section \ref{photo-prod} consists of a discussion on the photo-production in QGP within the aforementioned ${\cal M}$-theoretic QGP setup. In section \ref{Aniso+Gen-TOV} we discuss the generalized EoS, for the respective $\cal M$-theoretic QGP, which consists of an interesting interplay of anisotropic plasma, phantom energy and stable wormholes depending on the temperature range. Section \ref{Summary} consists of the summary of the results obtained.

\section{Type IIB/IIA Dual of Large-$N$ Thermal QCD-Like Theories, its ${\cal M}$-Theory Uplift and the MQGP Limit}
\label{MQGP limit}

The holographic dual of thermal QCD-like theories at finite coupling was successfully constructed in \cite{MQGP,NPB}. Finite gauge coupling on the gauge theory side would correspond to strong coupling limit of string theory, i.e. ${\cal M}$-theory, to be consistent with gauge - gravity duality. The same was effected via the "MQGP \footnote{Short for ${\cal M}$-theoretic Quark Gluon Plasma (QGP) - essentially implying study of QGP-like thermal QCD systems at intermediate/finite coupling, holographically, from ${\cal M}$ theory.} limit" defined as \cite{MQGP,NPB}: 
\begin{equation}
\label{MQGP_limit}
g_s^{-1}\equiv {\cal O}(1)-{\cal O}(10); N_f, M \equiv {\cal O}(1), N \gg1, \frac{\left(g_s M^2\right)^{m_1}\left(g_s N_f\right)^{m_2}}{N}\ll1, m_{1,2}\in\mathbb{Z}_+\cup\left\{0\right\},
\end{equation}
where $(g_s, N, M, N_f) \equiv$(string coupling, number of color $D3$-branes, number of fractional $D3$-branes/$D5$-branes wrapping the vanishing $S^2$ of a resolved conifold, number of flavor $D7$-branes) in the type IIB dual \cite{metrics} of thermal QCD-like theories at high (i.e. above the deconfinement) temperature.

In this work,  we use the specific values of $g_s, M, N_f$ as given in Table \tcb{2} (as also given in \cite{MQGP-chaos}) which is {\it purely motivated by the desire the theory makes contact with real QCD} as well as to work with {\it intermediate $N$ duals of thermal QCD-like theories}. However, it should be noted that {\bf all results pertaining to the $G_2$ structure, Almost Contact 3-Structures and the resultant transverse $SU(3)$ structures of this work are valid $\forall (g_s, M, N_f)$ satisfying (\ref{MQGP_limit})}.
\begin{table}[h]
\label{Parameters-real-QCD}
\begin{center}
\begin{tabular}{|c|c|c|c|} \hline
S. No. & Parameterc & Value chosen consistent with (\ref{MQGP_limit}) & Physics reason \\ \hline
1. & $g_s$ & 0.1 & {\footnotesize QCD fine structure constant (EW scale)} \\ 
\hline
2. & $M$ & 3 & {\footnotesize Number of colors in the IR after a } \\ 
&&& {\footnotesize Seiberg-like duality cascade} \\
&&& {\footnotesize to match real QCD} \\ \hline
3. & $N_f$ & 2 or 3 & {\footnotesize u, d (and s) quarks} \\
&&& {\footnotesize - the light quarks of QCD} \\  \hline
\end{tabular}
\end{center}
\caption{Values of $g_s, M, N_f$ in the IR motivated by realistic QCD}
\end{table}
We will see that for the values of $g_s, M, N_f$ as given in Table \tcb{2} [even though this table appears in the published \cite{MQGP-chaos} with both the authors as co-authors, chronologically, it first appeared in a preliminary version of this paper that appeared as arXiv:2211.13186 [hep-th], v1],  {\bf $N=100\pm{\cal O}(1)$ is the value of $N$ picked out to obtain explicit Contact 3-Structures (and the associated transverse $SU(3)$ 3-structures)}; a different choice of $(g_s, M = {\cal O}(1), N_f = {\cal O}(1))$ would pick out another (intermediate) $N$. 

The ${\cal M}$-theory uplift of the type IIB string dual \cite{metrics} of thermal QCD-like theories, was obtained by first constructing its  type IIA Strominger-Yau-Zaslow (SYZ) type IIA mirror [via triple T duality along a delocalized special Lagrangian (sLag) $T^3$ resolved/deformed conifold which could be identified with the $T^2$-invariant sLag of \cite{M.Ionel and M.Min-OO (2008)} with a large base ${\cal B}(r,\theta_1,\theta_2)$ \cite{NPB,MQGP}], and then uplifted to ${\cal M}$-theory.  As regards delocalization, (as summarized in \cite{OR4} \footnote{This is explained in \cite{delocalized-mirror-global-D5-resolved-S2}. A resolved warped {\it deformed} conifold (in the type IIB gravity dual (See Fig. \tcb{1})) does not possess an isometry along $\psi$. Therefore, to construct its type IIA SYZ mirror and its subsequent ${\cal M}$-theory uplift, to begin with, one works in the delocalized limit $\psi = \langle\psi\rangle$ wherein one replaces $S^2(\theta_{1,2},\phi_{1,2})$ by $T^2(\theta_{1,2},x/y)$ via (\ref{xyz-defs}). Then, similar to \cite{delocalized-mirror-global-D5-resolved-S2} in the context of $D5$-branes wrapped around the resolved squashed $S^2$ of a resolved conifold, it can be shown that freeing the uplift of the delocalization generates a $G_2$ structure, and therefore the ${\cal M}$-theory uplift and thus its type IIA descendant, are both free of delocalization.}) the ${\cal M}$-theory uplift (excluding the ${\cal O}(R^4)$ corrections) of the type IIB holographic dual of \cite{metrics} of our manuscript, was constructed in the MQGP limit in references \cite{MQGP}, \cite{NPB}, by first constructing the delocalized SYZ type IIA mirror (wherein a pair of squashed $S^2$s are replaced by a pair of $T^2$s, and the correct T-duality coordinates are identified). Analogous to \cite{delocalized-mirror-global-D5-resolved-S2}, the ${\cal M}$-theory uplift corresponds to a bona-fide $G_2$ structure satisfying the EOMs even if one removes the delocalization, i.e., take the uplift to be valid for all angles $\theta_{1,2}, \psi$.  Further, working in the aforementioned vanishing-Ouyang-embedding’s-modulus limit (essentially limiting to the first-generation quarks[+s quark]), it is evident that one will have to work near small values of $\theta_{1,2}$. {\it As an example}, we work in the neighborhood of 
\begin{eqnarray}
\label{alpha_theta_12}
& & (\theta_1, \theta_2) = \left(\frac{\alpha_{\theta_1}}{N^{1/5}}, \frac{\alpha_{\theta_2}}{N^{3/10}}\right),\nonumber\\
& & \alpha_{\theta_{1,2}}\equiv{\cal O}(1);
\end{eqnarray}
the slightly different powers of $N$ in the delocalized $\theta_{1,2}$ is also to remind us that in the pair of squashed $S^2$’s, the vanishing $S^2(\theta_1,\phi_1)$ and resolved $S^2(\theta_2,\phi_2)$ are not on the “same footing”.  At the level of on-shell action, the results up to ${\cal O}(\frac{1}{N})$ are made independent of the delocalization (as explained in \cite{OR4}) by replacing the ${\cal O}(1)$ delocalization parameters $\alpha_{\theta_{1,2}}$ respectively by $N^{1/5}\sin\theta_1$ or $N^{3/10}\sin\theta_2$. One can then choose a different delocalization by then replacing $\sin\theta_{1,2}$ by
\begin{eqnarray}
\label{alpha_theta_12_prime} 
& & \left(\frac{\tilde{\alpha}_{\theta_1}}{N^{\gamma_{\theta_1}}}, \frac{\tilde{\alpha}_{\theta_2}}{N^{\gamma_{\theta_2}}}\right), \gamma_{\theta_1}\neq\frac{1}{5}, \gamma_{\theta_2}\neq\frac{3}{10};\
 \tilde{\alpha}_{\theta_{1,2}}\equiv{\cal O}(1).
\end{eqnarray}
The results pertaining to  $G_2$-structure torsion classes of the closed $M_7$  and the existence of (Almost)Contact(3)(Metric)Structures  and transverse $SU(3)$ structures, remain unchanged and independent of delocalization.

The UV-complete (unlike \cite{SS}) Type IIB string dual of \cite{metrics}, involves $N$ color $D3$-branes placed at the tip of a resolved conifold, $M$ $D5$-branes and $\overline{D5}$-branes both wrapping the vanishing $S^2$ but at antipodal points of the resolved $S^2$, and $N_f$ flavor $D7$- and $\overline{D7}$-branes ``wrapping" a non-compact four cycle $\mathbb{R}_{>0}\times S^3$ involving the vanishing $S^2$ but at antipodal points of the resolved conifold. 
%\footnote{$S^2_a(\theta_2,\phi_2)$ is blown up $S^2$ in the conifold geometry, NP and SP stand for north pole and south pole of the resolved $S^2$, and $a$ is the resolution parameter.}. 

SYZ mirror symmetry is triple T-duality along three isometry directions ($\phi_1,\phi _2, \psi$). By performing first T-duality along $\psi$ direction, one obtains $N$ $D4$ branes which are wrapping $\psi$ direction and $M$ $D4$-branes straddling a pair of orthogonal NS5-branes. Further, from T-dualities along $\phi _1$ and $\phi _2$, one obtains a pair of Taub-Nut spaces and $N$ $D6$ branes. Effect of triple T-dualities on the flavor $D7$ branes is that $D7$ branes are replaced by $D6$-branes. The ${\cal M}$-theory mirror of the type IIA mirror yields KK monopoles (variants of Taub-NUT spaces). Therefore, we can see that there are no branes in ${\cal M}$-theory uplift and we have ${\cal M}$-theory on a $G_2$-structure manifold with fluxes. This is summarized in Fig. \tcb{1}.
\begin{figure}
\begin{center}
\includegraphics[width=0.83\textwidth]{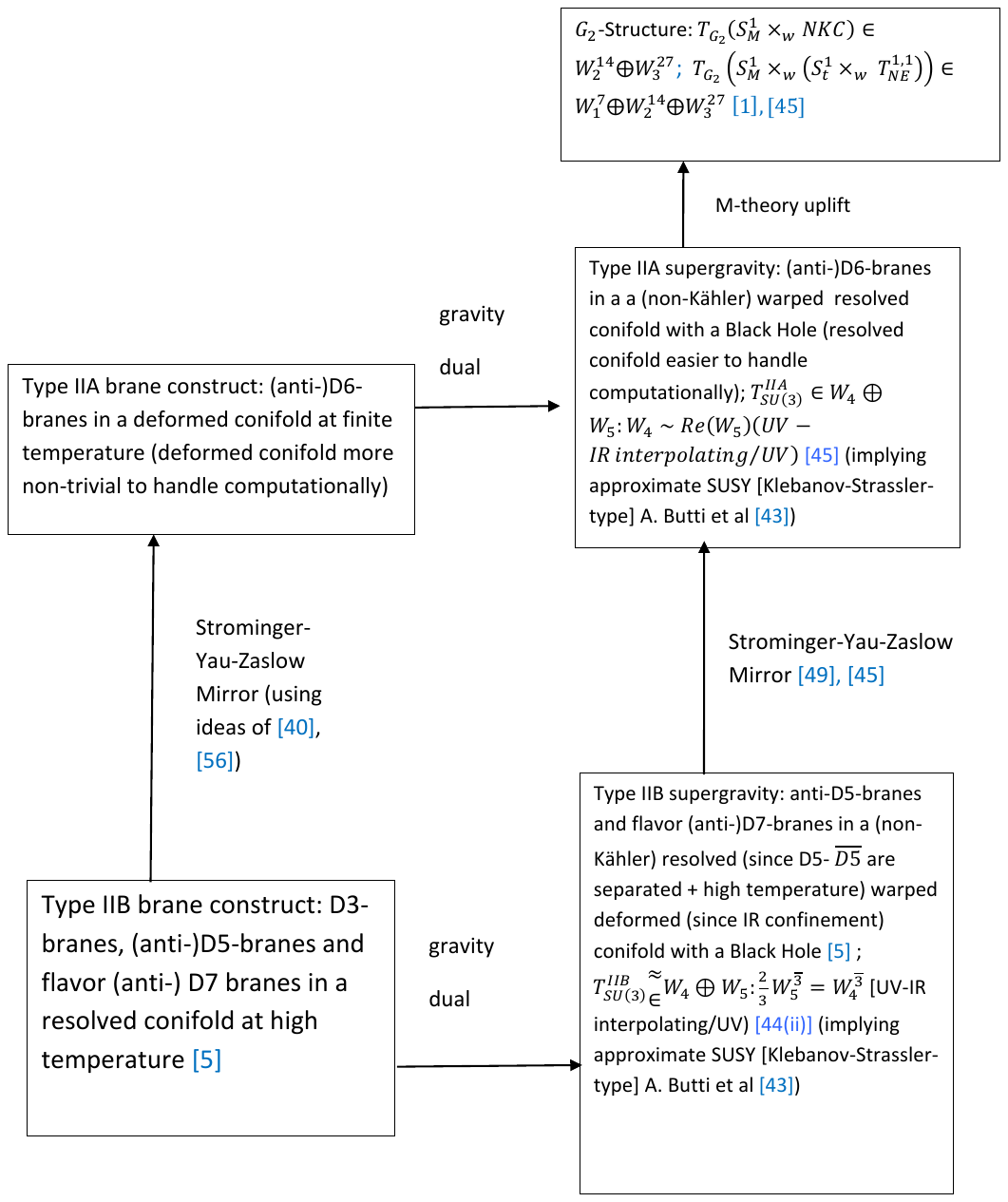}
\end{center}
\vskip -0.4in
\caption{The Status of the Type IIB/IIA/${\cal M}$-theory dual of large-N QCD at high temperature \cite{metrics}, \cite{MQGP}, \cite{NPB} inclusive of ${\cal O}(R^4)$ M-theory corrections \cite{OR4}; $T^{1,1}_{NE}$ denotes a non-Einsteinian deformation of $T^{1,1}$}
\label{Flowchart}
\end{figure}

\begin{tcolorbox}[enhanced,width=6.8in,center upper,size=fbox,
    %fontupper=\large\bfseries,
    drop shadow southwest,sharp corners]
    \begin{flushleft}
 (As explained in, e.g., \cite{Vikas+Gopal+Aalok}) After application of repeated Seiberg-like dualities at finite temperature, $N\ D3$-branes are cascaded away in the IR yielding an $SU(M)$ gauge theory that is UV-conformal, IR-confining wherein the quarks transform in the fundamental representation of flavor and color groups. As $M$ then gets identified with the number of colors in the IR , in the MQGP-limit (\ref{MQGP_limit}) $M$ can not only be taken to ${\cal O}(1)$ but in fact even the realistic-QCD-inspired value of 3. Further, the type IIB dual of \cite{metrics} is valid at  all temperatures.
\end{flushleft}
\end{tcolorbox}

\section{${\cal O}(l_p^6)$ Corrections and When to Go Beyond}
\label{O(R4)}

In this section, via two subsections, we talk about some aspects of ${\cal N}=1, D=11$ supergravity action up to terms quartic in curvature in subsection \ref{LOR4} and a competition between IR-enhancement and large-$N$ suppression thereby answering the question when one would require to go beyond the quartic-in-curvature corrections, in \ref{IR-large-N}.

\subsection{${\cal O}(l_p^6)$ terms in ${\cal N}=1, D=11$ Supergravity Action}
\label{LOR4}

The ${\cal N}=1, D=11$ supergravity action inclusive of ${\cal O}(l_p^6)$ terms is pretty well known, and has been summarized in several previous publications from our group, e.g., \cite{OR4}. Apart from the Einstein-Hilbert, the boundary Gibbons-Hawking-York and flux terms at the leading order, the higher derivative corrections start at terms quartic in the curvature (as well as terms which are cubic in curvature and quadratic in the four-form flux $G_4=d{\cal C}_3$ with ${\cal C}_3$ being the ${\cal M}$-theory three-form potential). The ${\cal O}(R^4)$ terms come in three varieties - the "$J_0 = t_8^2 R^4$" (see, e.g., \cite{OR4} for the definition of the $t_8$ tensor), the eleven-dimensional generalization of the eight-dimensional Eulerian density "$E_8 = \epsilon_{11}^2R^4$" ($\epsilon_{11}$ being the 11-dimensional Levi-Civita symbol) and "$X_8$" given in terms of the second and the square of the first Pontryagin classes of the 11-fold (relevant to anomaly inflow); $X_8$ was shown to vanish in \cite{MQGP} (See \cite{O(R^3G^2)} and \cite{OR4} for a discussion on a completion of the 1-loop ${\cal O}(R^4)$ in the presence of NS-NS $B$ in type IIA compatible with T duality and its ${\cal M}$-theory uplift). As in \cite{OR4}, the ${\cal O}(R^4)$ corrections are $\beta\sim l_p^6$($l_p$ being the Planckian length)-suppressed.

Now, the ${\cal M}$-theory uplift corresponding to high temperatures in QCD is given as follows \cite{MQGP}, \cite{OR4}:
\begin{eqnarray}
\label{TypeIIA-from-M-theory-Witten-prescription-T>Tc}
\hskip -0.1in ds_{11}^2 & = & e^{-\frac{2\phi^{\rm IIA}}{3}}\Biggl[\frac{1}{\sqrt{h(r,\theta_{1,2})}}\left(-g(r) dt^2 + \left(dx^1\right)^2 +  \left(dx^2\right)^2 +\left(dx^3\right)^2 \right)
\nonumber\\
& & \hskip -0.1in+ \sqrt{h(r,\theta_{1,2})}\left(\frac{dr^2}{g(r)} + ds^2_{\rm IIA}(r,\theta_{1,2},\phi_{1,2},\psi)\right)
\Biggr] + e^{\frac{4\phi^{\rm IIA}}{3}}\left(dx^{11} + A_{\rm IIA}^{F_1^{\rm IIB} + F_3^{\rm IIB} + F_5^{\rm IIB}}\right)^2,
\end{eqnarray} 
where $A_{\rm IIA}^{F^{\rm IIB}_{i=1,3,5}}$ correspond to the RR Type IIA one-form gauge field generated from the type IIB $F_{1,3,5}^{\rm IIB}$ via the SYZ mirror of the type IIB string dual \cite{metrics}. Also, $g(r) = 1 - \frac{r_h^4}{r^4}$.

The $D=11$ action  is holographically renormalizable by the construction of appropriate counter terms ${\cal S}^{\rm ct}$.  It was shown in \cite{Gopal+Vikas+Aalok} that inclusive of ${\cal O}(R^4)$-corrections, the bulk on-shell $D=11$ supergravity action  is given by:
\begin{equation}
\label{on-shell-D=11-action-up-to-beta}
\hskip -0.3in S_{D=11}^{\rm on-shell} = -\frac{1}{2}\Biggl[-2S_{\rm EH}^{(0)} + 2 S_{\rm GHY}^{(0)}+ \beta\left(\frac{20}{11}S_{\rm EH}^{(1)} - 2\int_{M_{11}}\left(\sqrt{-g}\right)^{(1)}R^{(0)}
+ 2 S_{\rm GHY}^{(1)} - \frac{2}{11}\int_{M_{11}}\sqrt{-g^{(0)}}g_{(0)}^{MN}\frac{\delta J_0}{\delta g_{(0)}^{MN}}\right)\Biggr],
\end{equation}
where the superscripts "(0)" and "(1)" refer to the contributions of the relevant term at ${\cal O}(\beta^0)$ and ${\cal O}(\beta)$ respectively. The UV divergences of the on-shell action of (\ref{on-shell-D=11-action-up-to-beta}) are of the following types:
\begin{eqnarray}
\label{UV_divergences}
& &\left. \int_{M_{11}}\sqrt{-g}R\right|_{\rm UV-divergent},\ \left.\int_{\partial M_{11}}\sqrt{-h}K\right|_{\rm UV-divergent} \sim r_{\rm UV}^4 \log r_{\rm UV},\nonumber\\
& & \left.\int_{M_{11}} \sqrt{-g}g^{MN}\frac{\delta J_0}{\delta g^{MN}}\right|_{\rm UV-divergent} \sim
\frac{r_{\rm UV}^4}{\log r_{\rm UV}}.
\end{eqnarray}
It was shown in \cite{Gopal+Vikas+Aalok} that a certain linear combination of the boundary terms: $\left.\int_{\partial M_{11}}\sqrt{-h}K\right|_{r=r_{\rm UV}}$ and $\left.\int_{\partial M_{11}}\sqrt{-h}h^{mn}\frac{\partial J_0}{\partial h^{mn}}\right|_{r=r_{\rm UV}}$ serves as the appropriate counter terms to cancel the UV divergences as given in (\ref{UV_divergences}).

Now, it was shown in \cite{OR4} that if one makes an ansatz:
\begin{eqnarray}
\label{ansaetze}
& & \hskip -0.8ing_{MN} = g_{MN}^{(0)} +\beta g_{MN}^{(1)},\nonumber\\
& & \hskip -0.8in{\cal C}_{MNP} = {\cal C}^{(0)}_{MNP} + \beta {\cal C}_{MNP}^{(1)},
\end{eqnarray}
to be substituted into the equations of motion, one can self-consistently set ${\cal C}_{MNP}^{(1)}=0$. Further, as proved in \cite{OR4} (as Lemma 1), in the neighborhood of the Ouyang embedding  of flavor $D7$-branes  (see \cite{metrics}) (that figure in the type IIB string dual of thermal QCD-like theories  at high temperatures \cite{metrics}) effected by working in the neighborhood of small $\theta_{1,2}$ (assuming a vanishingly small Ouyang embedding parameter), in the MQGP limit (\ref{MQGP_limit}), $\lim_{N\rightarrow\infty}\frac{E_8}{J_0}=0, \lim_{N\rightarrow\infty} \frac{t_8t_8G^2R^3}{E_8}=0$. Therefore, $E_8$ and $t_8^2G^2R^3$-contributions (were) are disregarded (in \cite{OR4}).

\subsection{When ${\cal O}(l_p^6)$ Is (Not) Enough}
\label{IR-large-N}

Based on the results of this paper and its applications as discussed in detail in \cite{Vikas+Gopal+Aalok}, \cite{Gopal+Vikas+Aalok}, we now address the question when it becomes necessary to go beyond ${\cal O}(R^4)$ corrections in ${\cal M}$-theory.

An extremely important lesson that we learn from the ${\cal O}(R^4)$ ${\cal M}$-theory corrections obtained in \cite{OR4}, can be abstracted from Table \ref{IR-enhancement-vs-Large-N-suppression}.
\begin{table}[h]
\begin{center}
\begin{tabular}{|c|c|c|c|} \hline
S. No. & $G^{\cal M}_{MN}$ & IR-Enhancement Factor & $N$-Suppression \\
& &  $\frac{\left(\log {\cal R}_h\right)^m}{{\cal R}_h^n}, m,n\in\mathbb{Z}^+$ & Factor \\
& & in the ${\cal O}(R^4)$ Correction & in the ${\cal O}(R^4)$ Correction   \\ \hline
1 & $G^{\cal M}_{\mathbb{R}^{1,3}}$ & $\log {\cal R}_h$ & ${N^{-\frac{9}{4}}}$ \\  \hline
2 & $G^{\cal M}_{rr, \theta_1x}$ & 1 & ${N^{-\frac{8}{15}}}$ \\  \hline
3 & $G^{\cal M}_{\theta_1z,\theta_2x}$ & ${{\cal R}_h^{-5}}$ & ${N^{-\frac{7}{6}}}$ \\  \hline
4 &  $G^{\cal M}_{\theta_2y}$ & $\log {\cal R}_h$ & ${N^{-\frac{7}{5}}}$ \\ \hline
5 &  $G^{\cal M}_{\theta_2z}$ & $\log {\cal R}_h$ & ${N^{-\frac{7}{6}}}$
\\ \hline
6 &  $G^{\cal M}_{xy}$ & $\log {\cal R}_h$ & ${N^{-\frac{21}{20}}}$ \\  \hline
7 &  $G^{\cal M}_{xz}$ &  $\left(\log {\cal R}_h\right)^3$ & ${N^{-\frac{5}{4}}}$ \\ \hline
8 &  $G^{\cal M}_{yy}$ & $\log {\cal R}_h$ & ${N^{-\frac{7}{4}}}$ \\  \hline
9 &  $G^{\cal M}_{yz}$ & $\frac{\log {\cal R}_h}{{\cal R}_h^7}$ & ${N^{-\frac{29}{12}}}$ \\ \hline
10 &  $G^{\cal M}_{zz}$ & $\log {\cal R}_h$ & ${N^{-\frac{23}{20}}}$ \\ \hline
11  &  $G^{\cal M}_{x^{10}x^{10}}$ & $\frac{\log {\cal R}_h^3}{{\cal R}_h^2}$ & ${N^{-\frac{5}{4}}}$ \\ \hline
\end{tabular}
\end{center}
\caption{IR Enhancement vs. large-$N$ Suppression in ${\cal O}(R^4)$-Corrections in the M-theory Metric in the $\psi=2n\pi, n=0,1,2$ Patches; ${\cal R}_h \equiv \frac{r_h}{{\cal R}_{D5/\overline{D5}}}<1$, ${\cal R}_{D5/\overline{D5}}$ being the $D5-\overline{D5}$ separation}
\label{IR-enhancement-vs-Large-N-suppression}
\end{table}

In Table 2, the delocalized $T^3(x, y, z)$ coordinates $x, y, z$ are defined near $r=\langle  r \rangle\in$IR and $\langle \theta_{1,2} \rangle$ close to the Ouyang embedding of the flavor $D7$-branes in the parent type IIB dual \cite{metrics}, as \cite{MQGP} \footnote{As explained in \cite{Knauf-thesis}, the $T^3$-valued $(x, y, z)$ are defined via:
 {\footnotesize
\begin{eqnarray*}
\label{xyz-definitions}
& & \phi_1 = \langle \phi_1 \rangle + \frac{x}{\sqrt{h_2}\left[h(\langle r \rangle,\langle \theta_{1,2} \rangle)\right]^{\frac{1}{4}} \sin\langle \theta_{1} \rangle\ \langle r \rangle},\nonumber\\
& & \phi_2 = \langle \phi_2 \rangle + \frac{y}{ \sqrt{h_4}
\left[h( \langle r \rangle,\langle \theta_{1,2} \rangle)\right]^{\frac{1}{4}}\sin\langle \theta_{2} \rangle\ \langle r \rangle}\nonumber\\
& & \psi = \langle \psi \rangle + \frac{z}{\sqrt{h_1} \left[h( \langle r \rangle,\langle \theta_{1,2} \rangle)
\right]^{\frac{1}{4}}\ \langle r \rangle}.
\end{eqnarray*}
}
In the IR, it was shown \cite{theta0-theta} that the delocalized $\langle \theta_{1,2} \rangle$ can be promoted to global $\theta_{1,2}$; we do so in all the results in the paper.}:
\begin{eqnarray}
\label{xyz-defs}
& &   dx = \sqrt{h_2}\biggl[h\biggl(\langle r \rangle,\langle \theta_{1,2} \rangle\biggr)\biggr]^{\frac{1}{4}} \sin\langle \theta_{1} \rangle\ \langle r \rangle  d\phi_1,\nonumber\\
& & dy = \sqrt{h_4}
\biggl[h\biggl( \langle r \rangle,\langle \theta_{1,2} \rangle\biggr)\biggr]^{\frac{1}{4}}\sin\langle \theta_{2} \rangle\ \langle r \rangle d\phi_2,\nonumber\\
& &  dz = \sqrt{h_1} \biggl[h\biggl( \langle r \rangle,\langle \theta_{1,2} \rangle\biggr)\biggr]^{\frac{1}{4}}\ \langle r \rangle d\psi,\nonumber\\
& &
\end{eqnarray}
$h(\langle r \rangle,\langle \theta_{1,2} \rangle)$ being the delocalized warp factor \cite{metrics}:
{\footnotesize
\begin{eqnarray}
\label{eq:h}
&& h(\langle r \rangle,\langle \theta_{1,2} \rangle) =\frac{L^4}{\langle r \rangle^4}\Bigg[1+\frac{3g_sM_{\rm eff}^2}{2\pi N}{\rm log}\langle r \rangle\left\{1+\frac{3g_sN^{\rm eff}_f}{2\pi}\left({\log} \langle r \rangle+\frac{1}{2}\right)+\frac{g_sN^{\rm eff}_f}{4\pi}{\rm log}\left({\sin}\frac{\langle \theta_{1} \rangle}{2}
{\rm sin}\frac{\langle \theta_{2} \rangle}{2}\right)\right\}\Bigg],\nonumber\\
\end{eqnarray}
}
wherein $L\equiv 4\pi g_s N \alpha'^2$, with the effective number of fractional $D3$-branes, $M_{\rm eff}$,  and the effective number of flavor $D7$-branes, $N_f^{\rm eff}$, defined, e.g., in \cite{Vikas+Gopal+Aalok}.  The squashing factors are defined below \cite{metrics}:
\begin{equation}
\label{h_{1,2,4}-defs}
h_1 = \frac{1}{9} + {\cal O}\left(\frac{g_sM^2}{N}\right),\ h_2 = \frac{1}{6} + {\cal O}\left(\frac{g_sM^2}{N}\right),\
h_4 = h_2 + \frac{4a^2}{\langle r\rangle^2},
\end{equation}
($a$ being the radius of the blown-up $S^2$).

 One notes that in the IR: $r = \chi r_h, \chi\equiv {\cal O}(1)$, and up to ${\cal O}(\beta)$:
\begin{equation}
\label{IR-beta-N-suppressed-logrh-rh-neg-exp-enhanced}
f_{MN} \sim \beta\frac{\left(\log {\cal R}_h\right)^{m}}{{\cal R}_h^n N^{\beta_N}},\ m\in\left\{0,1,3\right\},\ n\in\left\{0,2,5,7\right\},\
\beta_N>0.
\end{equation}
Note $|{\cal R}_h|\ll1$ and as estimated in \cite{Bulk-Viscosity-McGill-IIT-Roorkee}, 
\begin{equation}
\label{AbsLogCal Rh}
|\log {\cal R}_h|\sim \kappa_{r_h}N^{\frac{1}{3}}, 0<\kappa_{r_h} = \frac{1}{3(6\pi)^{1/3}\left(g_s N_f\right)^{2/3}\left(g_s M^2\right)^{1/3}}<1.
\end{equation}
 This implies Planckian and large-$N$ suppression, and infra-red enhancement arising from $m,n\neq0$ in (\ref{IR-beta-N-suppressed-logrh-rh-neg-exp-enhanced}), are mutually competing effects. As shown in \cite{OR4}, choosing a hierarchy: $\beta\sim e^{-\gamma_\beta N^{\gamma_N}}$ \cite{MQGP-Page}, $\gamma_\beta,\gamma_N>0: \gamma_\beta N^{\gamma_N}>7\kappa_{r_h}N^{\frac{1}{3}} + \left(\frac{m}{3} - \beta_N\right)\log N$, ensures that the IR-enhancement does not dominate over Planckian suppression. Hence, if $\gamma_\beta N^{\gamma_N}\sim7\kappa_{r_h}N^{\frac{1}{3}}$, one would have to go to a higher order in $\beta$.

Thus, for any $\alpha>0$, $\frac{E_8}{J_0}\sim\frac{1}{N^\alpha}$. It was shown in \cite{Gopal+Vikas+Aalok}that one obtains the hierarchy $t_8^2G^2R^3<E_8<J_0$ in the MQGP limit. Hence, we will consider only ``$J_0$'' term for the calculation purpose in this paper similar to \cite{Vikas+Gopal+Aalok,Gopal+Vikas+Aalok,Gopal-Tc-Vorticity,Gopal+Aalok,ACMS,Aalok+Gopal-Mesino}.

In the remainder of the paper it will be understood that use would have been made of the following for simplifying expression. In the ${\cal M}$-theory $C_3^\beta=0$-truncation, it was shown in \cite{OR4} that $|{\cal C}_{\theta_1x}^{\rm bh}|\ll1, \left(-{\cal C}_{zz}^{\rm bh} + 2 {\cal C}_{\theta_1z}^{\rm bh} - 3 {\cal C}_{\theta_1x}^{\rm bh}\right)=0$ where ${\cal C}_{MN}^{\rm bh}$ corresponds to the constant of integration appearing in solutions of the EOMs of the ${\cal O}(R^4)$ ${\cal M}$-theory metric $g_{MN}^{\cal M}$.

\section{Almost Contact 3-Structures Arising from $G_2$ Structure in the ${\cal M}$ Uplift in the Limit (\ref{MQGP_limit})}
\label{ACM3S-basics}

Due to non-trivial ${\cal M}$-theory four-form fluxes, $M_7$ is generically not Ricci-flat and hence does not possess $G_2$ holonomy, but usually possesses $G_2$ structure \footnote{If V is a seven-dimensional real vector space, then a three-form $\Phi$ is said to be positive if it lies in the $GL (7; \mathbb{R})$ orbit of $\Phi_0$, where $\Phi_0$ is a three-form on $\mathbb{R}^7$ which is preserved by $G_2$-subgroup of $GL (7; \mathbb{R})$. The pair $(\Phi; g)$ for a positive 3-form $\Phi$ and the corresponding metric $g$, constitute a $G_2$-structure.}. Given that the adjoint of $SO(7)$ decomposes under $G_2$ as ${\bf 21}\rightarrow{\bf 7}\oplus{\bf 14}$ where ${\bf 14}$ is the adjoint representation of $G_2$, one obtains the following four $G_2$-structure torsion classes:
\begin{equation}
\label{T-G2}
\tau \in \Lambda^1 \otimes g_2^\perp = W_1 \oplus W_7 \oplus W_{14} \oplus W_{27} = \tau_0 \oplus \tau_1 \oplus \tau_2 \oplus \tau_3,
\end{equation}
$g_2^\perp$ representing the orthogonal complement of $g_2$; the subscript $a$ in $W_a$ denotes the dimensionality of the torsion class $W_a$, and $p$ in $\tau_p$ denotes the rank of the associated differential form. The four intrinsic $G_2$-structure torsion classes are defined, e.g. in \cite{J. G. J. Held's thesis [2012]}.

The $G_2$-structure torsion classes $\tau_p$'s of the seven-fold $M_7=S^1_{\cal M}\times_w\left(S^1_{\rm thermal}\times_w M_5\right)$, $M_5$ being a non-Einsteinian generalization of $T^{1,1}$, and close to the Ouyang embedding\\ $\left(r^6 + 9 a^2 r^4\right)^{1/4}\sin\left(\frac{\theta_1}{2}\right)\sin\left(\frac{\theta_2}{2}\right) = \mu, \mu$ being the Ouyang embedding parameter, of the flavor $D7$-branes in the parent type IIB dual in the limit of very-small-Ouyang-embedding parameter limit ($|\mu_{\rm Ouyang}|\ll1$) were worked out in \cite{ACMS}:
\begin{equation}
\label{G2-torsion}
\tau\left(M_7\right) = \tau_1\oplus\tau_2\oplus\tau_3.
\end{equation}
It was also  shown in \cite{ACMS} that in the $N\gg1$-MQGP limit (\ref{MQGP_limit}) and the intermediate-$N$ MQGP limit ( (\ref{MQGP_limit})[``$\gg/\ll$''$\rightarrow>/<$]), the aforementioned closed $M_7$ supports Almost Contact 3-Structures [Lemma 2 of \cite{ACMS}]. But $M_7$ supports Contact 3-Structures  only in the latter limit (\ref{MQGP_limit}) [Lemma 4 of \cite{ACMS}].

The main result of \cite{ACMS} is that the four-parameter space ${\cal X}_{G_2}(g_s, M, N_f; N)$ [$g_s\in(0,1)$ and varying continuously; $M_{\rm UV}, N_f^{\rm UV} N$ varying in steps of 1 such that $M, N_f$ are ${\cal O}(1)$ and $\frac{1}{N}\ll1$] of $M_7$ supporting $G_2$ structures and relevant to the aforementioned ${\cal M}$-theory uplift of thermal QCD-like theories, is not $N$-path connected with reference to Contact Structures in the IR, i.e., the $N\gg1$ Almost Contact 3-Structures arising from the $G_2$ structure in the $N\gg1$ MQGP limit (footnote \tcb{1}), do not connect to a Contact 3-Structures (in the IR) which is shown to exist only for an appropriate intermediate $N$ effected by the intermediate-$N$ MQGP limit ( (\ref{MQGP_limit})[``$\gg/\ll$''$\rightarrow>/<$]) and, e.g., by the QCD-inspired parameters $M_{\rm UV}, N_f^{\rm UV} g_s$ of Table \tcb{1}.

\section{$D6$-Branes' Gauge Fields in the Absence of Magnetic Fields in the IR/UV}
\label{gauge-fields-B0}

In this section, we work out the gauge fields that can supported on the world volume of the type IIA flavor $D6$-branes in the absence of an external magnetic field in the IR (\ref{Amu-IR-beta0} - \ref{AmuB0beta}) and in the UV \ref{AmuB0UV}.

Here we will consider the DBI action for $N_f$ flavor $D6$-branes, 
\begin{equation}\label{D6DBI}
S_{D6}=-T_{D6}N_f\int d^{7}\xi~ e^{-\phi_{IIA}}\sqrt{-\det{\{i^*(g+B)+F\}}},
\end{equation}
with $2\pi\alpha^{\prime}=1$, $i:\Sigma_{D6}\hookrightarrow M_{10}$ defines the embedding of the $D6$-brane world volume in the ten-dimensional type IIA gravity dual involving a non-K\"{a}hler resolved conifold, and $\{t,x^1,x^2,x^3,Z,\theta_2,\tilde{y}\}$ are the coordinates of the worldvolume directions of the $D6$-branes with $\{t,x^1,x^2,x^3 \}$ are usual Minkowski coordinates. Here the radial coordinate is redefined as $r=r_{h}e^{Z}$, where $r$ is the radial coordinate and $\theta_{2}$, $\tilde{y}$ are angular coordinates. The $U(1)$ gauge field strength is $F_{\mu\nu}-\partial_{\mu}A_{\nu}-\partial_{\nu}A_{\mu}$, and $\phi_{IIA}$ is the type-IIA dilaton (triple T-dual of type-IIB dilaton). In this section, we will consider the aforementioned DBI action with vanishing magnetic field.

\subsection{$A_\mu, \mu = t, \rho, \phi, Z, x^3$ in the IR Up to ${\cal O}(\beta^0)$}
\label{Amu-IR-beta0}

We will work in the gauge $A_Z(\rho, Z, x^3)=0$. One can decompose the gauge field as, $A_\mu(t, \rho, \phi, Z, x^3) = A_\mu(\rho, \phi, Z, x^3)^{\beta^0} + \beta A_\mu(\rho, Z, x^3)^\beta$, where,  $ A_\mu(\rho, \phi, Z, x^3)^{\beta^0}$ are gauge fields without considering the ${\cal O}(R^4)$ corrections, and $A_\mu(\rho, \phi, Z, x^3)^{\beta}$ are the fields which encodes the ${\cal O}(R^4)$  corrections. First, we work out $A_\mu(\rho, Z, x^3)^{\beta^0}$  in the IR region.

Now,
{\footnotesize
\begin{eqnarray}
\label{SDBIuptoObeta-3}
& & \hskip -0.8in e^{-\phi^{\rm IIA}} = \frac{3 \left({g_s} {N_f}
   \log \left(9 b^2+e^{2 Z}\right)+6 {g_s} {N_f} \log ({r_h})+4 {g_s} {N_f} Z-8 \pi \right)}{8 \pi 
   {g_s}} + \frac{9 b (8 \pi  b+3) {g_s} M^2 {N_f} ({c_1}+{c_2} \log ({r_h}))}{4 \pi  N \left(9 b^2+e^{2
   Z}\right)}\nonumber\\
& & \hskip -0.8in   +\frac{243 b^{10} \left(9 b^2+1\right)^3 \beta  M \left(\frac{1}{N}\right)^{5/4} \left(19683 \sqrt{6}
   \alpha _{\theta _1}^6+6642 \alpha _{\theta _2}^2 \alpha _{\theta _1}^3-40 \sqrt{6} \alpha _{\theta _2}^4\right)
   \left(e^Z-2\right) e^{2 Z} \log ^3({r_h}) }{32 \pi ^2 \left(3 b^2-1\right)^5 \left(6
   b^2+1\right)^3 {g_s} (\log N)^4 {N_f} {r_h} \alpha _{\theta _2}^3}\nonumber\\  
& & \hskip -0.8in \times \left({g_s} {N_f} \log \left(9 b^2+e^{2 Z}\right)+6 {g_s}
   {N_f} \log ({r_h})+4 {g_s} {N_f} Z-8 \pi \right).   
\end{eqnarray}
}
In the IR,
{\footnotesize
\begin{eqnarray}
\label{sqrtGplusF}
& & \hskip -0.8in {\sqrt{-(i^*G+F)}}^{B=0,\ \beta^0}\nonumber\\
& & \hskip -0.8in = \frac{\sqrt{2} N^{3/5} {r_h} e^Z \sqrt{
(\partial_{x^3} A_t^{\beta^0}) ^2 \left((\partial_Z A_\rho^{\beta^0}) ^2+
(\partial_Z A_\phi^{\beta^0}) ^2\right) - 2
   (\partial_Z A_\rho^{\beta^0})   (\partial_\rho A_t^{\beta^0})  
   (\partial_{x^3} A_t^{\beta^0})  (\partial_Z A_{x^3}^{\beta^0}) + (\partial_\rho A_t^{\beta^0})^2 \left((\partial_Z A_{x^3})^2+(\partial_Z A_\phi)^2\right)}}{3 \sqrt[3]{3} \alpha _{\theta _2}^2}.\nonumber\\
& &  
\end{eqnarray}   
}
Assuming $\partial_{x^3}A_t^{\beta^0}=0$, (\ref{sqrtGplusF}) reduces to:
{\footnotesize
\begin{eqnarray}
\label{ExpZrhrhoLDBI-i}
& & \rho e^Z r_h {\cal L}_{\rm DBI}^{B=0}
= \frac{\rho  \sqrt{(\partial_\rho A_t^{\beta^0})  ^2 \left((\partial_Z A_{x^3}^{\beta^0})^2+(\partial_Z A_\phi^{\beta^0}) ^2\right)}
  }{139968 \sqrt{2} 3^{5/6} \pi  {g_s} \alpha _{\theta _2}^5}  \left(34992 \sqrt{3} N^{3/5} {r_h}^2 e^{2 Z} \alpha _{\theta _2}^3\right)\nonumber\\  
& & \times  \left(\frac{2 \left(9+8 \sqrt{3} \pi \right)
   {g_s}^2 M^2 {N_f} (c_1+c_2 \log ({r_h}))}{\sqrt{3} N \left(e^{2 Z}+3\right)}+{g_s}
   {N_f} \log \left(e^{2 Z}+3\right)+6 {g_s} {N_f} \log ({r_h})+4 {g_s} {N_f} Z-8 \pi \right).   
\end{eqnarray}
}

\noindent {\bf $A_\phi$ EOM}

As $\frac{\delta {\cal L}_{\rm DBI}^{B=0,\ {\rm IR}}}{\delta (\partial_{\rho}A_\phi^{\ \beta^0})} = \frac{\delta {\cal L}_{\rm DBI}^{B=0,\ {\rm IR}}}{\delta (\partial_{x^3}A_\phi^{\ \beta^0})}  = 0$, the ${A_\phi^{\ \beta^0}}$ EOM is $\frac{\delta {\cal L}_{\rm DBI}^{B=0,\ {\rm IR}}}{\delta (\partial_{Z}A_\phi^{\ \beta^0})} = {\cal C}_\phi^{\rho x^3}$. Assuming $\partial_{x^3}A_t=0$, and the following ansatz:
\begin{eqnarray}
\label{Amu-ansatz}
& & {A_\phi^{\ \beta^0}}(Z, \rho, Z) = {a_\phi^\rho}(\rho) {a_\phi^{x^3}}({x^3}) {a_\phi^Z}(Z),\nonumber\\
& & {A_\rho^{\ \beta^0}}(Z, \rho, Z) = {a_\rho^\rho}(\rho) {a_\phi^{x^3}}({x^3}) {a_\rho^{Z}}(Z),\nonumber\\
& &  {A_{x^3}^{\ \beta^0}}(Z, \rho, Z) = {a_{x^3}^\rho}(\rho ) {a_{x^3}^{x^3}}({x^3}) {a_{x^3}^Z}(Z),\nonumber\\
& & A_{t}^{\beta^0}\ (Z, \rho, Z) = {a_t^\rho}(\rho ) {a_{t}^{x^3}}({x^3}) {a_{t}^{Z}}(Z),
\end{eqnarray} 
this implies:
{\footnotesize
\begin{eqnarray}
\label{AphiEOM-i}
& & -\frac{N^{3/5} \rho  {r_h}^2 e^{2 Z} (\partial_\rho A_t^{\beta^0})  ^2 
(\partial_Z A_{\phi}^{\beta^0}) \left(6 {g_s} {N_f} \log ({r_h})-\frac{\left(9+8 \sqrt{3} \pi
   \right) {g_s}^2 M^2 {N_f} \left(Z^2+2 Z-4\right) (c_1+c_2 \log ({r_h}))}{8 \sqrt{3}
   N}\right)}{4 \sqrt{2} \sqrt[3]{3} \pi  {g_s} \alpha _{\theta _2}^2 \sqrt{-
   (\partial_\rho A_{t}^{\beta^0})^2 \left((\partial_Z A_{x^3}^{\beta^0})^2
   + (\partial_Z A_{\phi}^{\beta^0})^2\right)}}\nonumber\\
& & = -\frac{N^{3/5} \rho  {r_h}^2 e^{2 Z} {a_t^{x^3}}({x^3}) {a_t^Z}(Z) {a_\phi^\rho }(\rho ) {a_\phi^{x^3}}({x^3}) {a_t^\rho}'(\rho ) {a_\phi^Z}'(Z) \left(6 {g_s} {N_f} \log
   ({r_h})-\frac{\left(9+8 \sqrt{3} \pi \right) {g_s}^2 M^2 {N_f} \left(Z^2+2 Z-4\right)
   (c_1+c_2 \log ({r_h}))}{8 \sqrt{3} N}\right)}{4 \sqrt{2} \sqrt[3]{3} \pi  {g_s} \alpha _{\theta
   _2}^2 \sqrt{{a_{x^3}^\rho}(\rho )^2 {a_{x^3}^{x^3}}({x^3})^2 {a_{x^3}^Z}'(Z)^2+{a_\phi^\rho }(\rho )^2
   {a_\phi^{x^3}}({x^3})^2 {a_\phi^Z}'(Z)^2}}\nonumber\\
 & &   ={\cal C}_{\phi}^{\rho x^3,\ B=0}(\rho, x^3).
\end{eqnarray}
}   
Now, (\ref{AphiEOM-i}) simplifies to:
\begin{eqnarray}
\label{AphiEOM-ii}
& & \frac{3^{2/3} N^{3/5} {N_f} \rho  {r_h}^2 e^{2 Z} {a_t^{x^3}}({x^3}) {a_t^Z}(Z) \log ({r_h})
   {a_t^\rho}'(\rho )}{2 \pi  \alpha _{\theta _2}^2 \sqrt{2 
   {\cal C}_{x^3}\ ^{\rho\phi,\ B=0}\ ^2 {\cal C}_{x^3}\ ^{\phi,\ B=0}\ ^2
   {\cal C}_{x^3}\ ^{\phi Z,\ B=0}\ ^2+2}}\nonumber\\
& &    -\frac{\left(\frac{1}{N}\right)^{2/5} \left(\left(3 \sqrt{3}+8 \pi \right) {g_s} M^2
   {N_f} \rho  {r_h}^2 e^{2 Z} \left(Z^2+2 Z-4\right) {a_t^{x^3}}({x^3}) {a_t^Z}(Z) {at\rho
 }'(\rho ) (c_1+c_2 \log ({r_h}))\right)}{32 \left(\sqrt[3]{3} \pi  \alpha _{\theta _2}^2 \sqrt{2
   {\cal C}_{x^3}^{\rho\phi,\ B=0}\ ^2 {\cal C}_{x^3}\ ^{\phi,\ B=0}\ ^2 {\cal C}_{x^3}\ ^{\phi Z,\ B=0}\ ^2+2}\right)}\nonumber\\
& &    + {\cal O}\left(\left(\frac{1}{N}\right)^{6/5}\right)={\cal C}_{\phi}^{\rho x^3,\ B=0}(\rho, x^3).
\end{eqnarray}
One hence sees that (\ref{AphiEOM-ii}) implies:
\begin{eqnarray}
\label{AphiEOM-iii}
& & {a_t^Z}(Z)=-\frac{{\cal C}_\phi^{t Z,\ B=0}\  e^{-2 Z}}{729 \sqrt{3} \pi  {N_f} {r_h}^2 \alpha _{\theta _2}^3
   \left(\left(3 \sqrt{3}+8 \pi \right) c_2 {g_s} M^2 \left(\frac{1}{N}\right)^{2/5} \left(Z^2+2
   Z-4\right)-48 N^{3/5}\right)}\nonumber\\
& & = \frac{{\cal C}_\phi^{t Z,\ B=0}\  \left(\frac{1}{N}\right)^{3/5} (1 - 2 Z + 2 Z^2)}{34992 \sqrt{3} \pi  {N_f} {r_h}^2 \alpha
   _{\theta _2}^3}+{\cal O}\left(\left(\frac{1}{N}\right)^{6/5}\right),   
\end{eqnarray}
and
\begin{eqnarray}
\label{AphiEOM-iv}
& & \hskip -0.4in \frac{{\cal C}_\phi^{t Z,\ B=0}\  \rho  {a_t^{x^3}}({x^3}) \log ({r_h}) {a_t^\rho}'(\rho )}{23328\ 3^{5/6} \pi ^2
   \alpha _{\theta _2}^5 \sqrt{2 {\cal C}_{x^3}^{\rho\phi,\ B=0}\ ^2 {\cal C}_{x^3}\ ^{\phi,\ B=0}\ ^2 {\cal C}_{x^3}\ ^{\phi Z,\ B=0}\ ^2+2}}={\cal C}_{\phi}^{\rho x^3,\ B=0}(\rho, x^3) \equiv {\rm Constant},
\end{eqnarray}
which yields:
\begin{eqnarray}
\label{AphiEOM-v}
& & {a_t^\rho}(\rho )={\cal C}_\rho^{B=0}\    \log (\rho )+c_1,\nonumber\\
& & {a_t^{x^3}}({x^3})=\frac{23328\ 3^{5/6} \pi ^2 \alpha _{\theta _2}^5 \sqrt{2 {\cal C}_{x^3}^{\rho\phi,\ B=0}\ ^2
   {\cal C}_{x^3}\ ^{\phi,\ B=0}\ ^2 {\cal C}_{x^3}\ ^{\phi Z,\ B=0}\ ^2+2} {\cal C}_\phi^{\rho x^3,\ B=0}(\rho, x^3)}{{\cal C}_\phi^{t Z,\ B=0}\  {\cal C}_\rho^{B=0}\    \log ({r_h})}.
\end{eqnarray}

We have assumed $ \partial_{x^3}A_{t}^{\beta^0}\   = 0$.

\noindent{\bf $A_{\rho}$ EOM}

Now, assuming $ \partial_{x^3}A_{t}^{\beta^0}\   = 0$, one can show that the $A_\rho$-EOM is identically satisfied.

\noindent{\bf $A_t$ EOM}

As  $\frac{\delta {\cal L}_{\rm DBI}^{B=0,\ {\rm IR}}}{\delta \partial_{Z}A_{t}^{\beta^0}} = \frac{\delta {\cal L}_{\rm DBI}^{B=0,\ {\rm IR}}}{\delta \partial_{\rho}A_{t}^{\beta^0}}  = 0$, the $A_{t}^{\beta^0}\ $ EOM is $\frac{\delta {\cal L}_{\rm DBI}^{B=0,\ {\rm IR}}}{\delta \partial_{x^3}A_{t}^{\beta^0}} = {\cal C}_{t}^{\rho Z}$. This implies:
\begin{eqnarray}
\label{AtEOM-1}
& & \frac{N^{3/5} \rho  {r_h}^2 e^{2 Z} \left(6 {g_s} {N_f} \log ({r_h})-\frac{\left(9+8 \sqrt{3} \pi
   \right) {g_s}^2 M^2 {N_f} \left(Z^2+2 Z-4\right) (c_1+c_2 \log ({r_h}))}{8 \sqrt{3}
   N}\right) }{4 \sqrt{2} \sqrt[3]{3} \pi  {g_s} \alpha _{\theta
   _2}^2}\nonumber\\
& & \times \sqrt{{a_{x^3}^\rho}(\rho )^2 {a_{x^3}^{x^3}}({x^3})^2 {a_{x^3}^Z}'(Z)^2+{a_\phi^\rho }(\rho )^2
   {a_\phi^{x^3}}({x^3})^2 {a_\phi^Z}'(Z)^2} = {\cal C}_t^{x^3Z,\ B=0}(x^3, Z)
\end{eqnarray}
that is satisfied by:
\begin{eqnarray}
\label{AtEOM-2}
& & {a_{x^3}^\rho}(\rho )={\cal C}_{x^3}^{\rho\phi,\ B=0}\  {a_\phi^\rho }(\rho),
\nonumber\\
& & {a_{x^3}^{x^3}}({x^3})={\cal C}_{x^3}\ ^{\phi,\ B=0}\  {a_\phi^{x^3}}({x^3}),\nonumber\\
& & {a_{x^3}^Z}'(Z)={\cal C}_{x^3}\ ^{\phi Z,\ B=0} {a_\phi^Z}'(Z).
\end{eqnarray}
Hence, 
\begin{eqnarray}
\label{AtEOM-3}
& & \frac{{\cal C}_{\phi\rho}^{B=0}\  N^{3/5} {r_h}^2 \left(2 Z^2+2 Z+1\right) {a_\phi^Z}'(Z) \left(6 {g_s}
   {N_f} \log ({r_h})-\frac{\left(9+8 \sqrt{3} \pi \right) {g_s}^2 M^2 {N_f} \left(Z^2+2 Z-4\right)
   (c_1+c_2 \log ({r_h}))}{8 \sqrt{3} N}\right)}{4 \sqrt{2} \sqrt[3]{3} \pi  {g_s} \alpha _{\theta
   _2}^2}\nonumber\\
& &    = {\cal C}_{\phi Z}^{B=0}\ .
\end{eqnarray}
Defining,
{\footnotesize
\begin{eqnarray}
\label{a012-defs}
& & \hskip -0.6in {a_0}=17496 \left(9+8 \sqrt{3} \pi \right) {g_s}^2 M^2 \left(\frac{1}{N}\right)^{2/5} {N_f} {r_h}^2
   \alpha _{\theta _2}^3 (c_1+c_2 \log ({r_h}))+209952 \sqrt{3} {g_s} N^{3/5} {N_f} {r_h}^2
   \alpha _{\theta _2}^3 \log ({r_h}),\nonumber\\
& & \hskip -0.6in {a_1}=26244 \sqrt{3} \left(3 \sqrt{3}+8 \pi \right) {g_s}^2 M^2 \left(\frac{1}{N}\right)^{2/5} {N_f}
   {r_h}^2 \alpha _{\theta _2}^3 (c_1+c_2 \log ({r_h}))+419904 \sqrt{3} {g_s} N^{3/5} {N_f}
   {r_h}^2 \alpha _{\theta _2}^3 \log ({r_h}),\nonumber\\
& & \hskip -0.6in {a_2}=13122 \sqrt{3} \left(3 \sqrt{3}+8 \pi \right) {g_s}^2 M^2 \left(\frac{1}{N}\right)^{2/5} {N_f}
   {r_h}^2 \alpha _{\theta _2}^3 (c_1+c_2 \log ({r_h}))+419904 \sqrt{3} {g_s} N^{3/5} {N_f}
   {r_h}^2 \alpha _{\theta _2}^3 \log ({r_h}),      
\end{eqnarray}
}
and using 
\begin{eqnarray}
\label{a012-defs}
& & \int \frac{dZ}{{a_0}+{a_1} Z+{a_2} Z^2} = 
\frac{2  \tan ^{-1}\left(\frac{{a_1}+2 {a_2} Z}{\sqrt{4 {a_0}
   {a_2}-{a_1}^2}}\right)}{\sqrt{4 {a_0} {a_2}-{a_1}^2}},
\end{eqnarray}
one obtains:
\begin{eqnarray}
\label{Aphisol}
& & {a_\phi^Z}(Z)=\frac{i \sqrt{2} \pi  {\cal C}_{\phi Z}^{B=0}\  \left(\frac{1}{N}\right)^{3/5} \alpha _{\theta _2}^2
   (\log ((1-i)-2 i Z)-\log (2 i Z+(1+i)))}{3^{2/3} {\cal C}_{\phi\rho}^{B=0}\  {N_f} {r_h}^2 \log ({r_h})},\nonumber\\
& & {a_\phi^{x^3}}({x^3})=\frac{\sqrt{{\cal C}_{x^3}^{\rho\phi,\ B=0}\ ^2 {\cal C}_{x^3}\ ^{\phi,\ B=0}\ ^2 {\cal C}_{x^3}\ ^{\phi Z, B=0}\ ^2+1} {\cal C}_t^{x^3Z,\ B=0}(x^3, Z)}{{\cal C}_{\phi Z}^{B=0}\ }.   
\end{eqnarray}

\noindent {\bf $A_{x^3}$ EOM}

As $\frac{\delta {\cal L}_{\rm DBI}^{B=0,\ {\rm IR}}}{\delta \partial_\rho A_{x^3}^{\beta^0}} = \frac{\delta {\cal L}_{\rm DBI}^{B=0,\ {\rm IR}}}{\delta \partial_{x^3} A_{x^3}^{\beta^0}}  = 0$, the ${A_\phi^{\ \beta^0}}$ EOM is $\frac{\delta {\cal L}_{\rm DBI}^{B=0,\ {\rm IR}}}{\delta \partial_Z A_{x^3}^{\beta^0}} = {\cal C}_{x^3}^{\rho x^3}$, i. e.,
\begin{eqnarray}
\label{Ax3EOM-i}
& & -\frac{N^{3/5} \rho  {r_h}^2 e^{2 Z} {a_t^{x^3}}({x^3}) {a_t^Z}(Z) {a_{x^3}^\rho}(\rho )
   {a_{x^3}^{x^3}}({x^3}) {a_t^\rho}'(\rho ) {a_{x^3}^Z}'(Z) }{4 \sqrt{2} \sqrt[3]{3} \pi  {g_s} \alpha _{\theta
   _2}^2 \sqrt{{a_{x^3}^\rho}(\rho )^2 \left(-{a_{x^3}^{x^3}}({x^3})^2\right) {a_{x^3}^Z}'(Z)^2-{a\phi \rho
   }(\rho )^2 {a_\phi^{x^3}}({x^3})^2 {a_\phi^Z}'(Z)^2}}\nonumber\\
& & \times \left(6 {g_s} {N_f} \log
   ({r_h})-\frac{\left(9+8 \sqrt{3} \pi \right) {g_s}^2 M^2 {N_f} \left(Z^2+2 Z-4\right)
   (c_1+c_2 \log ({r_h}))}{8 \sqrt{3} N}\right)   
 \nonumber\\
 & & = {\cal C}_{x^3}^{\rho\phi,\ B=0}\  {\cal C}_{x^3}\ ^{\phi,\ B=0}\  {\cal C}_{x^3}\ ^{\phi Z,\ B=0} {\cal C}_\phi^{\rho x^3,\ B=0}(\rho, x^3)\nonumber\\
& &  -\frac{\left(3 \sqrt{3}+8 \pi \right) c_2 {\cal C}_{x^3}^{\rho\phi,\ B=0}\  {\cal C}_{x^3}\ ^{\phi,\ B=0}\ 
   {\cal C}_{x^3}\ ^{\phi Z,\ B=0} {g_s} M^2 \left(Z^2+2 Z-4\right) {\cal C}_{\phi}^{\rho x^3,\ B=0}(\rho, x^3)}{48 N}\nonumber\\  
& &    ={\cal C}_{x^3}\ ^{x^3\rho,\ B=0}\ ,
\end{eqnarray}
such that ${\cal C}_{x^3}^{\rho\phi,\ B=0}\  {\cal C}_{x^3}\ ^{\phi,\ B=0}\  {\cal C}_{x^3}\ ^{\phi Z,\ B=0}\sim N^{-x},x>0$.

The constants ${\cal C}_\phi^{t Z,\ B=0}, {\cal C}_\phi^{t Z,\ B=0}, {\cal C}_\rho^{B=0}, {\cal C}_{x^3}^{\rho\phi,\ B=0}, {\cal C}_{x^3}\ ^{\phi,\ B=0},
{\cal C}_{x^3}\ ^{\phi Z,\ B=0}, {\cal C}_\phi^{\rho x^3,\ B=0}(\rho, x^3),
{\cal C}_\phi^{t Z,\ B=0},\\ {\cal C}_\rho^{B=0}, {\cal C}_{x^3}\ ^{\phi,\ B=0},
{\cal C}_{\phi Z}^{B=0}, {\cal C}_{\phi\rho}^{B=0}, {\cal C}_t^{x^3Z,\ B=0}(x^3, Z) $ that figure in (\ref{AphiEOM-i}) - (\ref{Ax3EOM-i}), are the constants of integration appearing in the solutions of the EOMs for the type IIA flavor $D6$-branes in the IR and in the absence of an external magnetic field.

\subsection{Non-Renormalization in the IR Up to ${\cal O}(R^4)$ of $A_\mu, \mu=t, \rho, \phi, Z, x^3$}
\label{AmuB0beta}

In the $A_Z(\rho, \phi, Z, x^3)=0$ gauge for static solutions, we show here that $A_{\mu = \rho, \phi, Z, x^3}(\rho, Z, x^3)^\beta=0$ is a consistent truncation of the $A_\mu$ EOMs up to ${\cal O}(R^4)$.

Writing $A_\mu = A_\mu^{\beta^0} + \beta A_\mu^\beta \delta^t_\mu$, and $b = \frac{1}{\sqrt{3}} + \epsilon, A_t^\beta(Z, \rho, x^3) = a_t^{Z,\ \beta}(Z)a_t^{\rho,\ \beta^0}(\rho)a_t^{x^3,\ \beta^0}(x^3)$, one obtains:
{\scriptsize
\begin{eqnarray}
\label{ExpZrhrhoLDBI-ii}
& & \hskip -0.8in \rho e^Z r_h {\cal L}_{\rm DBI}^{B=0,\ \beta} = 
\frac{1}{139968 \sqrt{2} 3^{5/6} \pi  {g_s} \alpha
   _{\theta _2}^5}\Biggl\{\rho  \sqrt{(\partial_ZA_{x^3}^{\beta^0})^2+
(\partial_ZA_{\phi}^{\beta^0})^2}\nonumber\\
& & \hskip -0.8in \left(\frac{2 \left(9+8 \sqrt{3} \pi \right) {g_s}^2 M^2 {N_f} (c_1+c_2 \log
   ({r_h}))}{\sqrt{3} N \left(e^{2 Z}+3\right)}+{g_s} {N_f} \log \left(e^{2 Z}+3\right)+6 {g_s}
   {N_f} \log ({r_h})+4 {g_s} {N_f} Z-8 \pi \right)\nonumber\\
   & & \hskip -0.8in \left(34992 \sqrt{3} N^{3/5} {r_h}^2 e^{2 Z}
   \alpha _{\theta _2}^3 (\partial_\rho A_t^\beta)-\frac{2 \log r_h ^3 M
   \left(\frac{1}{N}\right)^{13/20} \left(35 Z^2+20 Z+4\right) \left(19683 \sqrt{6} \alpha _{\theta _1}^6+6642 \alpha
   _{\theta _2}^2 \alpha _{\theta _1}^3-40 \sqrt{6} \alpha _{\theta _2}^4\right) 
  (\partial_\rho A_t^{\beta^0})}{\pi  \epsilon ^5 (\log N)^4 {N_f} (3 Z+1)}\right)\Biggr\}\nonumber\\
& & \hskip -0.8in = \frac{\rho  \left(\frac{2 \left(9+8 \sqrt{3} \pi \right) {g_s}^2 M^2 {N_f} (c_1+c_2 \log
   ({r_h}))}{\sqrt{3} N \left(e^{2 Z}+3\right)}+{g_s} {N_f} \log \left(e^{2 Z}+3\right)+6 {g_s}
   {N_f} \log ({r_h})+4 {g_s} {N_f} Z-8 \pi \right) 
   }{209952 \sqrt{3} {g_s} \alpha
   _{\theta _2}^5}\nonumber\\
& & \hskip -0.8in \times \sqrt{\frac{\left(\frac{1}{N}\right)^{6/5} \alpha
   _{\theta _2}^4 \left({\cal C}_{x^3}^{\rho\phi,\ B=0}\ ^2 {\cal C}_{x^3}\ ^{\phi,\ B=0}\ ^2 {\cal C}_{x^3}\ ^{\phi Z,\ B=0}\ ^2+1\right)^2
   {\cal C}_t^{x^3Z,\ B=0}(x^3, Z)^2}{{N_f}^2 \rho ^2 {r_h}^4 \left(2 Z^2+2 Z+1\right)^2 \log ^2({r_h})}}\nonumber\\ 
   & & \hskip -0.8in \times \Biggl(34992 \sqrt{3} N^{3/5} {r_h}^2 e^{2 Z} \alpha _{\theta _2}^3 {a_t^{Z,\ \beta}}(Z) {a_t^{x^3,\ \beta^0}}({x^3})
   {a_t^{\rho, \beta^0}}'(\rho )\nonumber\\
& & \hskip -0.8in -\frac{4 {(\log r_h)}^3 M \left(\frac{1}{N}\right)^{5/4} \left(2 Z^2-2 Z+1\right)
   \left(35 Z^2+20 Z+4\right) \alpha _{\theta _2}^2 \left(19683 \sqrt{6} \alpha _{\theta _1}^6+6642 \alpha _{\theta
   _2}^2 \alpha _{\theta _1}^3-40 \sqrt{6} \alpha _{\theta _2}^4\right) {\cal C}_{\phi}^{\rho x^3,\ B=0}(\rho, x^3)}{3^{2/3} \epsilon
   ^5 (\log N)^4 {N_f}^2 \rho  {r_h}^2 (3 Z+1) \log ({r_h})}\Biggr)\nonumber\\
& & \hskip -0.8in \times  \sqrt{2 {\cal C}_{x^3}^{\rho\phi,\ B=0}\ ^2
   {\cal C}_{x^3}\ ^{\phi,\ B=0}\ ^2 {\cal C}_{x^3}\ ^{\phi Z,\ B=0}\ ^2+2}\nonumber\\ 
& & \hskip -0.8in = \left({\cal C}_{x^3}^{\rho\phi,\ B=0}\ ^2 {\cal C}_{x^3}\ ^{\phi,\ B=0}\ ^2 {\cal C}_{x^3}\ ^{\phi Z,\ B=0}\ ^2+1\right)
   {\cal C}_t^{x^3Z,\ B=0}(x^3, Z)\nonumber\\
   & & \hskip -0.8in \left(-\frac{\left(9+8 \sqrt{3} \pi \right) c_2 {g_s} M^2 \left(-Z^2-2
   Z+4\right) {a_t^{Z,\ \beta}}(Z) {a_t^{x^3,\ \beta}}({x^3}) {a_t^{\rho,\ \beta^0}}'(\rho )}{48 \sqrt{3}
   N}-\frac{{a_t^\beta}(Z) {a_t^{x^3,\ \beta^0}}({x^3}) {a_t^{\rho,\ \beta^0}}'(\rho ) (6 {g_s} {N_f} \log
   ({r_h})-8 \pi )}{6 {g_s} {N_f} \log ({r_h})}\right).      
\end{eqnarray}
}
One thus sees that (\ref{ExpZrhrhoLDBI-ii}) can be made arbitrarily negligible provided:
\begin{eqnarray}
\label{negligibleLDBIbeta}
& & \left({\cal C}_{x^3}^{\rho\phi,\ B=0}\ ^2 {\cal C}_{x^3}\ ^{\phi,\ B=0}\ ^2 {\cal C}_{x^3}\ ^{\phi Z,\ B=0}\ ^2+1\right)
   {\cal C}_t^{x^3Z,\ B=0}(x^3, Z)\sim N^{-(1+y)}, y>0.
\end{eqnarray} 
We hence can set $a_t^{Z,\ \beta}=0$ implying $A_{\mu = \rho, \phi, Z, x^3}(\rho, Z, x^3)^\beta=0$ in the IR.

\subsection{$A_\mu, \mu = t, \rho, \phi, Z, x^3$ in the UV Up to ${\cal O}(\beta^0)$}
\label{AmuB0UV}

Here, in the gauge $A_Z(\rho, Z, x^3)=0$, we will work out the background gauge fluctuations $A_\mu(\rho, Z, x^3)^{\beta^0}$  in the UV region. Now,
{\scriptsize
\begin{eqnarray}
\label{LDBI_UV_B=0}
& & \hskip -0.8in {\cal L}^{\rm DBI}_{\rm UV,\ B=0} =  -\frac{\sqrt{2} e^{4Z} \rho  {r_h}
   \sqrt{-\frac{N^{6/5} {r_h}^2 \left(\partial_{x^3}A_t ^2 \left({A\rho}^{(0,1,0,0)}(t,Z,\rho
   ,{x3})^2+\partial_Z A_\phi   ^2+2
   {r_h}^2\right)-2 \partial_Z A_\rho   
   \partial_\rho A_t  
   \partial_{x^3} A_t 
  \partial_ZA_{x^3}+\partial_\rho A_t  ^2
   \left(\partial_Z A_{x^3}^2+\partial_Z A_\phi   ^2+2
   {r_h}^2\right)\right)}{\alpha _{\theta _2}^4}}}{\sqrt[3]{3}
   {g_s}}\nonumber\\
& & \hskip -0.8in +   \frac{\beta  e^{5Z} \log r_h ^3 M^{\rm UV}
   \left(\frac{1}{N}\right)^{13/20} N^{3/5} \rho  {r_h} \left(19683
   \sqrt{6} \alpha _{\theta _1}^6+6642 \alpha _{\theta _2}^2 \alpha
   _{\theta _1}^3-40 \sqrt{6} \alpha _{\theta _2}^4\right)
  }{2187
   \sqrt{2} 3^{5/6} \pi  \epsilon ^5 {g_s} \log N ^4 {N_f^{\rm UV}}
   \alpha _{\theta _2}^7 \sqrt{-\frac{N^{6/5} {r_h}^2
   \left(\partial_{x^3} A_t ^2 \left(\partial_Z A_\rho   ^2+\partial_Z A_\phi   ^2+2 {r_h}^2\right)-2
   \partial_ZA_\rho  
   \partial_\rho A_t  
   \partial_{x^3} A_t 
   \partial_Z A_{x^3}+\partial_\rho A_t  ^2
   \left(\partial_Z A_{x^3}^2+\partial_Z A_\phi   ^2+2
   {r_h}^2\right)
   \right)}{e^{2Z} \alpha _{\theta_2}\ ^4}}}\nonumber\\
& & \hskip -0.8in \times  \left(\partial_{x^3} A_t ^2 \left(\partial_Z A_\rho   ^2+\partial_Z A_\phi   ^2+4 {r_h}^2\right)-2
   \partial_ZA_\rho  
   \partial_\rho A_t  
   \partial_{x^3} A_t 
   \partial_Z A_{x^3}+\partial_\rho A_t  ^2
   \left(\partial_Z A_{x^3}^2+\partial_Z A_\phi   ^2+4 {r_h}^2\right)\right).   
\end{eqnarray}
}
Assuming $\partial_{x^3} A_t =0$, here are the EOMs.

\noindent {\bf $A_{x^3}$ EOM}

The $A_{x^3}$ EOM: $\partial_\mu\left(\frac{\delta {\cal L}_{\rm DBI}^{\rm UV,\ B=0}}{\partial_\mu A_{x^3}}\right) = \frac{\delta {\cal L}_{\rm DBI}^{\rm UV,\ B=0}}{\delta A_{x^3}}$  yields:
$\frac{\delta {\cal L}_{\rm DBI}^{\rm UV,\ B=0}}{\partial_Z A_{x^3}}=$ constant. Now, using an ansatz similar to (\ref{Amu-ansatz}),
\begin{eqnarray}
\label{dLoverddZAx3}
& & \hskip -0.3in \frac{\delta {\cal L}_{\rm DBI}^{\rm UV,\ B=0}}{\partial_Z A_{x^3}} = \frac{\sqrt{2} N^{3/5} \rho  {r_h}^2 e^{4 Z} a_t^{x^3}(x^3) a_t^Z(Z)
   a_{x^3}^\rho(\rho) a_{x^3}^{x^3}(x^3) a_t^\rho\ '(\rho )
   a_{x^3}^Z\ '(Z)}{\sqrt{a_{x^3}^\rho(\rho)^2 a_{x^3}^{x^3}(x^3)^2
   a_{x^3}^Z\ '(Z)^2+a_\phi^\rho(\rho)^2 a_\phi^{x^3}(x^3)^2 {a\phi
   Z}'(Z)^2+2 {r_h}^2}}.
\end{eqnarray}
Assuming $a_{x^3}^Z\ '(Z)={\cal C}_{x3}^{\phi Z}\    a_\phi^Z\  '(Z) , a_{x^3}^\rho(\rho)={\cal C}_{a_{x3}^{\rho\phi}}^{B=0,\ {\rm UV}}\  {a_\phi^\rho}(\rho )$, one hence obtains:
\begin{eqnarray}
\label{Ax3EOMansatz2}
& & \frac{a_t^{x^3}(x^3)
   a_{x^3}^{x^3}(x^3)}{\sqrt{{\cal C}_{a_{x3}^{\rho\phi}}^{B=0,\ {\rm UV}}\ ^2
   {\cal C}_{x3}^{\phi Z}\   ^2 a_{x^3}^{x^3}(x^3)^2+a_\phi^{x^3}(x^3)^2}}={\cal C}_{a_{x3}^Z(x3)}\ ^{B=0,\ {\rm UV}}\  ,\nonumber\\
& & \rho  a_t^\rho\ '(\rho )={\cal C}_{a_{x3}^Z{\rho}}^{B=0,\ {\rm UV}}\ (\rho),\nonumber\\
& & e^{4 Z} a_t^Z(Z)={\cal C}_{a_{x3}^Z(Z)}^{B=0,\ {\rm UV}}\  ,
\end{eqnarray}
which are solved to yield:
\begin{eqnarray}
\label{ax2x3+atrho+atZ}
& & a_{x^3}^{x^3}(x^3)=\frac{{\cal C}_{a_{x3}^Z(x3)}\ ^{B=0,\ {\rm UV}}\  {a_\phi^{x^3}}(x^3)}{\sqrt{a_t^{x^3}(x^3)^2-{\cal C}_{a_{x3}^Z(x3)}^{B=0,\ {\rm UV}}\ ^2 
{\cal C}_{a_{x3}^{\rho\phi}}^{B=0,\ {\rm UV}}\ ^2 {\cal C}_{x3}^{\phi Z}\   ^2}},\nonumber\\
& & a_t^\rho(\rho)=\int _1^{\rho }\frac{{\cal C}_{a_{x3}^Z(\rho)}^{B=0,\ {\rm UV}}\ (\rho )}{\rho }d\rho +c_1,\nonumber\\
& & a_t^Z(Z) = {\cal C}_{a_{x3}^Z(Z)}^{B=0,\ {\rm UV}}\   e^{-4 Z}.
\end{eqnarray}

\noindent {\bf $A_t$ EOM}

The $A_t$ EOM: $\partial_\mu\left(\frac{\delta {\cal L}_{\rm DBI}^{\rm UV,\ B=0}}{\partial_\mu A_t}\right) = \frac{\delta {\cal L}_{\rm DBI}^{\rm UV,\ B=0}}{\delta A_t}$ yields:
$\frac{\delta {\cal L}_{\rm DBI}^{\rm UV,\ B=0}}{\partial_\rho A_t}=$ constant. Now, at ${\cal O}(\beta^0)$,
\begin{eqnarray}
\label{dLoverddrhoAt}
& & \frac{\delta {\cal L}_{\rm DBI}^{\rm UV,\ B=0}}{\partial_\rho A_t} 
% & & = \frac{\beta  M^{\rm UV} \left(\frac{1}{N}\right)^{13/20} \rho  e^{6 Z} \left(-19683 \sqrt{6}    \alpha _{\theta _1}^6-6642 \alpha _{\theta _2}^2 \alpha _{\theta _1}^3+40 \sqrt{6} \alpha_{\theta _2}^4\right) \log ^3({r_h}) \sqrt{\partial_ZA_{x^3}^2+\partial_Z A_\phi    ^2}}{2187 \sqrt{2} 3^{5/6} \pi     \epsilon ^5 {g_s} {N_f^{\rm UV}} \alpha _{\theta _2}^5 \log ^4(N)}\nonumber\\
= -\frac{\sqrt{2} N^{3/5}
   \rho  {r_h}^2 e^{5 Z} \sqrt{\partial_Z A_{x^3}    ^2+\partial_Z A_\phi   ^2}}{\sqrt[3]{3} {g_s} \alpha _{\theta _2}^2}.
\end{eqnarray}
Making an ansatz similar to (\ref{Amu-ansatz}),  one is motivated to assume:
\begin{eqnarray}
\label{aphiZ-i}
& & e^{5 Z} a_\phi^Z\  '(Z)=-5 {\cal C}_{a_t^{\phi}(Z)}^{B=0,\ {\rm UV}}\    ,
\end{eqnarray}
i.e.,
\begin{eqnarray}
\label{aphiZ-ii}
& & a_\phi^Z\  (Z)={\cal C}_{a_t^{\phi}(Z)}^{B=0,\ {\rm UV}}\     e^{-5 Z}+c_1.
\end{eqnarray}
Also, 
\begin{eqnarray}
\label{aphirho}
& & a_\phi^\rho(\rho) = \frac{{\cal C}_{a_t^{\phi}(\rho)}^{B=0,\ {\rm UV}}\   }{\rho }.
\end{eqnarray}
One hence also obtains:
\begin{eqnarray}
\label{aphix3}
& & a_\phi^{x^3}(x^3)=\frac{{\cal C}_{a_t^{x3}(x3)}^{B=0,\ {\rm UV}}\   
   \sqrt{a_t^{x^3}(x^3)^2-{\cal C}_{a_{x3}^Z(x3)}\ ^{B=0,\ {\rm UV}}\  ^2
   {\cal C}_{a_{x3}^{\rho\phi}}^{B=0,\ {\rm UV}}\ ^2 {\cal C}_{a_{x3}^{\phi Z}}^{B=0,\ {\rm UV}}\ ^2}}{a_t^{x^3}(x^3)}.
\end{eqnarray}

\noindent{\bf $A_{\rho}$ EOM}

Now, assuming $ \partial_{x^3}A_{t}^{\beta^0}\   = 0$, one can show that the $A_\rho$-EOM is identically satisfied.

\noindent{\bf $A_\phi$ EOM}

The $A_{\phi}$ EOM: $\partial_\mu\left(\frac{\delta {\cal L}_{\rm DBI}^{\rm UV,\ B=0}}{\partial_\mu A_{\phi}}\right) = \frac{\delta {\cal L}_{\rm DBI}^{\rm UV,\ B=0}}{\delta A_{\phi}}$  yields:
$\frac{\delta {\cal L}_{\rm DBI}^{\rm UV,\ B=0}}{\partial_Z A_\phi}=$ constant. Now, 
\begin{eqnarray}
\label{dLoverddZAphi}
& & \frac{\delta {\cal L}_{\rm DBI}^{\rm UV,\ B=0}}{\partial_Z A_\phi} = \frac{\rho  e^{4 Z} \partial_\rho A_t  
   \partial_ZA_\phi }{\sqrt{\left(\partial_ZA_{x^3}\right)^2+\partial_Z A_\phi    ^2}}.
\end{eqnarray}
Using results from $A_{x^3, t}$ EOMs' solutions, one is then required to impose the following constraint:
\begin{eqnarray}
\label{const-Aphi-EOM}
& & 4 {\cal C}_{a_{x3}^Z(Z)}^{B=0,\ {\rm UV}}\  
   \sqrt{{\cal C}_{a_t^{x3}(x3)}\ ^{B=0,\ {\rm UV}}\ ^2-{\cal C}_{a_{x3}^Z(x3)}\ ^{B=0,\ {\rm UV}}\  ^2
   {\cal C}_{a_{x3}^{\rho\phi}}^{B=0,\ {\rm UV}}\ ^2 {\cal C}_{x3}^{\phi Z}\   ^2}=1.
\end{eqnarray}
The constants ${\cal C}_{x3}^{\phi Z},
{\cal C}_{a_{x3}^{\rho\phi}}^{B=0,\ {\rm UV}},
{\cal C}_{a_{x3}^Z(x3)}\ ^{B=0,\ {\rm UV}},
{\cal C}_{a_{x3}^Z(\rho)}^{B=0,\ {\rm UV}},
{\cal C}_{a_{x3}^Z(Z)}^{B=0,\ {\rm UV}},
{\cal C}_{a_t^{\phi}(Z)}^{B=0,\ {\rm UV}},
{\cal C}_{a_t^{\phi}(\rho)}^{B=0,\ {\rm UV}},
{\cal C}_{a_t^{x3}(x3)}^{B=0,\ {\rm UV}},\\
{\cal C}_{a_{x3}^{\phi Z}}^{B=0,\ {\rm UV}}
$ appearing in (\ref{Ax3EOMansatz2}) - (\ref{const-Aphi-EOM}), are constants of integration appearing in the solutions of the EOMs for the type IIA flavor $D6$-branes in the UV and in the absence of an external magnetic field.

\subsection{Non-Renormalization in the UV of $A_{\mu = t, \rho, \phi, Z, x^3}$ up to ${\cal O}(R^4)$ in the Self-Consistent truncation $A^\beta_{\mu = \rho, \phi, Z, x^3}=0$}
\label{B0-i}

Now, we will show that the $U(1)$ gauge field $A_{\mu}$ in the UV and in the absence of an external magnetic field, is non-renormalized at $\cal O(\beta)$, in the self-consistent truncation: i.e., $A_{\mu = \rho, \phi, Z, x^3}^\beta=0$.

Now,
{\scriptsize
\begin{eqnarray}
\label{SDBI-UV-beta}
& & \hskip -0.8in {\cal L}_{\rm DBI} = -\frac{N^{3/5} \rho  {r_h}^2 e^{4 Z} \sqrt{-2 \partial_{x^3} A_t ^2 \left(\partial_Z A_\rho   ^2+\partial_Z A_\phi    ^2+2 {r_h}^2\right)+4
   \partial_ZA_\rho   \partial_\rho A_t  
   \partial_{x^3} A_t  \partial_Z A_{x^3}    -2
   \partial_\rho A_t  ^2 \left(\partial_Z A_{x^3}    ^2+(\partial_ZA_\phi)^2+2 {r_h}^2\right)}}{\sqrt[3]{3} {g_s} \alpha _{\theta _2}^2}\nonumber\\
& & \hskip -0.8in + \frac{M^{\rm UV} \left(\frac{1}{N}\right)^{13/20} \rho  e^{6 Z} \left(19683 \sqrt{6} \alpha _{\theta _1}^6+6642 \alpha
   _{\theta _2}^2 \alpha _{\theta _1}^3-40 \sqrt{6} \alpha _{\theta _2}^4\right) \log ^3({r_h})
   (\partial_\rho A_t^{\beta^0})^2 \left((\partial_ZA_{x^3}^{\beta^0})^2+(\partial_ZA_\phi^{\beta^0}\ )^2+4 {r_h}^2\right)}{2187 \sqrt{2} 3^{5/6} \pi
    \epsilon ^5 {g_s} \log N ^4 {N_f^{\rm UV}} \alpha _{\theta _2}^5 \sqrt{-(\partial_\rho A_t^{\beta^0}\ )^2 \left(\partial_ZA_{x^3,\ \beta^0}^2+(\partial_ZA_\phi^{\beta^0}\ )^2+2 {r_h}^2\right)}}.    
\end{eqnarray}
}
Assuming $\partial_{x^3}A_t=0, A_{\mu = \rho, \phi, Z, x^3}^\beta=0$,
replacing $\alpha_{\theta_{2}}\rightarrow N^{3/10}\sin\theta_{2}$ and integrating out $\theta_{1,2}, \phi_{2}$, the second term in (\ref{SDBI-UV-beta}) is proportional to the larger of :
\begin{eqnarray}
\label{SDBIbeta0UVB0}
& & \frac{\beta  {\cal C}_{a_t^{\rho}}^{B=0,\ {\rm UV}}\    {\cal C}_{a_t^{x3}(x3)}^{B=0,\ {\rm UV}}\    
    {\cal C}_{a_{x3}^Z(Z)}^{B=0,\ {\rm UV}}\    M^{\rm UV}  {r_h}
   e^{2{Z_{\rm UV}}} ({g_s} N)^{1/4}}{g_s \epsilon ^5 {g_s} N^{19/20}N_f^{\rm UV} \rho \log ^4(N)},
\end{eqnarray}
(${Z_{\rm UV}}$ defined in (\ref{ZUV}) and (\ref{N>})) and
{\footnotesize
\begin{eqnarray}
\label{SDBIbetaUVB0}
& & \frac{i \beta  {\cal C}_{a_t^{\rho}}^{B=0,\ {\rm UV}}\    {\cal C}_{a_t^{x3}(x3)}^{B=0,\ {\rm UV}}\    {\cal C}_{a_t^{\phi}(\rho)}^{B=0,\ {\rm UV}}\   
   {\cal C}_{a_t^{\phi}(Z)}^{B=0,\ {\rm UV}}\     {\cal C}_{a_{x3}^Z(Z)}^{B=0,\ {\rm UV}}\   {\cal C}_{a_t^{x3}(x3)}\ ^{B=0,\ {\rm UV}}\  M^{\rm UV} (\log r_h)^3
   ({g_s} N)^{1/4}}{\pi  \epsilon ^5 {g_s} {N_f^{\rm UV}} \rho ^2N^{27/10} \log ^4(N)},
\end{eqnarray}
}
subject to (\ref{const-Aphi-EOM}). Taking the lare-$N$ limit first and dropping terms of ${\cal O}(\frac{\beta}{N^{1+\alpha_N}}), \alpha_N>0$, and assuming: 
\begin{equation}
\label{MUV+NfUV}
\frac{M^{\rm UV}}{\epsilon^5N_f^{\rm UV}\ }\ll1,
\end{equation}
and $|{\cal C}_{a_t^{x3}(x3)}^{B=0,\ {\rm UV}}\    {\cal C}_{a_t^{\phi}(Z)}^{B=0,\ {\rm UV}}|\ll1$, and that $r_h\sim e^{-0.3 N^{1/3}}$ \cite{Bulk-Viscosity-McGill-IIT-Roorkee} for QCD-inspired values $(g_s, M, N_f) = (0.1, 3, 3)$, one can effect the vanishing of (\ref{SDBIbeta0UVB0}) and (\ref{SDBIbetaUVB0}).

The $A_t^\beta$-EOM will hence be:
\begin{equation}
\label{Atbeta-EOM}
\frac{e^Z\rho\sqrt{\left(2r_h^2 + \left(\partial_ZA_{x^2}\ ^{\beta^0}\right)^2 + \left(\partial_ZA_\phi^{\beta^0}\right)^2\right)}}{g_s}\equiv{\rm Constant}.
\end{equation}
The LHS of (\ref{Atbeta-EOM}) is the larger of $\frac{{\cal C}_{a_t^{x3}(x3)}^{B=0,\ {\rm UV}}\ {\cal C}_{a_t^{\phi}(\rho)}^{B=0,\ {\rm UV}}\   {\cal C}_{a_t^{\phi}(Z)}^{B=0,\ {\rm UV}}\   r_h^2 e^{-Z_{\rm UV}}}{g_s}$ and $\frac{e^{4Z_{\rm UV}}r_h^3\rho}{g_s}$, which for reasons stated in the previous paragraph as well as the exponential-in-$N^{1/3}$-suppression in $r_h$ (\ref{AbsLogCal Rh}), are negligible each and we can hence choose the Constant in the RHS of (\ref{Atbeta-EOM})  to be vanishing. We hence can set $a_t^{Z,\ \beta}=0$ implying $A_{\mu = \rho, \phi, Z, x^3}(\rho, Z, x^3)^\beta=0$ in the UV.

\section{$D6$-Branes' Gauge Fields in the  Presence of Strong Magnetic Fields}
\label{gauge-fields-B}

In this section, we will work out the gauge fields supported on the world volume of the flavor $D6$-branes in the presence of strong magnetic field (in $e=1$-units)  $B\ll T_c^2\sim0.02$ GeV$^2$. First, considering the gauge field in the IR region (later in the UV region) we will obtain their equations of motion via the standard method of variation of action, and then will obtain the respective gauge fluctuations up to $\cal O(\beta^{0})$. Then we will derive the non-renormalization of gauge fields in a self-consistent truncation of gauge fluctuations at $\cal O(\beta)$.

\subsection{In the IR}
\label{IR}
Here we will work out the gauge field up to $\cal O(\beta^{0})$ in the static gauge $A_{Z}=0$ in the presence of a strong magnetic field by varying the DBI action of D$6$-flavor brane, in the IR region. 

\subsubsection{$A_\mu,\ \mu=t, x^{1, 2, 3}, Z$ in the IR up to ${\cal O}(\beta^0)$}

Consider the DBI Lagrangian for flavor $D6$-branes, in the large $B$ limit, assuming $F_{tx^3}={\cal C}_{tx^3}$ and $F_{t\phi}=0$:  
{\footnotesize
\begin{eqnarray}
\label{SDBIuptoObeta}
& & \hskip -0.8in  {\cal L}_{\rm DBI}^{\beta^0 + \beta,\ B\neq0}\nonumber\\
& & \hskip -0.8in = \sqrt{ \left(2 Z^2-2 Z+1\right)
   \left((\partial_Z A_{x^3})^2 \left(\rho ^2 \Xi(Z) (\partial_\rho A_{t})^2-4 B^2 Z (2 Z+1)\right)+2
   B \Xi(Z) (\partial_\rho A_{t})
   (\partial_Z A_t) (\partial_Z A_\phi)+\Xi(Z) 
   (\partial_\rho A_t)^2 (\partial_Z A_\phi)^2+B^2\Xi(Z) (\partial_Z A_t)^2\right)}\nonumber\\
& & \hskip -0.8in \Biggl[\frac{N^{3/5} \rho  {r_h} \left(Z^2+2 Z+2\right) \left(48 {g_s} {N_f}
   \log ({r_h})+{g_s} {N_f} \left(3 Z^2+36 Z+8 \log (4)\right)-64 \pi
   \right) {r_h}}{64 \alpha_{\theta _2}^2\sqrt{2} \sqrt[3]{3} \pi  {g_s}}\nonumber\\
& & \hskip -0.8in -\frac{\beta  
   \log(r_h)^3 M \left(\frac{1}{N}\right)^{13/20} \rho  \left(Z^2-1\right)
   \left(-19683 \sqrt{3} \alpha _{\theta _1}^6-3321 \sqrt{2} \alpha _{\theta _2}^2
   \alpha _{\theta _1}^3+40 \sqrt{3} \alpha _{\theta _2}^4\right) \left(48
   {g_s} {N_f} \log ({r_h})+{g_s} {N_f} \left(3 Z^2+36 Z+8 \log
   (4)\right)-64 \pi \right) }{157464\ 3^{5/6} \pi ^2 \epsilon ^5 {g_s} (\log N)^4
   {N_f}  \alpha _{\theta _2}^5}\Biggr],\nonumber\\
& & 
\end{eqnarray}
}
where $\Xi(Z)\equiv \left(8 Z^2+4 Z+1\right)$. In (\ref{SDBIuptoObeta}), it is understood that $A_\mu = A_\mu^{\beta^0} + \beta A_\mu^\beta$ where $\mu = t, \rho, \phi, Z, x^3$. We will first work with $A_\mu^{\beta^0}$ in ${\cal L}_{\rm DBI}^{\beta^0}$.

In the large-$B$ limit,
{\scriptsize
\begin{eqnarray}
\label{LDBIlargeB}
& & {\cal L}_{\rm DBI}^{\beta^0} = \frac{3^{2/3} N^{3/5} {N_f} \rho  {r_h}^2 \log ({r_h})
   \left(\left(9 Z^2+6 Z+2\right) (\partial_\rho A_t)(\partial_Z A_t) 
   (\partial_Z A_\phi)+B \left(9 Z^2+6 Z+2\right)
   (\partial_Z A_t)^2-8 B Z (Z+1)
   (\partial_Z A_{x^3})^2\right)}{4 \pi  \alpha
   _{\theta _2}^2 \sqrt{\left(4 Z^2+4 Z+2\right)
   (\partial_Z A_t)^2-8 Z
   (\partial_Z A_{x^3})^2}}.\nonumber\\
& & 
\end{eqnarray}
}
\noindent {\bf $A_\phi$ EOM}

Assuming $\phi $ -independence of $A_{\mu = t, \phi, \rho, Z, x^3}$, given that $B = \frac{1}{\rho}\left(A_\phi + \rho\partial_\rho A_\phi\right)$,
\begin{eqnarray}
\label{Aphi-EOM-i}
& & \partial_Z\left(\frac{\delta {\cal L}_{\rm DBI}^{\beta^0}}{\delta \partial_Z A_\phi^{\beta^0}}\right) + \partial_\rho\left(\frac{\delta {\cal L}_{\rm DBI}^{\beta^0}}{\delta \partial_\rho A_\phi^{\beta^0}}\right) = \frac{\delta {\cal L}_{\rm DBI}^{\beta^0}}{\delta A_\phi},
\end{eqnarray}
or, equivalently,
\begin{eqnarray}
\label{Aphi-EOM-ii}
& & \partial_Z\left(\frac{\delta {\cal L}_{\rm DBI}^{\beta^0}}{\delta \partial_Z A_\phi^{\beta^0}}\right) + \partial_\rho\left(\frac{\delta {\cal L}_{\rm DBI}^{\beta^0}}{\delta B}\right) = \frac{1}{\rho}\frac{\delta {\cal L}_{\rm DBI}^{\beta^0}}{\delta B}.
\end{eqnarray}
Substituting (\ref{Amu-ansatz}) into (\ref{Aphi-EOM-ii}), its LHS:
{\scriptsize
\begin{eqnarray}
\label{Aphi-EOM-iii}
& & \frac{3^{2/3} N^{3/5} {N_f} {r_h}^2 \log ({r_h})}{4
   \sqrt{2} \pi  \alpha _{\theta _2}^2 \left(\left(2 Z^2+2 Z+1\right)
   {a_t^\rho}(\rho )^2 {a_t^{x^3}}({x^3})^2 {a_t^Z}'(Z)^2-4
   Z {a_{x^3}^\rho}(\rho )^2 {a_{x^3}\ ^{x^3}}({x^3})^2
   {a_{x^3}^Z}'(Z)^2\right)^{3/2}}\nonumber\\
& & \times  \Biggl\{\rho 
   {a_t^\rho}(\rho ) {a_t^{x^3}}({x^3})^2 {a_t^Z}(Z)
   {a_t^\rho}'(\rho )\Xi_1-4 \rho  Z \left(19 Z^2+12
   Z+4\right) {a_t^\rho}(\rho ) {a_t^{x^3}}({x^3})^2
   {a_{x^3}^\rho}(\rho )^2 {a_{x^3}\ ^{x^3}}({x^3})^2 a_t^\rho\ '(\rho ) {a_t^Z}'(Z)^2 {a_{x^3}^Z}'(Z)^2\nonumber\\
& & +2 \rho  \left(25 Z^2+10
   Z+2\right) {a_t^\rho}(\rho )^3 {a_t^{x^3}}({x^3})^4
   {a_t^\rho}'(\rho ) {a_t^Z}'(Z)^4-4 Z {a_t^\rho}(\rho
   )^2 {a_t^{x^3}}({x^3})^2 {a_{x^3}^\rho}(\rho )
   {a_{x^3}\ ^{x^3}}({x^3})^2 {a_t^Z}'(Z)^2 {a_{x^3}^Z}'(Z)^2\nonumber\\
& & \times    \left(\rho  \left(7 Z^2+6 Z+2\right) {a_{x^3}^\rho}'(\rho
   )+\left(17 Z^2+12 Z+4\right) {a_{x^3}^\rho}(\rho )\right)+\left(25
   Z^2+10 Z+2\right) {a_t^\rho}(\rho )^4 {a_t^{x^3}}({x^3})^4
   {a_t^Z}'(Z)^4+32 Z^2 (Z+1) {a_{x^3}^\rho}(\rho )^3
   {a_{x^3}\ ^{x^3}}({x^3})^4\nonumber\\
& & \times \left(\rho  {a_{x^3}^\rho}'(\rho
   )+{a_{x^3}^\rho}(\rho )\right) {a_{x^3}^Z}'(Z)^4\Biggr\},
\end{eqnarray}
}
where
{\scriptsize
\begin{eqnarray}
\label{Xi1-def}
& & \Xi_1 \equiv  \left(27 Z^2+20 Z+4\right)
   {a_t^\rho}(\rho )^2 {a_t^{x^3}}({x^3})^2 {a_t^Z}'(Z)^3+2
   {a_{x^3}^\rho}(\rho )^2 {a_{x^3}\ ^{x^3}}({x^3})^2 {a_t^Z}'(Z)
   {a_{x^3}^Z}'(Z)\nonumber\\
& &  \left(\left(-27 Z^2-6 Z+2\right) {a_{x^3}^Z}'(Z)+2 Z
   \left(9 Z^2+6 Z+2\right) {a_{x^3}^Z}''(Z)\right)-4 Z \left(9 Z^2+6
   Z+2\right) {a_{x^3}^\rho}(\rho )^2 {a_{x^3}\ ^{x^3}}({x^3})^2
   {a_t^Z}''(Z) {a_{x^3}^Z}'(Z)^2.\nonumber\\
& & 
\end{eqnarray}
}
Similarly, the RHS of (\ref{Aphi-EOM-ii}) yields,
{\scriptsize
\begin{eqnarray}
\label{Aphi-EOM-iv}
& &  \frac{3^{2/3} N^{3/5} {N_f} {r_h}^2 \log ({r_h})
   \left(\left(9 Z^2+6 Z+2\right) {a_t^\rho}(\rho )^2
   {a_t^{x^3}}({x^3})^2 {a_t^Z}'(Z)^2-8 Z (Z+1) a_{x^3}^\rho\ (\rho )^2 {a_{x^3}\ ^{x^3}}({x^3})^2 {a_{x^3}^Z}'(Z)^2\right)}{4 \pi
    \alpha _{\theta _2}^2 \sqrt{\left(4 Z^2+4 Z+2\right) a_t^\rho\ (\rho )^2 {a_t^{x^3}}({x^3})^2 {a_t^Z}'(Z)^2-8 Z
   {a_{x^3}^\rho}(\rho )^2 {a_{x^3}\ ^{x^3}}({x^3})^2
   {a_{x^3}^Z}'(Z)^2}}.
\end{eqnarray}
}
We make the following ansatz:
\begin{eqnarray}
\label{Aphi-EOM-v}
& & {a_{x^3}^\rho\ }(\rho )={\cal C}_{x^3}^{t\rho,\ B}\  {a_t^\rho\ }(\rho )\nonumber\\
& & a_{x^3}\ ^{x^3}({x^3})={{\cal C}_{x^3t}^B} {a_t^{x^3}}({x^3})\nonumber\\
& & {a_{x^3}^Z}\ '(Z)={\cal C}_{x^3}^{tZ,\ B}\  {a_t^Z}'(Z).
\end{eqnarray}
Hence,
{\scriptsize
\begin{eqnarray}
\label{Aphi-EOM-vi}
& & \hskip -0.8in \frac{1}{{a_t^\rho}(\rho )
   {a_t^Z}'(Z)^2 \left(Z \left(2-4 {{\cal C}_{x^3t}^B}^2
   {\cal C}_{x^3}^{t\rho,\ B}\ ^2 {\cal C}_{x^3}^{tZ,\ B}\ ^2\right)+2 Z^2+1\right)
   \left(Z^2 \left(8 {{\cal C}_{x^3t}^B}^2 {\cal C}_{x^3}^{t\rho,\ B}\ ^2
   {\cal C}_{x^3}^{tZ,\ B}\ ^2-9\right)+Z \left(8 {{\cal C}_{x^3t}^B}^2
   {\cal C}_{x^3}^{t\rho,\ B}\ ^2 {\cal C}_{x^3}^{tZ,\ B}\ ^2-6\right)-2\right)}\nonumber\\
& & \hskip -0.8in \times\Biggl\{\rho  \Biggl(2 {a_t^Z}'(Z)^2 \Biggl[{a_t^\rho}'(\rho )
   \left(Z^2 \left(24 {{\cal C}_{x^3t}^B}^2 {\cal C}_{x^3}^{t\rho,\ B}\ ^2
   {\cal C}_{x^3}^{tZ,\ B}\ ^2-25\right)+2 Z \left(4 {{\cal C}_{x^3t}^B}^2
   {\cal C}_{x^3}^{t\rho,\ B}\ ^2 {\cal C}_{x^3}^{tZ,\ B}\ ^2-5\right)-2\right)\nonumber\\
& & \hskip -0.8in+2
   {{\cal C}_{x^3t}^B}^2 {\cal C}_{x^3}^{t\rho,\ B}\  {\cal C}_{x^3}^{tZ,\ B}\ ^2 Z
   {a_{x^3}^\rho}'(\rho ) \left(Z^2 \left(7-8 {{\cal C}_{x^3t}^B}^2
   {\cal C}_{x^3}^{t\rho,\ B}\ ^2 {\cal C}_{x^3}^{tZ,\ B}\ ^2\right)+Z \left(6-8
   {{\cal C}_{x^3t}^B}^2 {\cal C}_{x^3}^{t\rho,\ B}\ ^2
   {\cal C}_{x^3}^{tZ,\ B}\ ^2\right)+2\right)\Biggr]\nonumber\\
& & \hskip -0.8in +{a_t^Z}(Z)
   {a_t^\rho}'(\rho ) \left(8 {{\cal C}_{x^3t}^B}^2
   {\cal C}_{x^3}^{t\rho,\ B}\ ^2 {\cal C}_{x^3}^{tZ,\ B}\  Z (3 Z+1)
   \left({\cal C}_{x^3}^{tZ,\ B}\ 
   {a_t^Z}''(Z)-{a_{x^3}^Z}''(Z)\right)+{a_t^Z}'(Z) \left(2
   {{\cal C}_{x^3t}^B}^2 {\cal C}_{x^3}^{t\rho,\ B}\ ^2 {\cal C}_{x^3}^{tZ,\ B}\ ^2
   \left(27 Z^2+6 Z-2\right)-27 Z^2-20
   Z-4\right)\right)\Biggr)\nonumber\\
& & \hskip -0.8in +{a_t^\rho}(\rho ) {a_t^Z}'(Z)^2
   \left(Z^2 \left(-32 {{\cal C}_{x^3t}^B}^4 {\cal C}_{x^3}^{t\rho,\ B}\ ^4
   {\cal C}_{x^3}^{tZ,\ B}\ ^4+48 {{\cal C}_{x^3t}^B}^2 {\cal C}_{x^3}^{t\rho,\ B}\ ^2
   {\cal C}_{x^3}^{tZ,\ B}\ ^2-25\right)+2 Z \left(8 {{\cal C}_{x^3t}^B}^2
   {\cal C}_{x^3}^{t\rho,\ B}\ ^2
   {\cal C}_{x^3}^{tZ,\ B}\ ^2-5\right)-2\right)\Biggr\}=1,
\end{eqnarray}
}
which can be satisfied by:
\begin{eqnarray}
\label{Aphi-EOM-vii}
& & \frac{\rho  {a_t^\rho}'(\rho )}{{a_t^\rho}(\rho
   )}={\cal C}_B^{t\rho}\ ,\nonumber\\
& & \frac{\rho  {a_{x^3}^\rho}'(\rho )}{{a_t^\rho}(\rho
   )}={\cal C}_B^{x^3\rho}\ .\nonumber\\
\end{eqnarray}
One therefore obtains:
{\footnotesize
\begin{eqnarray}
\label{Aphi-EOM-ix}
& &  {a_t^Z}'(Z) \Biggl[{\cal C}_B^{t\rho}\  \left(Z^2 \left(24
   {{\cal C}_{x^3t}^B}^2 {\cal C}_{x^3}^{t\rho,\ B}\ ^2
   {\cal C}_{x^3}^{tZ,\ B}\ ^2-25\right)+2 Z \left(4 {{\cal C}_{x^3t}^B}^2
   {\cal C}_{x^3}^{t\rho,\ B}\ ^2 {\cal C}_{x^3}^{tZ,\ B}\ ^2-5\right)-2\right)\nonumber\\
& &  +2
   {{\cal C}_{x^3t}^B}^2 {\cal C}_{x^3}^{t\rho,\ B}\  {\cal C}_{x^3}^{tZ,\ B}\ ^2 Z
   \left({\cal C}_B^{x^3\rho}\  (6 Z+2)-8 {\cal C}_B^{x^3\rho}\ 
   {{\cal C}_{x^3t}^B}^2 {\cal C}_{x^3}^{t\rho,\ B}\ ^2 {\cal C}_{x^3}^{tZ,\ B}\ ^2
   Z\right)\Biggr]\nonumber\\
& & +{\cal C}_B^{t\rho}\  {a_t^Z}(Z) \left(2
   {{\cal C}_{x^3t}^B}^2 {\cal C}_{x^3}^{t\rho,\ B}\ ^2 {\cal C}_{x^3}^{tZ,\ B}\ ^2
   \left(27 Z^2+6 Z-2\right)-27 Z^2-20 Z-4\right)=0.
\end{eqnarray}
}
As
{\scriptsize
\begin{eqnarray}
\label{Aphi-EOM-x}
& & -\int dZ\frac{{\cal C}_B^{t\rho}\  \left(2 \mathbb{X} \left(27 Z^2+6
   Z-2\right)-27 Z^2-20 Z-4\right)}{2 \left({\cal C}_B^{t\rho}\ 
   \left(Z^2 \left(24\mathbb{X}-25\right)+2 Z \left(4 \mathbb{X}-5\right)-2\right)+2
   {{\cal C}_{x^3t}^B}^2 {\cal C}_{x^3}^{t\rho,\ B}\  {\cal C}_{x^3}^{tZ,\ B}\ ^2 Z
   \left({\cal C}_B^{x^3\rho}\  (6 Z+2)-8 {\cal C}_B^{x^3\rho}\ 
   {{\cal C}_{x^3t}^B}^2 {\cal C}_{x^3}^{t\rho,\ B}\ ^2 {\cal C}_{x^3}^{tZ,\ B}\ ^2
   Z\right)\right)}\nonumber\\
& & =  \sum_{n=0}^2\kappa_n Z^n + {\cal O}(Z^3),
\end{eqnarray}
}
where $\mathbb{X}\equiv  {{\cal C}_{x^3t}^B}^2 {\cal C}_{x^3}^{t\rho,\ B}\ ^2
   {\cal C}_{x^3}^{tZ,\ B}\ ^2$, therefore, one can show:
\begin{eqnarray}
\label{Aphi-EOM-xi}
& & a_t(Z) = {\cal C}_{a_t^Z}^B\   e^{\kappa_0}+{\cal C}_{a_t^Z}^B\  
   e^{\kappa_0} \kappa_1 Z+\frac{3}{2} {\cal C}_{a_t^Z}^B\  
   e^{\kappa_0} \kappa_1^2 Z^2 + {\cal O}\left(Z^3\right),
\end{eqnarray}
where, e.g.,
{\footnotesize
\begin{eqnarray}
\label{kappa0-def}
& & \hskip -0.8in \kappa_0 \equiv \frac{1}{2 \left(16
   {\cal C}_B^{x^3\rho}\  {{\cal C}_{x^3t}^B}^4 {\cal C}_{x^3}^{t\rho,\ B}\ ^3 {\cal C}_{x^3}^{tZ,\ B}\ ^4-12 {{\cal C}_{x^3t}^B}^2
   {\cal C}_{x^3}^{t\rho,\ B}\  {\cal C}_{x^3}^{tZ,\ B}\ ^2 ({\cal C}_{x^3\rho}^B\ +2 {\cal C}_{x^3}^{t\rho,\ B}\ )+25\right)^2}\nonumber\\
& & \hskip -0.8in\times\Biggl\{\log (2) \Biggl(96 {\cal C}_B^{x^3\rho}\  {{\cal C}_{x^3t}^B}^6
   {\cal C}_{x^3}^{t\rho,\ B}\ ^5 {\cal C}_{x^3}^{tZ,\ B}\ ^6+4
   {{\cal C}_{x^3t}^B}^4 {\cal C}_{x^3}^{t\rho,\ B}\ ^3 {\cal C}_{x^3}^{tZ,\ B}\ ^4
   (18 {\cal C}_{x^3}^{t\rho,\ B}\ -31 {\cal C}_B^{x^3\rho}\ )+6
   {{\cal C}_{x^3t}^B}^2 {\cal C}_{x^3}^{t\rho,\ B}\  {\cal C}_{x^3}^{tZ,\ B}\ ^2
   (11 {\cal C}_B^{x^3\rho}\ +2 {\cal C}_{x^3}^{t\rho,\ B}\ )-115\Biggr)\nonumber\\
& & \hskip -0.8in-\frac{1}{\sqrt{-4 {{\cal C}_{x^3t}^B}^4 {\cal C}_{x^3}^{t\rho,\ B}\ ^2 {\cal C}_{x^3}^{tZ,\ B}\ ^4 ({\cal C}_B^{x^3\rho}\ -2
   {\cal C}_{x^3}^{t\rho,\ B}\ )^2-4 {{\cal C}_{x^3t}^B}^2
   {\cal C}_{x^3}^{t\rho,\ B}\  {\cal C}_{x^3}^{tZ,\ B}\ ^2 ({\cal C}_{x^3\rho}^B\ +2 {\cal C}_{x^3}^{t\rho,\ B}\ )+25}}\nonumber\\
& & \hskip -0.8in\times\Bigl\{8 {{\cal C}_{x^3t}^B}^2 {\cal C}_{x^3}^{t\rho,\ B}\ 
   {\cal C}_{x^3}^{tZ,\ B}\ ^2 \Biggl(-79 {\cal C}_B^{x^3\rho}\ +128
   {\cal C}_B^{x^3\rho}\ ^2 {{\cal C}_{x^3t}^B}^8 {\cal C}_{x^3}^{t\rho,\ B}\ ^7 {\cal C}_{x^3}^{tZ,\ B}\ ^8-8 {{\cal C}_{x^3t}^B}^4
   {\cal C}_{x^3}^{t\rho,\ B}\ ^5 {\cal C}_{x^3}^{tZ,\ B}\ ^4 \left(14
   {\cal C}_B^{x^3\rho}\ ^2 {{\cal C}_{x^3t}^B}^2
   {\cal C}_{x^3}^{tZ,\ B}\ ^2-27\right)\nonumber\\
& & \hskip -0.8in-2 {{\cal C}_{x^3t}^B}^2
   {\cal C}_{x^3}^{t\rho,\ B}\ ^3 {\cal C}_{x^3}^{tZ,\ B}\ ^2 \left(29
   {\cal C}_B^{x^3\rho}\ ^2 {{\cal C}_{x^3t}^B}^2
   {\cal C}_{x^3}^{tZ,\ B}\ ^2+279\right)-480 {\cal C}_B^{x^3\rho}\ 
   {{\cal C}_{x^3t}^B}^6 {\cal C}_{x^3}^{t\rho,\ B}\ ^6
   {\cal C}_{x^3}^{tZ,\ B}\ ^6\nonumber\\
& & \hskip -0.8in+728 {\cal C}_B^{x^3\rho}\  {{\cal C}_{x^3t}^B}^4
   {\cal C}_{x^3}^{t\rho,\ B}\ ^4 {\cal C}_{x^3}^{tZ,\ B}\ ^4-109
   {\cal C}_B^{x^3\rho}\  {{\cal C}_{x^3t}^B}^2 {\cal C}_{x^3}^{t\rho,\ B}\ ^2 {\cal C}_{x^3}^{tZ,\ B}\ ^2+{\cal C}_{x^3}^{t\rho,\ B}\  \left(39
   {\cal C}_B^{x^3\rho}\ ^2 {{\cal C}_{x^3t}^B}^2
   {\cal C}_{x^3}^{tZ,\ B}\ ^2+342\right)\Biggr)\nonumber\\
& & \hskip -0.8in \times \tan ^{-1}\left(\frac{5-2
   {{\cal C}_{x^3t}^B}^2 {\cal C}_{x^3}^{t\rho,\ B}\  {\cal C}_{x^3}^{tZ,\ B}\ ^2
   ({\cal C}_B^{x^3\rho}\ +2 {\cal C}_{x^3}^{t\rho,\ B}\ )}{\sqrt{-4
   {{\cal C}_{x^3t}^B}^4 {\cal C}_{x^3}^{t\rho,\ B}\ ^2 {\cal C}_{x^3}^{tZ,\ B}\ ^4
   ({\cal C}_B^{x^3\rho}\ -2 {\cal C}_{x^3}^{t\rho,\ B}\ )^2-4
   {{\cal C}_{x^3t}^B}^2 {\cal C}_{x^3}^{t\rho,\ B}\  {\cal C}_{x^3}^{tZ,\ B}\ ^2
   ({\cal C}_B^{x^3\rho}\ +2 {\cal C}_{x^3}^{t\rho,\ B}\ )+25}}\right)\Bigr\}\Biggr\}.
\end{eqnarray}
}

\noindent {\bf $A_{x^3}$ EOM}

{\scriptsize
\begin{eqnarray}
\label{Ax3-EOM-i}
& & \hskip -0.8in -\frac{3^{2/3} N^{3/5} {N_f} \rho  {r_h}^2 Z \log ({r_h})
   (\partial_Z A_{x^3}) \left(-\left(9 Z^2+6
   Z+2\right) (\partial_\rho A_{t})
   (\partial_Z A_t) (\partial_Z A_\phi)+B \left(8 Z^3+7 Z^2+6 Z+2\right)
   (\partial_Z A_t)^2-8 B Z (Z+1)
   (\partial_Z A_{x^3})^2\right)}{\sqrt{2} \pi 
   \alpha _{\theta _2}^2 \left(\left(2 Z^2+2 Z+1\right)
   (\partial_Z A_t)^2-4 Z
   (\partial_Z A_{x^3})^2\right)^{3/2}}\nonumber\\
& & \hskip -0.8in = {\cal C}_{x^3}^B(\rho, x^3),
\end{eqnarray}
}
implying
\begin{eqnarray}
\label{Ax3-EOM-ii}
& & {a_\phi^\rho}(\rho )={\cal C}_{\phi\rho}^B\  \rho,\nonumber\\
& & {\cal C}_{x^3}^B(\rho, x^3)=\rho  {\cal C}_{x^3}^B(x^3).
\end{eqnarray}
One hence obtains:
{\scriptsize
\begin{eqnarray}
\label{Ax3-EOM-iii}
& &\frac{3^{2/3} {{\cal C}_{x^3t}^B} {\cal C}_{x^3}^{t\rho,\ B}\  {\cal C}_{x^3}^{tZ,\ B}\ 
   N^{3/5} {N_f} {r_h}^2 Z \log ({r_h})}{4 \sqrt{2} \pi  \kappa_1^2 \alpha _{\theta
   _2}^2 (3 \kappa_1 Z+1)^2 \left(Z \left(2-4 {{\cal C}_{x^3t}^B}^2
   {\cal C}_{x^3}^{t\rho,\ B}\ ^2 {\cal C}_{x^3}^{tZ,\ B}\ ^2\right)+2
   Z^2+1\right)^{3/2}}\nonumber\\
& &\times  \Biggl[2
   {\cal C}_{\phi\rho}^B\  {\cal C}_{\phi x^3}^B\  \kappa_1 \left(9
   Z^2+6 Z+2\right) \left(3 \kappa_1^2 Z^2+2 \kappa_1
   Z+2\right) (3 \kappa_1 Z+1) a_\phi(Z)'(Z)+32 B
   {{\cal C}_{x^3t}^B}^2 {\cal C}_{x^3}^{t\rho,\ B}\ ^2 {\cal C}_{x^3}^{tZ,\ B}\ ^2
   \kappa_1^2 Z (Z+1) (3 \kappa_1 Z+1)^2\nonumber\\
& & -B \left(8 Z^3+7 Z^2+6
   Z+2\right) \left(3 \kappa_1^2 Z^2+2 \kappa_1
   Z+2\right)^2\Biggr] = {\cal C}_{x^3}^B(x^3).
\end{eqnarray}
}
The solution of (\ref{Ax3-EOM-iii}) is:
{\scriptsize
\begin{eqnarray}
\label{Ax3-EOM-iv}
& & \hskip -0.8in a_\phi(Z) = \frac{1}{4860
   \sqrt{2} 3^{2/3} {{\cal C}_{x^3t}^B} {\cal C}_{x^3}^{t\rho,\ B}\ 
   {\cal C}_{x^3}^{tZ,\ B}\  {\cal C}_{\phi\rho}^B\  {\cal C}_{\phi x^3}^B\ 
   \kappa_1 {N_f} {r_h}^2 \log ({r_h})}\nonumber\\
& & \hskip -0.8in\Biggl\{\frac{2160 \sqrt{2} 3^{2/3} B {{\cal C}_{x^3t}^B}^3
   {\cal C}_{x^3}^{t\rho,\ B}\ ^3 {\cal C}_{x^3}^{tZ,\ B}\ ^3 \kappa_1 \left(3
   \kappa_1^3-14 \kappa_1^2+15 \kappa_1-18\right)
   {N_f} {r_h}^2 \log ({r_h}) \log \left(3 \kappa_1^2
   Z^2+2 \kappa_1 Z+2\right)}{\kappa_1^4-2 \kappa_1^3+2
   \kappa_1^2-6 \kappa_1+9} -\frac{1}{2
   \kappa_1^6-6 \kappa_1^5+9 \kappa_1^4-18
   \kappa_1^3+32 \kappa_1^2-24 \kappa_1+9}\nonumber\\
& & \hskip -0.8in \times\Biggl\{40 \sqrt{2} 3^{2/3} B
   {{\cal C}_{x^3t}^B} {\cal C}_{x^3}^{t\rho,\ B}\  {\cal C}_{x^3}^{tZ,\ B}\  {N_f}
   {r_h}^2 \log ({r_h}) \log \left(9 Z^2+6 Z+2\right) \Biggl(3
   \kappa_1^7 \left(72 {{\cal C}_{x^3t}^B}^2 {\cal C}_{x^3}^{t\rho,\ B}\ ^2
   {\cal C}_{x^3}^{tZ,\ B}\ ^2+1\right)\nonumber\\
& & \hskip -0.8in-4 \kappa_1^6 \left(81
   {{\cal C}_{x^3t}^B}^2 {\cal C}_{x^3}^{t\rho,\ B}\ ^2
   {\cal C}_{x^3}^{tZ,\ B}\ ^2+4\right) +\kappa_1^5 \left(62-864
   {{\cal C}_{x^3t}^B}^2 {\cal C}_{x^3}^{t\rho,\ B}\ ^2
   {\cal C}_{x^3}^{tZ,\ B}\ ^2\right)\nonumber\\
& & \hskip -0.8in +\kappa_1^4 \left(1350
   {{\cal C}_{x^3t}^B}^2 {\cal C}_{x^3}^{t\rho,\ B}\ ^2
   {\cal C}_{x^3}^{tZ,\ B}\ ^2-131\right)+\kappa_1^3 \left(201-864
   {{\cal C}_{x^3t}^B}^2 {\cal C}_{x^3}^{t\rho,\ B}\ ^2
   {\cal C}_{x^3}^{tZ,\ B}\ ^2\right)+6 \kappa_1^2 \left(27
   {{\cal C}_{x^3t}^B}^2 {\cal C}_{x^3}^{t\rho,\ B}\ ^2
   {\cal C}_{x^3}^{tZ,\ B}\ ^2-58\right)+450 \kappa_1-189\Biggr)\Biggr\}\nonumber\\
& &\hskip -0.8in +\frac{4320\
   3^{2/3} \sqrt{10} B {{\cal C}_{x^3t}^B}^3 {\cal C}_{x^3}^{t\rho,\ B}\ ^3
   {\cal C}_{x^3}^{tZ,\ B}\ ^3 \kappa_1^2 \left(3 \kappa_1^2-2
   \kappa_1-3\right) {N_f} {r_h}^2 \log ({r_h}) \tan
   ^{-1}\left(\frac{3 \kappa_1
   Z+1}{\sqrt{5}}\right)}{\kappa_1^4-2 \kappa_1^3+2
   \kappa_1^2-6 \kappa_1+9}\nonumber\\
& & \hskip -0.8in-\frac{80 \sqrt{2} 3^{2/3} B
   {{\cal C}_{x^3t}^B} {\cal C}_{x^3}^{t\rho,\ B}\  {\cal C}_{x^3}^{tZ,\ B}\  {N_f}
   {r_h}^2 \log ({r_h}) \tan ^{-1}(3 Z+1) }{2
   \kappa_1^6-6 \kappa_1^5+9 \kappa_1^4-18
   \kappa_1^3+32 \kappa_1^2-24 \kappa_1+9}\nonumber\\
& & \hskip -0.8in \times \Biggl[\kappa_1^7
   \left(432 {{\cal C}_{x^3t}^B}^2 {\cal C}_{x^3}^{t\rho,\ B}\ ^2
   {\cal C}_{x^3}^{tZ,\ B}\ ^2+11\right)-\kappa_1^6 \left(1188
   {{\cal C}_{x^3t}^B}^2 {\cal C}_{x^3}^{t\rho,\ B}\ ^2
   {\cal C}_{x^3}^{tZ,\ B}\ ^2+37\right)+54 \kappa_1^5 \left(38
   {{\cal C}_{x^3t}^B}^2 {\cal C}_{x^3}^{t\rho,\ B}\ ^2
   {\cal C}_{x^3}^{tZ,\ B}\ ^2+1\right)\nonumber\\
& & \hskip -0.8in -2 \kappa_1^4 \left(1215
   {{\cal C}_{x^3t}^B}^2 {\cal C}_{x^3}^{t\rho,\ B}\ ^2
   {\cal C}_{x^3}^{tZ,\ B}\ ^2+56\right) + 7 \kappa_1^3 \left(216
   {{\cal C}_{x^3t}^B}^2 {\cal C}_{x^3}^{t\rho,\ B}\ ^2
   {\cal C}_{x^3}^{tZ,\ B}\ ^2+31\right)-9 \kappa_1^2 \left(54
   {{\cal C}_{x^3t}^B}^2 {\cal C}_{x^3}^{t\rho,\ B}\ ^2
   {\cal C}_{x^3}^{tZ,\ B}\ ^2+19\right)+90 \kappa_1-108\Biggr]\nonumber\\
& &\hskip -0.8in +\frac{30
   \sqrt{2} 3^{2/3} B {{\cal C}_{x^3t}^B} {\cal C}_{x^3}^{t\rho,\ B}\ 
   {\cal C}_{x^3}^{tZ,\ B}\  \left(28 \kappa_1^2+5 \kappa_1+40\right)
   {N_f} {r_h}^2 Z \log ({r_h})}{\kappa_1}\nonumber\\
& & \hskip -0.8in +\frac{50
   \sqrt{2} 3^{2/3} B {{\cal C}_{x^3t}^B} {\cal C}_{x^3}^{t\rho,\ B}\ 
   {\cal C}_{x^3}^{tZ,\ B}\  \left(54 \kappa_1^3-54 \kappa_1^2+21
   \kappa_1-8\right) {N_f} {r_h}^2 \log ({r_h}) \log (3
   \kappa_1 Z+1)}{\kappa_1^2 \left(2 \kappa_1^2-2
   \kappa_1+1\right)}\nonumber\\
& & \hskip -0.8in+720 \sqrt{2} 3^{2/3} B {{\cal C}_{x^3t}^B}
   {\cal C}_{x^3}^{t\rho,\ B}\  {\cal C}_{x^3}^{tZ,\ B}\  \kappa_1 {N_f}
   {r_h}^2 Z^3 \log ({r_h})+45 \sqrt{2} 3^{2/3} B
   {{\cal C}_{x^3t}^B} {\cal C}_{x^3}^{t\rho,\ B}\  {\cal C}_{x^3}^{tZ,\ B}\  (5
   \kappa_1+8) {N_f} {r_h}^2 Z^2 \log ({r_h})\Biggr\} + {\cal O}\left(1/N^{3/5}\right)\nonumber\\
& & \hskip -0.8in = \frac{1}{243 {\cal C}_{\phi\rho}^B\  {\cal C}_{\phi x^3}^B\  \kappa_1 \left(2 \kappa_1^2-2 \kappa_1+1\right)
   \left(\kappa_1^4-2 \kappa_1^3+2 \kappa_1^2-6
   \kappa_1+9\right)}\nonumber\\
& & \hskip -0.8in\Biggl\{B \Biggl[\pi  \Biggl(\kappa_1^7 \left(-\left(432
   {{\cal C}_{x^3t}^B}^2 {\cal C}_{x^3}^{t\rho,\ B}\ ^2
   {\cal C}_{x^3}^{tZ,\ B}\ ^2+11\right)\right)+\kappa_1^6 \left(1188
   {{\cal C}_{x^3t}^B}^2 {\cal C}_{x^3}^{t\rho,\ B}\ ^2
   {\cal C}_{x^3}^{tZ,\ B}\ ^2+37\right)-54 \kappa_1^5 \left(38
   {{\cal C}_{x^3t}^B}^2 {\cal C}_{x^3}^{t\rho,\ B}\ ^2
   {\cal C}_{x^3}^{tZ,\ B}\ ^2+1\right)\nonumber\\
& & \hskip -0.8in+2 \kappa_1^4 \left(1215
   {{\cal C}_{x^3t}^B}^2 {\cal C}_{x^3}^{t\rho,\ B}\ ^2
   {\cal C}_{x^3}^{tZ,\ B}\ ^2+56\right)-7 \kappa_1^3 \left(216
   {{\cal C}_{x^3t}^B}^2 {\cal C}_{x^3}^{t\rho,\ B}\ ^2
   {\cal C}_{x^3}^{tZ,\ B}\ ^2+31\right)+9 \kappa_1^2 \left(54
   {{\cal C}_{x^3t}^B}^2 {\cal C}_{x^3}^{t\rho,\ B}\ ^2
   {\cal C}_{x^3}^{tZ,\ B}\ ^2+19\right)-90 \kappa_1+108\Biggr)\nonumber\\
& & \hskip -0.8in -216
   {{\cal C}_{x^3t}^B}^2 {\cal C}_{x^3}^{t\rho,\ B}\ ^2 {\cal C}_{x^3}^{tZ,\ B}\ ^2
   \left(2 \kappa_1^2-2 \kappa_1+1\right) \kappa_1
   \Biggl[\kappa_1^4 \log (2)-\kappa_1^3 \left(\log (4)+3
   \sqrt{5} \tan
   ^{-1}\left(\frac{1}{\sqrt{5}}\right)\right)+\kappa_1^2
   \left(\log (4)+2 \sqrt{5} \tan
   ^{-1}\left(\frac{1}{\sqrt{5}}\right)\right)\nonumber\\
& & \hskip -0.8in+3 \kappa_1
   \left(\sqrt{5} \tan ^{-1}\left(\frac{1}{\sqrt{5}}\right)-\log
   (4)\right)+\log (512)\Biggr]-2 \kappa_1^7 \log (8)+32
   \kappa_1^6 \log (2)-124 \kappa_1^5 \log (2)+262
   \kappa_1^4 \log (2)-402 \kappa_1^3 \log (2)+696
   \kappa_1^2 \log (2)\nonumber\\
& & \hskip -0.8in -900 \kappa_1 \log (2)+378 \log
   (2)\Biggr]
+243 {\cal C}_{\phi\rho}^B\  {\cal C}_{\phi x^3}^B\  {\cal C}_{\phi Z}^B\ 
   \kappa_1 \left(2 \kappa_1^6-6 \kappa_1^5+9
   \kappa_1^4-18 \kappa_1^3+32 \kappa_1^2-24
   \kappa_1+9\right)\Biggr\}+\frac{B Z}{{\cal C}_{\phi\rho}^B\ 
   {\cal C}_{\phi x^3}^B\  \kappa_1}\nonumber\\ 
& & \hskip -0.8in-\frac{Z^2 \left(2 B
   {{\cal C}_{x^3t}^B}^2 {\cal C}_{x^3}^{t\rho,\ B}\ ^2 {\cal C}_{x^3}^{tZ,\ B}\ ^2
   \kappa_1+B\right)}{{\cal C}_{\phi\rho}^B\  {\cal C}_{\phi x^3}^B\ } + {\cal O}\left(Z^3\right) \equiv \alpha_1
 {\cal C}_{\phi Z}^B\  + B ( \alpha_2  + \alpha_3 Z + \alpha_4 Z^2) + \frac{\pi  \kappa_1
   \left(\frac{1}{N}\right)^{3/5} \alpha _{\theta _2}^2
   {\cal C}_{x^3}^B(x^3) \log (Z)}{\sqrt{2} 3^{2/3} {{\cal C}_{x^3t}^B}
   {\cal C}_{x^3}^{t\rho,\ B}\  {\cal C}_{x^3}^{tZ,\ B}\  {\cal C}_{\phi\rho}^B\ 
   {\cal C}_{\phi x^3}^B\  {N_f} {r_h}^2 \log ({r_h})}.
\end{eqnarray}
}
One also obtains:
\begin{equation}
\label{atrho}
a_t^\rho(\rho) = {\cal C}_t^{\rho,\ B}\  \rho.
\end{equation}

\noindent {\bf $A_t$ EOM}

\begin{eqnarray}
\label{At-EOM-i}
& & \partial_Z\left(\frac{\delta {\cal L}_{\rm DBI}^{\beta^0}}{\delta \partial_Z A_t^{\beta^0}}\right) + \partial_\rho\left(\frac{\delta {\cal L}_{\rm DBI}^{\beta^0}}{\delta \partial_\rho A_t^{\beta^0}}\right) = 0.
\end{eqnarray}
The numerator of (\ref{At-EOM-i}) in the IR and for large $B$ is proportional to:
{\scriptsize
\begin{eqnarray}
\label{At-EOM-ii}
& & \hskip -0.8in 2 (\partial_Z A_t)^3
   (\partial_Z A_{x^3}) \left(\left(-81 Z^2-10
   Z+2\right) (\partial_Z A_{x^3})+4 Z (5 Z+1)
   (\partial_Z^2 A_{x^3})\right)+16 Z
   (\partial_Z A_t)
   (\partial_Z A_{x^3})^3 \left(\left(15 Z^2+3
   Z-1\right) (\partial_Z A_{x^3})-2 Z (3 Z+1)
   (\partial_Z^2 A_{x^3})\right)\nonumber\\
& & \hskip -0.8in+32 Z^2
   (\partial_Z^2 A_{t})
   (\partial_Z A_{x^3})^4-8 Z (5 Z+1)
   (\partial_Z^2 A_{t})
   (\partial_Z A_t)^2
   (\partial_Z A_{x^3})^2+\left(75 Z^2+28
   Z+4\right) (\partial_Z A_t)^5\nonumber\\
& & \hskip -0.8in = {a_t^\rho}(\rho ) {a_t^{x^3}}({x^3}) \Biggl(-8 Z (5 Z+1)
   {a_t^\rho}(\rho )^2 {a_t^{x^3}}({x^3})^2 a_{x^3}^\rho\ (\rho )^2 {a_{x^3}^Z}(Z)^2 {a_t^Z}'(Z)^2 {a_t^Z}''(Z)
   {a_{x^3}\ ^{x^3}}'({x^3})^2+2 {a_t^\rho}(\rho )^2
   {a_t^{x^3}}({x^3})^2 {a_{x^3}^\rho}(\rho )^2
   {a_{x^3}\ ^{x^3}}({x^3})^2 {a_t^Z}'(Z)^3 {a_{x^3}^Z}'(Z)\nonumber\\
& & \hskip -0.8in \times
   \left(\left(-81 Z^2-10 Z+2\right) {a_{x^3}^Z}'(Z)+4 Z (5 Z+1)
   {a_{x^3}^Z}''(Z)\right)+\left(75 Z^2+28 Z+4\right) a_t^\rho\ (\rho )^4 {a_t^{x^3}}({x^3})^4 {a_t^Z}'(Z)^5+16 Z
   {a_{x^3}^\rho}(\rho )^4 {a_{x^3}\ ^{x^3}}({x^3})^4
   {a_t^Z}'(Z) {a_{x^3}^Z}'(Z)^3\nonumber\\
& & \hskip -0.8in \times \left(\left(15 Z^2+3 Z-1\right)
   {a_{x^3}^Z}'(Z)-2 Z (3 Z+1) {a_{x^3}^Z}''(Z)\right)+32 Z^2
   {a_{x^3}^\rho}(\rho )^4 {a_{x^3}\ ^{x^3}}({x^3})^4
   {a_t^Z}''(Z) {a_{x^3}^Z}'(Z)^4\Biggr).
\end{eqnarray}
}
Assuming,
\begin{equation}
\label{At-EOM-iii}
{a_t^{x^3}}({x^3})={\cal C}_{tx^3}^B\ ,
\end{equation}
and using (\ref{Aphi-EOM-v}), (\ref{At-EOM-ii}) yields:
{\scriptsize
\begin{eqnarray}
\label{At-EOM-iv}
& & \hskip -0.8in {\cal C}_{t\rho}^B\ ^4 \rho ^4 {a_t^{x^3}}({x^3})^4
   {a_t^Z}'(Z)^4 \Biggl({a_t^Z}'(Z) \left(16 {{\cal C}_{x^3t}^B}^4
   {\cal C}_{x^3}^{t\rho,\ B}\ ^4 {\cal C}_{x^3}^{tZ,\ B}\ ^4 Z \left(15 Z^2+3
   Z-1\right)-2 {{\cal C}_{x^3t}^B}^2 {\cal C}_{x^3}^{t\rho,\ B}\ ^2
   {\cal C}_{x^3}^{tZ,\ B}\ ^2 \left(81 Z^2+10 Z-2\right)+75 Z^2+28 Z+4\right)\nonumber\\
& & \hskip -0.8in+8
   {{\cal C}_{x^3t}^B}^2 {\cal C}_{x^3}^{t\rho,\ B}\ ^2 {\cal C}_{x^3}^{tZ,\ B}\ ^2 Z
   {a_t^Z}''(Z) \left(-12 {{\cal C}_{x^3t}^B}^2 {\cal C}_{x^3}^{t\rho, B}\ ^2 {\cal C}_{x^3}^{tZ,\ B}\ ^2 Z^2+5 Z+1\right)\Biggr)\nonumber\\
& & \hskip -0.8in = 4 {\cal C}_{a_t^Z}^B\   e^{\kappa_0} \kappa_1
   \left({{\cal C}_{x^3t}^B}^2 {\cal C}_{x^3}^{t\rho,\ B}\ ^2
   {\cal C}_{x^3}^{tZ,\ B}\ ^2+1\right)+4 {\cal C}_{a_t^Z}^B\   e^{\kappa_0}
   \kappa_1 Z \left({{\cal C}_{x^3t}^B}^2 {\cal C}_{x^3}^{t\rho,\ B}\ ^2
   {\cal C}_{x^3}^{tZ,\ B}\ ^2 (9 \kappa_1-5)-4 {{\cal C}_{x^3t}^B}^4
   {\cal C}_{x^3}^{t\rho,\ B}\ ^4 {\cal C}_{x^3}^{tZ,\ B}\ ^4+3
   \kappa_1+7\right)\nonumber\\
& &\hskip -0.8in -3 Z^2 \left({\cal C}_{a_t^Z}^B\  
   e^{\kappa_0} \kappa_1 \left(16 {{\cal C}_{x^3t}^B}^4
   {\cal C}_{x^3}^{t\rho,\ B}\ ^4 {\cal C}_{x^3}^{tZ,\ B}\ ^4 (\kappa_1-1)+2
   {{\cal C}_{x^3t}^B}^2 {\cal C}_{x^3}^{t\rho,\ B}\ ^2 {\cal C}_{x^3}^{tZ,\ B}\ ^2
   (27-10 \kappa_1)-28
   \kappa_1-25\right)\right) + {\cal O}\left(Z^3\right).
\end{eqnarray}
}
One hence sees that (\ref{At-EOM-i}) is identically satisfied in the IR and in the large-$B$ limit provided 
\begin{equation}
\label{At-EOM-v}
\kappa_0<0,\ |\kappa_0|\gg1.
\end{equation}

The constants $ {\cal C}_{x^3}^{t\rho,\ B},
{\cal C}_{x^3t}^B,
{\cal C}_{x^3}^{tZ,\ B},
{\cal C}_B^{t\rho},
{\cal C}_B^{x^3\rho},
{\cal C}_{a_t^Z}^B,
{\cal C}_{\phi\rho}^B,
{\cal C}_{x^3}^B(\rho, x^3),
{\cal C}_{x^3}^B(x^3),
{\cal C}_t^{\rho,\ B},
{\cal C}_{tx^3}^B,
{\cal C}_{\phi Z}^B,
{\cal C}_{x^3}^B(x^3),\\
{\cal C}_{\phi x^3}^B$ appearing in  (\ref{Aphi-EOM-v}) - (\ref{At-EOM-iv}), are constants of integration appearing in the solutions to the EOMs for the type IIA flavor $D6$-branes in the UV and in the presence of a strong magnetic field ($>T_c^2$ in $e=1$ units).

\subsubsection{Non-Renormalization in the IR of $A_{\mu = t, \rho, \phi, Z, x^3}$ in the Self-Consistent truncation $A_{\mu = \rho, \phi, Z, x^3}^\beta=0$}
\label{nonrenormalization-i}

Here we show the non-renormalization of the background gauge fluctuations $A_{\mu}$ in the self-consistent truncation $A_{\mu = \rho, \phi, Z, x^3}^\beta=0$ in the IR region. One can show:
\begin{eqnarray}
\label{Amu-beta-i}
& & \hskip -0.8in {\cal L}_{\rm DBI}^\beta \in \frac{N^{3/5} \rho  {r_h} \left(Z^2+2 Z+2\right) \left(48 {g_s}
   {N_f} \log ({r_h})+{g_s} {N_f} \left(3 Z^2+36 Z+8 \log
   (4)\right)-64 \pi \right)}{64 \sqrt{2} \sqrt[3]{3} \pi  {g_s}}.\nonumber\\
& & \hskip -0.8in \times  \sqrt{\Lambda(Z)}, 
\end{eqnarray}
where
{\scriptsize
\begin{eqnarray}
\label{Lambda-def}
& & \hskip -0.8in \Lambda(Z) \equiv \frac{1}{\alpha _{\theta
   _2}^4}\Biggl\{{r_h}^2 \left(2 Z^2-2 Z+1\right)  \Biggl[(\partial_Z A_{x^3})^2 \left(\rho ^2 \left(8
   Z^2+4 Z+1\right) (\partial_\rho A_{t})^2-4 B^2 Z (2
   Z+1)\right)+2 B \Xi(Z) (\partial_\rho A_t) (\partial_Z A_t) (\partial_Z A_\phi)\nonumber\\
& & \hskip -0.8in   +\Xi(Z)
   (\partial_\rho A_{t})^2 (\partial_Z A_\phi)^2+B^2 \Xi(Z)
   (\partial_Z A_t)^2\Biggr]\Biggr\}.
\end{eqnarray}
}
Now, once again, substituting $A_\mu = A_\mu^{\beta^0} + \beta A_\mu^\beta$ into (\ref{Amu-beta-i}), under the truncation $A_{\mu = \rho, \phi, Z, x^3}^\beta=0$, yields:
{\scriptsize
\begin{eqnarray}
\label{Amu-beta-ii}
& & \hskip -0.8in {\cal L}_{\rm DBI}^\beta = \beta\frac{1}{26244\ 3^{5/6} \pi ^2 \alpha _{\theta _2}^5
   \sqrt{\left(4 Z^2+4 Z+2\right) (\partial_Z A_t)^2-8
   Z (\partial_Z A_{x^3}\ ^{\beta^0})^2}}\Biggl\{\rho  \log ({r_h}) \Biggl(19683 \sqrt{3} \pi  N^{3/5} {N_f}
   {r_h}^2 \left(9 Z^2+6 Z+2\right) \alpha _{\theta _2}^3
   (\partial_Z A_t)(\partial_\rho A_t^\beta) (\partial_Z A_\phi\ ^{\beta^0})\nonumber\\
  &  & \hskip -0.8in +\frac{2
   \sqrt{2} \log(r_h)^3 M \left(\frac{1}{N}\right)^{13/20} \left(Z^2+2
   Z-2\right) \Sigma_1
   \left(\left(9 Z^2+6 Z+2\right) (\partial_Z A_t)
  (\partial_\rho A_t\ ^{\beta^0}) (\partial_Z A_\phi\ ^{\beta^0})+B \left(9 Z^2+6 Z+2\right)
   (\partial_Z A_t)^2-8 B Z (Z+1)
   (\partial_Z A_{x^3}\ ^{\beta^0})^2\right)}{\epsilon ^5
   (\log N)^4}\Biggr)\Biggr\}\nonumber\\
& & \hskip -0.8in = \frac{1}{26244\ 3^{5/6} \pi ^2 \alpha _{\theta
   _2}^5 \sqrt{\left(4 Z^2+4 Z+2\right) {a_t^\rho}(\rho )^2
   {a_t^{x^3}}({x^3})^2 {a_t^Z}'(Z)^2-8 Z a_{x^3}^\rho\ (\rho )^2 {a_{x^3}\ ^{x^3}}({x^3})^2 {a_{x^3}^Z}'(Z)^2}}\nonumber\\
   & & \hskip -0.8in \Biggl\{\rho  \log ({r_h}) \Biggl(19683 \sqrt{3} \pi  N^{3/5}
   {N_f} {r_h}^2 \left(9 Z^2+6 Z+2\right) \alpha _{\theta
   _2}^3 {a_t^\rho}(\rho ) {a_t^{x^3}}({x^3}) {a_{\phi}^\rho}(\rho ) {a_\phi^{\ x3}}({x^3}) {a_t^Z}'(Z)
   {a_\phi^{\ Z}}'(Z) (\partial_\rho A_t^\beta)-\frac{8 \sqrt{2} \log(r_h)^3 M
   \left(\frac{1}{N}\right)^{13/20} \Sigma_1}{\epsilon ^5
   (\log N)^4}\Biggr)\Biggr\}\nonumber\\
& & \hskip -0.8in \times \left((Z+1)^2 {a_t^{x^3}}({x^3}) \left({a_t^Z}(Z)
   {a_\phi^\rho }(\rho ) {a_\phi^{\ x3}}({x^3})
   {a_t^\rho}'(\rho ) a_\phi(Z)'(Z)
   (\partial_Z A_t)+B a_t^\rho\ (\rho )^2 {a_t^{x^3}}({x^3}) {a_t^Z}'(Z)^2\right)+4 B Z
   {a_{x^3}^\rho}(\rho )^2 {a_{x^3}\ ^{x^3}}({x^3})^2
   {a_{x^3}^Z}(Z)^2\right)   
\end{eqnarray}
}
which for $e^{-\kappa_0}\rightarrow0$ implies:
{\scriptsize
\begin{eqnarray}
\label{Amu-beta-iii}
& & \hskip -0.8in {\cal L}_{\rm DBI}^\beta =  \beta \Biggl[(\partial_\rho A_t\ ^{\beta})\frac{N^{3/5} \rho ^2 \left(9 Z^2+6 Z+2\right) \left(6 B
   {{\cal C}_{x^3t}^B} {\cal C}_{x^3}^{t\rho,\ B}\  {\cal C}_{x^3}^{tZ,\ B}\ 
   {\cal C}_{\phi\rho}^B\  {\cal C}_{\phi x^3}^B\  {N_f}
   {r_h}^2 Z \log ({r_h}) (\alpha_3+2 \alpha_4
   Z)+\sqrt{2} \sqrt[3]{3} \pi  \kappa_1
   \left(\frac{1}{N}\right)^{3/5} \alpha _{\theta _2}^2
   {\cal C}_{x^3}^B(x^3)\right)}{8 \sqrt[3]{3} \pi 
   {{\cal C}_{x^3t}^B} {\cal C}_{x^3}^{t\rho,\ B}\  {\cal C}_{x^3}^{tZ,\ B}\  Z
   \alpha _{\theta _2}^2 \sqrt{Z \left(4-8 {{\cal C}_{x^3t}^B}^2
   {\cal C}_{x^3}^{t\rho,\ B}\ ^2 {\cal C}_{x^3}^{tZ,\ B}\ ^2\right)+4 Z^2+2}}\nonumber\\
& & \hskip -0.8in + \frac{8 B {\cal C}_{t\rho}^B\  {\cal C}_{tx^3}^B\ 
   {{\cal C}_{x^3t}^B}^2 {\cal C}_{x^3}^{t\rho,\ B}\ ^2 {\cal C}_{x^3,\ 2}^{tZ,\ B}\ ^2
   e^{-\kappa_0} \log(r_h)^3 M
   \left(\frac{1}{N}\right)^{13/20} \rho ^2 Z \left(-19683 \sqrt{3}
   \alpha _{\theta _1}^6-3321 \sqrt{2} \alpha _{\theta _2}^2 \alpha
   _{\theta _1}^3+40 \sqrt{3} \alpha _{\theta _2}^4\right) \log
   ({r_h})}{6561\ 3^{5/6} \pi ^2 {\cal C}_{a_t^Z}^B\   \epsilon ^5
   \kappa_1 (\log N)^4 \alpha _{\theta _2}^5 (3
   \kappa_1 Z+1) \sqrt{Z \left(2-4 {{\cal C}_{x^3t}^B}^2
   {\cal C}_{x^3}^{t\rho,\ B}\ ^2 {\cal C}_{x^3}^{tZ,\ B}\ ^2\right)+2 Z^2+1}} \Biggr].
\end{eqnarray}
}
At ${\cal O}(\beta)$, the $A_t^\beta$ EOM: $\partial_\rho\left(\frac{\delta {\cal L}_{\rm DBI}^\beta}{\delta \partial_\rho A_t^\beta}\right)=0$ implies:
{\scriptsize
\begin{eqnarray}
\label{Amu-beta-iv}
& & \hskip -0.8in \frac{N^{3/5} \rho ^2 \left(9 Z^2+6 Z+2\right) \left(6 B {{\cal C}_{x^3t}^B}
   {\cal C}_{x^3}^{t\rho,\ B}\  {\cal C}_{x^3}^{tZ,\ B}\  {\cal C}_{\phi\rho}^B\  {\cal C}_{\phi x^3}^B\  {N_f} {r_h}^2 Z \log ({r_h}) (\alpha_3+2 \alpha_4
   Z)+\sqrt{2} \sqrt[3]{3} \pi  \kappa_1 \left(\frac{1}{N}\right)^{3/5} \alpha
   _{\theta _2}^2 {\cal C}_{x^3}^B(x^3)\right)}{8 \sqrt[3]{3} \pi  {{\cal C}_{x^3t}^B}
   {\cal C}_{x^3}^{t\rho,\ B}\  {\cal C}_{x^3}^{tZ,\ B}\  Z \alpha _{\theta _2}^2 \sqrt{Z \left(4-8
   {{\cal C}_{x^3t}^B}^2 {\cal C}_{x^3}^{t\rho,\ B}\ ^2 {\cal C}_{x^3}^{tZ,\ B}\ ^2\right)+4 Z^2+2}}
\nonumber\\
& & \hskip -0.8in = \frac{\kappa_1 \rho ^2 {\cal C}_{x^3}^B(x^3)}{4 {{\cal C}_{x^3t}^B}
   {\cal C}_{x^3}^{t\rho,\ B}\  {\cal C}_{x^3}^{tZ,\ B}\  Z} + {\cal O}(Z^0,r_h^2)\nonumber\\
& & = {\cal C}_t^\beta(Z,x^3).
\end{eqnarray}
}
Assuming $|\kappa_1 {\cal C}_{x^3}^B(x^3)|\ll1, |{\cal C}_t^\beta(Z,x^3)|\ll1$, the $A_t^\beta$-EOM is identically satisfied. One can hence set $A_\mu^\beta=0$.

\subsubsection{Log-Gravitational-DBI IR Renormalization}
Here we will derive the renormalization of DBI action by working out the DBI action at the boundary, in the IR region.

Given that:
{\scriptsize
\begin{eqnarray}
\label{detRicci6}
& & \left.{\rm det}\left({\rm Ricci}_{\Sigma^{D6}}\right)\right|_{{\rm fixed}\ Z\in {\rm IR}}
\sim \frac{10^{-15}{g_s}   ^{7/2} M^8 {N_f} ^4 {r_h} ^8 e^{4 Z} \left(e^{4Z}-1\right) (5.8 \beta {\cal C}_{zz}-4.1) (\cos (2 \theta_2)-3)^6
   \csc ^{11}(\theta_2) \log ^4\left({r_h} e^Z\right)}{\kappa_2^5 N^{17/2}},\nonumber\\
& & 
\end{eqnarray}
}
one sees that:
{\scriptsize
\begin{eqnarray}
\label{ExpminusPhiIIA-detRicci6}
& & e^{-\phi^{IIA}}\left.\log\left({\rm det}({\rm Ricci}_{\Sigma^{D6}}\right)\right|_{{\rm fixed}\ Z\in {\rm IR}}\sim\frac{3 \left(\frac{\left(9+8 \sqrt{3} \pi \right) {g_s}   ^2
   M^2 {N_f}  (c_1+c_2 \log ({r_h} ))}{2
   \sqrt{3} N}+6 {g_s}    {N_f}  \log
   ({r_h} )+{g_s}    {N_f}  \log (4)-8 \pi \right)
  }{8 \pi  {g_s}   }\nonumber\\
& & \times  \left[\log \left(\frac{10^{-13}{g_s}   ^{7/2} M^{16} {N_f} ^4
   {r_h} ^8 \left(\sqrt{2} \beta 
  {\cal C}_{zz}-2.\right)^2 \log ^4({r_h} ) (\cos (2
   \theta_2)-3)^6 \csc
   ^{11}(\theta_2)}{25\kappa_2^5
   N^{17/2}}\right)+\log (Z)\right].
\end{eqnarray}
}
As,
{\scriptsize
\begin{eqnarray}
\label{inttheta2CT}
& & {\cal I}(\theta_2)\equiv\int d\theta_2\log\left[(- 3 + \cos(2\theta_2))^6\csc^{11}\theta_2\right]\nonumber\\
& & = -\frac{11}{2} i \left(\theta_2^2+\text{Li}_2\left(e^{2 i
   \theta_2}\right)\right)+3 i \left[\text{Li}_2\left(\left(3-2
   \sqrt{2}\right) e^{2 i \theta_2}\right)+\text{Li}_2\left(\left(3+2
   \sqrt{2}\right) e^{2 i \theta_2}\right)\right]\nonumber\\
& & +6 i \theta_2^2+11
   \theta_2 \log \left(1-e^{2 i \theta_2}\right)-6 \log
   \left(1+\left(2 \sqrt{2}-3\right) e^{2 i \theta_2}\right)
   \left(\theta_2-i \sinh ^{-1}(1)\right)-6 \log \left(1-\left(3+2
   \sqrt{2}\right) e^{2 i \theta_2}\right) \left(\theta_2+i \sinh^{-1}(1)\right)\nonumber\\
& & -12 \sinh ^{-1}(1) \tan ^{-1}\left(\sqrt{2} \tan
   (\theta_2)\right)+\theta_2 \log \left((\cos (2 \theta_2)-3)^6
   \csc ^{11}(\theta_2)\right).
\end{eqnarray}
}
Hence,
{\scriptsize
\begin{eqnarray}
& & \hskip -0.8in {\cal I}(\theta_2=\epsilon_2\rightarrow0) = \frac{1}{2} i \epsilon_2 \left(-2 i \log \left(\frac{1}{32
   \epsilon_2^{11}}\right)-22 i \log (-2 i \epsilon_2)+22 i \log
   (\epsilon_2)+55 \pi -22 i\right)\nonumber\\
& & \hskip -0.8in -\frac{1}{12} i \left(-36
   \text{Li}_2\left(3-2 \sqrt{2}\right)-36 \text{Li}_2\left(3+2
   \sqrt{2}\right)+11 \pi ^2-72 i \pi  \log \left(3+2 \sqrt{2}\right)+72 i \pi 
   \sinh ^{-1}(1)-72 \log \left(3-2 \sqrt{2}\right) \sinh ^{-1}(1)\right);\nonumber\\
& & \hskip -0.8in {\cal I}(\theta_2 = \pi - \epsilon_2\rightarrow\pi) = \frac{1}{12} i \Biggl[-12 i \pi  \log
   \left(\frac{1}{\epsilon_2^{11}}\right)-132 i \pi  \log (2 i
   \epsilon_2)+36 \text{Li}_2\left(3+2 \sqrt{2}\right)+36
   \text{Li}_2\left(3-2 \sqrt{2}\right)+67 \pi ^2+72 i \pi  \log \left(3+2
   \sqrt{2}\right)+72 i \pi  \log \left(1+\sqrt{2}\right)\nonumber\\
& &\hskip -0.8in +72 i \pi  \log
   \left(\sqrt{2}-1\right)+12 i \pi  \log (64)+72 i \pi  \sinh ^{-1}(1)-72 \log
   \left(1+\sqrt{2}\right) \sinh ^{-1}(1)+72 \log \left(\sqrt{2}-1\right) \sinh
   ^{-1}(1)\Biggr]\nonumber\\
& & \hskip -0.8in-\frac{1}{2} i \epsilon_2 \left(-2 i \log
   \left(\frac{1}{32 \epsilon_2^{11}}\right)-22 i \log (2 i
   \epsilon_2)+22 i \log (\epsilon_2)+57 \pi -22
   i\right)+O\left(\epsilon_2^2\right)\sim-11\pi\log\epsilon_2.   
\end{eqnarray}
}
One can hence see that:
{\scriptsize
\begin{eqnarray}
\label{SDBI-IR}
& & \hskip -0.8in S_{\rm DBI}(Z\sim0)\sim \frac{\sqrt{2} \pi ^2 {\cal C}_{a_t^Z}^B\   {\cal C}_t^{\rho,\ B}\ 
   {\cal C}_{tx^3}^B\  e^{\kappa_0} \kappa_1 N^{9/20} \rho
   ^2 \Gamma \left(\frac{3}{4}\right)^2 \alpha _{\theta _1}
   {\cal C}_{x^3}^B(x^3) (-6 {g_s}    {N_f}  \log
   ({r_h} )-{g_s}    {N_f}  \log (4)+8 \pi )}{3
   {\cal C}_{x^3t}^B\  {\cal C}_{x^3}^{t\rho,\  B}\  {\cal C}_{x^3}^{tZ,\  B}\ 
   \sqrt{{g_s}   } {N_f}  {r_h}   | \log ({r_h} )| }\log Z.
\end{eqnarray}
}
We hence obtain:
{\scriptsize
\begin{eqnarray}
\label{GravSDBI-CT-IR-i}
& & \int_{\Sigma_{D6}({\rm IR})} e^{-\phi^{IIA}}\log\left({\rm det}({\rm Ricci}_{\Sigma^{D6}}\right) = \frac{4 \pi ^{5/2} \sqrt{N} \log (Z) \left(\frac{11}{16} {g_s}    {N_f} 
   \log (\epsilon_2) \left(-\frac{\left(3 \sqrt{3}+8 \pi \right) {g_s}   
   M^2 (c_1+c_2 \log ({r_h} ))}{N}-12 \log
   ({r_h} )\right)-\frac{3 \beta {\cal C}_{zz}}{8
   \sqrt{2}}\right)}{\sqrt{{g_s}   } {r_h} }.\nonumber\\
& & 
\end{eqnarray}
}
Now, the IR-divergent DBI action's IR-divergent contribution is given by:
{\scriptsize
\begin{eqnarray}
\label{SDBI_IR_div}
& & \hskip -0.8in {\cal S}_{\rm DBI}^{\rm IR-div} \sim \frac{ \pi ^2 {\cal C}_{a_t^Z}^B\   {\cal C}_t^{\rho, B} {\cal C}_{tx^3}^B\  e^{\kappa_0} \kappa_1 N^{9/20} \rho ^2 \Gamma
   \left(\frac{3}{4}\right)^2 \alpha _{\theta _1} {\cal C}_{x^3}^B(x^3) \log
   (Z) (-6 {g_s}    {N_f}  \log ({r_h} )-{g_s}    {N_f}  \log (4)+8
   \pi )}{ {\cal C}_{x^3t}^B\  {\cal C}_{x^3}^{t\rho,\  B}\  {\cal C}_{x^3}^{tZ,\  B}\ 
   \sqrt{{g_s}   } {N_f}  {r_h}  | \log ({r_h} )| }.\nonumber\\
& & 
\end{eqnarray}
}
Further, around some $\theta_1=\theta_{10}$
\begin{eqnarray}
\label{detFplusBconstIRZ}
& & {\rm det}\left.\left(F+i^*B_{\rm NS-NS}^{\rm IIA}\right)\right|_{Z\in{\rm IR}}\sim \left(B L F_{tx^3}\right)^2,
\end{eqnarray}
with $F_{tx^3}=\epsilon_{tx^3}\rightarrow0: B L \epsilon_{tx^3}$ is finite in the large $B$ and MQGP limit. Therefore, the IR boundary gravitational-DBI counter term will be given by:
\begin{eqnarray}
\label{GravSBI-CT-IR-ii}
& & {\cal S}_{\rm IR}^{\rm ct} \sim -
%\frac{
%{\cal C}_{a_t^Z}^B\   {\cal C}_t^{\rho,\ B}\     {\cal C}_{tx^3}^B\  \kappa_1 \rho ^2 \Gamma    \left(\frac{3}{4}\right)^2 \alpha _{\theta _1}   {\cal C}_{x^3}^B(x^3) e^{\kappa_0}
   \frac{\int_{\Sigma_{D6}(Z=0)} e^{-\phi^{IIA}}\sqrt{{\rm det}\left(F+i^*B_{\rm NS-NS}^{\rm IIA}\right)}\log\left(\sqrt{\frac{{\rm det}\left({\rm Ricci}_{\Sigma^{D6}}\right)}{{\rm det}\left(F+i^*B_{\rm NS-NS}^{\rm IIA}\right)}}\right) }{B L \epsilon_{tx^3}}, 
%   (-6 {g_s}       {N_f}  \log ({r_h} )-{g_s}    {N_f}  \log (4)+8    \pi ) \left(\sqrt{2} \beta {\cal C}_{zz}-44 {g_s}       {N_f}  \log (\epsilon_2) \log    ({r_h} )\right)}{ {\cal C}_{x^3t}^B\    {\cal C}_{x^3}^{t\rho,\  B}\  {\cal C}_{x^3}^{tZ,\  B}\  {g_s}   ^2   \sqrt[20]{N} {N_f} ^3 \log ^2(\epsilon_2) \log    ^2({r_h} ) | \log ({r_h} )| }.   
\end{eqnarray}
with $\epsilon_{tx^3}\sim\frac{1}{B L}$.

\subsection{In the UV at ${\cal O}(\beta^0)$}
\label{UV-beta0}

Following the footsteps of the previous sub-section for IR, here we will derive the EOMs and their solution for background gauge field fluctuation, $A_{\mu}$, but now in the UV region.

Consider the DBI action of flavor $D6$-branes in the UV region,
\begin{eqnarray}
\label{SDBIUV-i}
& &\rho r_h e^Z {\cal L}_{\rm DBI} \sim -\frac{N_f N^{3/5} \rho  {r_h}^2 e^{2 Z}
   \sqrt{\partial_Z A_{x^3}    ^2 \left(\rho ^2
   \partial_\rho A_t  ^2-B^2\right)+2 B
   \partial_\rho A_t  
   \partial_Z A_t      \partial_Z A_\phi   +\partial_\rho A_t^2 \partial_Z A_\phi    ^2+B^2
   \partial_Z A_t     ^2}}{ {g_s}
   \alpha _{\theta _2}^2},\nonumber\\
& & 
\end{eqnarray}
which in the large-$B$ limit is given by:
\begin{eqnarray}
\label{SDBIUV-ii}
& & -\frac{N_f N^{3/5} \rho  {r_h}^2 e^{2 Z}
   \left(\partial_\rho A_t  
   \partial_Z A_t      \partial_Z A_\phi     +B \partial_Z A_t ^2-B 
   \partial_Z A_{x^3}^2\right)}{ {g_s} \alpha _{\theta _2}^2
   \sqrt{\partial_Z A_{t}^2-\partial_Z A_{x^3}    ^2}}.
\end{eqnarray}

\noindent {\bf $A_t$ EOM}

One can show that $\partial_\rho\left(\frac{\delta {\cal L}_{\rm DBI}}{\delta \partial_\rho A_t}\right) +\partial_Z\left(\frac{\delta {\cal L}_{\rm DBI}}{\delta \partial_Z A_t}\right) = 0$, in the large-$B$ limit yields:
\begin{eqnarray}
\label{SDBIUV-iii}
& & \partial_Z A_t     
   \partial_Z A_{x^3}     \left(\rho 
   \left(\partial_Z\partial_\rho A_{x^3}   \partial_Z A_\phi   +B \partial_Z^2 A_{x^3}   \right)-\partial_Z A_{x^3}    
   \left(\rho  \partial_Z\partial_\rho A_{\phi}   +\partial_Z A_\phi    +2 B \rho
   \right)\right)\nonumber\\
& & -\rho  \partial_Z A_{x^3}    ^2
   \left(\partial_Z\partial_\rho A_t     \partial_Z A_\phi   +B \partial_Z^2 A_{t}\right)+\partial_Z A_t     ^3
   \left(\rho  \left(\partial_Z\partial_\rho A_{\phi}+2
   B\right)+\partial_Z A_\phi   \right)=0.
\end{eqnarray}
Assuming $\phi $ -independence of $A_\rho$,
\begin{equation}
\label{B}
B = \frac{\rho  \partial_\rho A_{\phi}+{A\phi
   }(Z,\rho ,{x3})}{\rho }.
\end{equation}
Using (\ref{B}), (\ref{SDBIUV-iii}) simplifies to:
\begin{eqnarray}
\label{SDBIUV-iv}
& & \hskip -0.8in \rho  \Biggl(\partial_Z A_t     
   \partial_Z A_{x^3}     \left(-\frac{2 \left(\rho 
   \partial_\rho A_{\phi}+A_{\phi}\right) \partial_Z A_{x^3}    }{\rho
   }+\partial_Z\partial_\rho A_{x^3}   \partial_Z A_{\phi}+\frac{\left(\rho  
   \partial_\rho A_{\phi}+A_{\phi}\right) \partial_Z^2A_{x^3}}{\rho}\right)\nonumber\\
& & \hskip -0.8in -\partial_Z A_{x^3}    ^2
   \left(\partial_Z\partial_\rho A_t    \partial_Z A_{\phi}+\frac{\left(\rho  
   \partial_\rho A_{\phi}+A_{\phi}\right) 
   \partial_Z^2 A_t}{\rho}\right)+\frac{2 \left(\rho  
   \partial_\rho A_\phi+{A\phi }(Z,\rho ,{x3})\right)
   \partial_Z A_t     ^3}{\rho }\Biggr)=0.\nonumber\\
& & 
\end{eqnarray}
Using the ansatz (\ref{Amu-ansatz}), (\ref{SDBIUV-iv}) can be rewritten as:
{\scriptsize
\begin{eqnarray}
\label{SDBIUV-v}
& & \hskip -0.8in a_t^{x^3}(x^3) \Biggl(-a_{x^3}^\rho(\rho)^2
   a_{x^3}^{x^3}(x^3)^2 a_{x^3}^Z\ '(Z)^2 \left(\frac{a_t^\rho(\rho) {a_t}''(Z) \left(\rho  
   \partial_\rho A_\phi+A_\phi\right)}{\rho }+a_\phi^\rho(\rho) 
a_\phi^{x^3}(x^3) a_t^\rho\ '(\rho ) a_t^Z\ '(Z) {a\phi Z}'(Z)\right)\nonumber\\
& & \hskip -0.8in +\frac{2 a_t^\rho(\rho)^3
   a_t^{x^3}(x^3)^2 a_t^Z\ '(Z)^3 \left(\rho  \partial_{x^3}A_\phi+A_\phi\right)}{\rho }\nonumber\\
& & \hskip -0.8in +a_t^\rho(\rho)a_{x^3}^\rho(\rho) a_{x^3}^{x^3}(x^3)^2 a_t^Z\ '(Z) a_{x^3}^Z\ '(Z)
   \left(\frac{a_{x^3}^\rho(\rho) \left(a_{x^3}^Z\ ''(Z)-2
   a_{x^3}^Z\ '(Z)\right) \left(\rho  \partial_{x^3}A_\phi+{A\phi }(Z,\rho ,{x3})\right)}{\rho
   }+a_\phi^\rho(\rho) a_\phi^{x^3}(x^3)
   a_{x^3}^\rho\ '(\rho) a_{x^3}^Z\ '(Z) {a\phi
   Z}'(Z)\right)\Biggr)=0.\nonumber\\
& & 
\end{eqnarray}
}
We see that (\ref{SDBIUV-v}) implies:
\begin{eqnarray}
\label{SDBI-vi}
& & a_t^{x^3}(x^3)={\cal C}_{a_t^{x3}}^{B\ {\rm UV}}\  a_{x^3}^{x^3}(x^3),\nonumber\\
& & a_\phi^{x^3}(x^3)={\cal C}_{a_\phi^{x3}}^{B\ {\rm UV}}\ ,\nonumber\\
& & a_t^\rho(\rho)={\cal C}_{a_t^\rho}^{B\ {\rm UV}}\  a_{x^3}^\rho(\rho),\nonumber\\
& & a_\phi^\rho(\rho)=\frac{{\cal C}_{a_t^\rho}^{B\ {\rm UV}}\ ^2 a_{x^3}^\rho(\rho)}{a_{x^3}^\rho\ '(\rho)},
\end{eqnarray}
and
{\scriptsize
\begin{eqnarray}
\label{SDBI-vii}
& & \hskip -0.8in B {\cal C}_{a_t^\rho}^{B\ {\rm UV}}\  a_{x^3}^\rho(\rho)^3
   a_{x^3}^{x^3}(x^3)^2 \left(a_t^Z\ '(Z) a_{x^3}^Z\ '(Z)
   \left(a_{x^3}^Z\ ''(Z)-2
   a_{x^3}^Z\ '(Z)\right)-{atZ}''(Z) a_{x^3}^Z\ '(Z)^2+2
   {\cal C}_{a_t^\rho}^{B\ {\rm UV}}\ ^2 {\cal C}_{a_t^{x3}}^{B\ {\rm UV}}\ ^2
   a_t^Z\ '(Z)^3\right)=0,\nonumber\\
& & 
\end{eqnarray}
}
satisfied by:
\begin{eqnarray}
\label{SDBI-viii}
& & a_t^Z(Z)={\cal C}_{tZ,1}^B\ +({\cal C}_{tZ,2}^B\ +{\epsilon_1})
   e^{-\kappa_{a_t^Z} Z},\nonumber\\
& & a_{x^3}^Z\ (Z)=\kappa_{a_x3^{Z,1}}+(\kappa_{a_{x3}^{Z,2}}+{\epsilon_2})
   e^{-\kappa_{a_x3^Z} Z},\nonumber\\
& & {\cal C}_{a_t^\rho}^{B\ {\rm UV}}\  =  \pm   \frac{\kappa_{a_{x3}^{Z,2}}}{{\cal C}_{tZ,2}^B\  {\cal C}_{a_t^{x3}}^{B\ {\rm UV}}\ },\nonumber\\
& & |{\cal C}_{tZ,1}^B\ |\ll1, |\kappa_{a_{x3}^{Z,2}}|\ll1.
\end{eqnarray}

\noindent {\bf $A_\phi$ EOM}

One can show that (\ref{Aphi-EOM-ii}) would yield:
{\scriptsize
\begin{eqnarray}
\label{SDBIUV-ix}
& & \hskip -0.8in -\frac{\sqrt{2} N^{3/5} \rho  {r_h}^2 e^{2 Z}
   \left(-\partial_Z A_t     ^2
   \partial_Z A_{x^3}    
   \partial_Z\partial_\rho A_{x^3}+\partial_Z A_t 
   \partial_Z A_{x^3}    
   \left(\partial_\rho A_t  
   \partial_Z^2A_{x^3}-2
   \partial_Z\partial_\rho A_t    
  \partial_Z^2A_{x^3} \right)+\partial_Z A_{x^3}    ^2
   \left(\partial_Z A_{x^3}    
   \partial_Z\partial_\rho A_{x^3}-\partial_\rho A_t  
   \partial_Z^2A_t \right)+2
   \partial_Z\partial_\rho A_t    
   \partial_Z A_t     ^3\right)}{\sqrt[3]{3}
   {g_s} \alpha _{\theta _2}^2 \left(\partial_Z A_{t}^2-\partial_Z A_{x^3}^2\right)^{3/2}}\nonumber\\
& & = 0.
\end{eqnarray}
}
It turns out that in the $\epsilon_2\rightarrow0$-limit, the LHS of (\ref{SDBIUV-ix}) is given by:
\begin{equation}
\label{LHS-Aphi-EOM-UV}
-\frac{\left(\kappa_{a_{x3}^{Z,2}}\right)^4e^{-4Z}({\cal C}_{tZ,2}^B\ +{\cal C}_{tZ,1}^B\ e^Z)}{{\cal C}_{tZ,2}^B\ }
\left(a_{x^3}\ ^{x^3}(x^3)\right)^4\left(a_{x^3}\ ^\rho(\rho)\right)^3a_{x^3}\ ^\rho\ '(\rho),
\end{equation}
which is at least $e^{-3Z_{UV}}$-suppressed in the UV.

\noindent {\bf $A_\rho$ EOM} is identically satisfied

\noindent {\bf $A_{x^3}$ EOM}

\begin{eqnarray}
\label{SDBIUV-x}
& & -\frac{\sqrt{2} e^{2Z} N^{3/5} \rho  {r_h}^2
   \partial_Z A_{x^3}    
   \left(\partial_\rho A_t  
   \partial_Z A_t      \partial_Z A_\phi -B
   \partial_Z A_t     ^2+B
   \partial_Z A_{x^3}^2\right)}{\sqrt[3]{3} {g_s} \alpha _{\theta
   _2}^2 \left(\partial_Z A_{t}^2-\partial_Z A_{x^3}^2\right)^{3/2}}={\cal C}_{x^3}^B(x^3, \rho),
\end{eqnarray}
which is equivalent to:
\begin{eqnarray}
\label{SDBIUV-xi}
& & \frac{(-\kappa_{a_{x3}^{Z,2}})^{7/2} {\cal C}_{a_\phi^{x3}}^{B\ {\rm UV}}\  N^{3/5}
   \rho  {r_h}^2 e^{2 Z} a_\phi^Z\  '(Z)
   \left({\cal C}_{tZ,1}^B\  e^Z+{\cal C}_{tZ,2}^B\ \right)}{2
   \sqrt[3]{3} {\cal C}_{tZ,2}^B\ ^3 {\cal C}_{a_t^{x3}}^{B\ {\rm UV}}\ ^2
   {\epsilon_2}^{3/2} {g_s} \alpha _{\theta
   _2}^2}={\cal C}_{x^3}^B(x^3, \rho),
\end{eqnarray}
solved by:
\begin{eqnarray}
\label{SDBIUV-xi}
& & a_\phi^Z\  (Z) = -\frac{{\epsilon_2}^{3/2}
   \kappa_\phi^Z  \left(2 {\cal C}_{tZ,1}^B\ ^2 \log
   \left({\cal C}_{tZ,1}^B\  e^Z+{\cal C}_{tZ,2}^B\ \right)-2
   {\cal C}_{tZ,1}^B\ ^2 Z-2 {\cal C}_{tZ,1}^B\  {\cal C}_{tZ,2}^B\ 
   e^{-Z}+{\cal C}_{tZ,2}^B\ ^2 e^{-2 Z}\right)}{2
   {\cal C}_{tZ,2}^B\ ^3}+c_1,\nonumber\\
& & \frac{\kappa_{a_{x3}^{Z,2}}^{7/2} {\cal C}_{a_\phi^{x3}}^{B\ {\rm UV}}\ 
   \kappa_\phi^Z  N^{3/5} \rho  {r_h}^2}{2 \sqrt[3]{3}
   {\cal C}_{tZ,2}^B\ ^3 {\cal C}_{a_t^{x3}}^{B\ {\rm UV}}\ ^2 {g_s} \alpha
   _{\theta _2}^2}={\cal C}_{x^3}^B(x^3, \rho)={\cal C}_{x^3}^B\  \rho.
\end{eqnarray}

The constants ${\cal C}_{a_t^{x3}}^{B\ {\rm UV}},
{\cal C}_{a_\phi^{x3}}^{B\ {\rm UV}},
{\cal C}_{a_t^\rho}^{B\ {\rm UV}},
{\cal C}_{a_t^\rho}^{B\ {\rm UV}},
{\cal C}_{tZ}^B,
{\cal C}_{tZ}^B,
{\cal C}_{x^3}^B(x^3, \rho)={\cal C}_{x^3}^B\  \rho
$ appearing in (\ref{SDBI-vi}) - (\ref{SDBIUV-xi}), are constants of integration in the solutions to the EOMs of the type IIA flavor $D6$-branes in the UV and in the presence of a strong magnetic field $B\gg T_c^2$ in $e=1$-units.

\subsection{In the UV at ${\cal O}(\beta)$ and Non-Renormalization of $A_{\mu = t, \rho, \phi, Z, x^3}$ in the Self-Consistent truncation $A_{\mu = \rho, \phi, Z, x^3}^\beta=0$}
\label{Strong-B-i}

The DBI integrand at ${\cal O}(\beta)$ assuming only $A_t^\beta(Z,\rho,x^3)=a_t^{Z,\ \beta}(Z)a_t^{\rho,\ \beta^0}(\rho)a_t^{x^3,\ \beta^0}(x^3)$, up to leading order in $\epsilon_{1,2}$, is given by:
{\footnotesize
\begin{eqnarray}
\label{LDBIbeta}
& & \hskip -0.8in {\cal L}_{\rm DBI}^\beta = -\frac{2\beta B^2 e^{-Z} {\cal C}_{a_{x3}^Z}\ ^{{\rm UV},  B}\ ^2 a_{x3}^\rho (\rho )^2
  a_{x^3}\ ^{x^3} ({x^3})^2 a_t^{Z,\ \beta}\ '(Z)}{{\cal C}_{tZ,2}^B\  }\nonumber\\
  & & \hskip -0.8in -\frac{8 \sqrt{\epsilon_2} \left(\sqrt{2} \beta  B
   \sqrt{-{\cal C}_{a_{x3}^Z}\ ^{{\rm UV},  B}\ } M^{\rm UV} \left(\frac{1}{N}\right)^{13/20} \rho 
   e^{3 Z} \left(19683 \sqrt{3} \alpha _{\theta _1}^6+3321 \sqrt{2} \alpha
   _{\theta _2}^2 \alpha _{\theta _1}^3-40 \sqrt{3} \alpha _{\theta _2}^4\right)
   a_{x3}^\rho (\rho )a_{x^3}\ ^{x^3} ({x^3}) \log
   ^3({r_h})\right)}{19683 \left(3^{5/6} \pi  \epsilon ^5 {g_s}
   N_f^{\rm UV}\  \alpha _{\theta _2}^5 \log
   ^4(N)\right)}.
\end{eqnarray}
}
Using (\ref{MUV+NfUV}) and in the $\epsilon_2\rightarrow0$-limit,  one obtains:
\begin{equation}
\label{atbeta}
a_t^{Z,\ \beta}\ '(Z)\sim0.
\end{equation}
We may choose $a_t^{Z,\ \beta}=0$.

\section{$T_c$ in the presence of a strong $B$}
\label{Tc-large-B}

It is known that the deconfinement temperature decreases when one turns on a magnetic field \cite{decrease-Tc-B}. We will demonstrante this in our type IIA dual.

One can show that the on-shell DBI action per-unit $\mathbb{R}^3$-coordinate volume, for the type IIA flavor $D6$-branes in the IR for $T>T_c$, is given by:
\begin{eqnarray}
\label{SDBI-on-shell-IR-bh}
\label{large-B-more-than-Tc-IR}
& & \frac{S_{\rm DBI}^{\rm IR, B}}{{\cal V}_3} = \kappa_{\rm IR}^{\rm B}\frac{\alpha_3 {\cal C}_{\phi x^3}^B {\cal C}_{\phi\rho}^B N^{3/5}r_h^2\rho  N_f |\log r_h|}{|\kappa_1|} B,
\end{eqnarray}
which is positive if one chooses  $\alpha_3 {\cal C}_{\phi x^3}^B {\cal C}_{\phi\rho}^B>0$. In the UV,
\begin{eqnarray}
\label{large-B-more-than-Tc-UV}
& & \lim_{\epsilon_{1,2}\rightarrow0}\frac{S_{\rm DBI}^{\rm UV, B}}{{\cal V}_3} = 0.
\end{eqnarray}
For the thermal background, in exactly an analogous way as was done in the black-hole background dual to $T>T_c$, one can show that one obtains the following background gauge field in the IR:
\begin{eqnarray}
\label{A-mu-IR-th}
& & a_\phi^\rho(\rho) = \frac{\left({\cal C}_{a_t^{x3}}\ ^{\rm IR, B}\right)^2\rho^{3/2}(2\sqrt{\rho}{\cal C}_{\rho}^{1, {\rm IR}} + {\cal C}_{\rho}^{2, {\rm IR}})}{B \left({\cal C}_{a_\phi^{\beta^0}}\ ^{\rm IR}\right)^2{\cal C}_{a_\phi^{x3}\ ^{\rm B}}^{\rm IR}},\nonumber\\
& & a_\phi^Z(Z) = {\cal C}_{\phi}^{1, {\rm IR}} + \frac{1}{2}(1 + 2 Z) {\cal C}_{Z}^{1, {\rm IR}} - \frac{1}{3}(1 + 3 Z){\cal C}_{\rho}^{2, {\rm IR}},\nonumber\\
& & a_\phi^{x^3}(x^3) = {\cal C}_{a_\phi^{x3}\ ^{\rm B}}^{\rm IR};\nonumber\\
& & a_t^Z(Z) = - (1 - Z) {\cal C}_{Z}^{1, {\rm IR}} + {\cal C}_{Z}^{2, {\rm IR}},\nonumber\\
& & a_t^\rho(\rho) = 2\sqrt{\rho}{\cal C}_{\rho}^{1, {\rm IR}} + {\cal C}_{\rho}^{2, {\rm IR}},\nonumber\\
& & a_t^{x^3}(x^3) = {\cal C}_{a_t^{x3}}\ ^{\rm IR, B}.
\end{eqnarray}
One similarly obtains the following background gauge field in the UV:
\begin{eqnarray}
\label{A-mu-UV-th}
& & a_\phi^\rho(\rho) = \frac{\left({\cal C}_{a_t^{x3}}\ ^{\rm UV, B}\right)^2\rho^{3/2}(2\sqrt{\rho}{\cal C}_{\rho}^{1, {\rm UV}} + {\cal C}_{\rho}^{2, {\rm UV}})}{B \left({\cal C}_{a_\phi^{\beta^0}}\ ^{\rm UV}\right)^2{\cal C}_{a_\phi^{x3}\ ^{\rm B}}^{\rm UV}},\nonumber\\
& & a_\phi^Z(Z) = {\cal C}_{\phi}^{1, {\rm UV}} + \frac{1}{2}e^{2Z} {\cal C}_{Z}^{1, {\rm UV}} - \frac{1}{3}e^{3Z}{\cal C}_{\rho}^{2, {\rm UV}},\nonumber\\
& & a_\phi^{x^3}(x^3) = {\cal C}_{a_\phi^{x3}\ ^{\rm B}}^{\rm UV};\nonumber\\
& & a_t^Z(Z) = - e^{-Z} {\cal C}_{Z}^{1, {\rm UV}} + {\cal C}_{Z}^{2, {\rm UV}},\nonumber\\
& & a_t^\rho(\rho) = 2\sqrt{\rho}{\cal C}_{\rho}^{1, {\rm UV}} + {\cal C}_{\rho}^{2, {\rm UV}},\nonumber\\
& & a_t^{x^3}(x^3) = {\cal C}_{a_t^{x3}}\ ^{\rm UV, B}.
\end{eqnarray}
In (\ref{A-mu-IR-th}) and (\ref{A-mu-UV-th}), ${\cal C}_{a_t^{x3}}\ ^{\rm IR/UV, B}, 
{\cal C}_{\rho}^{1/2, {\rm UV/IR}}, {\cal C}_{a_\phi^{x3}\ ^{\rm B}}^{\rm IR/UV}, {\cal C}_{a_\phi^{\beta^0}}\ ^{\rm IR/UV}, {\cal C}_{\phi}^{1/2, {\rm UV/IR}}, {\cal C}_{Z}^{1/2, {\rm UV/IR}}, {\cal C}_{\rho}^{1/2, {\rm UV/IR}}$ are constants of integration appearing in the solutions to the EOMs of the gauge fields in the IR/UV and in the presence of a strong magnetic field.
 
Using
{\footnotesize
\begin{eqnarray}
\label{Z-int-IR-th-B}
& & \int dZ e^{2Z}\sqrt{a_1 + b_1 Z}(a_2 + b_2 Z) = \frac{e^{-\frac{2a_1}{b_1}}}{32\sqrt{b_1}}\left(4\sqrt{b_1}e^{2\left(\frac{a_1}{b_1}+Z\right)}\sqrt{a_1 + b_1 Z}\left(4 a_2 + b_2(-3 + 4 Z\right)\right)\nonumber\\
& &  + \left(- 4 a_2 b_1 + 4 a_1 b_2 + 3 b_1 b_2\right)
\sqrt{2\pi} {\rm Erf}\left(\frac{\sqrt{2}\sqrt{a_1 + b_1 Z}}{\sqrt{b_1}}\right),
\end{eqnarray}
}
($Z_{D5/\overline{D5}} = \frac{\sqrt{3}g_s M^2 (c_1^{\rm th} + c_2^{\rm th} \log r_h)}{N} + \sqrt{3}\epsilon, \epsilon \sim r_0^2\left(|\log r_0|\right)^{9/2}N^{-9/10 - \alpha}, \alpha>0 $ \cite{OR4}) one can then show that in the IR, assuming $\left|\frac{{\cal C}_{\rho}^{2, {\rm IR}}}{{\cal C}_{\rho}^{2, {\rm IR}}}\right|\ll1$,
\begin{eqnarray}
\label{SDBI-th-IR-B-I}
& & \frac{S_{\rm DBI}^{\rm th, IR, B}}{{\cal V}_3} =\frac{1}{3} \left(g_s N\right)^{1/4}\lim_{\delta_{1,2}\rightarrow0}\int_0^{\beta_{\rm th}} dt\int_{\delta_1}^{\pi-\delta_2}d\theta_2 d\psi \int_{Z=0}^{Z_{D5/\overline{D5}}}dZ {\cal L}_{\rm DBI}^{\rm th, B}\nonumber\\
& & \sim \kappa_{\rm IR}^{\rm th, B} g_s^{3/4}\log r_0 r_0^2 \frac{M N_f^{3/2}}{N^{1/4}}\sqrt{- 2 + 3 \left({\cal C}_{a_t^{x3}\ ^{\rm IR, B}}\right)^2{\cal C}_{\rho}^{2, {\rm IR}}\ ^2} (\log N - 3 \log r_0)\log\left(\delta_1\delta_2\right)\epsilon B.\nonumber\\
& & 
\end{eqnarray}
Using the regularization:
\begin{equation}
\label{ang-reg-kappa-theta2-R4}
 \lim_{\delta_{1,2},\epsilon\rightarrow0}\log\left(\delta_1\delta_2\right)\epsilon = \kappa_{\theta_2}\ ^{(R^4)}<0,
\end{equation}
one obtains:
\begin{eqnarray}
\label{SDBI-th-IR-B-II}
& & \frac{S_{\rm DBI}^{\rm th, IR, B}}{{\cal V}_3}=\kappa_{\rm IR}^{\rm th, B}\beta_{\rm th} g_s^{3/4}\log r_0 r_0^2 \frac{M N_f^{3/2}}{N^{1/4}}\sqrt{- 2 + 3 \left({\cal C}_{a_t^{x3}\ ^{\rm IR, B}}\right)^2{\cal C}_{\rho}^{2, {\rm IR}}\ ^2} (\log N - 3 \log r_0)|\kappa_{\theta_2}\ ^{(R^4)}|.\nonumber\\
& & 
\end{eqnarray}
Using,
{\footnotesize
\begin{eqnarray}
\label{SDBI-th-UV-B-I}
& & \int dZ e^{2Z}\sqrt{-2 + 3 \left( {\cal C}_{a_t^{x3}}\ ^{\rm UV, B} {\cal C}_{Z}^{1, {\rm UV}}\right)^2 e^{-2Z}\left(2\sqrt{\rho}{\cal C}_{\rho}^{1, {\rm UV}} + {\cal C}_{\rho}^{2, {\rm UV}}\right)^2}=\nonumber\\
& & \frac{1}{4} {\cal C}_{a_t^{x3}}\ ^{\rm UV, B} {\cal C}_{Z}^{1, {\rm UV}} \left(2
   \sqrt{\rho } {\cal C}_{\rho}^{1, {\rm UV}}+{\cal C}_{\rho}^{2, {\rm UV}}\right) \Biggl(12 \sqrt{3} g_s N_f
   e^Z \,
   _3F_2\left(\frac{1}{2},\frac{1}{2},\frac{1}{2};\frac{3
   }{2},\frac{3}{2};\frac{2 e^{2 Z}}{3
   \left( {\cal C}_{a_t^{x3}}\ ^{\rm UV, B} {\cal C}_{Z}^{1, {\rm UV}}\right)^2 \left(2
   \sqrt{\rho }  {\cal C}_{\rho}^{1, {\rm UV}}+{\cal C}_{\rho}^{2, {\rm UV}}\right){}^2}\right)\nonumber\\
   & & +(2 g_s N_f \log
   (N)-6 g_s N_f \log (\text{rh})-6 g_s
   N_f Z+3 g_s N_f+8 \pi )\nonumber\\
   & & \times
   \Biggl[\frac{2 e^Z \sqrt{-2 e^{2 Z}+3
   \left( {\cal C}_{a_t^{x3}}\ ^{\rm UV, B} {\cal C}_{Z}^{1, {\rm UV}}\right)^2 \left(2
   \sqrt{\rho }  {\cal C}_{\rho}^{1, {\rm UV}}+{\cal C}_{\rho}^{2, {\rm UV}}\right){}^2}}{{\cal C}_{a_t^{x3}}\ ^{\rm UV, B} {\cal C}_{Z}^{1, {\rm UV}}
   \left(2 \sqrt{\rho } {\cal C}_{\rho}^{1, {\rm UV}}+{\cal C}_{\rho}^{2, {\rm UV}}\right)}+3 \sqrt{2}
  {\cal C}_{a_t^{x3}}\ ^{\rm UV, B} {\cal C}_{Z}^{1, {\rm UV}}
   \left(2 \sqrt{\rho } {\cal C}_{\rho}^{1, {\rm UV}}+{\cal C}_{\rho}^{2, {\rm UV}}\right)\nonumber\\
& &  \times  \sin ^{-1}\left(\frac{\sqrt{\frac{2}{3}} e^Z}{2
  {\cal C}_{a_t^{x3}}\ ^{\rm UV, B} \sqrt{\rho }  {\cal C}_{\rho}^{1, {\rm UV}}
   {\cal C}_{Z}^{1, {\rm UV}}+{\cal C}_{a_t^{x3}}\ ^{\rm UV, B} {\cal C}_{\rho}^{2, {\rm UV}} {\cal C}_{Z}^{1, {\rm UV}}}\right)\Biggr]\Biggr),
\end{eqnarray}
}
and $Z_{\rm UV} = \log\left(\kappa_{Z_{\rm UV}}\left\{4\pi g_s N\right\}^{1/4}\right)$, one obtains:
\begin{eqnarray}
\label{SDBIUVthB}
& & \frac{S_{\rm DBI}^{\rm th, UV, B}}{{\cal V}_3} =\frac{1}{3} \left(g_s N\right)^{1/4}\lim_{\delta_{1,2}\rightarrow0}\int_0^{\beta_{\rm th}} dt\int_{\delta_1}^{\pi-\delta_2}d\theta_2 d\psi \int_{Z_{D5/\overline{D5}}}^{Z_{\rm UV}}dZ {\cal L}_{\rm DBI}^{\rm th, B}\nonumber\\
& & \sim \kappa_{Z_{\rm UV}}B g_s^{7/4}r_0^2\log r_0 M_{\rm UV} \left(N_f^{\rm UV}\right)^{3/2}\left(\log N - 3 \left\{4\log(4\pi g_s(\kappa_{Z_{\rm UV}})^4) - 12\log r_h\right\}\right)N^{1/4}.\nonumber\\
& & 
\end{eqnarray}
As $M_{\rm UV}, N_f^{\rm UV}\rightarrow0$ hence upon comparison of (\ref{SDBI-th-IR-B-I}) and (\ref{SDBIUVthB}), one can drop the latter as compared to the former.

Now, if $\beta_{\rm BH,Th}$ are respectively the periodicities of the thermal circle in the black and thermal ${\mathscr {M}}$-theory backgrounds then at $r={\cal R}_{\rm UV}$, $\beta_{\rm BH}\sqrt{ G^{\rm BH}_{tt}} = \beta_{\rm Th} \sqrt{G^{\rm Th}_{tt}}$. Now at $T=T_c$ \cite{Witten-Hawking-Page-Tc}, 
\begin{eqnarray}
& & \beta_{\rm BH}\slashed{\int}_{M_{11}}\left({\cal L}^{\rm BH}_{\rm EH} + {\cal L}^{\rm BH}_{\rm GHY}\delta(r-{\cal R}_{\rm UV}) + {\cal L}_{\rm DBI, BH}^{\rm up\ to\ {\cal O}(R^4)}\delta\left(\theta_1 - \frac{\alpha_{\theta_1}}{N^{1/5}}\right)\delta(\tilde{x})\delta\left(\tilde{z} - {\cal C}\frac{\pi}{2}\right)
+ {\cal L}^{\rm BH}_{{\cal O}(R^4)}\right) \nonumber\\
& &  = \beta_{\rm Th}\slashed{\int}_{M_{11}}\left({\cal L}^{\rm Th}_{\rm EH} + {\cal L}^{\rm Th}_{\rm GHY}\delta(r-{\cal R}_{\rm UV}) + {\cal L}_{\rm DBI, th}^{\rm up\ to\ {\cal O}(R^4)}\delta\left(\theta_1 - \frac{\alpha_{\theta_1}}{N^{1/5}}\right)\delta(\tilde{x})\delta\left(\tilde{z} - {\cal C}\frac{\pi}{2}\right) + {\cal L}^{\rm Th}_{{\cal O}(R^4)}\right),\nonumber\\
& & 
\end{eqnarray}
where $\slashed{\int}$ excludes the coordinate integral with respect to  $x^{0}$,
 implying: 
\begin{eqnarray}
\label{actions-equal-Tc-ii}
& & \left(1+\frac{r_h^4}{2{\cal R}_{\rm UV}^4}\right)\int_{M_{10}}\left({\cal L}^{\rm BH}_{\rm EH}+{\cal L}^{\rm BH}_{\rm GHY}\delta(r-{\cal R}_{\rm UV}) +  {\cal L}_{\rm DBI, BH}^{\rm up\ to\ {\cal O}(R^4)}\delta\left(\theta_1 - \frac{\alpha_{\theta_1}}{N^{1/5}}\right)\delta(\tilde{x})\delta\left(\tilde{z} - {\cal C}\frac{\pi}{2}\right) + {\cal L}^{\rm BH}_{{\cal O}(R^4)}\right) 
\nonumber\\
& & = \int_{\tilde{M}_{10}}\left({\cal L}^{\rm Th}_{\rm EH} + {\cal L}^{\rm Th}_{\rm GHY}\delta(r-{\cal R}_{\rm UV})  + {\cal L}_{\rm DBI, th}^{\rm up\ to\ {\cal O}(R^4)}\delta\left(\theta_1 - \frac{\alpha_{\theta_1}}{N^{1/5}}\right)\delta(\tilde{x})\delta\left(\tilde{z} - {\cal C}\frac{\pi}{2}\right) + {\cal L}^{\rm Th}_{{\cal O}(R^4)}
+ {\cal L}^{\rm Th}_{{\cal O}(R^4)}\right).\nonumber\\
& & 
\end{eqnarray}   
One can show that (\ref{actions-equal-Tc-ii}) excluding the ${\cal O}(R^4)$-corrections, yields the following:
{\footnotesize  
\begin{eqnarray}
\label{rh-r0-beta0-i}
& & \frac{2 \kappa_{\rm GHY}^{\rm bh} {M_{\rm UV}} {r_h}^4 \log \left(\frac{{{\cal R}_{\rm UV}}}{{\cal R}_{D5/\overline{D5}}^{\rm bh}}\right)}{{g_s}^{9/4} {{\cal R}_{D5/\overline{D5}}^{\rm bh}}^4} + \kappa_{\rm IR}^{\rm B}\frac{\alpha_3 {\cal C}_{\phi x^3}^B {\cal C}_{\phi\rho}^B N^{3/5}r_h^2\rho  N_f |\log r_h|}{|\kappa_1|} B \nonumber\\
& & = \frac{2 \kappa_{\rm GHY}^{{\rm th},\ \beta^0} {M_{\rm UV}} {r_0}^4 \log \left(\frac{{{\cal R}_{\rm UV}}}{{\cal R}_{D5/\overline{D5}}^{\rm th}}\right)}{{g_s^{\rm UV}}^{9/4}
    {{\cal R}_{D5/\overline{D5}}^{\rm th}}^4} + \kappa_{\rm IR}^{\rm th, B} B g_s^{5/4}\log r_0 r_0^2 \frac{M N_f^{3/2}}{N^{1/4}}\sqrt{- 2 + 3 \left({\cal C}_{a_t^{x3}\ ^{\rm IR, B}}\right)^2{\cal C}_{\rho}^{2, {\rm IR}}\ ^2} (\log N - 3 \log r_0)|\kappa_{\theta_2}\ ^{(R^4)}|,\nonumber\\
& & {\rm or}\nonumber\\
& & \mathbb{A}_+r_h^4 + \mathbb{B}_+B r_h^2 \log r_h = \mathbb{C}_+ r_0^4 + \mathbb{D}_+ r_0^2\log r_0(\log N - 3 \log r_0).     
    \end{eqnarray}
}
As in \ref{different-EoS-scenarios}, $e^{2\kappa_0}\kappa_1^2 \left({\cal C}_{tx^3}^B {\cal C}_{a_t^Z}^B {\cal C}_t^{\rho,\ B}\right)^2$ is taken to be negative and $e^{2\kappa_0}\ll1$, thus for ${\cal C}_{tx^3}^B {\cal C}_{a_t^Z}^B {\cal C}_t^{\rho,\ B}\equiv {\cal O}(1)$ and real, $\Re e (\kappa_1)=0, |\kappa_1|\gg1$. So, assuming 
$B\sim N^{\gamma_B>0}, |\kappa_1|\sim N^{\gamma_{\kappa_1}}, \gamma_{\kappa_1}-(\gamma_B + 3/5)>1, $ (\ref{rh-r0-beta0-i}) simplifies to:
\begin{equation}
\label{rh-r0-beta0-ii}
 \mathbb{A}_+r_h^4  = \mathbb{C}_+ r_0^4 + \mathbb{D}_+ r_0^2\log r_0(\log N - 3 \log r_0),
\end{equation}
which implies:
{\footnotesize
\begin{eqnarray}
\label{Tc-B}
& & T_c(B) = T_c(B=0) - B\frac{\mathbb{D}_+{\cal R}_{D5/\overline{D5}}^{\rm th}\left(\log N +\left|\log\left(\frac{r_0}{{\cal R}_{D5/\overline{D5}}^{\rm th}}\right)\right|\right)\left|\log\left(\frac{r_0}{{\cal R}_{D5/\overline{D5}}^{\rm th}}\right)\right|}{2 g_s^2\sqrt{N}\pi^{3/2}r_0\mathbb{C}_+^{3/4}\mathbb{A}_+^{1/4}}<T_c(B=0),
\end{eqnarray}
 }
as in lattice QCD\cite{decrease-Tc-B}.

\section{Paramagnetic Pressure/Energy-Anisotropic Plasma and Generalized TOV Equations}
\label{Aniso+Gen-TOV}

In this section, we are interested in deciphering the features of the generalized EoS obtained for the M-theoretic uplift of thermal QCD-like theories at temperatures above the deconfinement temperature. The free energy $F$ is obtained by the regularized DBI action providing the pressure ($P$) via the relation $P=-F$. The energy density $E$ at finite temperature and chemical potential can be obtained via the relation, $E=T\frac{\partial P}{\partial T}+\mu \frac{\partial P}{\partial\mu}-P$. Then one obtains the generalized EOS by relating the pressure with energy density. We then demonstrate that the holographic dual, in principle, could correspond to several scenarios above $T_c$. These include stable wormholes, a stable wormhole transitioning via a smooth crossover to exotic matter as the universe cools (the converse being prohibited in our setup),  and a paramagnetic pressure/energy-anisotropic plasma. Given that above $T_c$ QGP is expected to be paramagnetic \cite{Bali et al-magnetic-chi}, the third possibility appears to be the preferred one. Generalizing the TOV equations to include angular mass/pressure/energy profiles, we show that it is not possible that the anisotropic plasma leads to the formation of a compact star. En route, we show that the IR renormalization of the DBI action requires a boundary Log-determinant-of-Ricci-tensor counter term.

\subsection{Pressure/Energy anisotropy, non-analyticity with respect to complexified gauge coupling, and Almost Contact 3-Structures}
\label{P+E+AC3S}

The UV-finite DBI action per unit $\mathbb{R}^3$-coodinate density of the type IIA flavor $D6$-branes, near, e.g. $\theta_1= \frac{\alpha_{\theta_1}}{N^{1/5}}$, is:
\begin{eqnarray}
\label{SDBIUVfin-i}
& & \hskip -0.6in {\cal S}_{\rm UV-finite} \sim -\frac{N_f^{\rm UV}B \sqrt[5]{N}  \left(\alpha
   _{\theta _2} \left(6 a^2 {r_h} ^2 e^{2 Z_{D5/\overline{D5}}\ }-2
   {r_h} ^4 e^{4 Z_{D5/\overline{D5}}\ }\right)+99 \pi  a^4
   Z_{D5/\overline{D5}}\ \right)}{ \sqrt{{g_s}   } {r_h} },
\end{eqnarray}
(the ``$\sim$'' is indicative of the fact that multiplicative numerical factors have been disregarded) which upon substituting $e^{Z_{{\cal R}_{D5/\overline{D5}}}} = \frac{\sqrt{3}a}{r_h}$ yields:
\begin{eqnarray}
\label{SDBIUVfin-ii}
& & {\cal S}_{\rm UV-div} \sim -\frac{N_f^{\rm UV}\pi ^{5/2} B \sqrt{N} \rho {r_h} ^3 e^{4
   Z_{\rm UV}}}{\sqrt{{g_s}   }}.
\end{eqnarray}
As, the UV boundary cosmological constant term near $\theta_1=\frac{\alpha_{\theta_1}}{N^{1/5}}$ with the embedding $i:\Sigma_{D6}\hookrightarrow M_{10}\cong \left(S^1_t\times\mathbb{R}^3\right)\times_w \mathbb{R}_{>0}\times_w{\cal T}^{1,1}_{\rm NE}$:
\begin{eqnarray}
\label{UVccCT-i}
& & \int_{\Sigma_{D6}(Z=Z_{\rm UV})}\sqrt{-{\rm det}(i^*g)}\sim \frac{e^{4 Z_{\rm UV}}r_h^3\alpha_{\theta_1}}{g_s^{1/4}N^{9/20}},
\end{eqnarray}
The UV counter term will be proportional to:
\begin{eqnarray}
\label{UVccCT-ii}
& &  \frac{B N N_f^{\rm UV}}{g_s^{1/4}} \int_{\Sigma_{D6}(Z=Z_{\rm UV})}\sqrt{-{\rm det}(i^*g)}.
\end{eqnarray}
Now,
{\footnotesize
\begin{eqnarray}
\label{pres-UV}
& &\hskip -0.8inP^{\rm UV}=\frac{11 \pi ^2 B \rho ^2 r_{h}   ^4 \left(3 c_{1}    g_{s}    M^2+3 c_{2}    g_{s}    M^2 \log (r_{h}   )+\sqrt{3}
   N\right)^4 \log \left(\frac{\sqrt{3} g_{s}    M^2 (c_{1}   +c_{2}    \log (r_{h}   ))}{N}+1\right)}{36 g_{s}   
   N^{43/10}}
\end{eqnarray}
}
utilizing $\mu = \frac{\kappa_{\mu}}{a^2}$ \cite{Aalok+Gopal-Mesino} 
implying:
{\footnotesize
\begin{eqnarray}
\label{mudPresUVoverdmu}
 \left.\mu\frac{d P^{\rm UV}}{d\mu}\right|_{\mu = \mu(a(r_h))}=\frac{99 \pi ^2 B \rho ^2 r_{h}   ^4 \left(\frac{g_{s}    M^2 (c_{1}   +c_{2}    \log
   (r_{h}   ))}{N}+\frac{1}{\sqrt{3}}\right)^4 \left(4 \log \left(\frac{\sqrt{3} g_{s}    M^2 (c_{1}   +c_{2}    \log
   (r_{h}   ))}{N}+1\right)+1\right)}{8 g_{s}    N^{3/10}}\nonumber\\
\end{eqnarray}
}
and
{\footnotesize
\resizebox{\textwidth}{!}{%
\begin{minipage}{\textwidth}
\begin{eqnarray}
\label{TdPresUVoverdmu}
\left. T \frac{dP^{\rm UV}}{dT} \right|_{a = a(r_h)}
 &=& 11 \pi ^2 B \rho ^2 r_{h}^4 \left(3 c_{1} g_{s} M^2 + 3 c_{2} g_{s} M^2 \log (r_{h}) + \sqrt{3} N\right)^3 \nonumber\\
&& \times \frac{1}{36 N^{53/10}} \Bigg( 
12 c_{2} M^2 N \log \left( \frac{\sqrt{3} g_{s} M^2 (c_{1} + c_{2} \log (r_{h}))}{N} + 1 \right) \nonumber\\
&& + \frac{ \sqrt{3} c_{2} M^2 \left(3 c_{1} g_{s} M^2 + 3 c_{2} g_{s} M^2 \log (r_{h}) + \sqrt{3} N \right) }
{ \frac{\sqrt{3} g_{s} M^2 (c_{1} + c_{2} \log (r_{h}))}{N} + 1 } \nonumber\\
&& + \frac{ 4 N \left(3 c_{1} g_{s} M^2 + 3 c_{2} g_{s} M^2 \log (r_{h}) + \sqrt{3} N \right) 
\log \left( \frac{ \sqrt{3} g_{s} M^2 (c_{1} + c_{2} \log (r_{h})) }{N} + 1 \right) }{g_{s}} 
\Bigg)
\end{eqnarray}
\end{minipage}
}
}
Similarly,

{\footnotesize
\begin{eqnarray}
\label{presIR}
& & \hskip -0.8in P^{\rm IR}=-\frac{\pi  (\mathcal{C}_{a_{t}^{Z}}^{B})^2 (\mathcal{C}_{t}^{\rho,B} ) ^2 (\mathcal{C}_{x^{3}t}^{B})^2 e^{2\kappa_{0}}
   (\kappa_{1}^2 \rho ^3 \log (3)\mathcal{C}_{x^{3}}^{B}(x^{3}) \left(\text{gs} N_{f}    \log
   \left(r_{h}^6 \left(9 \left(\frac{\text{gs} M^2 (\text{c1}+\text{c2} \log
   (r_{h}))}{N}+\frac{1}{\sqrt{3}}\right)^2+1\right)\right)-8 \pi \right)}{
   \sqrt{\frac{\left(\frac{\text{gs} M^2 (\text{c1}+\text{c2} \log
   (r_{h}))}{N}+\frac{1}{\sqrt{3}}\right)^2 \left(6 r_{h}^2-9 (\mathcal{C}_{a_{t}^{Z}}^{B})^2
   (\mathcal{C}_{t}^{\rho,B})^2 (\mathcal{C}_{x^{3}t}^{B})^2 e^{2\kappa_{0}} (\kappa_{1}^2 \rho
   ^2\right)-(\mathcal{C}_{a_{t}^{Z}}^{B})^2 (\mathcal{C}_{t}^{\rho,B})^2 (\mathcal{C}_{x^{3}t}^{B})^2 e^{2
  \kappa_{0}} (\kappa_{1}^2 \rho ^2+r_{h}^2}{9 \left(\frac{\text{gs} M^2
   (\text{c1}+\text{c2} \log (r_{h}))}{N}+\frac{1}{\sqrt{3}}\right)^2+1}}} \nonumber\\
   && \hskip -6.4in \times \frac{1}{6 \sqrt{2} 3^{2/3}
   \mathcal{C}_{tx^{3}}^{B}\mathcal{C}_{x^{3}}^{t\rho,B} \mathcal{C}_{x^{3}}^{tZ,B} g_{s}N_{f}    \log (r_{h})}
\end{eqnarray}
}

implying:

{\scriptsize
\begin{eqnarray}
\label{mudPIRoverdmu}
\left.\mu\frac{d P^{\rm IR}}{d\mu}\right|_{\mu = \mu(a(r_h))}= \pi  (\mathcal{C}_{a_{t}^{Z}}^{B})^2 (\mathcal{C}_{t}^{\rho,B} ) ^2 (\mathcal{C}_{x^{3}t}^{B})^2 e^{2\kappa_{0}} (\kappa_{1}^2 \rho ^3 \log
   (3)\mathcal{C}_{x^{3}}^{B}(x^{3})  \nonumber\\
&& \hskip-3.0in \times \frac{\left(6g_{s}N_{f}    \left(3 r_{h}^2-4 (\mathcal{C}_{a_{t}^{Z}}^{B})^2 (\mathcal{C}_{t}^{\rho,B} ) ^2 (\mathcal{C}_{x^{3}t}^{B})^2 e^{2 (\kappa_{0}} (\kappa_{1}^2 \rho ^2\right)+\text{gs} \text{Nf} r_{h}^2 \log \left(4
   r_{h}^6\right)-8 \pi  r_{h}^2\right)}{24 \sqrt{2} 3^{2/3} \mathcal{C}_{tx^{3}}^{B}\mathcal{C}_{x^{3}}^{t\rho,B}
   \mathcal{C}_{x^{3}}^{tZ,B} g_{s}\text{Nf} \log (r_{h}) \left(3 r_{h}^2-4 (\mathcal{C}_{a_{t}^{Z}}^{B})^2 (\mathcal{C}_{t}^{\rho,B} ) ^2 (\mathcal{C}_{x^{3}t}^{B})^2 e^{2 (\kappa_{0}} (\kappa_{1}^2 \rho ^2\right)^{3/2}}
\end{eqnarray}
}

and

{\footnotesize
\resizebox{\textwidth}{!}{%
\begin{minipage}{\textwidth}
\begin{eqnarray}
\label{TdpIRoverdT}
\left.T\frac{d P^{\rm IR}}{d T}\right|_{a = a(r_h)} &=& \frac{\pi  (\mathcal{C}_{a_{t}^{Z}}^{B})^2 (\mathcal{C}_{t}^{\rho,B})^2 (\mathcal{C}_{x^{3}t}^{B})^2 e^{2\kappa_{0}} (\kappa_{1}^2 \rho ^3 \log
   (3)\mathcal{C}_{x^{3}}^{B}(x^{3}) }{3 \sqrt{2} 3^{2/3} \mathcal{C}_{tx^{3}}^{B}\mathcal{C}_{x^{3}}^{t\rho,B} \mathcal{C}_{x^{3}}^{tZ,B} g_{s}N_{f}    \log ^2(r_{h}) \left(3 r_{h}^2-4 (\mathcal{C}_{a_{t}^{Z}}^{B})^2
   (\mathcal{C}_{t}^{\rho,B} ) ^2 (\mathcal{C}_{x^{3}t}^{B})^2 e^{2\kappa_{0}} (\kappa_{1}^2 \rho ^2\right)^{3/2}} \nonumber\\
  && \times \left(\left(8 \pi -g_s N_{f} \log \left(4 r_{h}^6\right)\right) \left(4
   (\mathcal{C}_{a_{t}^{Z}}^{B})^2 (\mathcal{C}_{t}^{\rho,B} ) ^2 (\mathcal{C}_{x^{3}t}^{B})^2 e^{2\kappa_{0}} (\kappa_{1}^2 \rho ^2-3
   r_{h}^2\right)\right.\nonumber\\
   && \left.-3 \log (r_{h}) \left(-8 (\mathcal{C}_{a_{t}^{Z}}^{B})^2 (\mathcal{C}_{t}^{\rho,B})^2 (\mathcal{C}_{x^{3}t}^{B})^2 g_s
   e^{2\kappa_{0}} (\kappa_{1}^2 N_{f} \rho ^2+6g_{s}N_{f} r_{h}^2-g_s N_{f} r_{h}^2
   \log \left(4 r_{h}^6\right)+8 \pi  r_{h}^2)\right)\right) \nonumber\\
\end{eqnarray}
\end{minipage}
}
}

Using standard KS(Klebanov-Strassler)-like RG-flow equations \cite{Klebanov:2000hb}, \cite{Bulk-Viscosity}, one sees that (similar to \cite{effec_kin_model_zetaovereta}):
{\footnotesize
\begin{equation}
\label{g(T)}
g^2(T/T_c) \sim \left.\left(e^{-\phi^{\rm IIA}}\int_{S^2}B^{\rm IIA}\right)^{-1}\right|_{r=r_h}
\sim\frac{1}{(g_s M) (g_s N_f)}\left[\left|\log \left(\frac{\sqrt{3}T}{2(1 + \sqrt{3}\varepsilon )T_c}\right)\right| \left(\log N - 3 \log \left(\frac{\sqrt{3}T}{2(1 + \sqrt{3}\varepsilon )T_c}\right)\right)\right]^{-1}. 
\end{equation}
}
One sees from (\ref{pres-UV}) - (\ref{TdpIRoverdT}) that near $\theta_1\sim\frac{1}{N^{1/5}}$, up to LO in $r_h, |\log r_h|(>1)$ and $N$:
\begin{eqnarray}
\label{no-log[rh]-dependence-E+p}
& & P^{\rm UV} \sim -\frac{B N^{1/5}r_h^3\rho^2\left(\frac{{\cal R}_{D5/\overline{D5}}}{r_h}\right)^4\log\left(\frac{{\cal R}_{D5/\overline{D5}}}{r_h}\right)}{\sqrt{g_s}},\nonumber\\
& & \mu\frac{\partial P^{\rm UV}}{\partial\mu} \sim -\frac{B N^{1/5}r_h^3\rho^2\left(\frac{{\cal R}_{D5/\overline{D5}}}{r_h}\right)^4\Biggl[1 + 4 \log\left(\frac{{\cal R}_{D5/\overline{D5}}}{r_h}\right)\Biggr]}{\sqrt{g_s}},\nonumber\\
& & T\frac{\partial P^{\rm UV}}{\partial\mu} \sim -\frac{N^{1/5} B r_h^3\rho^2\left(\frac{{\cal R}_{D5/\overline{D5}}}{r_h}\right)^3}{\sqrt{g_s}}\Biggl[\sqrt{3}c_2 \frac{g_sM^2}{N}\left(1 + 4 \log\left(\frac{{\cal R}_{D5/\overline{D5}}}{r_h}\right)\right) + 3 \frac{{\cal R}_{D5/\overline{D5}}}{r_h}\log\left(\frac{{\cal R}_{D5/\overline{D5}}}{r_h}\right)\Biggr],\nonumber\\
& & P^{\rm IR} \sim \frac{\sqrt{g_s N} \rho^2 \sqrt{{\cal C}_{\rm combo}}{\cal C}_{x^3}^B(x^3)}{r_h \sqrt{{\cal C}^{(2)}_{\rm combo}}},\nonumber\\
& & \mu\frac{\partial P^{\rm IR}}{\partial\mu} \sim -\frac{\sqrt{g_s N}r_h {\cal C}_{x^3}^B(x^3)}{\sqrt{{\cal C}_{\rm combo} {\cal C}^{(2)}_{\rm combo}} },\nonumber\\
& & T\frac{\partial P^{\rm IR}}{\partial T} \sim \frac{\sqrt{g_s N}\rho^2\sqrt{{\cal C}_{\rm combo}}{\cal C}_{x^3}^B(x^3)}{r_h  \sqrt{{\cal C}^{(2)}_{\rm combo}}}. 
\end{eqnarray}
We hence note the absence of an explicit $\log r_h$-dependence of the pressure and energy densities (\ref{no-log[rh]-dependence-E+p}). Hence, from (\ref{g(T)}), writing
$g^2 = -\frac{{\cal C}_{\rm gauge}}{g_s^2 M  N_f \log r_h\left(\log N - 3 \log r_h\right)}, {\cal C}_{\rm gauge}>0$, one sees that 
\begin{equation}
\label{beta-g}
\log r_h = \frac{1}{6}\Biggl[\log N - \frac{\sqrt{12 {\cal C}_{\rm gauge} + g_s^2\log^2N M N_f g^2\left(\frac{T}{T_c}\right)}}{g_s\sqrt{M N_f}g\left(\frac{T}{T_c}\right)}\Biggr].
\end{equation}
 Hence, one sees that if one were to complexify the expressions for pressure and energy densities, then there will be no non-analytic dependence (of pressure and energy densities) on the complexified $g\left(\frac{T}{T_c}\right)$. We also note the absence of ${\cal O}(R^4)$-corrections of the pressure and energy densities. 
 \begin{tcolorbox}[enhanced,width=7in,center upper,size=fbox,
    %fontupper=\large\bfseries,
    drop shadow southwest,sharp corners]
    \begin{flushleft}
 Using the results of \cite{ACMS}, and motivated by the conjecture of the connection between the non-analytic-complexified-gauge-coupling-dependence of the complexified bulk-to-shear-viscosity ratio and the existence of Contact 3-Structures for $(g_s, M, N_f)=(0.1, 3, 3)$ and $N=100$ in \cite{Shivam+Aalok_Bulk}, we now conjecture that the existence of Almost Contact 3-Structures for the same values of $(g_s, M, N_f)$ as above but for $N=200$  [$N=100$ resulting in complex free energy derivable from the renormalized DBI action for the type IIA flavor $D6$-branes]  maps to the absence of non-analytic dependence of the complexified pressure and energy on the complexified coupling constant.
\end{flushleft}
\end{tcolorbox}

\subsection{Various scenarios arising on the basis of Equations of State}
\label{different-EoS-scenarios}
From (\ref{p+E-angular-dependence}), it turns out that $p(r,\theta; r_h(T))$ and $E(r,\theta; r_h(T))$ depend on the following combinations of integrations of constant for the type IIA $D6$-world-volume gauge fields in the IR:
\begin{eqnarray}
\label{IR-combos}
& & {\cal C}_{\rm combo}\equiv e^{2\kappa_0}\kappa_1^2 \left({\cal C}_{tx^3}^B {\cal C}_{a_t^Z}^B {\cal C}_t^{\rho,\ B}\right)^2, {\cal C}^{(2)}_{\rm combo}\equiv {\cal C}_{x^3}^{tZ,\  B} {\cal C}_{x^3}^{t\rho,\ B} {\cal C}_{x^3t}^B\ , {\cal C}_{x^3}^B(x^3).
\end{eqnarray}
In the following, we will choose  $\varepsilon = 0.3,  c_1 = -109, c_2 = -10$ \cite{Shivam+Aalok_Bulk} and take $N=200$ (to ensure $p(r,\theta), E(r,\theta)\in\mathbb{R}$, unlike \cite{Shivam+Aalok_Bulk}), and $\rho = 1$. 

In the following, we will re-interpret the plots in terms of variations of pressure, energy, and equation-of-state parameter as a function of "boundary time", relevant to calculating the entanglement entropy of the relevant eternal black hole corresponding to the Hartman-Maldacena-like surface using Dong's formula \cite{Dong} for the computation of entanglement entropy in higher derivative theories. 
In ${\cal M}$-theory dual, Hartman-Maldacena-like surface is a co-dimension two surface which is located at $x^1=x_R$ and corresponds to the embedding $t = t(r)$. We can write an expression for the entanglement entropy for Hartman-Maldacena-like surface in the following form:
\begin{equation}
\label{HD-Entropy-HM}
S_{EE}=\int dr dx^2 dx^3  d\theta_1 d\theta_2dxdydz dx^{10} \sqrt{-g}\Biggl[\frac{\partial {\cal L}}{\partial R_{z\bar{z}z\bar{z}}}+\sum_{\alpha}\left(\frac{\partial^2 {\cal L}}{\partial R_{zizj}\partial R_{\bar{z}m\bar{z}l}}\right)_{\alpha} \frac{8 K_{zij}K_{\bar{z}ml}}{(q_{\alpha}+1)}\Biggr],
\end{equation}
where $g$ is the determinant of the induced metric  on the co-dimension two surface, $ z = x e^{i t}$ along the directions normal to the HM surface, $(i,j,k,l)$ are along tangential directions, $K_{zij}=\frac{1}{2}\partial_z G_{ij}$ and its trace is defined as $K_z=K_{zij}G^{ij}$; see \cite{Dong} for the definition of $q_\alpha$. The solution to the embedding equation $t=t(r)$ at $r=r_h$ with $t(r=r_h)=t_b$, i.e., the "boundary time" was shown in \cite{Gopal+Aalok} to yield:
\begin{equation}
\label{rh-tb-relation}
r_h=\frac{\left(\frac{2}{3}\right)^{2/3}}{\left(\frac{(c_2-{t_{b_0}}) \
N^{-\frac{3 n_{t_b}}{2}}}{c_1}\right)^{2/3}},
\end{equation}
where $c_{1,2}<0$ and $|c_2|\sim e^{|c_1|}$ (rendering a "Swiss-Cheese" structure to the HM entanglement entropy) and $n_{t_b}={\cal O}(1)$.

From (\ref{pres-UV}) - (\ref{TdpIRoverdT}), one sees that the pressure and energy per unit $\mathbb{R}^3$ coordinate volume, near $\theta_1\sim\frac{1}{N^{1/5}}$, have an angular dependence:
\begin{eqnarray}
\label{p+E-angular-dependence}
& & p(r,\theta; r_h(T)) =  r\sin\theta\left( \sqrt{N}f_p^{(1)}(g_s; r_h; \left\{{\cal C}\right\}) + \frac{1}{\sqrt{N}}f^{(2)}_p(g_s, M; r_h; \left\{{\cal C}\right\}) + \frac{B}{N^{4/5}}f^{(3)}_p(g_s, M; r_h; \left\{{\cal C}\right\})\right),\nonumber\\
& & E(r,\theta; r_h(T)) = r\sin\theta\left(\sqrt{N} f_E^{(1)}(g_s, M, N_f; r_h; \left\{{\cal C}\right\}) + B N^{1/5} f_E^{(2)}(g_s; r_h; \left\{{\cal C}\right\})
+ \frac{f_E^{(3)}(g_s, M; r_h; \left\{{\cal C}\right\})}{\sqrt{N}}\right),\nonumber\\
& & 
\end{eqnarray}
where $\left\{{\cal C}\right\}\equiv  c_1, c_2, \kappa_0,  {\cal C}_{a_t^Z}^B, {\cal C}_t^{\rho,\ B}, {\cal C}_{tx^3}^B, \kappa_1, {\cal C}_{a_t^Z}^B, {\cal C}_{x^3t}^B, 
{\cal C}_{x^3}^{tZ,\  B}, {\cal C}_{x^3}^{t\rho,\ B}, {\cal C}_{x^3}^B(x^3)$. We hence see a pressure/energy anisotropy. Introducing the magnetic field along a direction (say $z$-direction in three dimensional space) breaks the rotational invariance and gives rise to the pressure anisotropy, and is vital in proving the paramagnetic behaviour of QGP. The anisotropic pressure decomposes into longitudinal pressure along the direction of the external magnetic field and transverse pressure orthogonal to the direction of the external magnetic field. Our results of increased transverse pressure as compared to the longitudinal pressure in the presence of a strong uniform magnetic field is consistent with, e.g. \cite{Koothottil:2020riy} based on a quasi-particle model. 

Here are some possible scenarios determined by $p = p(r_h),  E = E(r_h)$ for $\rho=1$ where depending on the sign of the two we talk about stable wormhole, exotic matter and anisotropic plasma. 

\noindent{\bf (a) A smooth crossover from anisotrpioc plasma to more exotic matter e.g. phantom energy as the universe cools (in terms of the ``boundary time'' defined in (\ref{rh-tb-relation})) but stays above $T_c$}: $ {\cal C}_{\rm combo}\approx 0.33,  {\cal C}^{(2)}_{\rm combo}=-1, {\cal C}_{x^3}^B(x^3) = 1$; see Fig. \tcb{4}..

\begin{figure}[h!]
\begin{center}
\includegraphics[width=0.7\textwidth]{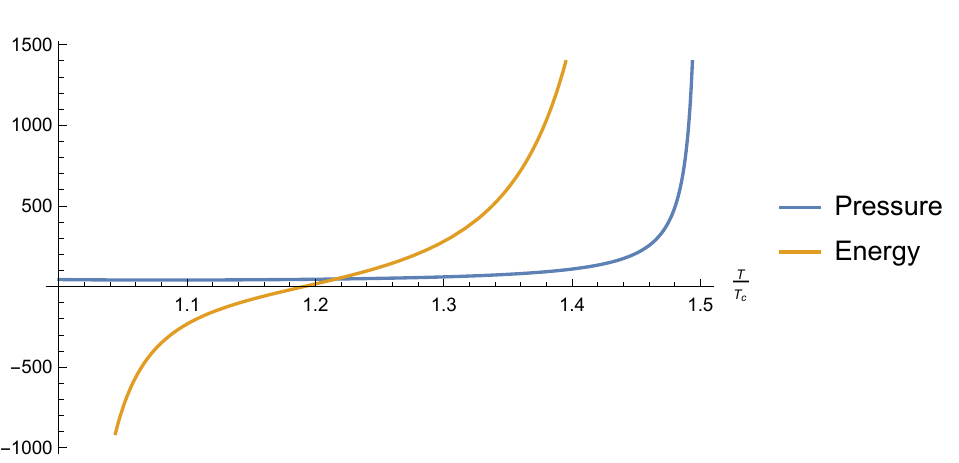}
\end{center}

\caption{A smooth crossover from anisotrpioc plasma to exotic matter as the universe cools (defined in terms of ``boundary time'' (\ref{rh-tb-relation})) but stays above $T_c$: $ {\cal C}_{\rm combo}\approx 0.33,  {\cal C}^{(2)}_{\rm combo}=-1, {\cal C}_{x^3}^B(x^3) = 1$}.
\end{figure}
\newpage

\noindent{\bf (b) Anisotropic Plasma}\\
I) $ {\cal C}_{\rm combo} =- 1,  {\cal C}^{(2)}_{\rm combo}=1, {\cal C}_{x^3}^B(x^3) = -0.001$; see Fig. \tcb{5} and \tcb{6} \\
\begin{figure}[h!]
\begin{subfigure}{.5\textwidth}
\centering
\includegraphics[width=1.1\textwidth]{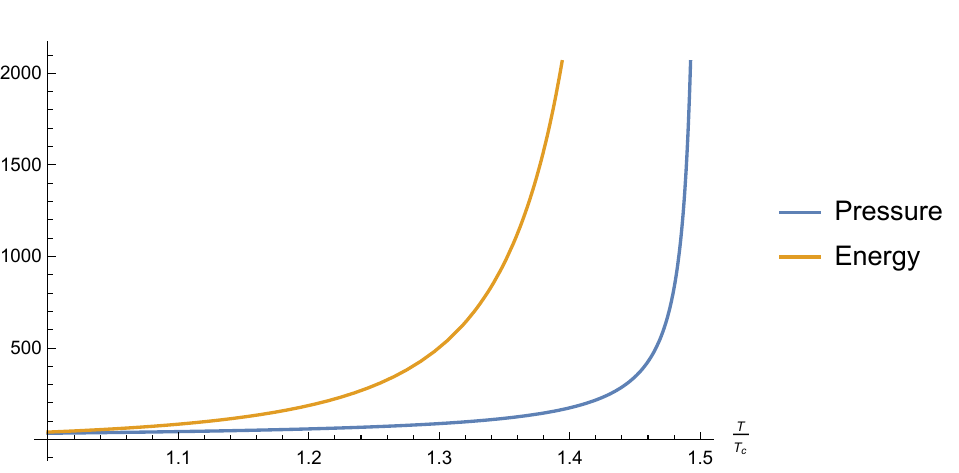}

\caption{Pressure/Energy density-vs-temperature}
\end{subfigure}
\hskip 0.1in
\begin{subfigure}{.5\textwidth}
\centering
\includegraphics[width=1\textwidth]{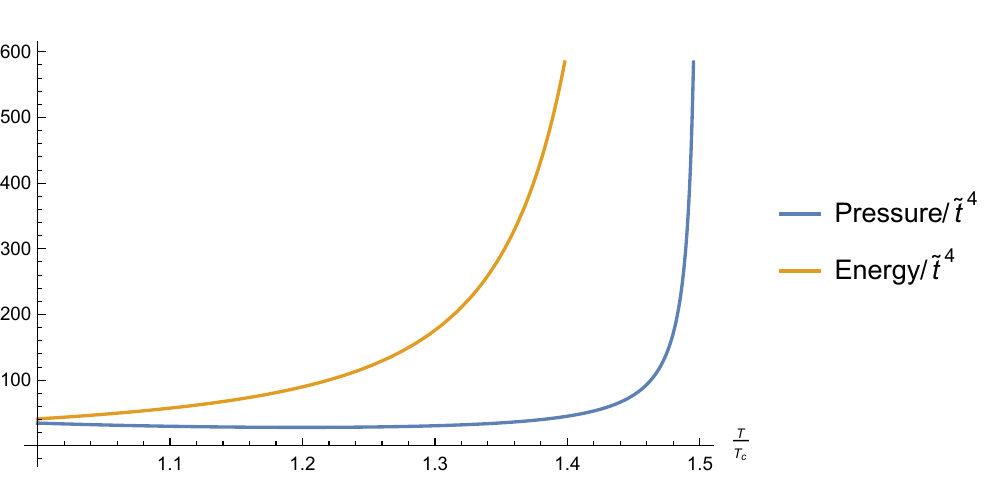}

\caption{$\frac{\rm Energy}{\tilde{t}^4}/\frac{\rm Pressure}{\tilde{t}^4}$-vs-temperature}
\end{subfigure}
\caption{Anisotropic plasma: $ {\cal C}_{\rm combo} =- 1,  {\cal C}^{(2)}_{\rm combo}=1, {\cal C}_{x^3}^B(x^3) = -0.001, B = 1 \left({\rm GeV}\right)^4;\ \tilde{t}\equiv\frac{T}{T_c}$}
\end{figure}

\begin{figure}[h!]
\begin{subfigure}{.5\textwidth}
\centering
\includegraphics[width=1.1\textwidth]{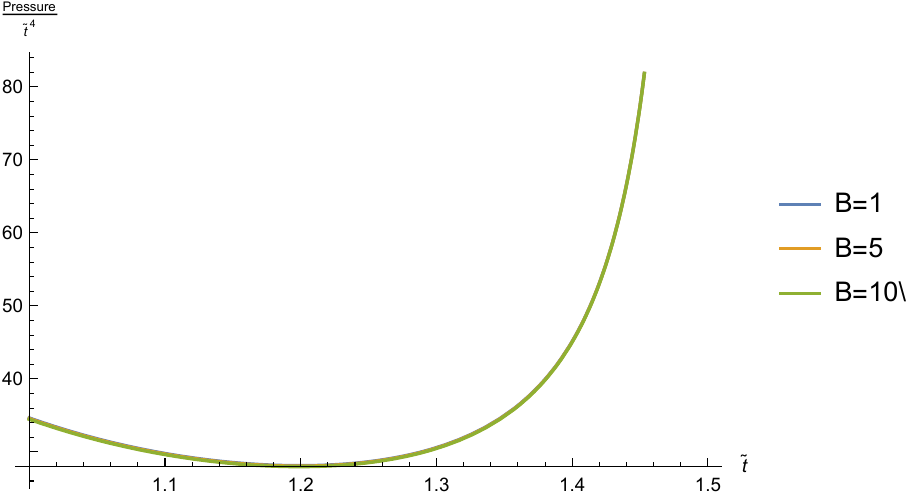}
\caption{$\frac{\rm Pressure}{\tilde{t}^4}$-vs-temperature for $B=0.2, 0.3, 0.5\left({\rm GeV}\right)^2$ (in $e=1$-units)}
\end{subfigure}

\begin{subfigure}{.5\textwidth}
\centering
\includegraphics[width=1\textwidth]{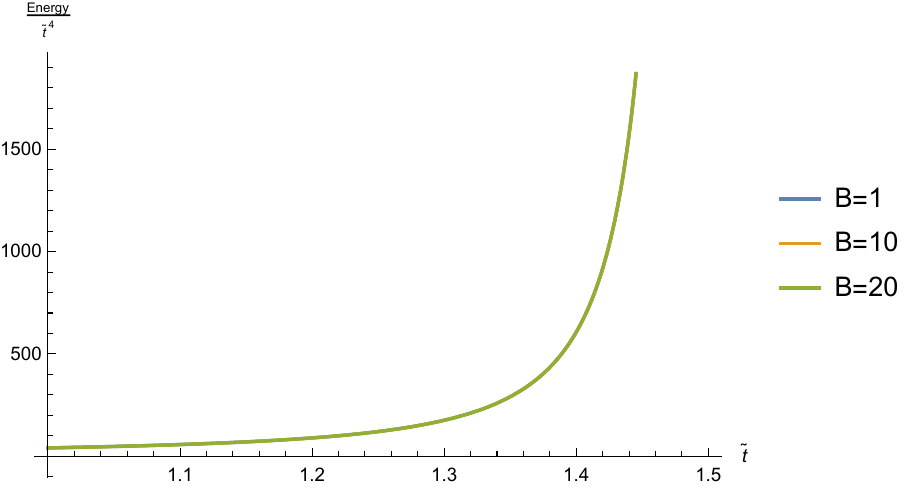}
\caption{$\frac{\rm Energy}{\tilde{t}^4}$-vs-temperature $T\in[T_c, 1.2T_c]$}
\end{subfigure}
\hskip 0.1in
\begin{subfigure}{.5\textwidth}
\centering
\includegraphics[width=1\textwidth]{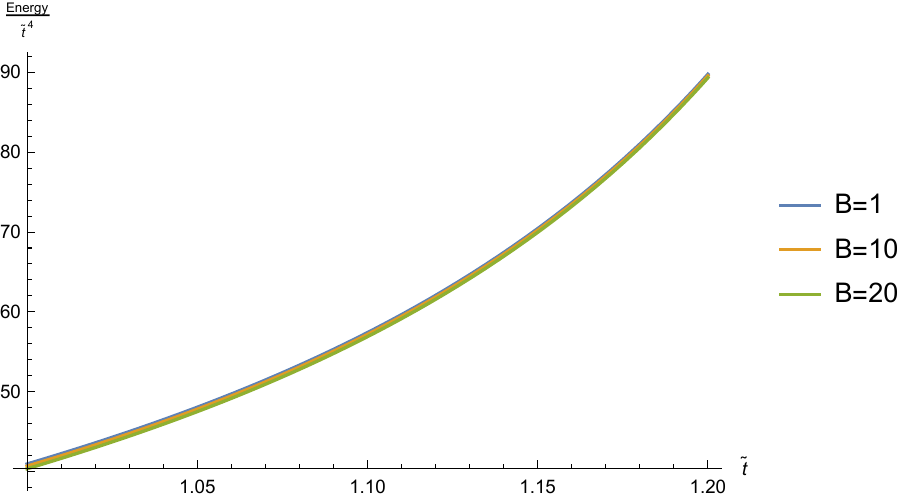}

\caption{$\frac{\rm Energy}{\tilde{t}^4}$-vs-temperature $T\in[1.2T_c, 1.5T_c]$}
\end{subfigure}
\caption{Anisotropic plasma: $ {\cal C}_{\rm combo} =- 1,  {\cal C}^{(2)}_{\rm combo}=1, {\cal C}_{x^3}^B(x^3) = -0.001;\ \tilde{t}\equiv\frac{T}{T_c}$}
\end{figure}

II) $ {\cal C}_{\rm combo} =- 1,  {\cal C}^{(2)}_{\rm combo}=100, {\cal C}_{x^3}^B(x^3) = -0.001$; see Figs. 7 and 8\\
\begin{figure}[h!]
\begin{subfigure}{.4\textwidth}
\centering
\includegraphics[width=1.1\textwidth]{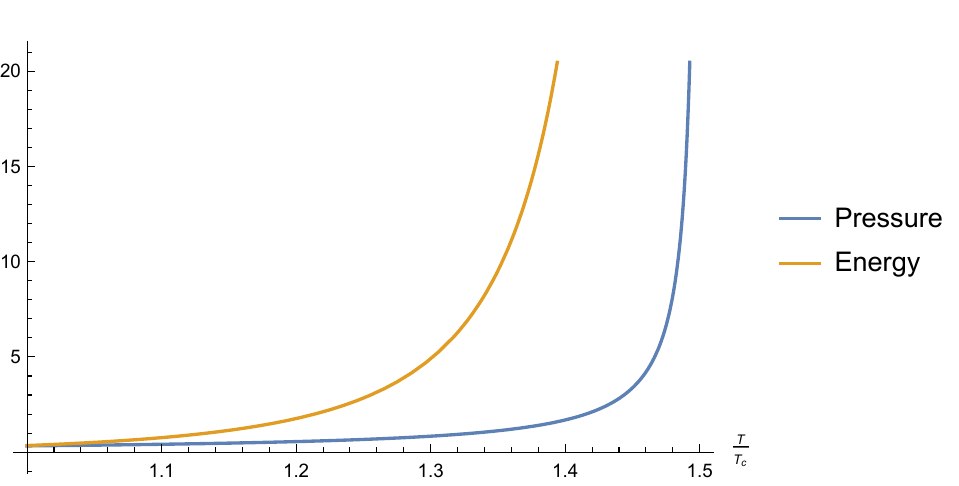}
\caption{Pressure/Energy density-vs-temperature}
\end{subfigure}
\hskip 0.4in
\begin{subfigure}{.4\textwidth}
\centering
\includegraphics[width=1\textwidth]{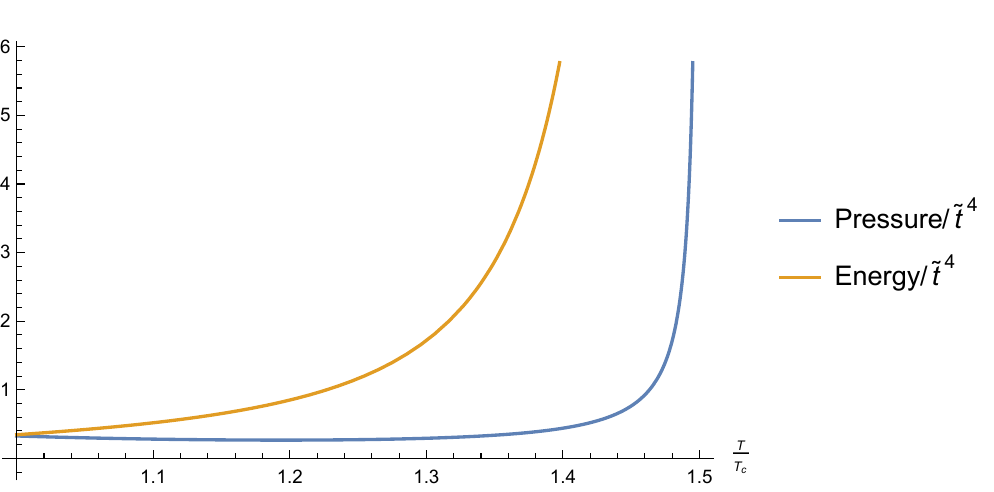}
\caption{$\frac{\rm Energy}{\tilde{t}^4}/\frac{\rm Pressure}{\tilde{t}^4}$-vs-temperature}
\end{subfigure}
\caption{Anisotropic plasma: $ {\cal C}_{\rm combo} =- 1,  {\cal C}^{(2)}_{\rm combo}=100, {\cal C}_{x^3}^B(x^3) = -0.001;\ \tilde{t}\equiv\frac{T}{T_c}$}
\end{figure}

\begin{figure}[h!]
\begin{subfigure}{.4\textwidth}
\centering
\includegraphics[width=1.1\textwidth]{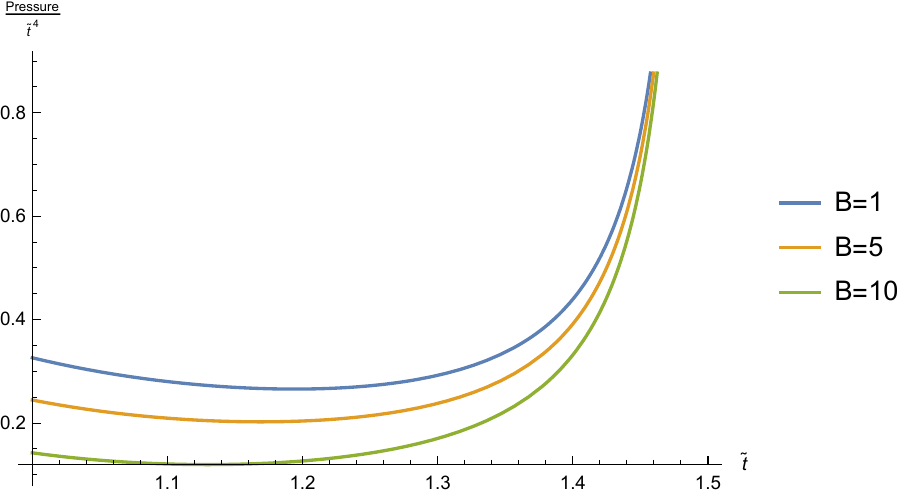}
\caption{$\frac{\rm Pressure}{\tilde{t}^4}$-vs-temperature for $B=0.2, 0.3, 0.5\left({\rm GeV}\right)^2$ (in $e=1$ units)}
\end{subfigure}
\hskip 0.2in
\begin{subfigure}{.5\textwidth}
\centering
\includegraphics[width=1\textwidth]{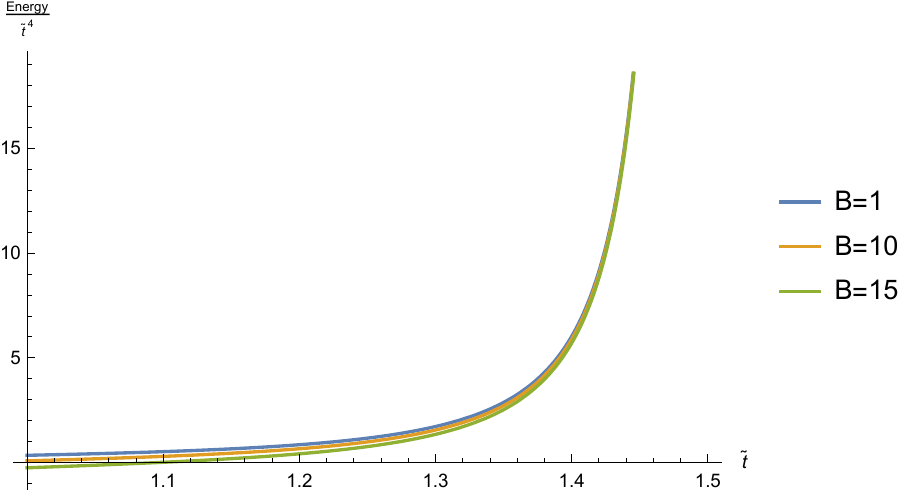}
\caption{$\frac{\rm Energy}{\tilde{t}^4}$-vs-temperature $T\in[T_c, 1.2T_c]$}
\end{subfigure}
\caption{Anisotropic plasma: $ {\cal C}_{\rm combo} =- 1,  {\cal C}^{(2)}_{\rm combo}=100, {\cal C}_{x^3}^B(x^3) = -0.001;\ \tilde{t}\equiv\frac{T}{T_c}$}
\end{figure}

\noindent{\bf (c) Stable Wormhole}: $ {\cal C}_{\rm combo}=-1,  {\cal C}^{(2)}_{\rm combo}=-1, {\cal C}_{x^3}^B(x^3) = 10^5$; see Fig. 7
\begin{figure}[h]
\begin{center}
\includegraphics[width=0.7\textwidth]{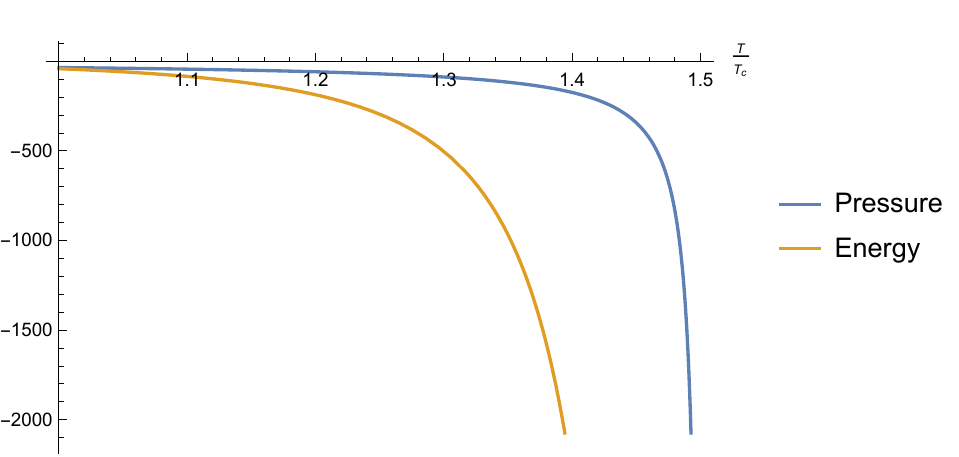}
\end{center}

\caption{A  Stable Wormhole: $ {\cal C}_{\rm combo}=-1,  {\cal C}^{(2)}_{\rm combo}=-1, {\cal C}_{x^3}^B(x^3) = 10^5$}
\end{figure}

To understand the relevance of half Ellis-like wormholes, consider the following.
\begin{itemize}
\item
The asymptotic $AdS_{1+1}$ metric written by delocalizing w.r.t. $\theta_{1,2}, \phi_{1,2},\psi, x^{1,2}$ and disregarding the $\frac{1}{N}$-suppressed non-conformal corrections is given by: $ds^2 = - g_{tt} dt^2 + g_{rr}dr^2$. Using the Eddington-Finkelstein coordinates $\chi_1 = t - \int\sqrt{\frac{g_{rr}}{g_{tt}}}dr, \chi_2 = t + \int\sqrt{\frac{g_{rr}}{g_{tt}}}dr$, one obtains $ds^2_{1+1} = -\frac{g_{tt}}{4}d\chi_1 d\chi_2$. As an example near $r=0$, $ds^2 \sim \frac{\left(\frac{r_h^4}{L^2}\right)^{1/3}d\chi_1 d\chi_2}{\left(\chi_1 - \chi_2\right)^{2/3}}$ where $\chi_1 = t - \frac{v^{3/2}L^2}{3 r_h^4},
\chi_2 = t + \frac{v^{3/2}L^2}{3r_h^4}$ where $v = r^2$. Defining $\kappa = (\chi_1 - \chi_2)^{1/3}$, one sees that
$ds^2_{1+1} \sim dt d\kappa $. Finally defining $T = (t-\kappa)/2, X = (t+\kappa)/2$, one obtains $ds_{1+1}^2 = - dT^2+dX^2$. Schematically, near $r=0$, one thus sees that the $AdS_{1+1}\times S^2_{\rm resolved}$-metric would be given as: $ds^2 = -dT^2 + dX^2 + ((X - T)^2 + a^2) ds^2_{S^2({\rm resolved conifold})}$, which is somewhat like half an  Ellis wormhole(as $X - T \sim \kappa \sim r >0$). 
\item
By writing $\log r = \frac{1}{2}\log r^2$ and approximating $\frac{\left(r + \# a^2\log r\right)}{r}$ by 1 (given that for QCD-inspired values of $(g_s, M, N_f)$ of Table 1, in the UV $\frac{a^2\log r}{r}\sim e^{-{\cal O}(1) N^{1/3}} \frac{\log N}{N^{1/4}}\ll1$ and in the IR, $\frac{a^2\log r}{r}\sim e^{-\frac{N^{1/3}}{{\cal O}(1)}}\frac{N^{1/3}}{{\cal O}(1)}<1$), in the type IIB/IIA ten-dimensional warp factor $h$, $B_2^{\rm NS-NS}, H_3^{\rm NS-NS}, F_3^{IIB, {\rm RR}}/F_2^{IIA, {\rm RR}}$ of the type IIB/IIA dual \cite{metrics} of thermal QCD-like theories at high temperatures, one can show that the type IIB/IIA/${\cal M}$-theory dual of \cite{metrics}, \cite{MQGP, NPB, OR4} have $r\rightarrow-r$ symmetry.
\end{itemize}

\subsection{Impossibility of obtaining a Compact Star}
Let us now discuss the possibility of whether one could obtain a quark star  composed of a perfect fluid whose energy-momentum tensor is given by $T_{\mu\nu} = (E + p) u_\mu u_\nu + p g_{\mu\nu}$ from our background gauge configuration on the world volume of the flavor $D6$-branes. Inside a static spherical (quark) star with the metric:
\begin{equation}
\label{radial-metric-TOV}
ds^2 = - e^{2\Phi(r)}dt^2 + \left(1 - \frac{2 G m(r)}{r}\right)dr^2 + r^2\left(d\theta^2 + \sin^2\theta d\phi^2\right)
\end{equation}
 (by Birkoff's theorem, the exterior will be given by the Schwarzschild metric). Respectively from the $tt, rr$ components and conservation of the energy-momentum tensor, one obtains the Tolman-Oppenheimer-Volkov equations:
\begin{eqnarray}
\label{TOV-equations}
& & P'(r) = - \frac{E(r) m(r)}{r^2}\left(1 + \frac{P(r)}{E(r)}\right)\frac{\left(1 + \frac{4\pi P(r) r^3}{m(r)}\right)}{\left(1 - \frac{2 m(r)}{r}\right)},\nonumber\\
& & \Phi'(r) = - \frac{\frac{2}{E(r)}P'(r)}{\left(1 + \frac{P(r)}{E(r)}\right)},\nonumber\\
& & m'(r) = 4\pi r^2 E(r).
\end{eqnarray}

Motivated by (\ref{p+E-angular-dependence}), (\ref{radial-metric-TOV}) is modified include a $\theta$ dependence in $\Phi$ as well as $m$, i.e., $\Phi = \Phi(r, \theta), m = m(r, \theta)$. With an ansatz, $m(r, \theta) = m_r(r) m_\theta(\theta), P(r,\theta) = P_r(r)P_\theta(\theta), E(r,\theta) = E_r(r)E_\theta(\theta)$, this results in the following modified TOV equations:
{\footnotesize
\begin{eqnarray}
\label{modified-TOV}
& & \hskip -0.8in (i) G_{tt} = \frac{\Sigma_1(r,\theta)}{r^2 (r-2 G
   \ m_r(r)\ m_\theta(\theta ))^2} = 8\pi T_{tt} = 8\pi E_r(r)E_\theta(\theta) e^{-\Phi(r,\theta)},\nonumber\\
& & \hskip -0.8in \Sigma_1(r,\theta) \equiv G e^{2 \Phi(r,\theta )} \Biggl(8 G^2 \ m_r(r)^2
  \ m_\theta(\theta )^3 \ m_r'(r)+2\ m_\theta(\theta ) \left(G
   \ m_r(r)^2 \left(\cot (\theta )\ m_\theta'(\theta
   )+ m_\theta''(\theta )\right)+r^2 \ m_r'(r)\right)-8 G r
   \ m_r(r)\ m_\theta(\theta )^2 \ m_r'(r)\nonumber\\
& & \hskip -0.8in -\ m_r(r)    \left(3 G \ m_r(r)\ m_\theta'(\theta )^2+r \cot (\theta )
  \ m_\theta'(\theta )+r\ m_\theta''(\theta )\right)\Biggr)  ;\nonumber\\
& & \hskip -0.8in (ii) G_{rr}=\frac{r \left(\frac{\partial\Phi}{\partial\theta}^2+\frac{\partial^2\Phi}{\partial \theta^2}+2 r \frac{\partial\Phi}{\partial r}+\cot (\theta ) \frac{\partial\Phi}{\partial\theta}\right)-2 G \ m_r(r)\ m_\theta(\theta ) \left(2 r
   \frac{\partial\Phi}{\partial r}+1\right)}{r^2 (r-2 G \ m_r(r)
  \ m_\theta(\theta ))} = 8\pi T_{rr} = 8\pi \left(1 - \frac{2 G m_r(r)m_\theta(\theta)}{r}\right)P_r(r)P_\theta(\theta);
\nonumber\\
& & \hskip -0.8in  (iii) G_{\theta\theta}=\frac{\Sigma_2(r,\theta)}{r (r-2 G \ m_r(r)\ m_\theta(\theta ))} = 8\pi T_{\theta\theta} = 8\pi\frac{P_r(r)P_\theta(\theta)}{r^2},\nonumber\\
& & \hskip -0.8in \Sigma_2(r,\theta) \equiv 2 G^2 \ m_r(r)^2\ m_\theta(\theta )^2 \left(2 r^2
   \frac{\partial\Phi}{\partial r}^2+2 r^2 \frac{\partial^2\Phi}{\partial r^2}+r
   \frac{\partial\Phi}{\partial r}-1\right)\nonumber\\
& & \hskip -0.8in +G r \ m_r(r) \left(2 G
  \ m_\theta(\theta )^2 \ m_r'(r) \left(r\frac{\partial\Phi}{\partial r}+1\right)+ m_\theta'(\theta ) \left(\frac{\partial\Phi}{\partial\theta}+\cot
   (\theta )\right)- m_\theta(\theta ) \left(4 r^2\frac{\partial\Phi}{\partial r}^2+4 r^2 \frac{\partial^2\Phi}{\partial r^2}+3 r \frac{\partial\Phi}{\partial r}+2 \cot
   (\theta ) \frac{\partial\Phi}{\partial\theta}-1\right)\right)\nonumber\\
& & \hskip -0.8in+r^2 \left(-G
  \ m_\theta(\theta ) \ m_r'(r) \left(r\frac{\partial\Phi}{\partial r}+1\right)+r \left(r \frac{\partial\Phi}{\partial r}^2+\Phi_r^{(1,0)}(r,\theta
   )+r \frac{\partial^2\Phi}{\partial r^2}\right)+\cot (\theta ) \frac{\partial\Phi}{\partial \theta}\right);\nonumber\\
& & \hskip -0.8in (iv)G_{\phi\phi}=\frac{\Sigma_3(r,\theta)}{r (r-2 G \ m_r(r)
  \ m_\theta(\theta ))^2}= 8\pi T_{\phi\phi} = 8\pi \frac{\csc^2\theta P_r(r)P_\theta(\theta)}{r^2},\nonumber\\
& & \hskip -0.8in \Sigma_3(r,\theta) \equiv \sin ^2(\theta ) \Biggl[G^2 r \ m_r(r)^2 \Biggl(-4 G
  \ m_\theta(\theta )^3 \ m_r'(r) \left(r\frac{\partial\Phi}{\partial r}+1\right)-2\ m_\theta(\theta ) \left( m_\theta'(\theta )
   \frac{\partial\Phi}{\partial\theta}+ m_\theta''(\theta )\right)\nonumber\\
& & \hskip -0.8in +4
  \ m_\theta(\theta )^2 \left(3 r^2 \frac{\partial\Phi}{\partial r}^2+3 r^2
   \frac{\partial^2\Phi}{\partial r^2}+2 r\frac{\partial\Phi}{\partial r}+\frac{\partial\Phi}{\partial\theta}^2+\frac{\partial^2\Phi}{\partial \theta^2}-1\right)+3
  \ m_\theta'(\theta )^2\Biggr)\nonumber\\
& & \hskip -0.8in -4 G^3 \ m_r(r)^3\ m_\theta(\theta
   )^3 \left(2 r^2 \frac{\partial\Phi}{\partial r}^2+2 r^2 \frac{\partial^2\Phi}{\partial ^2r}+r \frac{\partial\Phi}{\partial r}-1\right)+G r^2 \ m_r(r) \Biggl(4 G
  \ m_\theta(\theta )^2 \ m_r'(r) \left(r \Phi_r^{(1,0)}(r,\theta
   )+1\right)+ m_\theta'(\theta ) \Phi_r^{(0,1)}(r,\theta
   )\nonumber\\
& & \hskip -0.8in- m_\theta(\theta ) \left(6 r^2 \frac{\partial\Phi}{\partial r}^2+6 r^2
   \frac{\partial^2\Phi}{\partial r^2}+5 r \frac{\partial\Phi}{\partial r}+4
   \frac{\partial\Phi}{\partial\theta}^2+4 \Phi_r^{(0,2)}(r,\theta
   )-1\right)+ m_\theta''(\theta )\Biggr)\nonumber\\
& & \hskip -0.8in +r^3 \left(-G\ m_\theta(\theta
   ) \ m_r'(r) \left(r \frac{\partial\Phi}{\partial r}+1\right)+r^2
   \frac{\partial\Phi}{\partial r}^2+r^2 \frac{\partial^2\Phi}{\partial r^2}+r
   \frac{\partial\Phi}{\partial r}+\Phi_r^{(0,1)}(r,\theta
   )^2+\frac{\partial^2\Phi}{\partial \theta^2}\right)\Biggr];\nonumber\\
& & \hskip -0.8in (vi)\ {\rm We\ require}:\ G_{r\theta} = \frac{G \ m_r(r) \left( m_\theta'(\theta ) \left(r
   \frac{\partial\Phi}{\partial r}+1\right)+2\ m_\theta(\theta )
   \left(\frac{\partial\Phi}{\partial\theta} \left(r \frac{\partial\Phi}{\partial r}-1\right)+r \frac{\partial^2\Phi}{\partial r\partial\theta}\right)\right)+r
   \left(\frac{\partial\Phi}{\partial\theta} \left(1-r \frac{\partial\Phi}{\partial r}\right)-r\frac{\partial^2\Phi}{\partial r\partial\theta}\right)}{r (r-2 G \ m_r(r)
  \ m_\theta(\theta ))} = 8\pi T_{r\theta} = 0.\nonumber\\
& & 
\end{eqnarray}
}
\begin{itemize}
\item
Assuming
\begin{eqnarray}
\label{Phi}
& &\ m_\theta'(\theta )  = 0,\nonumber\\
& & \Phi(r,\theta) = \Phi_r(r) + \epsilon \Phi_r^{(2)}(r)\Phi_\theta(\theta),
\end{eqnarray}
one can show that $G_{r\theta} = {\cal O}(\epsilon^2)$, and hence is negligible. 

\item
$G_{tt} = 8\pi T_{tt}$, up to ${\cal O}(G)$ and approximating $e^{\Phi_r(r) + \epsilon \Phi_r^{(2)}(r)\Phi_\theta}\approx e^{\Phi_r(r)}$, would yield:
\begin{eqnarray}
\label{tt-a}
& & \frac{G e^{2\Phi_r(r)} \left(2 r\ m_\theta(\theta )
   \ m_r'(r)-\ m_r(r) \left(\cot (\theta )
  \ m_\theta'(\theta )+ m_\theta''(\theta
   )\right)\right)}{r} + {\cal O}\left(G^2\right) = 8 \pi  r  E_r(r) \ E_{\theta}(\theta ) e^{-2
  \Phi_r(r)},\nonumber\\
& & 
\end{eqnarray}
which implies:
\begin{eqnarray}
\label{tt-b}
& & \cot (\theta )\ m_\theta'(\theta
   )+ m_\theta''(\theta )={\cal C}_{m_\theta}\ 
  \ m_\theta(\theta ),\nonumber\\
& & \ E_{\theta}(\theta )={\cal C}_{m_\theta}^E\ 
  \ m_\theta(\theta ),
\end{eqnarray}
i.e.,
\begin{eqnarray}
\label{tt-c}
& &\ m_\theta(\theta )=c_1 P_{\frac{1}{2} \left(\sqrt{1-4
   {\cal C}_{m_\theta}\ }-1\right)}(\cos (\theta ))+c_2 Q_{\frac{1}{2}
   \left(\sqrt{1-4 {\cal C}_{m_\theta}\ }-1\right)}(\cos (\theta )).
\end{eqnarray}
Now, one of the simplest ways to ensure $ m_\theta'(\theta )  = 0$ is to have an infinitesimal  ${\cal C}_{m_\theta}\ $ and $c_2=0$ as $P_0(\cos\theta)=1$. One thus sees that (\ref{tt-b}) yields:
\begin{equation}
\label{tt-d}
 E_r(r)=\frac{G e^{4\Phi_r(r)} \left(2 r
   \ m_r'(r)-{\cal C}_{m_\theta}\  \ m_r(r)\right)}{8 \pi 
   {\cal C}_{m_\theta}^E\  r},\ |{\cal C}_{m_\theta}|\ll1.
\end{equation}
 
\item
$G_{rr} = 8\pi T_{rr}$ yields:
\begin{eqnarray}
\label{rr}
& &  P_r (r)=\frac{2 r^2\Phi_r'(r)-2 {\cal C}_{m_\theta}\  G
   \ m_r(r) \left(2 r
  \Phi_r'(r)+1\right)}{{\cal C}_{P_\theta}\  r (r-2
   {\cal C}_{m_\theta}\  G \ m_r(r))^2},\nonumber\\
& &\ P_\theta(\theta )={\cal C}_{P_\theta}\ .
\end{eqnarray}

\item
$G_{\theta\theta} = 8 \pi T_{\theta\theta}$ yields:
{\footnotesize
\begin{eqnarray}
\label{thetatheta-a}
& & \hskip -0.8in -{\cal C}_{m_\theta}\  G \ m_r'(r) \left(r
  \Phi_r'(r)+1\right)-\frac{{\cal C}_{m_\theta}\  G
   \ m_r(r) \left(2 r^2\Phi_r'(r)^2+2 r^2
  \Phi_r''(r)+r\Phi_r'(r)-1\right)}{r}+r \left(r
  \Phi_r'(r)^2+\Phi_r '(r)+r
  \Phi_r''(r)\right)\nonumber\\
& & \hskip -0.8in =\frac{8 \pi  \left(2 r^2
  \Phi_r'(r)-2 {\cal C}_{m_\theta}\  G \ m_r(r)
   \left(2 r\Phi_r'(r)+1\right)\right)}{r^3 (r-2
   {\cal C}_{m_\theta}\  G \ m_r(r))^2},
\end{eqnarray}
}
which if $ \ m_r(r) = r {\cal C}_m\ $, would imply
{\footnotesize
\begin{eqnarray}
\label{thetatheta-b}
& & \hskip -0.8in \left(r^2\Phi_r'(r)^2+\left(r-\frac{16 \pi }{r^3}\right)
  \Phi_r'(r)+r^2\Phi_r''(r)\right)-\frac{2 G
   \left({\cal C}_{m_\theta}\  {\cal C}_m\  \left(r^6
  \Phi_r'(r)^2+\left(r^5+16 \pi  r\right)
  \Phi_r'(r)+r^6\Phi_r''(r)-8 \pi
   \right)\right)}{r^4} + {\cal O}\left(G^2\right) = 0.\nonumber\\
& & 
\end{eqnarray}
}
Making an ansatz $\Phi_r (r) =\Phi_r^{(0)}\ (r) + G \Phi_r^{(1)}\ (r)$ and substituting into (\ref{thetatheta-b}) obtains:
\begin{eqnarray}
\label{thetatheta-c}
& &\Phi_r^{(0)}\ (r)=\log \left(-{Ei}\left[-\frac{4 \pi
   }{r^4}\right]+4 c_{\Phi_r^{0,1}}\right)+c_{\Phi_r^{0,2}}\nonumber\\
& & \Downarrow {\rm near}\ r=0\nonumber\\
& & \log (4 c_{\Phi_r^{0,1}})+c_{\Phi_r^{0,2}}.
\end{eqnarray}
Near $r=0$, one can show:
\begin{eqnarray}
\label{thetatheta-d}
& & r^2 \Phi_r^{(1)}\ '(r)^2+\left(r-\frac{16 \pi }{r^3}\right)
   \Phi_r^{(1)}\ '(r)+r^2 \Phi_r^{(1)}\ ''(r)=\frac{16 \pi 
   {\cal C}_{m_\theta}\  {\cal C}_m\ }{r^4},
\end{eqnarray}
that is solved to yield:
\begin{eqnarray}
\label{thetatheta-d}
& & \Phi_r^{(1)}\ (r)=\int^r-\frac{4 \pi 
   \left({\cal C}_{m_\theta}\  c_1 \,
   _1F_1\left(1-\frac{{\cal C}_{m_\theta}\ 
   {\cal C}_m\ }{4};2;-\frac{4 \pi }{\tilde{r}^4}\right)
   {\cal C}_m\ -4 G_{1,2}^{2,0}\left(\frac{4 \pi }{\tilde{r}^4}|
\begin{array}{c}
 \frac{{\cal C}_{m_\theta}\  {\cal C}_m\ }{4} \\
 -1,0 \\
\end{array}
\right)\right)}{\tilde{r}^5 \left(c_1 \,
   _1F_1\left(-\frac{{\cal C}_{m_\theta}\ 
   {\cal C}_m\ }{4};1;-\frac{4 \pi
   }{\tilde{r}^4}\right)+G_{1,2}^{2,0}\left(\frac{4 \pi }{\tilde{r}^4}|
\begin{array}{c}
 \frac{{\cal C}_{m_\theta}\  {\cal C}_m\ }{4}+1 \\
 0,0 \\
\end{array}
\right)\right)}d\tilde{r}+c_2.\nonumber\\
& & 
\end{eqnarray}
Near $r=0$, the integral in (\ref{thetatheta-d}) can be shown to be approximated by:
{\footnotesize
\begin{eqnarray}
\label{thetatheta-e}
& & \hskip -0.8in\Phi_r^{(1)}(r) = 2^{-{\cal C}_{m_\theta}\  {\cal C}_m\ } \pi
   ^{-\frac{{\cal C}_{m_\theta}\  {\cal C}_m\ }{2}}
   \Biggl[\frac{{\cal C}_{m_\theta}\ ^2 {\cal C}_m\ ^2 r^{2
   {\cal C}_{m_\theta}\  {\cal C}_m\ +4} \,
   _2F_1\left(1,\frac{{\cal C}_{m_\theta}\ 
   {\cal C}_m\ }{2}+1;\frac{{\cal C}_{m_\theta}\ 
   {\cal C}_m\ }{2}+2;-\frac{{\cal C}_{m_\theta}\ ^2 r^4
   {\cal C}_m\ ^2}{64 \pi }\right)}{32 \pi 
   {\cal C}_{m_\theta}\  {\cal C}_m\ +64 \pi }\nonumber\\
& & \hskip -0.8in-\frac{1}{2}
   r^{2 {\cal C}_{m_\theta}\  {\cal C}_m\ }+\frac{(4 \pi
   )^{{\cal C}_{m_\theta}\  {\cal C}_m\ } r^{1-2
   {\cal C}_{m_\theta}\  {\cal C}_m\ }}{2
   {\cal C}_{m_\theta}\  {\cal C}_m\ -1}\Biggr] = r^{2 {\cal C}_{m_\theta}\  {\cal C}_m\ }
   \left(\frac{2^{-{\cal C}_{m_\theta}\  {\cal C}_m\ -5}
   {\cal C}_{m_\theta}\ ^2 \pi ^{-\frac{{\cal C}_{m_\theta}\ 
   {\cal C}_m\ }{2}-1} {\cal C}_m\ ^2
   r^4}{{\cal C}_{m_\theta}\ 
   {\cal C}_m\ +2}+O\left(r^5\right)\right)\nonumber\\
& & \hskip -0.8in -2^{-{\cal C}_{m_\theta} {\cal C}_m\ -1} \pi ^{-\frac{{\cal C}_{m_\theta}\ 
   {\cal C}_m\ }{2}} r^{2 {\cal C}_{m_\theta}\ 
   {\cal C}_m\ }+\frac{2^{{\cal C}_{m_\theta}\ 
   {\cal C}_m\ } \pi ^{\frac{{\cal C}_{m_\theta}\ 
   {\cal C}_m\ }{2}} r^{1-2 {\cal C}_{m_\theta}\ 
   {\cal C}_m\ }}{2 {\cal C}_{m_\theta}\  {\cal C}_m\ -1}.
\end{eqnarray}
}
Assuming ${\cal C}_{m_\theta}\ {\cal C}_m\ =1$,  
\begin{equation}
\label{thetatheta-f}
\Phi_r(r\sim0) = \log (4 c_{\Phi_r^{0,1}})+c_{\Phi_r^{0,2}} + G \left( \frac{2\sqrt{\pi}}{r}\right).
\end{equation}

\item
$G_{\phi\phi} = 8\pi T_{\phi\phi}$ using (\ref{Phi}), implies:
{\footnotesize
\begin{eqnarray}
\label{phiphi-a}
& & \sin ^2(\theta ) \left(\epsilon ^2 \Phi_r^{(2)}(r)^2 \Phi_\theta '(\theta )^2+\epsilon  \Phi_r^{(2)}(r)
   \Phi_\theta ''(\theta )+r^2\Phi_r'(r)^2+r^2\Phi_r''(r)+r
  \Phi_r'(r)\right)\nonumber\\
& & -\frac{{\cal C}_{m_\theta}\  G \sin ^2(\theta ) \left(r \ m_r'(r) \left(r
  \Phi_r'(r)+1\right)+\ m_r(r) \left(2 r^2\Phi_r'(r)^2+2 r^2\Phi_r''(r)+r
  \Phi_r'(r)-1\right)\right)}{r} + {\cal O}\left(G^2\right) = 0.\nonumber\\
& &    
\end{eqnarray}
}
Assuming
\begin{equation}
\label{phiphi-b}
r^2\Phi_r'(r)^2+r^2\Phi_r''(r)+r\Phi_r '(r)=0
\end{equation}
(so that $G_{\phi\phi}$, like $T_{\phi\phi}$, can be written as a product of a $\theta$-dependent and an $r$-dependent functions), implying:
\begin{equation}
\label{phiphi-c}
\Phi_r (r)=\log (\log (r)+c_{\Phi_r^{(1)}\ })+c_{\Phi_r^{(2)}}.
\end{equation}
Assuming $r\rightarrow0$ as $r\sim N^{-\alpha_r}, \alpha_r\geq1$, $c_{c_{\Phi_r^{(2)}\ }}>0, c_{c_{\Phi_r^{(2)}\ }}>\log N$, one reconciles (\ref{phiphi-c}) with the ${\cal O}(G^0)$ result of (\ref{thetatheta-f}).
 
One hence obtains:
\begin{eqnarray}
\label{Tphiphi}
& & \hskip -0.8in T_{\phi\phi} = \frac{16 \pi  \csc ^2(\theta )}{r^4 (\log
   (r)+c_{\Phi_r^{(1)}\ })}-\frac{16 G \left(\pi 
   {\cal C}_{m_\theta}\ \ \ m_r(r) \csc
   ^2(\theta ) (c_{\Phi_r^{(1)}\ }+\log
   (r)-2)\right)}{r^5 (\log
   (r)+c_{\Phi_r^{(1)}\ })} + {\cal O}\left(G^2\right).\nonumber\\
& & 
\end{eqnarray}
Therefore,
\begin{eqnarray}
\label{phiphi-d}
& & \epsilon  \Phi_r^{(2)}(r) \sin ^2(\theta )
   \Phi_\theta ''(\theta )=\frac{16 \pi  \csc
   ^2(\theta )}{r^4 (\log (r)+c_{\Phi_r^{(1)}\ })},
\end{eqnarray}
implying
\begin{eqnarray}
\label{phiphi-e}
& & \Phi_r^{(2)}(r)=\frac{16 \pi }{\epsilon  r^4 (\log
   (r)+c_{\Phi_r^{(1)}\ })}.
\end{eqnarray}
Assuming $|c_{\Phi_r^{(1)}\ }| \gg  |\log r| \forall r$ assuming $r$ approaches any value like $N^{\kappa_r}$, $\epsilon c_{\Phi_r^{(1)}\ }= c_{\Phi_r^{\rm finite}}$= finite as $\epsilon \rightarrow 0$, implying $\Phi_r^{(2)}(r)=\frac{16 \pi }{r^4 c_{\Phi_r^{\rm finite}}}$.
Further, 
\begin{eqnarray}
\label{phiphi-f}
& & \sin ^2(\theta ) \Phi_\theta ''(\theta )=\csc ^2(\theta ),
\end{eqnarray}
which yields:
\begin{equation}
\label{phiphi-g}
\Phi_\theta (\theta )=c_{\Phi_\theta^1}\ +\frac{\csc
   ^2(\theta )}{6}-\frac{2}{3} \log (\sin (\theta )).
\end{equation}

\end{itemize}

The conservation of the energy-momentum tensor yields ($T^{t\mu}_{\ \ \ ;\mu} = T^{\phi\mu}_{\ \ \ ;\mu} \equiv0$):
{\scriptsize
\begin{eqnarray}
\label{conserv-EM-tensor}
& &\hskip -0.8in T^{r\mu}_{\ \  \ ;\mu} = \frac{ E_r(r) \ E_{\theta}(\theta )
   \frac{\partial\Phi}{\partial r} (r-2 G \ m_r(r)\ m_\theta(\theta
   ))}{r}+ P_r (r)\ P_\theta(\theta ) \frac{\partial\Phi}{\partial r} \left(1-\frac{2
   G \ m_r(r)\ m_\theta(\theta )}{r}\right)+ P_r (r)\ P_\theta(\theta
   ) \left(\frac{2 G \ m_r(r)\ m_\theta(\theta )}{r^2}-\frac{2 G
  \ m_\theta(\theta ) \ m_r'(r)}{r}\right)\nonumber\\
  & & \hskip -0.8in +\frac{2 G
  \ m_\theta(\theta )  P_r (r)\ P_\theta(\theta ) \left(r
   \ m_r'(r)-\ m_r(r)\right) \left(1-\frac{2 G \ m_r(r)
  \ m_\theta(\theta )}{r}\right)}{r (r-2 G \ m_r(r)
  \ m_\theta(\theta ))}+ P_\theta (\theta )  P_r '(r) \left(1-\frac{2 G
   \ m_r(r)\ m_\theta(\theta )}{r}\right)-\frac{ P_r (r)
  \ P_\theta(\theta ) (r-2 G \ m_r(r)\ m_\theta(\theta
   ))}{r^2}\nonumber\\
& &  \hskip -0.8in  +\frac{ P_r (r)\ P_\theta(\theta ) (2 G \ m_r(r)
  \ m_\theta(\theta )-r)}{r^2}+\frac{2  P_r (r)\ P_\theta(\theta )
   \left(1-\frac{2 G \ m_r(r)\ m_\theta(\theta )}{r}\right)}{r}\nonumber\\
& & \hskip -0.8in   = \left( E_r(r) \ E_{\theta}(\theta ) \frac{\partial\Phi}{\partial r}+P_\theta(\theta ) \left( P_r (r) \frac{\partial\Phi}{\partial r}+ P_r
   '(r)\right)\right)-\frac{2 G \left(\ m_r(r)\ m_\theta(\theta )
   \left( E_r(r) \ E_{\theta}(\theta ) \frac{\partial\Phi}{\partial r}+ P_\theta (\theta ) \left( P_r (r) \frac{\partial\Phi}{\partial r}+ P_r
   '(r)\right)\right)\right)}{r} + {\cal O}\left(G^2\right) \nonumber\\
& & \hskip -0.8in = \frac{2 \left(r\Phi_r'(r)^2-\Phi_r '(r)+r
  \Phi_r''(r)\right)}{r^2}+\frac{{\cal C}_{m_\theta}\ 
   G \left(r \ m_r'(r) \left(r \left(r^2 e^{4
  \Phi_r(r)}+16 \pi \right)\Phi_r'(r)-8 \pi
   \right)-24 \pi  \ m_r(r) \left(r
  \Phi_r'(r)-1\right)\right)}{4 \pi 
   r^4}+O\left(G^2\right);\nonumber\\
& & \hskip -0.8in T^{\theta\mu}_{\ \ \ ;\mu} = \frac{ E_r(r) \ E_{\theta}(\theta ) 
\frac{\partial\Phi}{\partial\theta}}{r^2}+\frac{ P_r (r)\ P_\theta(\theta ) \frac{\partial\Phi}{\partial\theta}}{r^2}+\frac{G \ m_r(r)  P_r (r)\ P_\theta(\theta )
  \ m_\theta'(\theta )}{r^2 (r-2 G \ m_r(r)\ m_\theta(\theta
   ))}-\frac{G \ m_r(r)  P_r (r)\ P_\theta(\theta )\ m_\theta'(\theta) \left(1-\frac{2 G \ m_r(r)\ m_\theta(\theta )}{r}\right)}{r (r-2 G
   \ m_r(r)\ m_\theta(\theta ))^2}+\frac{ P_r (r)\ P_\theta'(\theta
   )}{r^2}=0.\nonumber\\
& &    
\end{eqnarray} 
}
Now, $T^{\theta\mu}_{\ \ \ ;\mu}\equiv0$ as $T^{\theta\mu}_{\ \ \ ;\mu}\sim\epsilon \Phi_r^{(2)}(r)\Phi_\theta (\theta )$, which is negligible as $|\epsilon|\ll1$. Substituting $\ m_r(r)=r {\cal C}_m\ $ into
$T^{r\mu}_{\ \ \ ;\mu}=0$, writing $\Phi_r (r) =\Phi_r^{(0)}\ (r) + G \Phi_r^{(1)}\ (r)$, one obtains at ${\cal O}(G^0)$:
\begin{equation}
\label{Trmumu-ii}
r\Phi_r^{(0)}\ '(r)^2-\Phi_r^{(0)}\ '(r)+r\Phi_r^{(0)}\ ''(r)=0,
\end{equation}
which is solved to obtain:
\begin{equation}
\label{Trmumu-iii}
\Phi_r^{(0)}\ (r)=\log \left(r^2+2
   \tilde{c}_{\Phi_r^{0,1}}\right)+\tilde{c}_{\Phi_r^{0,2}}.
\end{equation}
Near $r=0$, one needs to solve:
\begin{equation}
\label{Trmumu-iv}
r \Phi_r^{(1)}\ '(r)^2-\Phi_r^{(1)}\ '(r)+r
   \Phi_r^{(1)}\ ''(r)=\frac{4 {\cal C}_{m_\theta}\ 
   {\cal C}_m\ }{r^3},
\end{equation}
which is solved to yield:
\begin{eqnarray}
\label{Trmumu-v}
& & \hskip -0.8in  \Phi_r^{(1)}\ (r) = \tilde{c}_{\Phi_r^{(2)}} + \int^r d\tilde{r} \frac{\sqrt{{\cal C}_{m_\theta}\ {\cal C}_m\ }}{\left(4 \sqrt{\pi } I_1\left(\frac{2
   \sqrt{{\cal C}_{m_\theta}\ 
   {\cal C}_m\ }}{\tilde{r}}\right)+K_1\left(\frac{2
   \sqrt{{\cal C}_{m_\theta}\  {\cal C}_m\ }}{\tilde{r}}\right)
   \tilde{c}_{\Phi_r^{(1)}}\right)
   \tilde{r}^2}\times\nonumber\\
& &  \hskip -0.8in \Biggl( \left[-4 \sqrt{\pi } I_0\left(\frac{2
   \sqrt{{\cal C}_{m_\theta}\ 
   {\cal C}_m\ }}{\tilde{r}}\right)-4 \sqrt{\pi }
   I_2\left(\frac{2 \sqrt{{\cal C}_{m_\theta}\ 
   {\cal C}_m\ }}{\tilde{r}}\right)+\left(K_0\left(\frac{2
   \sqrt{{\cal C}_{m_\theta}\ 
   {\cal C}_m\ }}{\tilde{r}}\right)+K_2\left(\frac{2
   \sqrt{{\cal C}_{m_\theta}\ 
   {\cal C}_m\ }}{\tilde{r}}\right)\right)
    \tilde{c}_{\Phi_r^{(1)}}\right]\nonumber\\
& & \hskip -0.8in   +4 \sqrt{\pi } I_1\left(\frac{2
   \sqrt{{\cal C}_{m_\theta}\  {\cal C}_m\ }}{\tilde{r}}\right)
   \tilde{r}+K_1\left(\frac{2 \sqrt{{\cal C}_{m_\theta}\ 
   {\cal C}_m\ }}{\tilde{r}}\right)\tilde{c}_{\Phi_r^{(1)}}
   \tilde{r}\Biggr).\nonumber\\
& & 
\end{eqnarray}
Setting $ \tilde{c}_{\Phi_r^{(1)}}=0$, and using:
\begin{eqnarray}
\label{Trmumu-vi}
& & \int^r\frac{\sqrt{{\cal C}_{m_\theta}\ 
   {\cal C}_m\ } \left(-4 \sqrt{\pi } I_0\left(\frac{2
   \sqrt{{\cal C}_{m_\theta}\ 
   {\cal C}_m\ }}{\tilde{r}}\right)-4 \sqrt{\pi }
   I_2\left(\frac{2 \sqrt{{\cal C}_{m_\theta}\ 
   {\cal C}_m\ }}{\tilde{r}}\right)\right)+4 \sqrt{\pi } \tilde{r}
   I_1\left(\frac{2 \sqrt{{\cal C}_{m_\theta}\ 
   {\cal C}_m\ }}{\tilde{r}}\right)}{4 \sqrt{\pi } \tilde{r}^2
   I_1\left(\frac{2 \sqrt{{\cal C}_{m_\theta}\ 
   {\cal C}_m\ }}{\tilde{r}}\right)}d\tilde{r}\nonumber\\
   & & =\log
   \left[I_1\left(\frac{2 \sqrt{{\cal C}_{m_\theta}\ 
   {\cal C}_m\ }}{r}\right)\right]+\log (r),
\end{eqnarray}
one obtains:
\begin{eqnarray}
\label{Trmumu-vii}
& &\Phi_r(r\sim0) = \log (2 \tilde{c}_{\Phi_0^{(1)}}\ )+\tilde{c}_{\Phi_0^{(2)}}\  + G \left(\frac{2 \sqrt{{\cal C}_{m_\theta}\ 
   {\cal C}_m\ }}{r}-\frac{1}{4} \log
   \left({\cal C}_{m_\theta}\ 
   {\cal C}_m\ \right)+\tilde{c}_{\Phi_1^{(2)}}\ +\frac{3 \log
   (r)}{2}\right).\nonumber\\
& & 
\end{eqnarray}
For ${\cal C}_{m_\theta}\     {\cal C}_m\ =1$, assuming $r\rightarrow0$ as $r=N^{-\alpha_r}, \alpha_r\geq1$ and $2N^{\kappa_r} = 2N^{\alpha_r}\sqrt{\pi}+\tilde{c}_{\Phi_1^{(2)}}\ -\frac{3\kappa_r \log N}{2}$,
 (\ref{Trmumu-vii}) and the ${\cal O}(G)$ result of (\ref{thetatheta-f}) are shown to be mutually consistent.  

Upon comparison of (\ref{tt-b}) and (\ref{rr}) with (\ref{p+E-angular-dependence}), one concludes that one can not have a compact-star solution. Instead, we have a pressure/energy anisotropic phase of plasma.

\section{Photoproduction}
\label{photo-prod}

In this section, we will derive the spectral density of photon production in the UV region, which is related to the differential photon production rate. First considering the case with zero magnetic field, and utilizing the solutions of the gauge field derived in section \ref{UV-beta0}, one can derive the gauge invariant gauge fluctuations,  $E_{\rm long}$, and $E_{\rm trans }$ [see appendix A of \cite{Holographic-Photoprod-B} for more details], the longitudinal and transverse directions here are considered based on photon's $4$-momentum, say $k^{\mu}=(\omega,\omega,0,0)$ in $\mathbb{R}^{1,3}( t,x^1,x^2,x^3)$ with a uniform and strong magnetic field along $x^3: {\bf B} = (0,0,B)$, where $x^2$-direction is the transverse direction. By solving EOMs for $E_{\rm trans }$ one can derive the spectral density of photon production. In the second part, we will repeat the same procedure in the presence of a strong magnetic field.   
\subsection{Photoproduction in the absence of magnetic field}
\label{photoprod-B_0}
Here we derive the EOM for $E_{\rm trans }$. After solving the EOM we derive the spectral density of photon production in the absence of a magnetic field in the UV region. Using results of gauge fluctuations in the UV for $B=0$ of (\ref{AmuB0UV}),
{\scriptsize
\begin{eqnarray}
\label{Ginverse}
& & \hskip -0.8in G^{x^2x^2}_{B=0,{\rm UV}}=\frac{2
   \sqrt{\pi } {\cal C}_{a_{x3}^Z(x3)}\ ^{B=0,\ {\rm UV}}\  ^2
   {\cal C}_{a_{x3}^{\rho\phi}}^{B=0,\ {\rm UV}}\ ^2 {\cal C}_{x3}^{\phi Z}\   ^2
   \sqrt{{g_s}} \sqrt{N} e^{-2 Z} \cos ^2(\phi
   )}{{r_h}^2 \left({\cal C}_{a_{x3}^Z(x3)}\ ^{B=0,\ {\rm UV}}\  ^2
   {\cal C}_{a_{x3}^{\rho\phi}}^{B=0,\ {\rm UV}}\ ^2 {\cal C}_{x3}^{\phi Z}\   ^2
   \left(\rho
   ^2-1\right)+{\cal C}_{a_t^{x3}(x3)}\ ^{B=0,\ {\rm UV}}\ ^2\right)}\nonumber\\
& & \hskip -0.8in+\frac{2 \sqrt{\pi } \beta 
   {\cal C}_{a_t^{x3}(x3)}\ ^{B=0,\ {\rm UV}}\ ^2 \sqrt{{g_s}} \sqrt{N}
   \rho ^2 e^{10 Z} \cos ^2(\phi ) ({\cal C}_{zz}-2
   {\cal C}_{\theta_1z}+2{\cal C}_{\theta_1x})
   \left({\cal C}_{a_t^{x3}(x3)}\ ^{B=0,\ {\rm UV}}\ ^2-
   {\cal C}_{a_{x3}^Z(x3)}\ ^{B=0,\ {\rm UV}}\  ^2 {\cal C}_{x3}^{\rho\phi}\  ^2 {\cal C}_{x3}^{\phi Z}\   ^2\right)}{25 {\cal C}_{a_t^{x3}(x3)}^{B=0,\ {\rm UV}}\   ^2
   {\cal C}_{a_t^{\phi}(\rho)}^{B=0,\ {\rm UV}}\   ^2
   {\cal C}_{a_t^{\phi}(Z)}^{B=0,\ {\rm UV}}\    ^2
   \left({\cal C}_{a_{x3}^Z(x3)}\ ^{B=0,\ {\rm UV}}\  ^2 {\cal C}_{x^3}^{\rho\phi}\ ^2 {\cal C}_{x3}^{\phi Z}\   ^2 \left(\rho
   ^2-1\right)+{\cal C}_{a_t^{x3}(x3)}\ ^{B=0,\ {\rm UV}}\ ^2\right)^2}\nonumber\\
& & \hskip -0.8in G^{ZZ}_{B=0,{\rm UV}}=-\frac{\sqrt{\frac{1}{N}} {r_h}^2
   e^{10 Z}}{32 \sqrt{\pi } {\cal C}_{a_{x3}^Z(Z)}^{B=0,\ {\rm UV}}\  ^2
   {\cal C}_{a_t^{x3}(x3)}\ ^{B=0,\ {\rm UV}}\ ^2 \sqrt{{g_s}}
   ({\cal C}_{a_{x3}^Z{\rho}}^{B=0,\ {\rm UV}}\  \log (\rho )+c_1){}^2}-\frac{\beta  \sqrt{\frac{1}{N}} \rho ^4
   {r_h}^4 e^{20 Z} ({\cal C}_{zz}-2
   {\cal C}_{\theta_1z}+2{\cal C}_{\theta_1x})}{512 \sqrt{\pi }
   {\cal C}_{a_t^{\rho}}^{B=0,\ {\rm UV}}\   ^4
   {\cal C}_{a_{x3}^Z(Z)}^{B=0,\ {\rm UV}}\  ^4 {\cal C}_{a_t^{x3}(x3)}\ ^{B=0,\ {\rm UV}}\ ^4
   \sqrt{{g_s}}}\nonumber\\
   & & G^{tt}_{B=0,\ {\rm UV}}=\frac{\sqrt{\frac{1}{N}}
   \rho ^2 {r_h}^2 e^{10 Z}}{2 \sqrt{\pi }
   {\cal C}_{a_{x3}^Z{\rho}}^{B=0,\ {\rm UV}}\ ^2
   {\cal C}_{a_{x3}^Z(Z)}^{B=0,\ {\rm UV}}\  ^2 {\cal C}_{a_t^{x3}(x3)}\ ^{B=0,\ {\rm UV}}\ ^2
   \sqrt{{g_s}}}\frac{\sqrt{\pi } \beta 
   \sqrt{{g_s}} \sqrt{N} \rho ^2 e^{8 Z}
   ({\cal C}_{zz}-2 {\cal C}_{\theta_1z}+2{\cal C}_{\theta_1x})}{8
   {\cal C}_{a_t^{\rho}}^{B=0,\ {\rm UV}}\   ^2
   {\cal C}_{a_{x3}^Z(Z)}^{B=0,\ {\rm UV}}\  ^2
   {\cal C}_{a_t^{x3}(x3)}\ ^{B=0,\ {\rm UV}}\ ^2},\nonumber\\
& & \hskip -0.8in {\cal L}_{\rm DBI}^{B=0,\ {\rm UV}}\  =\frac{5 i \beta 
   {\cal C}_{a_t^{\rho}}^{B=0,\ {\rm UV}}\    {\cal C}_{a_t^{x3}(x3)}^{B=0,\ {\rm UV}}\   
   {\cal C}_{a_t^{\phi}(\rho)}^{B=0,\ {\rm UV}}\   
   {\cal C}_{a_t^{\phi}(Z)}^{B=0,\ {\rm UV}}\     {\cal C}_{a_{x3}^Z(Z)}^{B=0,\ {\rm UV}}\  
   {\cal C}_{a_t^{x3}(x3)}\ ^{B=0,\ {\rm UV}}\  \log r_h ^3 M^{\rm UV}
   \left(\frac{1}{N}\right)^{13/20} e^{-3 Z}}{2187 \sqrt{2} 3^{5/6} \pi 
   \epsilon ^5 {g_s} \log N ^4 {N_f^{\rm UV}} \rho
   ^2 \alpha _{\theta _2}^5}\nonumber\\
& & \hskip -0.8in \times  \left(-19683
   \sqrt{6} \alpha _{\theta _1}^6-6642 \alpha _{\theta
   _2}^2 \alpha _{\theta _1}^3+40 \sqrt{6} \alpha
   _{\theta _2}^4\right)  -\frac{5 i \sqrt{2}
   {\cal C}_{a_t^{x3}(x3)}^{B=0,\ {\rm UV}}\    {\cal C}_{\phi^\rho}^{B,\ {\rm UV}}\  {\cal C}_{a_t^{\phi}(Z)}^{B=0,\ {\rm UV}}\    
   {\cal C}_{a_{x3}^Z{\rho}}^{B=0,\ {\rm UV}}\ 
   {\cal C}_{a_{x3}^Z(Z)}^{B=0,\ {\rm UV}}\   {\cal C}_{a_t^{x3}(x3)}\ ^{B=0,\ {\rm UV}}\ 
   N^{3/5} {r_h}^2 e^{-5 Z}}{\sqrt[3]{3} {g_s}
   \rho  \alpha _{\theta _2}^2}.\nonumber\\
& &       
\end{eqnarray}
}

Assume
\begin{equation}
\label{assumption_rho}
\rho  \left(\int _1^{\rho }\frac{{\cal C}_{a_{x3}^Z{\rho}}^{B=0,\ {\rm UV}}\ (\kappa)}{\kappa}d\kappa+c_1\right)={\cal C}_{a_t^\rho}^{B=0,\ {\rm UV}}\ ,
\end{equation}
to obtain a sensible $E_{\rm trans}$ EOM.

Writing $E_{\rm trans} = E_{\rm trans}^{\beta^0} + \beta E_{\rm trans}^\beta$, using standard techniques \cite{Holographic-Photoprod-B}, the EOM for $E_{\rm trans}^\beta\ (Z)$ turns to be:
{\scriptsize
\begin{eqnarray}
\label{EtransbetaB0UV}
& & \hskip -0.8in {\cal X}^{\beta^0} 
   E_{\rm trans}^\beta\ '(Z)+E_{\rm trans}^\beta\ ''(Z)+E_{\rm trans}^\beta(Z) \left(-\left(q^2 {\cal Y}_2^{\beta^0}+w^2
   {\cal Y}_1^{\beta^0}\right)\right)-E_{\rm trans}^{\beta^0}\ (Z) \left(q^2
   {\cal Y}_2^{\beta}+w^2 {\cal Y}_1^\beta \right)+{\cal X}^\beta 
   E_{\rm trans}^{\beta^0}\ (Z)=0,
\end{eqnarray}
}
wherein,
{\scriptsize
\begin{eqnarray}
\label{Y1Y2Xprime-defs-i}
& & {\cal X} \equiv \partial_Z \log\left({\cal L}_{\rm DBI}^{B=0,\ {\rm UV}} G^{ZZ}_{B=0,{\rm UV}} G^{x^2x^2}_{B=0,{\rm UV}}\right) = {\cal X}^{\beta^0} + \beta {\cal X}^{\beta},\nonumber\\
& & {\cal Y}_1 \equiv \frac{G^{tt}_{B=0,{\rm UV}}}{G^{ZZ}_{B=0,{\rm UV}}} = {\cal Y}_1^{B=0,\ \beta^0} + \beta {\cal Y}_1^{B=0,\ \beta},\nonumber\\
& & {\cal Y}_2 \equiv \frac{G^{x^1x^1}_{B=0,{\rm UV}}}{G^{ZZ}_{B=0,{\rm UV}}} = {\cal Y}_2^{B=0,\ \beta^0} + \beta {\cal Y}_2^{B=0,\ \beta},
\end{eqnarray}
}
where:
{\scriptsize
\begin{eqnarray}
\label{EOM_Etrans_B0_UV}
& & \hskip -0.8in {\cal Y}_1^{B=0,\ \beta^0}=-\frac{16 \rho ^2 ({\cal C}_{a_{x3}^Z{\rho}}^{B=0,\ {\rm UV}}\  \log (\rho )+c_1){}^2}{{\cal C}_{a_{x3}^Z{\rho}}^{B=0,\ {\rm UV}}\ ^2},\nonumber\\
& & \hskip -0.8in {\cal Y}_1^{B=0,\ \beta}=-\frac{4 \pi  \beta  {g_s} N \rho ^2 e^{-2 Z} ({\cal C}_{zz}-2 {\cal C}_{\theta_1z}+2{\cal C}_{\theta_1x})
   ({\cal C}_{a_{x3}^Z{\rho}}^{B=0,\ {\rm UV}}\  \log (\rho )+c_1){}^2}{{\cal C}_{a_t^{\rho}}^{B=0,\ {\rm UV}}\   ^2 {r_h}^2};\nonumber\\
& & \hskip -0.8in {\cal Y}_2^{B=0,\ \beta^0}=-\frac{64 \pi  {\cal C}_{a_{x3}^Z(x3)}\ ^{B=0,\ {\rm UV}}\  ^2 {\cal C}_{a_{x3}^Z(Z)}^{B=0,\ {\rm UV}}\  ^2 {\cal C}_{a_t^{x3}(x3)}\ ^{B=0,\ {\rm UV}}\ ^2
   {\cal C}_{x^3}^{\rho\phi}\ ^2 {\cal C}_{x3}^{\phi Z}\   ^2 {g_s} N e^{-12 Z} \cos ^2(\phi ) ({\cal C}_{a_{x3}^Z(\rho)}^{B=0,\ {\rm UV}}\  \log (\rho )+c_1){}^2}{{r_h}^4 \left({\cal C}_{a_{x3}^Z(x3)}\ ^{B=0,\ {\rm UV}}\  ^2 {\cal C}_{x^3}^{\rho\phi}\ ^2
   {\cal C}_{x3}^{\phi Z}\   ^2 \left(\rho ^2-1\right)+{\cal C}_{a_t^{x3}(x3)}\ ^{B=0,\ {\rm UV}}\ ^2\right)},\nonumber\\
& & \hskip -0.8in {\cal Y}_2^{B=0,\ \beta}\nonumber\\
& & \hskip -0.8in=-\frac{64 \pi  \beta  {\cal C}_{a_{x3}^Z(Z)}^{B=0,\ {\rm UV}}\  ^2 {\cal C}_{a_t^{x3}(x3)}\ ^{B=0,\ {\rm UV}}\ ^4 {g_s} N \rho ^2 \cos
   ^2(\phi ) ({\cal C}_{zz}-2 {\cal C}_{\theta_1z}+2{\cal C}_{\theta_1x}) ({\cal C}_{a_{x3}^Z{\rho}}^{B=0,\ {\rm UV}}\  \log (\rho )+c_1){}^2
   }{25 {\cal C}_{a_t^{x3}(x3)}^{B=0,\ {\rm UV}}\   ^2 {\cal C}_{a_{t}^\phi(\rho)}^{B=0,\ {\rm UV}}\ ^2 {\cal C}_{a_t^{\phi}(Z)}^{B=0,\ {\rm UV}}\    ^2
   {r_h}^2 \left({\cal C}_{a_{x3}^Z(x3)}\ ^{B=0,\ {\rm UV}}\  ^2 {\cal C}_{x^3}^{\rho\phi}\ ^2 {\cal C}_{x3}^{\phi Z}\   ^2 \left(\rho
   ^2-1\right)+{\cal C}_{a_t^{x3}(x3)}\ ^{B=0,\ {\rm UV}}\ ^2\right)^2}\nonumber\\
& & \hskip -0.8in \times  \left({\cal C}_{a_t^{x3}(x3)}\ ^{B=0,\ {\rm UV}}\ ^2-{\cal C}_{a_{x3}^Z(x3)}\ ^{B=0,\ {\rm UV}}\  ^2 {\cal C}_{x^3}^{\rho\phi}\ ^2 {\cal C}_{x3}^{\phi Z}\ ^2\right);\nonumber\\         
& & \hskip -0.8in {\cal X}^{B=0,\ \beta^0}=3,\nonumber\\
& & \hskip -0.8in  {\cal X}^{B=0,\ \beta}=\frac{12 {\cal C}_{a_t^{x3}(x3)}\ ^{B=0,\ {\rm UV}}\ ^2 \rho ^2 {r_h}^2 e^{12 Z} ({\cal C}_{zz}-2
   {\cal C}_{\theta_1z}+2{\cal C}_{\theta_1x}) \left({\cal C}_{a_t^{x3}(x3)}\ ^{B=0,\ {\rm UV}}\ ^2-{\cal C}_{a_{x3}^Z(x3)}\ ^{B=0,\ {\rm UV}}\  ^2 {\cal C}_{x3}^{\rho\phi}\  ^2 {\cal C}_{x3}^{\phi Z}\   ^2\right)}{25 {\cal C}_{a_t^{x3}(x3)}^{B=0,\ {\rm UV}}\   ^2 {\cal C}_{a_{t}^\phi(\rho)}^{B=0,\ {\rm UV}}\ ^2
   {\cal C}_{a_t^{\phi}(Z)}^{B=0,\ {\rm UV}}\    ^2 {\cal C}_{a_{x3}^Z(x3)}\ ^{B=0,\ {\rm UV}}\  ^2 {\cal C}_{x^3}^{\rho\phi}\ ^2 {\cal C}_{x3}^{\phi Z}\   ^2
   \left({\cal C}_{a_{x3}^Z(x3)}\ ^{B=0,\ {\rm UV}}\  ^2 {\cal C}_{x^3}^{\rho\phi}\ ^2 {\cal C}_{x3}^{\phi Z}\   ^2 \left(\rho
   ^2-1\right)+{\cal C}_{a_t^{x3}(x3)}\ ^{B=0,\ {\rm UV}}\ ^2\right)}.\nonumber\\
& &    
\end{eqnarray}
}

Making the following perturbative ansatz: $E_{\rm trans} = \sum_{n=0}^\infty w^{2n}E_{\rm trans}^{w^n}$, we come with the following EOMs for $E_{\rm trans}^{w^{2n}, \beta^0}$.
\begin{itemize}
\item ${\cal O}(w^0)$:
\begin{eqnarray}
\label{EtransB0betaEOM}
& & e^{7 Z} c_{\rm UV}^{2, B=0}+c_{\rm UV}^{1, B=0}+3 E_{\rm trans}^{w^0,\ \beta^0}\ '(Z)+E_{\rm trans}^{w^0,\ \beta^0}\ ''(Z)=0,
\end{eqnarray}
whose solution is given by:
\begin{eqnarray}
\label{EtransB0betaw0sol}
& & E_{\rm trans}^{w^0,\ \beta^0} = -\frac{Z c_{\rm UV}^{1, B=0}}{3}-\frac{1}{70} e^{7 Z}
   c_{\rm UV}^{2, B=0}-\frac{1}{3} c_1 e^{-3 Z}+c_2.
\end{eqnarray}

\item ${\cal O}(w^2)$:
\begin{eqnarray}
\label{EtransB0UVwsqEOM}
& & \frac{16 {\cal C}_{a_t^{\rho}}^{B=0,\ {\rm UV}}\   ^2 Z \left(c_{\rm UV}^{1, B=0}-e^{7 Z} c_{\rm UV}^{2, B=0}\right)}{7
   {\cal C}_{a_{x3}^Z{\rho}}^{B=0,\ {\rm UV}}\ ^2}+\frac{16 {\cal C}_{a_t^{\rho}}^{B=0,\ {\rm UV}}\   ^2 \left(-\frac{Z
   c_{\rm UV}^{1, B=0}}{3}-\frac{1}{70} e^{7 Z} c_{\rm UV}^{2, B=0}-\frac{1}{3} c_1 e^{-3
   Z}+c_2\right)}{{\cal C}_{a_{x3}^Z{\rho}}^{B=0,\ {\rm UV}}\ ^2}\nonumber\\
   & & +3 E_{\rm trans}^{w^2,\ \beta^0}\ '(Z)+E_{\rm trans}^{w^2,\ \beta^0}\ ''(Z)=0,
\end{eqnarray}
whose solution is given by:
{\scriptsize
\begin{eqnarray}
\label{EtransB0UVwsol}
& & \hskip -0.8in E_{\rm trans}^{w^2,\ \beta^0}\ (Z) = \frac{7840 {\cal C}_{a_t^{\rho}}^{B=0,\ {\rm UV}}\   ^2 Z (3 Z-2) c_{\rm UV}^{1, B=0}+216 {\cal C}_{a_t^\rho}^{B=0,\ {\rm UV}}\ ^2 e^{7 Z} (7 Z-1) c_{\rm UV}^{2, B=0}-1715 e^{-3 Z} }{46305 {\cal C}_{a_{x3}^Z{\rho}}^{B=0,\ {\rm UV}}\ ^2}\nonumber\\
& & \times \left(16 {\cal C}_{a_t^{\rho}}^{B=0,\ {\rm UV}}\   ^2 \left(3 c_1 Z+9 c_2
   e^{3 Z} Z+c_1\right)+9 {\cal C}_{a_{x3}^Z{\rho}}^{B=0,\ {\rm UV}}\ ^2 c_3\right)+c_4.
\end{eqnarray}
}
\item ${\cal O}(w^4)$:
{\scriptsize
\begin{eqnarray}
\label{EtransB0UVw4}
& & \hskip -1in \frac{128 {\cal C}_{a_t^{\rho}}^{B=0,\ {\rm UV}}\   ^4 Z \left((7 Z+2) c_{\rm UV}^{1, B=0}+e^{7 Z} (7 Z-2) c_{\rm UV}^{2, B=0}\right)}{343
   {\cal C}_{a_{x3}^Z{\rho}}^{B=0,\ {\rm UV}}\ ^4} +\frac{16 {\cal C}_{a_t^{\rho}}^{B=0,\ {\rm UV}}\   ^2 }{{\cal C}_{a_{x3}^Z{\rho}}^{B=0,\ {\rm UV}}\ ^2}\times\nonumber\\
 & & \hskip -1in   \Biggl[\frac{7840 {\cal C}_{a_t^\rho}^{B=0,\ {\rm UV}}\  ^2 Z (3 Z-2) c_{\rm UV}^{1, B=0}+216 {\cal C}_{a_t^{\rho}}^{B=0,\ {\rm UV}}\   ^2 e^{7 Z} (7 Z-1) c_{\rm UV}^{2, B=0}-1715 e^{-3
   Z}}{46305 {\cal C}_{a_{x3}^Z{\rho}}^{B=0,\ {\rm UV}}\ ^2}  \left(16 {\cal C}_{a_t^{\rho}}^{B=0,\ {\rm UV}}\   ^2 \left(3 c_1 Z+9 c_2 e^{3 Z} Z+c_1\right)+9 {\cal C}_{a_{x3}^Z(\rho)}^{B=0,\ {\rm UV}}\ ^2 c_3\right)   +c_4\Biggr]\nonumber\\
   & & \hskip -1in  +3
   E_{\rm trans}^{w^4,\ \beta}\ '(Z)+E_{\rm trans}^{w^4,\ \beta}\ ''(Z)=0,
\end{eqnarray}
}
whose solution is given by:
{\scriptsize
\begin{eqnarray}
\label{EtransB0UVwsqsol}
& & \hskip -0.8in E_{\rm trans}^{w^4,\ \beta}\ (Z) = c_6 -\frac{640 {\cal C}_{a_t^{\rho}}^{B=0,\ {\rm UV}}\   ^4 Z \left(777 Z^2-1284 Z+856\right)
   c_{\rm UV}^{1, B=0}+\frac{15552}{49} {\cal C}_{a_t^{\rho}}^{B=0,\ {\rm UV}}\   ^4 e^{7 Z} (2-7 Z)^2 c_{\rm UV}^{2, B=0}+1715 e^{-3 Z}
  }{416745 {\cal C}_{a_{x3}^Z{\rho}}^{B=0,\ {\rm UV}}\ ^4}\nonumber\\
& &\hskip -0.8in \times  \left(144 {\cal C}_{a_t^{\rho}}^{B=0,\ {\rm UV}}\   ^2 {\cal C}_{a_{x3}^Z{\rho}}^{B=0,\ {\rm UV}}\ ^2 \left(3 c_3 Z+9 c_4 e^{3 Z}
   Z+c_3\right)+128 {\cal C}_{a_t^{\rho}}^{B=0,\ {\rm UV}}\   ^4 \left(c_1 (3 Z+2)^2-9 c_2 e^{3 Z} Z (3 Z-2)\right)+81
   {\cal C}_{a_{x3}^Z{\rho}}^{B=0,\ {\rm UV}}\ ^4 c_5\right).\nonumber\\
& &     
\end{eqnarray}
}
\end{itemize}

We therefore obtain:
{\footnotesize
\begin{eqnarray}
\label{Etransbetasoluptow4}
& & E_{\rm trans}^\beta\ =w^4 \left(-\frac{64 {\cal C}_{a_t^{\rho}}^{B=0,\ {\rm UV}}\   ^4 e^{7 Z} Z^2 c_{\rm UV}^{2, B=0}}{1715
   {\cal C}_{a_{x3}^Z{\rho}}^{B=0,\ {\rm UV}}\ ^4}-\frac{c_5}{3}+c_6\right)+w^2 \left(\frac{8 {\cal C}_{a_t^{\rho}}^{B=0,\ {\rm UV}}\   ^2 e^{7 Z}
   Z c_{\rm UV}^{2, B=0}}{245 {\cal C}_{a_{x3}^Z{\rho}}^{B=0,\ {\rm UV}}\ ^2}+c_4\right)\nonumber\\
   & & -\frac{Z c_{\rm UV}^{1, B=0}}{3}-\frac{1}{70} e^{7
   Z} c_{\rm UV}^{2, B=0}+c_2.
\end{eqnarray}
}
So, finally,
{\scriptsize
\begin{eqnarray}
\label{EtranssolB0UV}
& & E_{\rm trans}^{B=0}=w^4 \Biggl(\frac{128 {\cal C}_{a_t^{\rho}}^{B=0,\ {\rm UV}}\   ^4 Z \left((7 Z+2) c_{\rm UV}^{1, B=0}+e^{7 Z}
   (7 Z-2) c_{\rm UV}^{2, B=0}\right)}{343 {\cal C}_{a_{x3}^Z{\rho}}^{B=0,\ {\rm UV}}\ ^4}   \nonumber\\
& &  +\beta  \left[-\frac{4736 {\cal C}_{a_t^{\rho}}^{B=0,\ {\rm UV}}\   ^4 Z^3 c_{\rm UV}^{1, B=0}}{3969
   {\cal C}_{a_{x3}^Z{\rho}}^{B=0,\ {\rm UV}}\ ^4}-\frac{64 {\cal C}_{a_t^{\rho}}^{B=0,\ {\rm UV}}\   ^4 e^{7 Z} Z^2 c_{\rm UV}^{2, B=0}}{1715
   {\cal C}_{a_{x3}^Z{\rho}}^{B=0,\ {\rm UV}}\ ^4}+\frac{128 {\cal C}_{a_t^{\rho}}^{B=0,\ {\rm UV}}\   ^4 c_2 Z^2}{9 {\cal C}_{3,\ Z}^{B=0,\ {\rm UV}}\  ^4}-\frac{16 {\cal C}_{a_t^{\rho}}^{B=0,\ {\rm UV}}\   ^2 c_4 e^{3 Z} Z}{3 {\cal C}_{3,\ Z}^{B=0,\ {\rm UV}}\  ^2}-\frac{c_5}{3}+c_6\right]  \Biggr)\nonumber\\
& &    +w^2 \left(\frac{16 {\cal C}_{a_t^{\rho}}^{B=0,\ {\rm UV}}\   ^2 Z \left(c_{\rm UV}^{1, B=0}-e^{7 Z} c_{2
   {B0UV}}\right)}{7 {\cal C}_{a_{x3}^Z{\rho}}^{B=0,\ {\rm UV}}\ ^2}+\beta  \left[\frac{8
   {\cal C}_{a_t^{\rho}}^{B=0,\ {\rm UV}}\   ^2 e^{7 Z} Z c_{\rm UV}^{2, B=0}}{245 {\cal C}_{3,\ Z}^{B=0,\ {\rm UV}}\  ^2}+c_4\right]\right)\nonumber\\
& &     +e^{7 Z} c_{\rm UV}^{2, B=0}+c_{\rm UV}^{1, B=0} + \beta  \left(-\frac{Z c_{\rm UV}^{1, B=0}}{3}-\frac{1}{70} e^{7 Z} c_{\rm UV}^{2, B=0}+ c_2\right).
\end{eqnarray}
}
Therefore, 
{\scriptsize
\begin{eqnarray}
\label{dEtransoverEtransbeta0w4}
& & \left(\frac{E_{\rm trans}^{B=0}\ '(Z)}{E_{\rm trans}^{B=0}(Z)}\right)^{\beta^0}\  = \frac{7 e^{7 Z} c_{\rm UV}^{2, B=0}}{e^{7 Z} c_{\rm UV}^{2, B=0}+c_{\rm UV}^{1, B=0}}\nonumber\\
& & +\frac{16 {\cal C}_{a_t^{\rho}}^{B=0,\ {\rm UV}}\   ^2
   w^2 \left(-14 e^{7 Z} Z c_{\rm UV}^{2, B=0} c_{\rm UV}^{1, B=0}-e^{14 Z} c_{2
   {B0UV}}{}^2+c_{\rm UV}^{1, B=0}{}^2\right)}{7 {\cal C}_{a_{x3}^Z{\rho}}^{B=0,\ {\rm UV}}\ ^2 \left(e^{7 Z} c_{2
   {B0UV}}+c_{\rm UV}^{1, B=0}\right){}^2}\nonumber\\
& &    +\frac{256 {\cal C}_{a_t^{\rho}}^{B=0,\ {\rm UV}}\   ^4 w^4 \left(e^{7 Z} \left(98 Z^2+14
   Z+1\right) c_{\rm UV}^{2, B=0} c_{\rm UV}^{1, B=0}{}^2-e^{14 Z} \left(98 Z^2-14 Z+1\right) c_{\rm UV}^{2, B=0}{}^2
   c_{\rm UV}^{1, B=0}-e^{21 Z} c_{\rm UV}^{2, B=0}{}^3+c_{\rm UV}^{1, B=0}{}^3\right)}{343 {\cal C}_{a_{x3}^Z{\rho}}^{B=0,\ {\rm UV}}\ ^4
   \left(e^{7 Z} c_{\rm UV}^{2, B=0}+c_{\rm UV}^{1, B=0}\right){}^3}+O\left(w^5\right).\nonumber\\
& & 
\end{eqnarray}
}
At ${\cal O}(\beta)$, assuming:
\begin{eqnarray}
\label{dEtransoverEtransbetauptow4assumptions}
& & c_{\rm UV}^{1, B=0}=e^{-{\alpha_{\rm UV}^{B=0}}  {Z_{\rm UV}}},\nonumber\\
& & c_{\rm UV}^{2, B=0}=e^{-{\alpha_{2,\ \rm UV}^{B=0}}  {Z_{\rm UV}}},\nonumber\\
& & {\alpha_{2,\ \rm UV}^{B=0}} -{\alpha_{\rm UV}^{B=0}} -7>0,\nonumber\\
& & {\alpha_{2,\ \rm UV}^{B=0}} -7<2 {\alpha_{\rm UV}^{B=0}} ,\nonumber\\
& & c_2 = \frac{1}{3} {Z_{\rm UV}} e^{-{\alpha_{\rm UV}^{B=0}}  {Z_{\rm UV}}}-\frac{1}{21} e^{{Z_{\rm UV}} ({\alpha_{2,\ \rm UV}^{B=0}} -2
   {\alpha_{\rm UV}^{B=0}} -7)}, 
\end{eqnarray}
one obtains:
{\scriptsize
\begin{eqnarray}
\label{dEtransoverEtransbetauptow4}
& &  \left(\frac{E_{\rm trans}^{B=0}\ '(Z)}{E_{\rm trans}^{B=0}(Z)}\right)  =
\frac{
   16 {\cal C}_{a_t^{\rho}}^{B=0,\ {\rm UV}}\   ^2 w^2 \left(-14
   e^{7 Z_{\rm UV}} Z_{\rm UV} c_{2 \text{B0UV}}
   c_{\rm UV}^{1, B=0}-e^{14 Z_{\rm UV}}  c_{\rm UV}^{2, B=0}{}^2+c_{\rm UV}^{1, B=0}{}^2\right)}{7
  {\cal C}_{a_{x3}^Z{\rho}}^{B=0,\ {\rm UV}}\ ^2 \left(e^{7
   Z_{\rm UV}} c_{\rm UV}^{2, B=0}+c_{\rm UV}^{1, B=0}\right){}^2}\nonumber\\
& &  +
\frac{256 {\cal C}_{a_t^{\rho}}^{B=0,\ {\rm UV}}\   ^4 w^4
   \left(98 e^{7 Z_{\rm UV}} Z_{\rm UV}^2 c_{\rm UV}^{2, B=0} c_{\rm UV}^{2, B=0}{}^2-98 e^{14
   Z_{\rm UV}} Z_{\rm UV}^2 c_{\rm UV}^{2, B=0}{}^2
   c_{\text{B0UV}}-e^{21 Z_{\rm UV}} c_{\rm UV}^{2, B=0}{}^3+c_{\rm UV}^{2, B=0}{}^3\right)}{34
   3{\cal C}_{a_{x3}^Z{\rho}}\ ^{B=0,\ {\rm UV}}\ ^4 \left(e^{7
   Z_{\rm UV}} c_{\rm UV}^{2, B=0}+c_{\rm UV}^{1, B=0}\right){}^3}\nonumber\\
& & + \frac{\beta  w^2 e^{{\alpha_{2,\ \rm UV}^{B=0}}  {Z_{\rm UV}}} \left(16 {\cal C}_{a_t^{\rho}}^{B=0,\ {\rm UV}}\   ^2 e^{{Z_{\rm UV}} (3
   {\alpha_{2,\ \rm UV}^{B=0}} -{\alpha_{\rm UV}^{B=0}} -7)}-1029 {\cal C}_{a_{x3}^Z{\rho}}^{B=0,\ {\rm UV}}\ ^2 c_4 e^{(3 {\alpha_{\rm UV}^{B=0}} +14)
   {Z_{\rm UV}}}\right)}{147 {\cal C}_{a_{x3}^Z{\rho}}^{B=0,\ {\rm UV}}\ ^2 \left(e^{{\alpha_{2,\ \rm UV}^{B=0}} 
   {Z_{\rm UV}}}+e^{({\alpha_{\rm UV}^{B=0}} +7) {Z_{\rm UV}}}\right)^3}\nonumber\\
& &   -\frac{128 \beta  {\cal C}_{a_t^{\rho}}^{B=0,\ {\rm UV}}\   ^4 w^4}{12005 {\cal C}_{a_{x3}^Z{\rho}}^{B=0,\ {\rm UV}}\ ^4}  -\frac{32 \beta 
   {\cal C}_{a_t^{\rho}}^{B=0,\ {\rm UV}}\   ^2 w^4 {Z_{\rm UV}} e^{-4 ({\alpha_{\rm UV}^{B=0}} +7) {Z_{\rm UV}}}}{9261 {\cal C}_{a_{x3}^Z{\rho}}^{B=0,\ {\rm UV}}\ ^4}
   \times  \Biggl[464 {\cal C}_{a_t^\rho}^{B=0,\ {\rm UV}}\  ^2 e^{{Z_{\rm UV}} (5 {\alpha_{2,\ \rm UV}^{B=0}} -{\alpha_{\rm UV}^{B=0}} -7)}\nonumber\\
& &    +3087 {\cal C}_{a_{x3}^Z{\rho}}^{B=0,\ {\rm UV}}\ ^2 c_4
   \left(e^{{Z_{\rm UV}} (3 {\alpha_{2,\ \rm UV}^{B=0}} +2 {\alpha_{\rm UV}^{B=0}} +10)}-2 e^{{Z_{\rm UV}} ({\alpha_{2,\ \rm UV}^{B=0}} +4
   {\alpha_{\rm UV}^{B=0}} +24)}\right)\Biggr].   
\end{eqnarray}
}
Using results of \cite{Holographic-Photoprod-B}
{\footnotesize
\begin{eqnarray}
\label{Chi2}
& & \chi_2= \Im m\left[\left(\frac{E_{\rm trans}^{B=0}\ '(Z)}{E_{\rm trans}^{B=0}(Z)}\right) 
   G^{x^2x^2}_{B=0,{\rm UV}} G^{ZZ}_{B=0,{\rm UV}}
   {\cal L}_{\rm DBI}^{B=0,\ {\rm UV}}\right] =  w^2(\chi_2^{w^2,\ \beta^0}\  + \beta \chi_2^{w^2,\ \beta}\ )
 + w^4(\chi_2^{w^4,\ \beta^0}\  + \beta \chi_2^{w^4,\ \beta}\ ),  
   \nonumber\\
\end{eqnarray}
}
where
{\scriptsize
\begin{eqnarray}
\label{Chi2w024uptobeta}
& & \hskip -0.8in \chi_2^{w^2,\ \beta^0}\ = \frac{320 i \sqrt{2} \pi 
   {\cal C}_{a_t^{\rho}}^{B=0,\ {\rm UV}}\   ^2
   {\cal C}_{a_t^{x3}(x3)}^{B=0,\ {\rm UV}}\    {\cal C}_{a_t^\phi(\rho)}^{B=0,\ {\rm UV}}\   {\cal C}_{a_t^{\phi}(Z)}^{B=0,\ {\rm UV}}\    
   {\cal C}_{a_{x3}^Z(x3)}\ ^{B=0,\ {\rm UV}}\  ^4
   {\cal C}_{a_{x3}^Z(Z)}^{B=0,\ {\rm UV}}\   {\cal C}_{a_t^{x3}(x3)}\ ^{B=0,\ {\rm UV}}\ 
   {\cal C}_{x^3}^{\rho\phi}\ ^4 {\cal C}_{x3}^{\phi Z}\ ^4 N^{8/5} w^2 e^{-9 {Z_{\rm UV}}} \cos
   ^4(\phi )}{7
   \sqrt[3]{3} {\cal C}_{a_{x3}^Z{\rho}}^{B=0,\ {\rm UV}}\  \rho
    {r_h}^2 \alpha _{\theta _2}^2 \left(e^{7
   {Z_{\rm UV}}} c_{2 {B0UV}}+c_{\rm UV}^{1, B=0}\right){}^2
   \left({\cal C}_{a_{x3}^Z(x3)}\ ^{B=0,\ {\rm UV}}\  ^2
   {\cal C}_{x^3}^{\rho\phi}\ ^2 {\cal C}{x3}^{\phi Z}\ ^2 \left(\rho
   ^2-1\right)+{\cal C}_{a_t^{x3}(x3)}\ ^{B=0,\ {\rm UV}}\ ^2\right)^2}\nonumber\\
& & \hskip -0.8in \times  \left(14 e^{7 {Z_{\rm UV}}} {Z_{\rm UV}}
   c_{\rm UV}^{2, B=0} c_{\rm UV}^{1, B=0}+e^{14
   {Z_{\rm UV}}} c_{2 {B0UV}}{}^2-c_{\rm UV}^{1, B=0}{}^2\right),\nonumber\\   
& & \hskip -0.8in \chi_2^{w^2,\ \beta}\ =\frac{20 i \sqrt{2} \pi  \beta 
   {\cal C}_{a_t^{x3}(x3)}^{B=0,\ {\rm UV}}\    {\cal C}_{a_t^\phi(\rho)}^{B=0,\ {\rm UV}}\   {\cal C}_{a_t^{\phi}(Z)}^{B=0,\ {\rm UV}}\    
   {\cal C}_{a_{x3}^Z(x3)}\ ^{B=0,\ {\rm UV}}\  ^4
   {\cal C}_{a_{x3}^Z(Z)}^{B=0,\ {\rm UV}}\   {\cal C}_{a_t^{x3}(x3)}\ ^{B=0,\ {\rm UV}}\ 
   {\cal C}_{x^3}^{\rho\phi}\ ^4 {\cal C}_{x3}^{\phi Z}\  ^4 N^{8/5} w^2 e^{-9 {Z_{\rm UV}}} \cos
   ^4(\phi )}{147 \sqrt[3]{3}
   {\cal C}_{a_{x3}^Z{\rho}}^{B=0,\ {\rm UV}}\  \rho 
   {r_h}^2 \alpha _{\theta _2}^2
   \left(e^{{\alpha_{2,\ \rm UV}^{B=0}} 
   {Z_{\rm UV}}}+e^{({\alpha_{\rm UV}^{B=0}} +7)
   {Z_{\rm UV}}}\right)^3
   \left({\cal C}_{a_{x3}^Z(x3)}\ ^{B=0,\ {\rm UV}}\  ^2
   {\cal C}_{x^3}^{\rho\phi}\ ^2 {\cal C}_{x3}^{\phi Z}\ ^2 \left(\rho
   ^2-1\right)+{\cal C}_{a_t^{x3}(x3)}\ ^{B=0,\ {\rm UV}}\ ^2\right)^2}\nonumber\\
   & & \hskip -0.8in \times  \left(16 {\cal C}_{a_t^\rho}^{B=0,\ {\rm UV}}\ ^2 e^{{Z_{\rm UV}} (4
   {\alpha_{2,\ \rm UV}^{B=0}} -{\alpha_{\rm UV}^{B=0}} -7)}-1029
   {\cal C}_{a_{x3}^Z{\rho}}^{B=0,\ {\rm UV}}\ ^2 c_4
   e^{{Z_{\rm UV}} ({\alpha_{2,\ \rm UV}^{B=0}} +3
   {\alpha_{\rm UV}^{B=0}} +14)}\right);\nonumber\\
& & \hskip -0.8in \chi_2^{w^4,\ \beta^0}\ =-\frac{5120 i \sqrt{2} \pi 
   {\cal C}_{a_t^{\rho}}^{B=0,\ {\rm UV}}\   ^4
   {\cal C}_{a_t^{x3}(x3)}^{B=0,\ {\rm UV}}\    {\cal C}_{a_t^\phi(\rho)}^{B=0,\ {\rm UV}}\   {\cal C}_{a_t^{\phi}(Z)}^{B=0,\ {\rm UV}}\    
   {\cal C}_{a_{x3}^Z(x3)}\ ^{B=0,\ {\rm UV}}\  ^4
   {\cal C}_{a_{x3}^Z(Z)}^{B=0,\ {\rm UV}}\   {\cal C}_{a_t^{x3}(x3)}\ ^{B=0,\ {\rm UV}}\ 
   {\cal C}_{x^3}^{\rho\phi}\ ^4 {\cal C}_{x3}^{\phi Z}\ ^4 N^{8/5} w^4 e^{-9 {Z_{\rm UV}}} \cos
   ^4(\phi )}{34
   3 \sqrt[3]{3} {\cal C}_{a_{x3}^Z{\rho}}^{B=0,\ {\rm UV}}\ ^3
   \rho  {r_h}^2 \alpha _{\theta _2}^2
   \left(e^{7 {Z_{\rm UV}}} c_{\rm UV}^{2, B=0}\ +c_{\rm UV}^{1, B=0}\right){}^3
   \left({\cal C}_{a_{x3}^Z(x3)}\ ^{B=0,\ {\rm UV}}\  ^2
   {\cal C}_{x^3}^{\rho\phi}\ ^2 {\cal C}_{x3}^{\phi Z}\ ^2 \left(\rho
   ^2-1\right)+{\cal C}_{a_t^{x3}(x3)}\ ^{B=0,\ {\rm UV}}\ ^2\right)^2}\nonumber\\
   & & \hskip -0.8in \times  \left(c_{\rm UV}^{1, B=0}-e^{7
   {Z_{\rm UV}}} c_{\rm UV}^{2, B=0}\right) \left(e^{7
   {Z_{\rm UV}}} \left(98 {Z_{\rm UV}}^2+1\right) c_{2 {B0UV}} c_{\rm UV}^{1, B=0}+e^{14 {Z_{\rm UV}}}
   c_{\rm UV}^{2, B=0}\ {}^2+c_{\rm UV}^{1, B=0}{}^2\right),\nonumber\\
& & \hskip -0.8in \chi_2^{w^4,\ \beta}\ =\frac{16 i \sqrt{2} \pi  \beta 
   {\cal C}_{a_t^{x3}(x3)}^{B=0,\ {\rm UV}}\    {\cal C}_{a_t^{\phi}(\rho)}^{B=0,\ {\rm UV}}\   {\cal C}_{a_t^{\phi}(Z)}^{B=0,\ {\rm UV}}\    
   {\cal C}_{a_{x3}^Z(x3)}\ ^{B=0,\ {\rm UV}}\  ^4
   {\cal C}_{a_{x3}^Z(Z)}^{B=0,\ {\rm UV}}\   {\cal C}_{a_t^{x3}(x3)}\ ^{B=0,\ {\rm UV}}\ 
   {\cal C}_{x^3}^{\rho\phi}\ ^4 {\cal C}_{x^3}^{\phi Z}\ ^4 N^{8/5} w^4 e^{-9 {Z_{\rm UV}}} \cos
   ^4(\phi )}{64827
   \sqrt[3]{3} {\cal C}_{a_{x3}^Z{\rho}}^{B=0,\ {\rm UV}}\ ^3
   \rho  {r_h}^2 \alpha _{\theta _2}^2
   \left({\cal C}_{a_{x3}^Z(x3)}\ ^{B=0,\ {\rm UV}}\  ^2
   {\cal C}_{x^3}^{\rho\phi}\ ^2 {\cal C}_{x3}^{\phi Z}\ ^2 \left(\rho
   ^2-1\right)+{\cal C}_{a_t^{x3}(x3)}\ ^{B=0,\ {\rm UV}}\ ^2\right)^2}\nonumber\\
& & \hskip -0.8in \times  \left(32 {\cal C}_{a_t^\rho}^{B=0,\ {\rm UV}}\ ^4 \left(4060 {Z_{\rm UV}} e^{5 {Z_{\rm UV}}
   ({\alpha_{2,\ \rm UV}^{B=0}} -{\alpha_{\rm UV}^{B=0}} -7)}+27\right)-864360 {\cal C}_{a_t^{\rho}}^{B=0,\ {\rm UV}}\   ^2
   {\cal C}_{a_{x3}^Z{\rho}}^{B=0,\ {\rm UV}}\ ^2 c_4 {Z_{\rm UV}}
   \left(2 e^{({\alpha_{2,\ \rm UV}^{B=0}} -4)
   {Z_{\rm UV}}}-e^{{Z_{\rm UV}} (3
   {\alpha_{2,\ \rm UV}^{B=0}} -2
   ({\alpha_{\rm UV}^{B=0}} +9))}\right)\right). \nonumber\\
& &            
\end{eqnarray}
}
Therefore,
\begin{eqnarray}
\label{chi2B0}
& & \frac{\chi}{w}\sim w \left(\chi_2^{w^2,\ \beta^0} e^{w^2
   \left(\frac{\beta  (\chi_2^{w^2,\ \beta^0}
   \chi_2^{w^4,\ \beta}-\chi_2^{w^2,\ \beta}
   \chi_2^{w^4,\ \beta^0})}{\chi_2^{w^2,\ \beta^0}\ ^2}+\frac{\chi_2^{w^4,\ \beta^0}}{\chi_2^{w^2,\ \beta^0}}\right
   )}+\beta  \chi_2^{w^2,\ \beta}
   e^{\frac{\chi_2^{w^4,\ \beta^0}
   w^2}{\chi_2^{w^2,\ \beta^0}}}\right).
\end{eqnarray}

\subsubsection{Matching with $D=5$ gauged SUGRA truncation over an $S^5$ of $D=10$ type IIB SUGRA for $B=0$ \cite{D5gaugedSUGRAtrunc}}
\label{Match-D5-gauged-SUGRA}

Utilizing the result obtained in the previous section for the spectral density of photon production in the absence of a magnetic field, we will see our results match nicely with those in gauged type IIB supergravity compactified on $S^5$ \cite{D5gaugedSUGRAtrunc}.

Now, (\ref{chi2B0}) can be rewritten as:
{\footnotesize
\begin{eqnarray}
\label{chioverwB0-i}
& & \hskip -0.8 in \frac{\chi}{w}\sim \frac{{\cal C}_{a_t^\rho}^{B=0, {\rm UV}}\ ^2 {\cal C}_{a_t^{x3}(x3)}^{B=0, {\rm UV}}\  {\cal C}_{a_t^\phi(\rho)}^{B=0, {\rm UV}}\ 
   {\cal C}_{a_t^\phi(Z)}^{B=0, {\rm UV}}\  {\cal C}_{a_{x3}^Z(x3)}^{B=0, {\rm UV}}\ ^4 {\cal C}_{a_{x3}^Z(Z)}^{B=0, {\rm UV}}\ 
   {\cal C}_{a_{t}^{x3}(x3)}^{B=0, {\rm UV}}\  {\cal C}_{a_{x3}^{\rho}(\phi)}^{B=0, {\rm UV}}\ ^4 {\cal C}_{a_{x3}^{\phi}(Z)}^{B=0, {\rm UV}}\ ^4 N  \left(\beta  e^{\frac{928 e^5 \beta 
   {\cal C}_{a_t^\rho}^{B=0, {\rm UV}}\ ^2 w^2}{1323 {\cal C}_{a_{x3}^Z(\rho)}^{B=0, {\rm UV}}\ ^2
   \xi_{\rm UV}^{B=0}\ }+1}-21\right)}{{\cal C}_{a_{x3}^Z(\rho)}^{B=0, {\rm UV}}\  {r_h}^2
   \left({\cal C}_{a_{x3}^Z(x3)}^{B=0, {\rm UV}}\ ^2 {\cal C}_{a_{x3}^{\rho}(\phi)}^{B=0, {\rm UV}}\ ^2 {\cal C}_{a_{x3}^\phi(Z)}^{B=0,\ {\rm UV}}\ ^2+{\cal C}_{a_{t}^{x3}(x3)}^{B=0, {\rm UV}}\ ^2\right)^2}\nonumber\\
& & \hskip -0.6in \times w e^{-\frac{16
   {\cal C}_{a_t^\rho}^{B=0, {\rm UV}}\ ^2 w^2 \left(\frac{58 e^5 \beta }{\xi_{\rm UV}^{B=0}\ }-27\right)}{1323
   {\cal C}_{a_{x3}^Z(\rho)}^{B=0, {\rm UV}}\ ^2}-\frac{9}{\xi_{\rm UV}^{B=0}\ }}.
\end{eqnarray}
}
Using \cite{Shivam+Aalok_Bulk}:
\begin{eqnarray}
\label{beta_Bulk_Viscosity_cs_paper}
& & \beta =\frac{16384 \pi ^7 \kappa_{\beta} \left(\frac{24 {g_s} M^2 {N_f}
   ({c_1}+{c_2} \log ({r_h})) \left(\frac{{g_s} M^2 ({c_1}+{c_2} \log
   ({r_h}))}{N}+\frac{1}{\sqrt{3}}\right)^2}{9 \left(\frac{{g_s} M^2 ({c_1}+{c_2}
   \log ({r_h}))}{N}+\frac{1}{\sqrt{3}}\right)^2+1}+\frac{3 N {N_f} (\log (N)-3 \log
   ({r_h}))}{4 \pi }\right)^3}{2187 {g_s}^4 N^3 {N_f}^7 (\log (N)-3 \log ({r_h}))^7},\nonumber\\
& & 
\end{eqnarray}
along with $r_h = e^{-\kappa_{r_h}(g_s, M, N_f)N^{1/3}}$, which for $g_s = 0.1, M=N_f=3, N=100$ and values of $c_{1,2}$ as obtained in \cite{Shivam+Aalok_Bulk} to match with $SU(3)$ Gluodynamics lattice results \cite{lattice-SU3_Glue}, yields $\beta=1.1$ (essentially implying that the higher derivative corrections can not be disregarded). Defining ${\cal R}=-\frac{{\cal C}_{a_t^\rho}^{B=0, {\rm UV}}\ ^2}{{\cal C}_{a_{x3}^Z(\rho)}^{B=0, {\rm UV}}\ ^2}$, one obtains:
\begin{eqnarray}
\label{chioverwB0-ii}
& & \frac{\chi}{w} = {\cal C} w e^{-\frac{16 {\cal R} w^2
   \left(27-\frac{9468.76}{\xi_{\rm UV}^{B=0}\ }\right)}{1323}-\frac{9}{\xi_{\rm UV}^{B=0}\ }} \left(1.1 e^{1-\frac{114.513 {\cal R}
   w^2}{\xi_{\rm UV}^{B=0}\ }}-21\right).
\end{eqnarray}
Now, $c_2, c_{0, {\rm UV}}^{B=0}, c_{2, {\rm UV}}^{B=0}$ have dimensions of $T$,
$c_4$ has dimensions of $1/T,  {\cal C}_{a_t^\rho}^{B=0,\ {\rm UV}} / {\cal C}_{a_{x3}^Z(\rho)}^{B=0,\ {\rm UV}}$ has dimensions of $1/T$;
$c_5, c_6$ must have dimensions of $1/T^3$. Hence, ${\cal R}\rightarrow  {\cal R}/T^2$ with now a dimensionless ratio $\tilde{w} \equiv w/(2 \pi T)$. Thus, with ${\cal C}\rightarrow\frac{{\cal C}}{4\pi^2N^2}$,
\begin{eqnarray}
\label{chioverwB0-iii}
& & \frac{\chi}{4 N^2 \tilde{w} T^2} = {\cal C} \tilde{w} e^{-\frac{16 {\cal R} \tilde{w}^2
   \left(27-\frac{9468.76}{\xi_{\rm UV}^{B=0}\ }\right)}{1323}-\frac{9}{\xi_{\rm UV}^{B=0}\ }} \left(1.1
   e^{1-\frac{114.513 {\cal R} \tilde{w}^2}{\xi_{\rm UV}^{B=0}\ }}-21\right).
\end{eqnarray}
Numerically, one obtains a reasonable match with \cite{D5gaugedSUGRAtrunc} for
\begin{eqnarray}
\label{Const+ratio+xiUV}
{\cal C} = 0.0064,\ \xi_{\rm UV}^{B=0}\  = -3.095, {\cal R} = 0.5727.
\end{eqnarray}
\begin{figure}
\begin{center}
\includegraphics[width=0.6\textwidth]{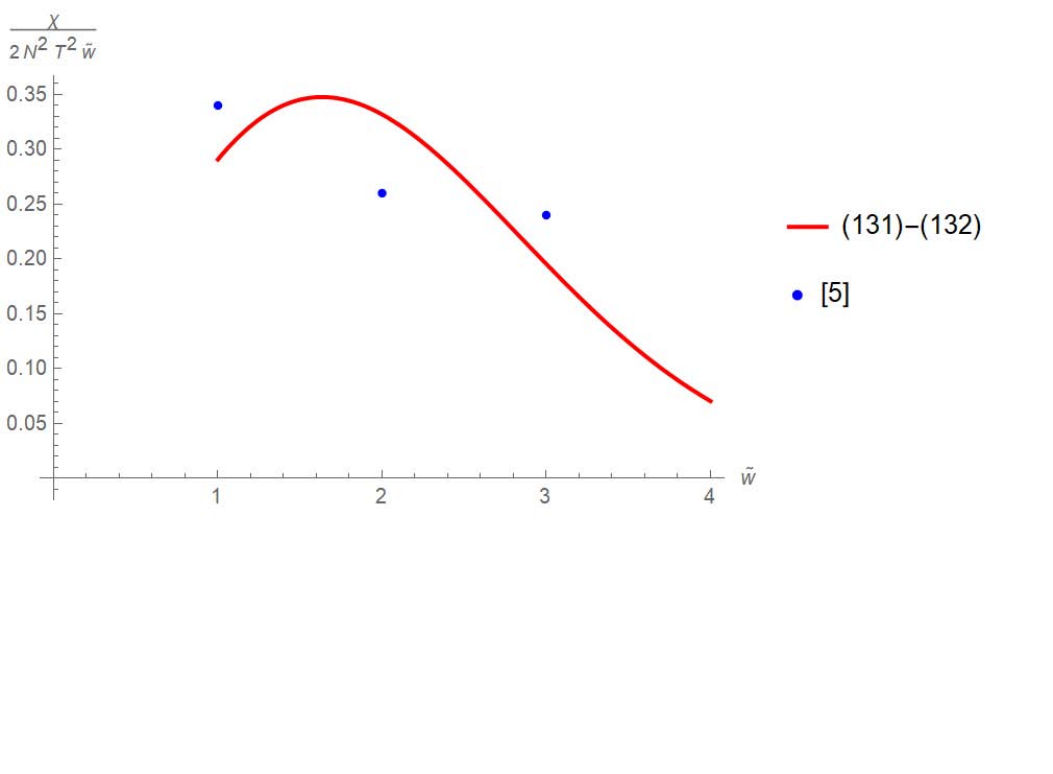}
\end{center}
\caption{$\frac{\chi}{2 N^2\tilde{w}T^2}$-vs-$\tilde{w}$;  the plot in red is based on (\ref{chioverwB0-iii})-(\ref{Const+ratio+xiUV}), and the plots points in blue are from \cite{D5gaugedSUGRAtrunc}; the comparison though  not entirely justified [the wave vector is orthogonal to the $x^3$-axis in (\ref{chioverwB0-iii})-(\ref{Const+ratio+xiUV}), but at $45^0$ relative to $x^3$ in \cite{D5gaugedSUGRAtrunc} for $B=0$], is just to illustrate that one can fit the parameters ${\cal C},\ \xi_{\rm UV}^{B=0}\  , {\cal R}$ to results of \cite{D5gaugedSUGRAtrunc}}
\end{figure}

\subsection{Photoproduction in the presence of strong magnetic field}
\label{photoBneq0}

In this section, we will generalize the procedure adopted in section \ref{photoprod-B_0} in the presence of a strong magnetic field and similarly will obtain the respective spectral density by utilizing the $E_{trans}$ in the presence of the strong magnetic field.

Now,
{\scriptsize
\begin{eqnarray}
\label{iG22+iGZZ+SDBI}
& & \hskip -0.8in G^{x^2x^2} = \frac{4 \sqrt{\pi } \beta  {\cal C}_{x^3x^3}^B\ ^2 \kappa_{a_{x3}^{Z,2}}^2 {\cal C}_{tZ}^B\  {\cal C}_{x^3\rho}^B\ ^2 \sqrt{{g_s}} \sqrt{N} e^{Z-2
   \rho } \cos ^2(\phi ) \left({\cal C}_{tZ,1}^B\  e^Z+{\cal C}_{tZ,2}^B\ \right)}{B^2 {\cal C}_{tZ,2}^B\ ^2 {r_h}^2}   +\frac{2 \sqrt{\pi }
   {\cal C}_{x^3x^3}^B\ ^2 \kappa_{a_{x3}^{Z,2}}^2 {\cal C}_{x^3\rho}^B\ ^2 \sqrt{{g_s}} \sqrt{N} e^{-2 \rho } \cos ^2(\phi )
   \left({\cal C}_{tZ,1}^B\  e^Z+{\cal C}_{tZ,2}^B\ \right)^2}{B^2 {\cal C}_{tZ,2}^B\ ^2 {r_h}^2},\nonumber\\
& & \hskip -0.8in G^{ZZ} = -\frac{\beta  \sqrt{\frac{1}{N}} {r_h}^4 e^{12 Z} \left(e^{4 Z}-1\right) \left(6 b^2+e^{2 Z}\right) ({\cal C}_{zz}-2 {\cal C}_{\theta_1z}+2
  {\cal C}_{\theta_1x})}{2 \sqrt{\pi } {\cal C}_{x^3x^3}^B\ ^4 \kappa_{a_{x3}^{Z,2}}^4 \sqrt{{g_s}} \left(9 b^2+e^{2 Z}\right) \left({\cal C}_{x^3\rho, 1}^B\ +{\cal C}_{x^3\rho}^B\  e^{-\rho }\right)^4}-\frac{\sqrt{\frac{1}{N}} {r_h}^2 e^{4 Z} \left(e^{4 Z}-1\right)}{2 \sqrt{\pi }
   {\cal C}_{x^3x^3}^B\ ^2 \kappa_{a_{x3}^{Z,2}}^2 \sqrt{{g_s}} \left({\cal C}_{x^3\rho,1}^B\ +{\cal C}_{x^3\rho}^B\  e^{-\rho }\right)^2},\nonumber\\
& & G^{tt}=\frac{2 \sqrt{\pi } \beta  \sqrt{{g_s}} \sqrt{N} \left(6 b^2+e^{2 Z}\right) e^{2 \rho +6 Z} ({\cal C}_{zz}-2 {\cal C}_{\theta_1z}+2{\cal C}_{\theta_1x})}{{\cal C}_{x^3x^3}^B\ ^2
   \kappa_{a_{x3}^{Z,2}}^2 \left(e^{4 Z}-1\right) \left(9 b^2+e^{2 Z}\right) \left({\cal C}_{x^3\rho,1}^B\  e^{\rho }+{\cal C}_{x^3\rho}^B\ \right)^2}+\frac{2 \sqrt{\pi } \sqrt{{g_s}} \sqrt{N} e^{2
   Z}}{{r_h}^2},\nonumber\\
& & \hskip -0.8in {\cal L}_{\rm DBI}^B = \frac{2 \beta  {\cal C}_{x^3x^3}^B\ ^4 {\cal C}_{tZ,1}^B\  \kappa_{a_{x3}^{Z,2}}^2 {\cal C}_{tZ}^B\ 
   {\cal C}_{x^3\rho}^B\ ^2 \rho ^2 (\kappa_{a_{x3}^{Z,2}}+{\epsilon_2})^2 e^{-4 \rho -2 Z}
   \left({\cal C}_{x^3\rho,1}^B\  e^{\rho }+{\cal C}_{x^3\rho,2}^B\ \right)^2}{{\cal C}_{tZ,2}^B\ ^2}\nonumber\\
& & \hskip -0.9in   -\frac{\sqrt{2} N^{3/5} \rho  {r_h}^2
   \sqrt{\frac{{\cal C}_{x^3x^3}^B\ ^2 e^{-4 \rho } \left({\cal C}_{x^3\rho,1}^B\  e^{\rho
   }+{\cal C}_{x^3\rho}^B\ \right)^2 \left({\cal C}_{x^3x^3}^B\ ^2 {\cal C}_{tZ,1}^B\ ^2
   \kappa_{a_{x3}^{Z,2}}^2 {\cal C}_{x^3\rho}^B\ ^2 \rho ^2
   (\kappa_{a_{x3}^{Z,2}}+{\epsilon_2})^2-B^2 e^{2 \rho } \left({\cal C}_{tZ,2}^B\ ^2
   {\epsilon_2} (2 \kappa_{a_{x3}^{Z,2}}+{\epsilon_2})-2 {\cal C}_{tZ,2}^B\ 
   \kappa_{a_{x3}^{Z,2}}^2 {\epsilon_1}-\kappa_{a_{x3}^{Z,2}}^2
   {\epsilon_1}^2\right)\right)}{{\cal C}_{tZ,2}^B\ ^2}}}{\sqrt[3]{3} {g_s} \alpha
   _{\theta _2}^2}.   \nonumber\\
& &      
\end{eqnarray}
}
The $E_{\rm trans}(Z)$ EOM hence is:
\begin{eqnarray}
\label{EtransBUVEOM}
& & E_{\rm trans}'(Z) \left(-\frac{2 e^{-Z}
   ({\cal C}_{tZ,2}^B\ +{\epsilon_1})}{{\cal C}_{tZ,1}^B\ }-2\right)+E_{\rm trans}''(Z)+\frac{4 \pi  {g_s} N w^2 e^{-2
   Z} E_{\rm trans}(Z)}{{r_h}^2}=0.
\end{eqnarray}
The solution of (\ref{EtransBUVEOM}) is given by:
{\scriptsize
\begin{eqnarray}
\label{EtransBUVsol-i}
& & = \exp \left(-\frac{e^{-Z} \left(\sqrt{{\cal C}_{tZ,2}^B\ ^2 {r_h}^2-4 \pi  {\cal C}_{tZ,1}^B\ ^2
   {g_s} N w^2}+{\cal C}_{tZ,2}^B\  {r_h}\right)}{{\cal C}_{tZ,1}^B\  {r_h}}\right)\nonumber\\
 & & \times  \Biggl[c_{1,\ \rm UV}^{B\neq0}U\left(\frac{9 \pi ^2 {\cal C}_{tZ,1}^B\ ^4 {g_s}^2 N^2 w^4}{{r_h}^4
   ({\cal C}_{tZ,2}^B\ +{\epsilon_1})^4}+\frac{3 \pi  {\cal C}_{tZ,1}^B\ ^2 {g_s} N
   w^2}{{r_h}^2 ({\cal C}_{tZ,2}^B\ +{\epsilon_1})^2}+3,3,-\frac{4 \pi ^2
   {\cal C}_{tZ,1}^B\ ^3 {g_s}^2 N^2 w^4 e^{-Z}}{{r_h}^4
   ({\cal C}_{tZ,2}^B\ +{\epsilon_1})^3}-\frac{4 \pi  {\cal C}_{tZ,1}^B\  {g_s} N w^2
   e^{-Z}}{{r_h}^2 ({\cal C}_{tZ,2}^B\ +{\epsilon_1})}+\frac{2 e^{-Z}
   ({\cal C}_{tZ,2}^B\ +{\epsilon_1})}{{\cal C}_{tZ,1}^B\ }\right)\Biggr]\nonumber\\
& & + c_{2,\ \rm UV}^{B\neq0} L_{-\frac{9 {\cal C}_{tZ,1}^B\ ^4 {g_s}^2 N^2 \pi ^2
   w^4}{({\cal C}_{tZ,2}^B\ +{\epsilon_1})^4 {r_h}^4}-\frac{3 {\cal C}_{tZ,1}^B\ ^2 {g_s}
   N \pi  w^2}{({\cal C}_{tZ,2}^B\ +{\epsilon_1})^2 {r_h}^2}-3}^2\left(-\frac{4
   {\cal C}_{tZ,1}^B\ ^3 e^{-Z} {g_s}^2 N^2 \pi ^2 w^4}{({\cal C}_{tZ,2}^B\ +{\epsilon_1})^3
   {r_h}^4}-\frac{4 {\cal C}_{tZ,1}^B\  e^{-Z} {g_s} N \pi 
   w^2}{({\cal C}_{tZ,2}^B\ +{\epsilon_1}) {r_h}^2}+\frac{2 e^{-Z}
   ({\cal C}_{tZ,2}^B\ +{\epsilon_1})}{{\cal C}_{tZ,1}^B\ }\right) \Biggr].\nonumber\\
& & = c_{1,\ \rm UV}^{B\neq0} e^{2 Z} \left(\frac{2 \pi ^2 {\cal C}_{tZ,1}^B\ ^6 {g_s}^2 N^2 w^4}{{\cal C}_{tZ,2}^B\ ^6
   {r_h}^4}+\frac{\pi  {\cal C}_{tZ,1}^B\ ^4 {g_s} N w^2}{2 {\cal C}_{tZ,2}^B\ ^4
   {r_h}^2}+\frac{{\cal C}_{tZ,1}^B\ ^2}{8 {\cal C}_{tZ,2}^B\ ^2}\right)\nonumber\\
& & + c_{2\ ,\rm UV}^{B\neq0} \left(\frac{177.653 {\cal C}_{tZ,1}^B\ ^4 {g_s}^2 N^2 w^4}{{\cal C}_{tZ,2}^B\ ^4
   {r_h}^4}-\frac{4 {\cal C}_{tZ,2}^B\ ^2 e^{-2 Z}}{{\cal C}_{tZ,1}^B\ ^2}+\frac{{g_s} N w^2
   e^{-Z} \left(\frac{2 \pi  {\cal C}_{tZ,1}^B\  {r_h}}{\sqrt{{\cal C}_{tZ,2}^B\ ^2
   {r_h}^2}}-\frac{12.5664
   {\cal C}_{tZ,1}^B\ }{{\cal C}_{tZ,2}^B\ }\right)}{{r_h}^2}+1\right).         
\end{eqnarray}   
}

Let us look at the EOM for $E_{\rm trans}^\beta\ (Z)$:
{\footnotesize
\begin{eqnarray}
\label{EtransbetaBUV}
& & {\cal X}^{\beta^0} 
   E_{\rm trans}^\beta\ '(Z)+E_{\rm trans}^\beta\ ''(Z)+E_{\rm trans}^\beta\ (Z) \left(-\left(q^2 {\cal Y}_2^{\beta^0}+w^2
   {\cal Y}_1^{\beta^0}\right)\right)-E_{\rm trans}^{\beta^0}\ (Z) \left(q^2
   {\cal Y}_2^{\beta}+w^2 {\cal Y}_1^\beta \right)+{\cal X}^\beta 
   E_{\rm trans}^{\beta^0}\ (Z)=0,\nonumber\\
& &    
\end{eqnarray}
}
wherein,
\begin{eqnarray}
\label{Y1Y2Xprime-defs-i}
& & {\cal X} \equiv \partial_Z \log\left({\cal L}_{\rm DBI}^B G^{ZZ} {G^{x^2x^2}}\right) = {\cal X}^{\beta^0} + \beta {\cal X}^{\beta},\nonumber\\
& & {\cal Y}_1 \equiv \frac{{G^{tt}}}{G^{ZZ}} = {\cal Y}_1^{\beta^0} + \beta {\cal Y}_2^{\beta},\nonumber\\
& &{\cal Y}_2 \equiv \frac{{G^{x^1x^1}}}{G^{ZZ}} = {\cal Y}_2^{\beta^0} + \beta {\cal Y}_2^{\beta};
\end{eqnarray}
\begin{eqnarray}
\label{Y1Y2Xprime-defs-ii}
& & {\cal Y}_1^{\beta^0}=-\frac{4 \pi  {\cal C}_{x^3x^3}^B\ ^2
   \kappa_{a_{x3}^{Z,2}}^2 {g_s} \sqrt{\frac{1}{N}} N^{3/2}
   e^{-2 \rho -6 Z} \left({\cal C}_{x^3\rho,1}^B\  e^{\rho
   }+{\cal C}_{x^3\rho}^B\ \right)^2}{{r_h}^4},\nonumber\\
& & {\cal Y}_1^\beta = -\frac{4 \pi  \beta  {g_s}
   N^{3/2} e^{-2 Z} ({\cal C}_{zz}-2 {\cal C}_{\theta_1z}+2
  {\cal C}_{\theta_1x})}{{r_h}^2};\nonumber\\
& & {\cal Y}_2^{\beta^0}=-\frac{4 \pi  {\cal C}_{x^3x^3}^B\ ^4
   {\cal C}_{tZ,1}^B\ ^2 \kappa_{a_{x3}^{Z,2}}^4 {\cal C}_{x^3\rho, 2}^B\ ^2 {g_s} N e^{-4 \rho -6 Z} \sin ^2(\phi )
   \left({\cal C}_{x^3\rho,1}^B\  e^{\rho }+{\cal C}_{x^3\rho, 2}^B\ \right)^2}{B^2 {\cal C}_{tZ,2}^B\ ^2 {r_h}^4},\nonumber\\
& & {\cal Y}_2^{\beta}=-\frac{8 \pi  \beta  {\cal C}_{x^3x^3}^B\ ^4
   {\cal C}_{tZ,1}^B\  \kappa_{a_{x3}^{Z,2}}^4 {\cal C}_{tZ}^B\ 
   {\cal C}_{x^3\rho}^B\ ^2 {g_s} N e^{-4 \rho -6 Z} \sin
   ^2(\phi ) \left({\cal C}_{x^3\rho,1}^B\  e^{\rho
   }+{\cal C}_{x^3\rho}^B\ \right)^2}{B^2 {\cal C}_{tZ,2}^B\ ^2
   {r_h}^4};\nonumber\\
& & {\cal X}^{\beta^0}=2-\frac{2 {\cal C}_{tZ,2}^B\ 
   e^{-Z}}{{\cal C}_{tZ,1}^B\ },\nonumber\\
& & {\cal X}^{\beta}=\frac{2 {\cal C}_{tZ,2}^B\  {\cal C}_{tZ}^B\ 
   e^{-Z}}{{\cal C}_{tZ,1}^B\ ^2}.
\end{eqnarray}
Now, making the following perturbative ansatz up to ${\cal O}(w^4)$,
\begin{equation}
\label{EtransuptoW4}
E_{\rm trans}^\beta\ (Z)=E_{\rm trans}^{w^0,\ \beta}\ (Z)+w^2
   E_{\rm trans}^{w^2,\ \beta}\ (Z)+w^4 E_{\rm trans}^{w^4,\ \beta}\ (Z),
\end{equation}
one obtains the following EOMs.
\begin{itemize}
\item
${\cal O}(w^0)$:\\
{\footnotesize
\begin{eqnarray}
\label{EOMw0}
& & \frac{2 {\cal C}_{tZ,2}^B\  {\cal C}_{tZ}^B\  c_2 e^{-3 Z}
   \left({\cal C}_{tZ,1}^B\ ^2 e^{2 Z}-4
   {\cal C}_{tZ,2}^B\ ^2\right)}{{\cal C}_{tZ,1}^B\ ^4}+\left(2-\frac{
   2 {\cal C}_{tZ,2}^B\  e^{-Z}}{{\cal C}_{tZ,1}^B\ }\right)
   E_{\rm trans}^{w^0,\ \beta}\ '(Z)+\frac{{\cal C}_{tZ}^B\  c_1 e^Z}{4
   {\cal C}_{tZ,2}^B\ }+E_{\rm trans}^{w^0,\ \beta}\ ''(Z) = 0,\nonumber\\
& &    
\end{eqnarray}
}
whose solution is given by:
{\footnotesize
\begin{eqnarray}
\label{EOMw0-sol}
& & \hskip -0.8in E_{\rm trans}^{w^0,\ \beta}\ (Z) = c_4 +  \frac{1}{12} \Biggl(\frac{24
   {\cal C}_{tZ,2}^B\ ^2 {\cal C}_{tZ}^B\  c_2 e^{-2
   Z}}{{\cal C}_{tZ,1}^B\ ^3}+\frac{{\cal C}_{tZ}^B\  (c_1+12 c_2)
   e^{-\frac{2 {\cal C}_{tZ,2}^B\  e^{-Z}}{{\cal C}_{tZ,1}^B\ }}
   \left({\cal C}_{tZ,1}^B\ +2 {\cal C}_{tZ,2}^B\  e^{-Z}\right)
   {Ei}\left(\frac{2 {\cal C}_{tZ,2}^B\ 
   e^{-Z}}{{\cal C}_{tZ,1}^B\ }\right)}{{\cal C}_{tZ,1}^B\ ^2}\nonumber\\
 & &   +\frac{3
   {\cal C}_{tZ,1}^B\  c_3 e^{-\frac{2 {\cal C}_{tZ,2}^B\ 
   e^{-Z}}{{\cal C}_{tZ,1}^B\ }-Z} \left({\cal C}_{tZ,1}^B\  e^Z+2
   {\cal C}_{tZ,2}^B\ \right)}{{\cal C}_{tZ,2}^B\ ^2}+\frac{12
   {\cal C}_{tZ}^B\  c_2
   Z}{{\cal C}_{tZ,1}^B\ }-\frac{{\cal C}_{tZ}^B\  c_1
   e^Z}{{\cal C}_{tZ,2}^B\ }\Biggr) \nonumber\\
& & \hskip -0.8in = \frac{-2 {\cal C}_{tZ,2}^B\ ^2 {\cal C}_{tZ}^B\  c_1 Z-24
   {\cal C}_{tZ,2}^B\ ^2 {\cal C}_{tZ}^B\  c_2 Z}{24
   {\cal C}_{tZ,1}^B\  {\cal C}_{tZ,2}^B\ ^2}-\frac{{\cal C}_{tZ}^B\ 
   c_1 e^Z}{12 {\cal C}_{tZ,2}^B\ }.
\end{eqnarray}
}
\item
${\cal O}(w^2)$:\\
{\scriptsize
\begin{eqnarray}
\label{EOMw2}
& & \hskip -0.8in \frac{1}{{r_h}^4}\Biggl\{4 \pi  {\cal C}_{x^3x^3}^B\ ^2 \kappa_{a_{x3}^{Z,2}}^2 {g_s}
   \sqrt{\frac{1}{N}} N^{3/2} e^{-2 (\rho +3 Z)}
   \left({\cal C}_{x^3\rho,1}^B\  e^{\rho }+{\cal C}_{x^3\rho, 2}^B\ \right)^2 \Biggr\}
   \nonumber\\
& & \hskip -0.8in \times \Biggl[\frac{1}{12} \Biggl\{\frac{24
   {\cal C}_{tZ,2}^B\ ^2 {\cal C}_{tZ}^B\  c_2 e^{-2
   Z}}{{\cal C}_{tZ,1}^B\ ^3}+\frac{{\cal C}_{tZ}^B\  (c_1+12 c_2)
   e^{-\frac{2 {\cal C}_{tZ,2}^B\  e^{-Z}}{{\cal C}_{tZ,1}^B\ }}
   \left({\cal C}_{tZ,1}^B\ +2 {\cal C}_{tZ,2}^B\  e^{-Z}\right)
   {Ei}\left(\frac{2 {\cal C}_{tZ,2}^B\ 
   e^{-Z}}{{\cal C}_{tZ,1}^B\ }\right)}{{\cal C}_{tZ,1}^B\ ^2}\nonumber\\
& & \hskip -0.8in   +\frac{3
   {\cal C}_{tZ,1}^B\  c_3 e^{-\frac{2 {\cal C}_{tZ,2}^B\ 
   e^{-Z}}{{\cal C}_{tZ,1}^B\ }-Z} \left({\cal C}_{tZ,1}^B\  e^Z+2
   {\cal C}_{tZ,2}^B\ \right)}{{\cal C}_{tZ,2}^B\ ^2}+\frac{12
   {\cal C}_{tZ}^B\  c_2
   Z}{{\cal C}_{tZ,1}^B\ }-\frac{{\cal C}_{tZ}^B\  c_1
   e^Z}{{\cal C}_{tZ,2}^B\ }\Biggr\}+c_4\Biggr]\nonumber\\
   & & \hskip -0.8in   +\frac{
   \pi  {\cal C}_{tZ,1}^B\ ^2 {\cal C}_{tZ}^B\  c_1 {g_s} N
   e^Z}{{\cal C}_{tZ,2}^B\ ^3 {r_h}^2}+\frac{4.
   {\cal C}_{tZ}^B\  c_2 {g_s} N e^{-2 Z} \left(\pi 
   \sqrt{{\cal C}_{tZ,2}^B\ ^2 {r_h}^2}-6.28319
   {\cal C}_{tZ,2}^B\  {r_h}\right)}{{\cal C}_{tZ,1}^B\ 
   {\cal C}_{tZ,2}^B\  {r_h}^3}+\left(2-\frac{2
   {\cal C}_{tZ,2}^B\  e^{-Z}}{{\cal C}_{tZ,1}^B\ }\right)
   E_{\rm trans}^{w^2,\ \beta}\ '(Z)+E_{\rm trans}^{w^2,\ \beta}\ ''(Z) = 0,\nonumber\\
& &    
\end{eqnarray}
}
whose solution is given by:
{\scriptsize
\begin{eqnarray}
\label{EOMw2-sol}
& & \hskip -0.8in E_{\rm trans}^{w^2,\ \beta}\ (Z)=c_3  + \frac{e^{-\frac{2 {\cal C}_{tZ,2}^B\ 
   e^{-Z}}{{\cal C}_{tZ,1}^B\ }-Z} }{12
   {\cal C}_{tZ,2}^B\ ^3 {r_h}^2}\nonumber\\
& & \hskip -0.8in \times \Biggl[{\cal C}_{tZ,1}^B\  \left(-4
   \pi  {\cal C}_{tZ,1}^B\  {\cal C}_{tZ}^B\  c_1 {g_s} N e^{2
   \left(\frac{{\cal C}_{tZ,2}^B\ 
   e^{-Z}}{{\cal C}_{tZ,1}^B\ }+Z\right)}+3 {\cal C}_{tZ,1}^B\ 
   {\cal C}_{tZ,2}^B\  c_2 {r_h}^2 e^Z+6 {\cal C}_{tZ,2}^B\ ^2 c_2
   {r_h}^2\right)\nonumber\\
& & \hskip -0.8in +4 \pi  {\cal C}_{tZ,2}^B\  {\cal C}_{tZ}^B\ 
   c_1 {g_s} N \left({\cal C}_{tZ,1}^B\  e^Z+2
   {\cal C}_{tZ,2}^B\ \right) {Ei}\left(\frac{2
   {\cal C}_{tZ,2}^B\  e^{-Z}}{{\cal C}_{tZ,1}^B\ }\right)\Biggr]  = -\frac{\pi  {\cal C}_{tZ,1}^B\ ^2 {\cal C}_{tZ}^B\  c_1 {g_s}
   N e^Z}{3 {\cal C}_{tZ,2}^B\ ^3 {r_h}^2}-\frac{\pi 
   {\cal C}_{tZ,1}^B\  {\cal C}_{tZ}^B\  c_1 {g_s} N Z}{3
   {\cal C}_{tZ,2}^B\ ^2 {r_h}^2}.
\end{eqnarray}
}
\item
${\cal O}(w^4)$:\\
\begin{eqnarray}
\label{EOMw4}
& & \hskip -0.8in \frac{4 \pi  {\cal C}_{x^3x^3}^B\ ^2 \kappa_{a_{x3}^{Z,2}}^2 {g_s}
   \sqrt{\frac{1}{N}} N^{3/2} e^{-2 (\rho +3 Z)}
   \left({\cal C}_{x^3\rho,1}^B\  e^{\rho }+{\cal C}_{x^3\rho, 2}^B\ \right)^2 \left(-\frac{\pi  {\cal C}_{tZ,1}^B\ 
   {\cal C}_{tZ}^B\  c_1 {g_s} N \left({\cal C}_{tZ,1}^B\ 
   e^Z+{\cal C}_{tZ,2}^B\  Z\right)}{3 {\cal C}_{tZ,2}^B\ ^3
   {r_h}^2}\right)(Z)}{{r_h}^4}\nonumber\\
& &  \hskip -0.8in  +\frac{4 \pi ^2
   {\cal C}_{tZ,1}^B\ ^4 {\cal C}_{tZ}^B\  c_1 {g_s}^2 N^2
   e^Z}{{\cal C}_{tZ,2}^B\ ^5 {r_h}^4}+\frac{355.306
   {\cal C}_{tZ,1}^B\ ^2 {\cal C}_{tZ}^B\  c_2 {g_s}^2 N^2
   e^{-Z}}{{\cal C}_{tZ,2}^B\ ^3 {r_h}^4}+\left(2-\frac{2
   {\cal C}_{tZ,2}^B\  e^{-Z}}{{\cal C}_{tZ,1}^B\ }\right)
   E_{\rm trans}^{w^4,\ \beta}\ '(Z)\nonumber\\
& &  \hskip -0.8in  + E_{\rm trans}^{w^4,\ \beta}\ ''(Z)=0,
\end{eqnarray}
whose solution is given by:
{\footnotesize
\begin{eqnarray}
\label{EOMw4-sol}
& & \hskip -0.8in E_{\rm trans}^{w^4,\ \beta}\ (Z) = c_3 + \frac{{\cal C}_{tZ,1}^B\  e^{-\frac{2
   {\cal C}_{tZ,2}^B\  e^{-Z}}{{\cal C}_{tZ,1}^B\ }-Z} }{12
   {\cal C}_{tZ,2}^B\ ^5 {r_h}^4}\nonumber\\
& & \hskip -0.8in \times \Biggl(-16 \pi
   ^2 {\cal C}_{tZ,1}^B\ ^3 {\cal C}_{tZ}^B\  c_1 {g_s}^2 N^2
   e^{2 \left(\frac{{\cal C}_{tZ,2}^B\ 
   e^{-Z}}{{\cal C}_{tZ,1}^B\ }+Z\right)}+3 {\cal C}_{tZ,1}^B\ 
   {\cal C}_{tZ,2}^B\ ^3 c_2 {r_h}^4 e^Z+16 \pi ^2
   {\cal C}_{tZ,1}^B\  {\cal C}_{tZ,2}^B\  {\cal C}_{tZ}^B\  c_1
   {g_s}^2 N^2 \left({\cal C}_{tZ,1}^B\  e^Z+2
   {\cal C}_{tZ,2}^B\ \right)\nonumber\\
& & \hskip -0.8in \times    {Ei}\left(\frac{2
   {\cal C}_{tZ,2}^B\  e^{-Z}}{{\cal C}_{tZ,1}^B\ }\right)+6
   {\cal C}_{tZ,2}^B\ ^4 c_2 {r_h}^4\Biggr) = -\frac{4 \pi ^2 {\cal C}_{tZ,1}^B\ ^4 {\cal C}_{tZ}^B\  c_1
   {e^Z} {g_s}^2 N^2}{3 {\cal C}_{tZ,2}^B\ ^5
   {r_h}^4}-\frac{4 \pi ^2 {\cal C}_{tZ,1}^B\ ^3
   {\cal C}_{tZ}^B\  c_1 {g_s}^2 N^2 Z}{3
   {\cal C}_{tZ,2}^B\ ^4 {r_h}^4}.
\end{eqnarray}
}
\end{itemize}
Thus,
{\footnotesize
\begin{eqnarray}
\label{dEtransoverEtransuptobetauptow4}
& & \hskip -0.8in \left(\frac{E_{\rm trans}\ '(Z)}{E_{\rm trans}(Z)}\right) = 2 + \frac{c_2 {g_s} N
   w^2 (\beta  {\cal C}_{tZ}^B\  (100.5\, -201.1 Z)+201.1
   {\cal C}_{tZ,1}^B\ )}{{\cal C}_{tZ,1}^B\  c_1 e^{2Z}
   {r_h}^2}-\frac{2842.4 {\cal C}_{tZ,1}^B\ ^2 c_2 {g_s}^2
   N^2 w^4}{{\cal C}_{tZ,2}^B\ ^2 c_1 e^{2Z}
   {r_h}^4}   +\frac{2 \beta  {\cal C}_{tZ,2}^B\ 
   {\cal C}_{tZ}^B\ }{3 {\cal C}_{tZ,1}^B\ ^2 e^Z}.
\end{eqnarray}
}
Hence,
{\scriptsize
\begin{eqnarray}
\label{Chi2}
& & \chi_2 = \frac{{\cal C}_{x^3x^3}^B\ ^2 {\cal C}_{tZ,1}^B\ ^2 \kappa_{a_{x3}^{Z,2}}^2
   {\cal C}_{x^3\rho}^B\ ^3 N^{3/5} \rho ^2 e^{8 Z-2 \rho } \cos ^2(\phi )
  }{B^2
   {\cal C}_{tZ,2}^B\ ^5 c_1 {g_s} {r_h}^2 \alpha _{\theta _2}^2
   \left({\cal C}_{x^3\rho,1}^B\  e^{\rho }+{\cal C}_{x^3\rho}^B\ \right)}\nonumber\\
& & \times  \left(-2787.15 {\cal C}_{tZ,1}^B\ ^3 c_2 {g_s}^2 N^2 w^4+{\cal C}_{tZ,1}^B\ 
   {\cal C}_{tZ,2}^B\ ^2 {r_h}^2 \left(197.191 c_2 {g_s} N w^2+1.96112 c_1
   {r_h}^2 e^{2 Z}\right)-197.191 \beta  {\cal C}_{tZ,2}^B\ ^2
   {\cal C}_{tZ}^B\  c_2 {g_s} N {r_h}^2 w^2 (Z-0.5)\right).\nonumber\\
& &    
\end{eqnarray}
}
Assuming 
\begin{equation}
\label{ZUV}
(4 \pi g_s N)^{1/N_{>4}} = \frac{r_h}{{\cal R}_{D5/\overline{D5}}} e^{Z_{\rm UV}},
\end{equation}
 and $r_h (g_s = 0.1, M = N_f = 3)\sim e^{-0.3 N^{1/3}} l_s$, and $w\sim \frac{r_h^{\kappa_w}}{l_s^2}$, one sees that $\frac{1.96 e^{2 Z_{\rm UV}} r_h^2 }{ 
 197.19 g_s N w^2 } \sim \frac{\left(l_s^2{\cal O} (1) g_s N)^{2/N_{>4}} r_h^2 \right)}{\left({\cal O} (10^2) g_s N r_h^{2 \kappa_w} \right)}$. Therefore, $\frac{{\cal O} (1) 
    (g_s N)^{2/N_{>4}} r_h^2}{({\cal O} (10^2) g_s N r_h^{2 \kappa_w})} \sim 
    (g_s N)^{2/N_{>4} - 1} e^{-0.6 (1 - \kappa_w) N^{1/3}}$. Now, $2/N_{>4} < 1/2$, hence $2/N_{>4} - 1 < 0$. Hence, in a large - $N$ limit, $\left| \frac{{\cal O} (w^0)}{{\cal O} (w)}\right| \ll 1$.

Writing 
\begin{equation}
\label{N>}
N^{\frac{1}{N_{>4} }}=\frac{\sqrt[4]{N}}{{\kappa_{Z_{\rm UV}}}}, N_{>4} \in\mathbb{Z}, N_{>4} >4,
\end{equation}
one obtains the following:
\begin{eqnarray}
\label{spectral-functionoverwNsq-B}
& & \hskip -0.8in \frac{{\chi_2}}{N^2  \tilde{w} T^2}\sim \frac{{\cal C}_{x^3x^3}^B\ ^2 {\cal C}_{tZ,1}^B\ ^2
   \kappa_{a_{x3}^{Z,2}}^2 c_2 N \tilde{w}
   \left({\cal C}_{x^3\rho}^B\ 
   ({\cal C}_{x^3\rho}^B\ -8. {\cal C}_{x^3\rho,1}^B\ )-8. {\cal C}_{x^3\rho,1}^B\ ^2
   {Li}_3\left(-\frac{{\cal C}_{x^3\rho,2}^B\ }{{\cal C}_{x^3\rho,1}^B\ }\right)\right)
   (\beta  {\cal C}_{tZ}^B\ 
   {Z_{\rm UV}}-{\cal C}_{tZ,1}^B\ ) }{{\alpha_{\rm ZUV}}^8 B^2 {\cal C}_{tZ,2}^B\ ^3
   c_1}\nonumber\\
& & \hskip -0.3in \times \ e^{-\frac{4
   {\cal C}_{tZ,1}^B\  \tilde{w}^2 ({\cal C}_{tZ,1}^B\ +\beta 
   {\cal C}_{tZ}^B\ 
   {Z_{\rm UV}})}{{\cal C}_{tZ,2}^B\ ^2
   }}  \equiv \frac{\kappa_1 \tilde{w} e^{-\kappa_2 \tilde{w}^2}}{\tilde{B}^2},     
\end{eqnarray} 
where $\tilde{B}\equiv\frac{B}{T^2}$. If one adds a term quadratic in $\tilde{w}^2$ in the numerator of the RHS of (\ref{spectral-functionoverwNsq-B}),
\begin{equation}
\label{extra-term-chi-over-wtilde}
\frac{{\chi_2}}{N^2  \tilde{w} T^2} = \frac{\kappa_1 \tilde{w} e^{-\kappa_2 \tilde{w}^2} + \kappa_3 \tilde{w}^2}{\tilde{B}^2},
\end{equation}
one obtains the comparative plot in Fig. \tcb{3}.
\begin{figure}
\begin{center}
\includegraphics[width=0.6\textwidth]{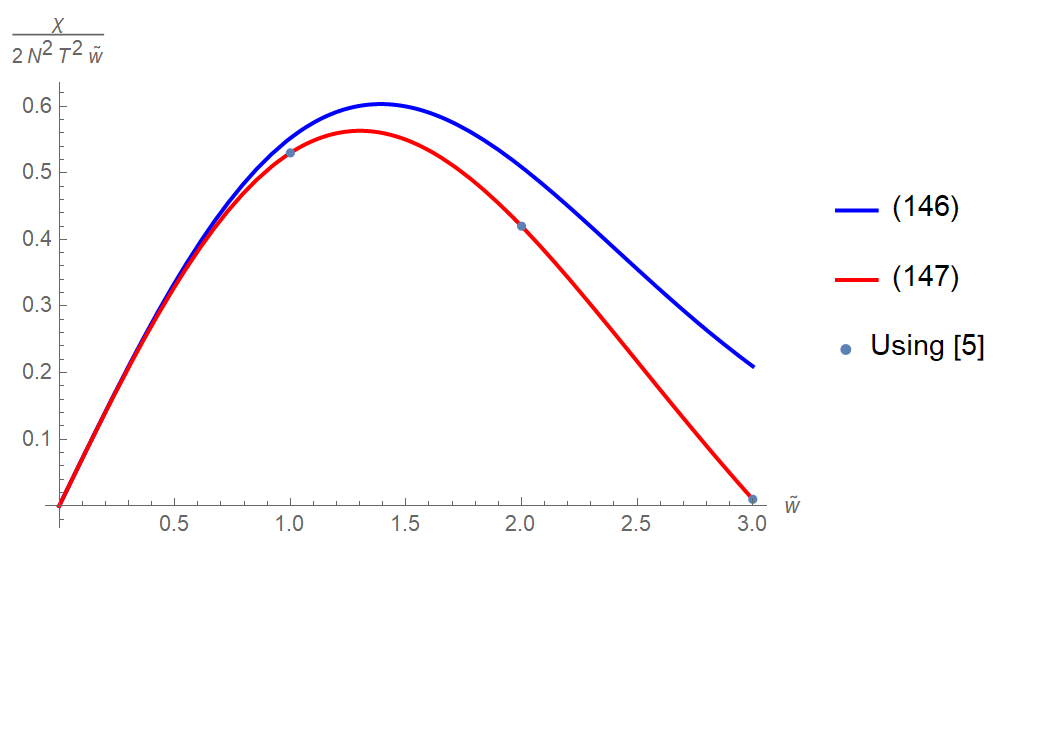}
\end{center}
\vskip-1in
\caption{$\frac{\chi}{2 N^2\tilde{w}T^2}$-vs-$\tilde{w}$ -  the plot in red is with an extra term quadratic in $\tilde{w}$ added by hand, the plot in thick blue is based on (\ref{spectral-functionoverwNsq-B}) and and the plot points in pale blue are from \cite{D5gaugedSUGRAtrunc} for $\frac{B}{T^2} = 11.24$}
\label{chi-over-w_vs_w}
\end{figure}

\section{Summary}
\label{Summary}

The study of thermal QCD-like theories in the presence of strong magnetic fields and at intermediate coupling from ${\cal M}$ theory, had been missing in the literature. We fill this gap by looking at the world-volume theory of the flavor $D6$-branes in the type IIA SYZ mirror of the parent type IIB dual \cite{metrics}, the former constructed in \cite{MQGP}, \cite{NPB} inclusive of ${\cal O}(R^4)$ corrections \cite{OR4}, in the presence of a large uniform magnetic field. 

 \begin{tcolorbox}[enhanced,width=7in,center upper,size=fbox,
    %fontupper=\large\bfseries,
    drop shadow southwest,sharp corners]
    \begin{flushleft}
The two issues we look at are photoproduction and Equation of State. In the process of doing so, in the ${\cal M}$-theory $C_3=0$-truncation, one sees that while the former receives ${\cal O}(R^4)$ corrections, the latter - motivated by the type IIB modular completion at ${\cal O}(R^4)$ \cite{Green and Gutperle} - receives no such corrections. In the language of differential geometry, this corresponds to the lack of $N$-connectedness of the parameter space corresponding to (Almost) Contact 3-Structures derivable from $G_2$-structures supported on a closed seven-fold \cite{ACMS} - a warped product of the ${\cal M}$-theory circle and a non-Kähler six-fold with the six-fold being a warped product of the thermal circle with a non-Einsteinian deformation of $T^{1,1}$. 
\end{flushleft}
\end{tcolorbox}

\subsection{$D6$-brane world-volume Gauge fields in the deconfined phase in the presence/absence of $B\gg T_c^2$ (in $e=1$-units)}

For the purpose of computation of the spectral density relevant to photoproduction and the free energy relevant to EoS via the pressure in the deconfined phase ($T>T_c$), we first obtained the following background gauge field configurations supported on the world-volume of the type IIA flavor $D6$-branes:
\begin{itemize}
\item
UV, $B=0$:
{\footnotesize
\begin{eqnarray}
\label{Amu-UV-B0}
& & a_{x^3}^Z\ '(Z)={\cal C}_{x3}^{\phi Z}\    a_\phi^Z\  '(Z) , a_{x^3}^\rho(\rho)={\cal C}_{a_{x3}^{\rho\phi}}^{B=0,\ {\rm UV}}\  {a_\phi^\rho}(\rho ),\nonumber\\
& & a_{x^3}^{x^3}(x^3)=\frac{{\cal C}_{a_{x3}^Z(x3)}\ ^{B=0,\ {\rm UV}}\  {a_\phi^{x^3}}(x^3)}{\sqrt{a_t^{x^3}(x^3)^2-{\cal C}_{a_{x3}^Z(x3)}^{B=0,\ {\rm UV}}\ ^2 
{\cal C}_{a_{x3}^{\rho\phi}}^{B=0,\ {\rm UV}}\ ^2 {\cal C}_{x3}^{\phi Z}\   ^2}},\nonumber\\
& & a_t^\rho(\rho)=\int _1^{\rho }\frac{{\cal C}_{a_{x3}^Z(\rho)}^{B=0,\ {\rm UV}}\ (\rho )}{\rho }d\rho +c_1,\nonumber\\
& & a_t^Z(Z) = {\cal C}_{a_{x3}^Z(Z)}^{B=0,\ {\rm UV}}\   e^{-4 Z}\nonumber\\
& & a_\phi^Z\  (Z)={\cal C}_{a_t^{\phi}(Z)}^{B=0,\ {\rm UV}}\     e^{-5 Z}+c_1,
\nonumber\\
& & a_\phi^\rho(\rho) = \frac{{\cal C}_{a_t^{\phi}(\rho)}^{B=0,\ {\rm UV}}\   }{\rho }\nonumber\\
& & a_\phi^{x^3}(x^3)=\frac{{\cal C}_{a_t^{x3}(x3)}^{B=0,\ {\rm UV}}\   
   \sqrt{a_t^{x^3}(x^3)^2-{\cal C}_{a_{x3}^Z(x3)}\ ^{B=0,\ {\rm UV}}\  ^2
   {\cal C}_{a_{x3}^{\rho\phi}}^{B=0,\ {\rm UV}}\ ^2 {\cal C}_{a_{x3}^{\phi Z}}^{B=0,\ {\rm UV}}\ ^2}}{a_t^{x^3}(x^3)}.
\end{eqnarray}
subject to the following constraint:
\begin{eqnarray}
\label{const-Aphi-EOM}
& & 4 {\cal C}_{a_{x3}^Z(Z)}^{B=0,\ {\rm UV}}\  
   \sqrt{{\cal C}_{a_t^{x3}(x3)}\ ^{B=0,\ {\rm UV}}\ ^2-{\cal C}_{a_{x3}^Z(x3)}\ ^{B=0,\ {\rm UV}}\  ^2
   {\cal C}_{a_{x3}^{\rho\phi}}^{B=0,\ {\rm UV}}\ ^2 {\cal C}_{x3}^{\phi Z}\   ^2}=1.
\end{eqnarray}
}

\item Strong $B\gg(0.15 GeV)^2$ in the UV:
{\footnotesize
\begin{eqnarray}
\label{Amu-UV-B}
& & a_t^{x^3}(x^3)={\cal C}_{a_t^{x3}}^{B\ {\rm UV}}\  a_{x^3}^{x^3}(x^3),\nonumber\\
& & a_\phi^{x^3}(x^3)={\cal C}_{a_\phi^{x3}}^{B\ {\rm UV}}\ ,\nonumber\\
& & a_t^\rho(\rho)={\cal C}_{a_t^\rho}^{B\ {\rm UV}}\  a_{x^3}^\rho(\rho),\nonumber\\
& & a_\phi^\rho(\rho)=\frac{{\cal C}_{a_t^\rho}^{B\ {\rm UV}}\ ^2 a_{x^3}^\rho(\rho)}{a_{x^3}^\rho\ '(\rho)},\nonumber\\
& & a_t^Z(Z)={\cal C}_{tZ,1}^B\ +({\cal C}_{tZ,2}^B\ +{\epsilon_1})
   e^{-\kappa_{a_t^Z} Z},\nonumber\\
& & a_{x^3}^Z\ (Z)=\kappa_{a_x3^{Z,1}}+(\kappa_{a_{x3}^{Z,2}}+{\epsilon_2})
   e^{-\kappa_{a_x3^Z} Z},\nonumber\\
& & {\cal C}_{a_t^\rho}^{B\ {\rm UV}}\  =  \pm   \frac{\kappa_{a_{x3}^{Z,2}}}{{\cal C}_{tZ,2}^B\  {\cal C}_{a_t^{x3}}^{B\ {\rm UV}}\ },\nonumber\\
& & |{\cal C}_{tZ,1}^B\ |\ll1, |\kappa_{a_{x3}^{Z,2}}|\ll1,\nonumber\\
& & a_\phi^Z\  (Z) = -\frac{{\epsilon_2}^{3/2}
   \kappa_\phi^Z  \left(2 {\cal C}_{tZ,1}^B\ ^2 \log
   \left({\cal C}_{tZ,1}^B\  e^Z+{\cal C}_{tZ,2}^B\ \right)-2
   {\cal C}_{tZ,1}^B\ ^2 Z-2 {\cal C}_{tZ,1}^B\  {\cal C}_{tZ,2}^B\ 
   e^{-Z}+{\cal C}_{tZ,2}^B\ ^2 e^{-2 Z}\right)}{2
   {\cal C}_{tZ,2}^B\ ^3}+c_1,\nonumber\\
& & \frac{\kappa_{a_{x3}^{Z,2}}^{7/2} {\cal C}_{a_\phi^{x3}}^{B\ {\rm UV}}\ 
   \kappa_\phi^Z  N^{3/5} \rho  {r_h}^2}{2 \sqrt[3]{3}
   {\cal C}_{tZ,2}^B\ ^3 {\cal C}_{a_t^{x3}}^{B\ {\rm UV}}\ ^2 {g_s} \alpha
   _{\theta _2}^2}={\cal C}_{x^3}^B(x^3, \rho)={\cal C}_{x^3}^B\  \rho,
\end{eqnarray}
}
\item
IR, $B=0$:
{\footnotesize
\begin{eqnarray}
\label{Amu-IR-B0}
& & {a_t^Z}(Z)=-\frac{{\cal C}_\phi^{t Z,\ B=0}\  e^{-2 Z}}{729 \sqrt{3} \pi  {N_f} {r_h}^2 \alpha _{\theta _2}^3
   \left(\left(3 \sqrt{3}+8 \pi \right) c_2 {g_s} M^2 \left(\frac{1}{N}\right)^{2/5} \left(Z^2+2
   Z-4\right)-48 N^{3/5}\right)}\nonumber\\
& & = \frac{{\cal C}_\phi^{t Z,\ B=0}\  \left(\frac{1}{N}\right)^{3/5} (1 - 2 Z + 2 Z^2)}{34992 \sqrt{3} \pi  {N_f} {r_h}^2 \alpha
   _{\theta _2}^3}+{\cal O}\left(\left(\frac{1}{N}\right)^{6/5}\right),\nonumber\\
& & {a_t^\rho}(\rho )={\cal C}_\rho^{B=0}\    \log (\rho )+c_1,\nonumber\\
& & {a_t^{x^3}}({x^3})=\frac{23328\ 3^{5/6} \pi ^2 \alpha _{\theta _2}^5 \sqrt{2 {\cal C}_{x^3}^{\rho\phi,\ B=0}\ ^2
   {\cal C}_{x^3}\ ^{\phi,\ B=0}\ ^2 {\cal C}_{x^3}\ ^{\phi Z,\ B=0}\ ^2+2} {\cal C}_\phi^{\rho x^3,\ B=0}(\rho, x^3)}{{\cal C}_\phi^{t Z,\ B=0}\  {\cal C}_\rho^{B=0}\    \log ({r_h})},\nonumber\\
& & {a_{x^3}^\rho}(\rho )={\cal C}_{x^3}^{\rho\phi,\ B=0}\  {a_\phi^\rho }(\rho),
\nonumber\\
& & {a_{x^3}^{x^3}}({x^3})={\cal C}_{x^3}\ ^{\phi,\ B=0}\  {a_\phi^{x^3}}({x^3}),\nonumber\\
& & {a_{x^3}^Z}'(Z)={\cal C}_{x^3}\ ^{\phi Z,\ B=0} {a_\phi^Z}'(Z),\nonumber\\
& & {a_\phi^Z}(Z)=\frac{i \sqrt{2} \pi  {\cal C}_{\phi Z}^{B=0}\  \left(\frac{1}{N}\right)^{3/5} \alpha _{\theta _2}^2
   (\log ((1-i)-2 i Z)-\log (2 i Z+(1+i)))}{3^{2/3} {\cal C}_{\phi\rho}^{B=0}\  {N_f} {r_h}^2 \log ({r_h})},\nonumber\\
& & {a_\phi^{x^3}}({x^3})=\frac{\sqrt{{\cal C}_{x^3}^{\rho\phi,\ B=0}\ ^2 {\cal C}_{x^3}\ ^{\phi,\ B=0}\ ^2 {\cal C}_{x^3}\ ^{\phi Z, B=0}\ ^2+1} {\cal C}_t^{x^3Z,\ B=0}(x^3, Z)}{{\cal C}_{\phi Z}^{B=0}\ } 
\end{eqnarray}
}

\item
IR, Strong $B\gg(0.15\ {\rm GeV})^2$:
{\footnotesize
\begin{eqnarray}
\label{Amu-IR-B}
& & {a_{x^3}^\rho\ }(\rho )={\cal C}_{x^3}^{t\rho,\ B}\  {a_t^\rho\ }(\rho )\nonumber\\
& & a_{x^3}\ ^{x^3}({x^3})={{\cal C}_{x^3t}^B} {a_t^{x^3}}({x^3})\nonumber\\
& & {a_{x^3}^Z}\ '(Z)={\cal C}_{x^3}^{tZ,\ B}\  {a_t^Z}'(Z)\nonumber\\
& & \frac{\rho  {a_t^\rho}'(\rho )}{{a_t^\rho}(\rho
   )}={\cal C}_B^{t\rho}\ ,\nonumber\\
& & \frac{\rho  {a_{x^3}^\rho}'(\rho )}{{a_t^\rho}(\rho
   )}={\cal C}_B^{x^3\rho}\nonumber\\
& & a_t(Z) = {\cal C}_{a_t^Z}^B\   e^{\kappa_0}+{\cal C}_{a_t^Z}^B\  
   e^{\kappa_0} \kappa_1 Z+\frac{3}{2} {\cal C}_{a_t^Z}^B\  
   e^{\kappa_0} \kappa_1^2 Z^2 + {\cal O}\left(Z^3\right),\nonumber\\
& & {a_\phi^\rho}(\rho )={\cal C}_{\phi\rho}^B\  \rho,\nonumber\\
& & {\cal C}_{x^3}^B(\rho, x^3)=\rho  {\cal C}_{x^3}^B(x^3),\nonumber\\
& & a_t^\rho(\rho) = {\cal C}_t^{\rho,\ B}\  \rho\nonumber\\
& & {a_t^{x^3}}({x^3})={\cal C}_{tx^3}^B,\nonumber\\
& &  a_\phi(Z) = \alpha_1
 {\cal C}_{\phi Z}^B\  + B ( \alpha_2  + \alpha_3 Z + \alpha_4 Z^2) + \frac{\pi  \kappa_1
   \left(\frac{1}{N}\right)^{3/5} \alpha _{\theta _2}^2
   {\cal C}_{x^3}^B(x^3) \log (Z)}{\sqrt{2} 3^{2/3} {{\cal C}_{x^3t}^B}
   {\cal C}_{x^3}^{t\rho,\ B}\  {\cal C}_{x^3}^{tZ,\ B}\  {\cal C}_{\phi\rho}^B\ 
   {\cal C}_{\phi x^3}^B\  {N_f} {r_h}^2 \log ({r_h})}.\nonumber\\
& &    
\end{eqnarray}
}

\end{itemize}

\begin{table}[h]
\centering
\begin{tabular}{|c|c|c|c|} 
\hline
S. No. & Constants of Integration & UV/IR & $B=0/B\gg\left(0.15\ {\rm GeV}\right)^2$ \\ 
\hline
1 & 
\begin{tabular}{c}
${\cal C}_{x3}^{\phi Z}, {\cal C}_{a_{x3}^{\rho\phi}}^{B=0,\ {\rm UV}},
{\cal C}_{a_{x3}^Z(x3)}\ ^{B=0,\ {\rm UV}}, {\cal C}_{a_{x3}^Z(\rho)}^{B=0,\ {\rm UV}}$\\
${\cal C}_{a_{x3}^Z(Z)}^{B=0,\ {\rm UV}}, {\cal C}_{a_t^{\phi}(Z)}^{B=0,\ {\rm UV}},
{\cal C}_{a_t^{\phi}(\rho)}^{B=0,\ {\rm UV}}, {\cal C}_{a_t^{x3}(x3)}^{B=0,\ {\rm UV}}, {\cal C}_{a_{x3}^{\phi Z}}^{B=0,\ {\rm UV}}$
\end{tabular} 
& UV & $B=0$ \\ 
\hline
2 & 
\begin{tabular}{c}
${\cal C}_{a_t^{x3}}^{B\ {\rm UV}}, {\cal C}_{a_\phi^{x3}}^{B\ {\rm UV}},
{\cal C}_{a_t^\rho}^{B\ {\rm UV}}, {\cal C}_{a_t^\rho}^{B\ {\rm UV}}, 
{\cal C}_{tZ}^B , {\cal C}_{tZ}^B, {\cal C}_{x^3}^B(x^3, \rho)={\cal C}_{x^3}^B\  \rho
$ 
\end{tabular} 
& UV & $B\gg\left(0.15\ {\rm GeV}\right)^2$ \\ 
\hline
3 & 
\begin{tabular}{c}
${\cal C}_\phi^{t Z,\ B=0}, {\cal C}_\phi^{t Z,\ B=0}, 
{\cal C}_\rho^{B=0}, {\cal C}_{x^3}^{\rho\phi,\ B=0} $\\
${\cal C}_{x^3}\ ^{\phi,\ B=0}, {\cal C}_{x^3}\ ^{\phi Z,\ B=0}, 
{\cal C}_\phi^{\rho x^3,\ B=0}(\rho, x^3), {\cal C}_\phi^{t Z,\ B=0}$\\
$
{\cal C}_\rho^{B=0}, {\cal C}_{x^3}\ ^{\phi,\ B=0}, 
{\cal C}_{\phi Z}^{B=0}, {\cal C}_{\phi\rho}^{B=0}, {\cal C}_t^{x^3Z,\ B=0}(x^3, Z)
$

\end{tabular} 
& IR & $B=0$ \\ 
\hline
4 & 
\begin{tabular}{c}
$ {\cal C}_{x^3}^{t\rho,\ B}, {\cal C}_{x^3t}^B , 
{\cal C}_{x^3}^{tZ,\ B}, {\cal C}_B^{t\rho}, {\cal C}_B^{x^3\rho}, 
{\cal C}_{a_t^Z}^B 
$ \\
${\cal C}_{\phi\rho}^B, {\cal C}_{x^3}^B(\rho, x^3), {\cal C}_{x^3}^B(x^3), 
{\cal C}_t^{\rho,\ B}, {\cal C}_{tx^3}^B, 
{\cal C}_{\phi Z}^B, {\cal C}_{x^3}^B(x^3), 
{\cal C}_{\phi x^3}^B
$
\end{tabular} 
& IR & $B\gg\left(0.15\ {\rm GeV}\right)^2$ \\ 
\hline
\end{tabular}
\caption{Constants of integration appearing in the solutions of the type IIA $D6$-brane world-volume for $T>T_c$, categorized by UV/IR behavior and magnetic field strength.}
\label{tab:constants}
\end{table}

\subsection{$D6$-brane world-volume gauge field in the confined phase in the presence of $B>T_c^2$ (in $e=1$-units)}

The world-volume gauge fields on the type IIA flavor $D6$-branes for $T<T_c$ in the presence of a strsong magnetic field $B$ were found to be given as under: 
\begin{eqnarray}
\label{A-mu-IR+UV-th}
& & a_\phi^\rho(\rho) = \frac{\left({\cal C}_{a_t^{x3}}\ ^{\rm IR/UV, B}\right)^2\rho^{3/2}(2\sqrt{\rho}{\cal C}_{\rho}^{1, {\rm IR/UV}} + {\cal C}_{\rho}^{2, {\rm IR/UV}})}{B \left({\cal C}_{a_\phi^{\beta^0}}\ ^{\rm IR/UV}\right)^2{\cal C}_{a_\phi^{x3}\ ^{\rm B}}^{\rm IR/UV}},\nonumber\\
& & a_\phi^Z(Z) = {\cal C}_{\phi}^{1, {\rm IR/UV}} + \frac{1}{2}(1 + 2 Z) {\cal C}_{Z}^{1, {\rm IR/UV}} - \frac{1}{3}(1 + 3 Z){\cal C}_{\rho}^{2, {\rm IR/UV}},\nonumber\\
& & a_\phi^{x^3}(x^3) = {\cal C}_{a_\phi^{x3}\ ^{\rm B}}^{\rm IR/UV};\nonumber\\
& & a_t^Z(Z) = - (1 - Z) {\cal C}_{Z}^{1, {\rm IR/UV}} + {\cal C}_{Z}^{2, {\rm IR/UV}},\nonumber\\
& & a_t^\rho(\rho) = 2\sqrt{\rho}{\cal C}_{\rho}^{1, {\rm IR/UV}} + {\cal C}_{\rho}^{2, {\rm IR/UV}},\nonumber\\
& & a_t^{x^3}(x^3) = {\cal C}_{a_t^{x3}}\ ^{\rm IR/UV, B},
\end{eqnarray}
where ${\cal C}_{a_t^{x3}}\ ^{\rm IR/UV, B}, 
{\cal C}_{\rho}^{1/2, {\rm UV/IR}}, {\cal C}_{a_\phi^{\beta^0}}\ ^{\rm IR/UV}, {\cal C}_{a_\phi^{x3}\ ^{\rm B}}^{\rm IR/UV}, {\cal C}_{\phi}^{1/2, {\rm UV/IR}}, {\cal C}_{Z}^{1/2, {\rm UV/IR}}, {\cal C}_{\rho}^{1/2, {\rm UV/IR}}$ are constants of integration appearing in the solutions to the EOMs of the gauge fields in the IR/UV.

\subsection{Effect of a strong magnetic field on deconfinement temperature}

As shown in (\ref{rh-r0-beta0-i}) - (\ref{Tc-B}), 
{\footnotesize
\begin{eqnarray}
& & T_c(B) = T_c(B=0) - B\frac{\mathbb{D}_+{\cal R}_{D5/\overline{D5}}^{\rm th}\left(\log N +\left|\log\left(\frac{r_0}{{\cal R}_{D5/\overline{D5}}^{\rm th}}\right)\right|\right)\left|\log\left(\frac{r_0}{{\cal R}_{D5/\overline{D5}}^{\rm th}}\right)\right|}{2 g_s^2\sqrt{N}\pi^{3/2}r_0\mathbb{C}_+^{3/4}\mathbb{A}_+^{1/4}}<T_c(B=0),
\end{eqnarray}
 }
where
{\footnotesize  
\begin{eqnarray}
& & \mathbb{A}_+ =  \frac{2 \kappa_{\rm GHY}^{{\rm bh},\ \beta^0} {M_{\rm UV}} \log \left(\frac{{{\cal R}_{\rm UV}}}{{\cal R}_{D5/\overline{D5}}^{\rm bh}}\right)}{{g_s}^{9/4} {{\cal R}_{D5/\overline{D5}}^{\rm bh}}^4},\nonumber\\
& & \mathbb{C}_+ = \frac{2 \kappa_{\rm GHY}^{{\rm th},\ \beta^0} {M_{\rm UV}}  \log \left(\frac{{{\cal R}_{\rm UV}}}{{\cal R}_{D5/\overline{D5}}^{\rm th}}\right)}{{g_s^{\rm UV}}^{9/4}
    {{\cal R}_{D5/\overline{D5}}^{\rm th}}^4},\nonumber\\
& & \mathbb{D}_+ =     \kappa_{\rm IR}^{\rm th, B} B g_s^{5/4} \frac{M N_f^{3/2}}{N^{1/4}}\sqrt{- 2 + 3 \left({\cal C}_{a_t^{x3}}\ ^{\rm IR, B}\right)^2{\cal C}_{\rho}^{2, {\rm IR}}\ ^2}|\kappa_{\theta_2}\ ^{(R^4)}|\nonumber\\ 
    \end{eqnarray}
}
($r_0$ being the IR cut-off in the thermal dual of $T<T_c$ QCD-like theories, $\kappa_{\theta_2}\ ^{(R^4)}$ defined in (\ref{ang-reg-kappa-theta2-R4}), $\kappa_{\rm GHY}^{{\rm th/bh},\ \beta^0}$ being numerical pre-factors in the contribution from the Gibbons-Hawking-York surface terms excluding the ${\cal O}(R^4)$-corrections, ${\cal R}_{D5/\overline{D5}}^{\rm th/bh}$ being the $D5-\overline{D5}$-brane separations in the parent type IIB thermal/black-hole gravity dual of thermal QCD-like theories respectively at $T<T_c, T>T_c$, and ${\cal R}_{\rm UV} = (4 \pi g_s N)^{1/N_{>4}}, \mathbb{Z}\ni N_{>4}>4$ is the UV cut-off but in the near-horizon limit, ${\cal R}_{\rm UV} < L \equiv \left(4\pi g_s N\right)^{1/4}$; ${\cal C}_{a_t^{x3}}\ ^{\rm IR, B}, {\cal C}_{\rho}^{2, {\rm IR}}$ being constants of integration appearing in the solutions to the type IIA $D6$-brane world-volume gauge fields' EOMs in the IR as in (\ref{A-mu-IR-th})), implying the reduction of the deconfinement temperature in the presence of a (strong) magnetic field, as in lattice QCD\cite{decrease-Tc-B}. 

\subsection{Photoproduction in QGP}
\label{photoproduction-summary}
In this part of the study, we explored the photon production through QGP with or without a strong magnetic field produced via background gauge field supported on the world-volume of the type IIA flavor $D6$-branes in the UV region. For this part, we derived the spectral density of photon production with or without a strong magnetic field. First, the spectral density of photons without a magnetic field in the UV was computed by deriving the EOM for the transverse gauge invariant field $E_{\rm trans }$, up to $\mathcal O(\omega^{4})$ in the absence of magnetic field:
{\footnotesize
\begin{eqnarray}
%\label{Chi2}
& & \chi_2= \Im m\left[\left(\frac{E_{\rm trans}^{B=0}\ '(Z)}{E_{\rm trans}^{B=0}(Z)}\right) 
   G^{x^2x^2}_{B=0,{\rm UV}} G^{ZZ}_{B=0,{\rm UV}}
   {\cal L}_{\rm DBI}^{B=0,\ {\rm UV}}\right] =  w^2(\chi_2^{w^2,\ \beta^0}\  + \beta \chi_2^{w^2,\ \beta}\ )
 + w^4(\chi_2^{w^4,\ \beta^0}\  + \beta \chi_2^{w^4,\ \beta}\ ),  
   \nonumber\\
\end{eqnarray}
}
where, $\chi_2^{w^2,\ \beta^0}$, $\chi_2^{w^2,\ \beta}$, $\chi_2^{w^4,\ \beta^0}$, and $\chi_2^{w^4,\ \beta}$ is given in eq(\ref{Chi2w024uptobeta}). From Table \tcb{1}, considering the QCD-inspired parameters $g_s=0.3$, $M=N_f=3$ along with $r_h=e^{-\kappa_{r_{h}}(g_{s},M,N_f)N^{1/3}}$, and the tunable parameters, $c_{1,2}$ obtained in \cite{Shivam+Aalok_Bulk}, we obtained the spectral density as:
\begin{eqnarray}
%\label{chioverwB0-iii}
& & \frac{\chi}{4 N^2 \tilde{w} T^2} = {\cal C} \tilde{w} e^{-\frac{16 {\cal R} \tilde{w}^2
   \left(27-\frac{9468.76}{\xi_{\rm UV}^{B=0}\ }\right)}{1323}-\frac{9}{\xi_{\rm UV}^{B=0}\ }} \left(1.1
   e^{1-\frac{114.513 {\cal R} \tilde{w}^2}{\xi_{\rm UV}^{B=0}\ }}-21\right).
\end{eqnarray}
Numerically, one obtains a reasonable match with \cite{D5gaugedSUGRAtrunc} for
\begin{eqnarray}
\label{Const+ratio+xiUV}
{\cal C} = 0.0064,\ \xi_{\rm UV}^{B=0}\  = -3.095, {\cal R} = 0.5727.
\end{eqnarray}

We have also obtained the spectral density of photon production in the presence of a strong magnetic field up to ${\cal O}(R^4)$, via a simple generalization of the procedure adopted in the absence of a magnetic field case:
\begin{eqnarray}
%\label{spectral-functionoverwNsq-B}
& & \hskip -0.8in \frac{{\chi_2}}{N^2  \tilde{w} T^2}\sim \frac{{\cal C}_{x^3x^3}^B\ ^2 {\cal C}_{tZ,1}^B\ ^2
   \kappa_{a_{x3}^{Z,2}}^2 c_2 N \tilde{w}
   \left({\cal C}_{x^3\rho}^B\ 
   ({\cal C}_{x^3\rho}^B\ -8. {\cal C}_{x^3\rho,1}^B\ )-8. {\cal C}_{x^3\rho,1}^B\ ^2
   {Li}_3\left(-\frac{{\cal C}_{x^3\rho,2}^B\ }{{\cal C}_{x^3\rho,1}^B\ }\right)\right)
   (\beta  {\cal C}_{tZ}^B\ 
   {Z_{\rm UV}}-{\cal C}_{tZ,1}^B\ ) }{{\alpha_{\rm ZUV}}^8 B^2 {\cal C}_{tZ,2}^B\ ^3
   c_1}\nonumber\\
& & \hskip -0.3in \times \ e^{-\frac{4
   {\cal C}_{tZ,1}^B\  \tilde{w}^2 ({\cal C}_{tZ,1}^B\ +\beta 
   {\cal C}_{tZ}^B\ 
   {Z_{\rm UV}})}{{\cal C}_{tZ,2}^B\ ^2
   }}  \equiv \frac{\kappa_1 \tilde{w} e^{-\kappa_2 \tilde{w}^2}}{\tilde{B}^2},     
\end{eqnarray} 
where $\tilde{B}\equiv\frac{B}{T^2}$. If one adds a term quadratic in $\tilde{w}^2$ in the numerator of the RHS of (\ref{spectral-functionoverwNsq-B}),
\begin{equation}
\frac{{\chi_2}}{N^2  \tilde{w} T^2} = \frac{\kappa_1 \tilde{w} e^{-\kappa_2 \tilde{w}^2} + \kappa_3 \tilde{w}^2}{\tilde{B}^2},
\end{equation}
one obtains a good match with results of type IIB gauged supergravity \cite{D5gaugedSUGRAtrunc}. The various constants of integration appearing in (\ref{spectral-functionoverwNsq-B}), are defined in (\ref{Amu-IR-B}).

Now, $\beta\sim l_p^6\sim\left(\frac{\sqrt{G_{x^{10}x^{10}}^{\cal M}}}{g_s^{2/3}}\right)^6$ where $\sqrt{G_{x^{10}x^{10}}^{\cal M}}$ is the size of the ${\cal M}$-theory circle near the $\psi=2n\pi, n=0, 1, 2$-patches. Now, near (\ref{alpha_theta_12}), \cite{Shivam+Aalok_Bulk}
\begin{eqnarray}
%\label{M-theory-circle-metric}
& & G_{x^{10}x^{10}}^{\cal M} \sim \frac{ \left(\frac{24 a^2 {g_s} M^2 N_f
   \left(c_1+c_2 \log \left(r_h\right)\right)}{9 a^2+r^2}+\frac{3 N N_f (\log (N)-3 \log (r))}{4 \pi
   }\right)}{ N (N_f (\log (N)-3 \log (r)))^{7/3}}.\nonumber\\
& &    
\end{eqnarray}
One hence sees that 
\begin{equation}
\label{interpolating-i}
\beta\sim l_p^6 \sim \frac{1}{|\log r_h|^4}\sim\left(\frac{\left(g_s N_f\right)^{2/3}\left(g_s M^2\right)^{1/3}}{N^{1/3}}\right)^4,
\end{equation}
which, using (\ref{beta-g}) implies that the complexified $\beta$ will not be analytic in the complexified coupling $g\left(\frac{T}{T_c}\right)$. The complexified spectral density pertaining to photoproduction, would hence have a non-analytic gauge-coupling-$g$ dependence due to a non-trivial ${\cal O}(R^4)$ contribution with the non-analyticity contained in the latter. For the purpose of comparison with the gauged supergravity results of \cite{D5gaugedSUGRAtrunc} consistent with \cite{Shivam+Aalok_Bulk} the latter having been shown to be consistent with $SU(3)$ Gluodynamics result of \cite{lattice-SU3_Glue}, apart from $(g_s, M, N_f) = (0.1, 3, 3)$ one needs to consider $N=100$, that as shown in \cite{ACMS}, corresponds to the existence of Contact 3-Structures.

\subsection{Study of generalized EoS}
\label{EoS-summary}

In this section, we summarize the results obtained for EoS  $P = P(E)$, which is obtained from the renormalized DBI action. With the embedding $i:\Sigma_{D6}\hookrightarrow M_{10}\cong \left(S^1_t\times\mathbb{R}^3\right)\times_w \mathbb{R}_{>0}\times_w{\cal T}^{1,1}_{\rm NE}$,  the UV boundary cosmological term near $\theta_1=\frac{\alpha_{\theta_1}}{N^{1/5}}$ given by eq(\ref{UVccCT-i}), the $Z_{\rm UV}$ counter term will be proportional to : 
\begin{eqnarray}
\label{UVccCT-ii}
& &  \frac{B N N_f^{\rm UV}}{g_s^{1/4}} \int_{\Sigma_{D6}(Z=Z_{\rm UV})}\sqrt{-{\rm det}(i^*g)},
\end{eqnarray}
where $r = r_h e^Z, Z_{\rm UV}\equiv$ UV cutoff. The DBI action also possesses logarithmic divergence in the IR - see (\ref{SDBI_IR_div}). One hence needs the following counter term: 
\begin{eqnarray}
\label{GravSBI-CT-IR-ii}
& & {\cal S}_{\rm IR}^{\rm ct} \sim -
%\frac{
%{\cal C}_{a_t^Z}^B\   {\cal C}_t^{\rho,\ B}\     {\cal C}_{tx^3}^B\  \kappa_1 \rho ^2 \Gamma    \left(\frac{3}{4}\right)^2 \alpha _{\theta _1}   {\cal C}_{x^3}^B(x^3) e^{\kappa_0}
   \frac{\int_{\Sigma_{D6}(Z=0)} e^{-\phi^{IIA}}\sqrt{{\rm det}\left(F+i^*B_{\rm NS-NS}^{\rm IIA}\right)}\log\left(\sqrt{\frac{{\rm det}\left({\rm Ricci}_{\Sigma^{D6}}\right)}{{\rm det}\left(F+i^*B_{\rm NS-NS}^{\rm IIA}\right)}}\right) }{B L \epsilon_{tx^3}}, 
%   (-6 {g_s}       {N_f}  \log ({r_h} )-{g_s}    {N_f}  \log (4)+8    \pi ) \left(\sqrt{2} \beta {\cal C}_{zz}-44 {g_s}       {N_f}  \log (\epsilon_2) \log    ({r_h} )\right)}{ {\cal C}_{x^3t}^B\    {\cal C}_{x^3}^{t\rho,\  B}\  {\cal C}_{x^3}^{tZ,\  B}\  {g_s}   ^2   \sqrt[20]{N} {N_f} ^3 \log ^2(\epsilon_2) \log    ^2({r_h} ) | \log ({r_h} )| }.   
\end{eqnarray}
with $\epsilon_{tx^3}\sim\frac{1}{B L}$.
Using the renormalized DBI action for flavor $D6$-brane we have computed the pressure and energy density.

We demonstrate from the EoS that the holographic dual, in principle, could correspond to several $T>T_c$ scenarios: stable wormhole, stable wormhole transitioning via a smooth crossover to exotic matter as the universe cools (in terms of ``boundary time'' defined in (\ref{rh-tb-relation})), and a paramagnetic pressure/energy-anisotropic plasma. Given that $T>T_c$ QGP is expected to be paramagnetic \cite{Bali et al-magnetic-chi}, the third possibility appears to be the preferred one. We also show that it is not possible that the anisotropic plasma leads to the formation of a compact star.

In the consistent trunction of only $A_t^\beta$ being the only non-trivial ${\cal O}(R^4)$-correction to the background $D6$-brane world-volume gauge field, it was shown that there are no ${\cal O}(R^4)$-corrections to the $D6$-brane world-volume gauge field supporting a constant magnetic field and hence no ${\cal O}(R^4)$-corrections to the free energy/pressure and energy densities, and hence no non-analytic-in-complexified-gauge-coupling dependence in complexified pressure/energy density. For the aforementioned values of $(g_s, M, N_f)$, it turns out that $N=100$ results in complex free energy/pressure and energy densities, but, e.g., for $N=200$, one obtains real free energy/pressure and energy densities. From \cite{ACMS}, this corresponds to the existence of Almost Contact 3-Structures, and not Contact 3-Structures.  Combined with the conjecture of \cite{Shivam+Aalok_Bulk}, namely the existence of Contact 3-Structures is mapped to the non-analytic-complexified-gauge-coupling dependence of the complexified bulk-to-shear-viscosity ratio on the temperature-dependent gauge coupling, we hence further conjecture the following.
\begin{tcolorbox}[enhanced,width=7in,center upper,size=fbox,
    %fontupper=\large\bfseries,
    drop shadow southwest,sharp corners]
    \begin{flushleft}
The failure of the space of AC3S to be $N$-path connected to C3S in the parameter space of such structures induced from $G_2$-structures on closed seven-folds that are a warped product of ${\cal M}$-theory circle and a non-Kähler six-fold with the six-fold being a warped product of the thermal circle with a non-Einsteinian deformation of $T^{1,1}$, conjectured earlier/above to be mapped to the existence of non-analytic(corresponding to C3S)/analytic(corresponding to AC3S)-dependence, is the differential geometric analog of the following pair of statements. In the presence of a strong magnetic field, (i) fluctuations in world-volume gauge fields (relevant to, e.g., holographic photoproduction spectral density computation)  can not be finite, unlike the finite background world-volume gauge field (relevant to, e.g., EoS); (ii) in the zero-instanton sector, (type-IIB modular-completion-inspired) ${\cal O}(R^4)$ non-renormalized gauge fields corresponding to AC3S produce ${\cal O}(R^4)$-corrected gauge fluctuations corresponding to C3S.
\end{flushleft}
\end{tcolorbox}
Lastly, motivated by the above conjecture and a pair of observations: (i) replacing $M$ and $N_f$ by the effective number of $D5$-branes and $D7$-branes respectively in the parent type IIB dual of \cite{metrics},  (\ref{interpolating-i}) implies that $\beta\stackrel{\rm UV}{\longrightarrow}0$ as there is no net effective $D5/D7$-brane charge in the UV validating the expected UV conformality, and (ii) the ${\cal O}(R^4)$-corrections to the ${\cal M}$-theory uplift of thermal QCD-like theories vanish in the UV \cite{Vikas+Gopal+Aalok}, we now conjecture:
\begin{tcolorbox}[enhanced,width=7in,center upper,size=fbox,
    %fontupper=\large\bfseries,
    drop shadow southwest,sharp corners]
    \begin{flushleft}
   \hskip 3 in C3S $\stackrel{\rm UV}{\longrightarrow}$ AC3S.
    \end{flushleft}
    \end{tcolorbox}
Note, the above is not a  contradiction of the lack of $N$-path connectedness in the parameter space of (A)C3S as noted in \cite{ACMS}. First, the aforementioned conjecture of \cite{ACMS} was in the IR. What the last conjecture above is based on is given that the effective number of $D5/D7$-branes can RG-flow continuously with $r$, the ${\cal O}(R^4)$ corrections become vanishingly small in the UV.   

During heavy nuclei collisions when QGP is produced, the strong magnetic field is observed for a short time \cite{Skokov:2009qp}. QGP is the plasma of charged particles, and hence, it is highly responsive to a strong magnetic fields. Defining the magnetization, $M=-\frac{\partial F}{\partial B}=\frac{\partial P}{\partial B}$, we see that $M = \frac{P^{\rm UV}}{B} + {\cal O}\left(\frac{1}{B^2}\right)$ [$P^{\rm UV}$ as defined in (\ref{different-EoS-scenarios})] for $B\gg \left(0.15\ {\rm GeV}\right)^2$, which  turns out to be positive for strong magnetic fields, implying the anisotropic plasma of scenario (b) in \ref{different-EoS-scenarios}, is paramagnetic for high temperatures above $T_c$. The magnetic susceptibility, $\chi_{m}=-\frac{\partial^2F}{\partial B^2} = {\cal O}\left(\frac{1}{B}\right)^3$ in our setup which in the large-$B$ limit, we drop and hence $\chi_m\approx0$ (as supported by the negligible value of $\chi_m$ obtained in "parton-hadron-string-dynamics transport approach" \cite{small-chi_m-i}, as well as lattice results \cite{Bali et al-magnetic-chi}). 
\begin{figure}[h!]
\begin{center}
\includegraphics[width=0.5\textwidth]{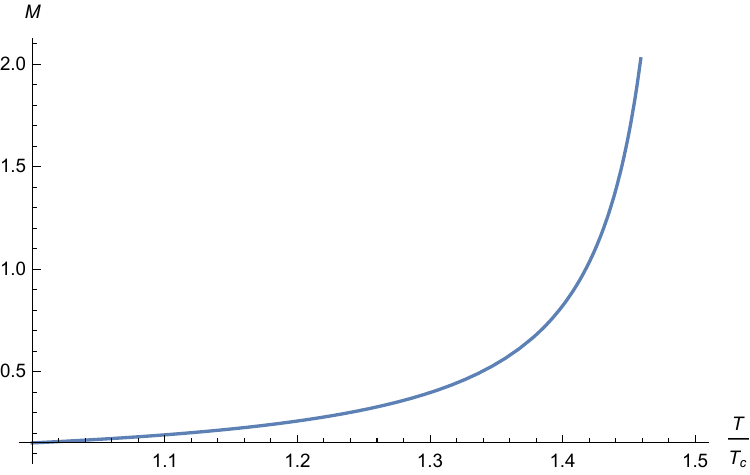}
\end{center}
\caption{Magnetization-vs-Temperature}
\end{figure}

%
%\begin{figure}
%\begin{subfigure}{.5\textwidth}
%\centering
%\includegraphics[width=1\textwidth]{cssq-DBI+SUGRA-I.jpg}
%\caption{$c_s^2$-vs-temperature for different $B=\left(1, 10, 100, 300\right)\left({\rm GeV}\right)^2$ for anisotropic plasma, scenario I)}
%\end{subfigure}
%\hskip 0.25in
%\begin{subfigure}{.5\textwidth}
%\centering
%\includegraphics[width=1\textwidth]{cssq-DBI+SUGRA-II.jpg}
%\caption{$c_s^2$-vs-temperature for different $B=\left(1, 10, 100, 300\right)\left({\rm %GeV}\right)^2$ for anisotropic plasma, scenario II)}
%\end{subfigure}
%\caption{Anisotropic plasma: $ {\cal C}_{\rm combo} =- 1,  {\cal C}^{(2)}_{\rm combo}=1 \ or\ 100, {\cal C}_{x^3}^B(x^3) = -0.001;\ \tilde{t}\equiv\frac{T}{T_c}$}
%\end{figure}

\section*{Acknowledgements}

SSK is supported by a Junior Research Fellowship (JRF) from the Ministry of Human Resource
and Development (MHRD), Govt. of India. AM is partly supported by a Core Research Grant
number SER-1829-PHY from the Science and Engineering Research Board, Govt. of India. We also thank Rajnish Kumar Jha and Ankit Patel for participation in some developments  pertaining to  the paper.

\end{document}